\documentclass[a4paper,12pt,reqno]{amsart}

\usepackage{a4}
\usepackage{verbatim}
\usepackage{comment}
\usepackage{color}
\usepackage{todonotes}
\usepackage{doi}
\usepackage{mathrsfs}

\usepackage{stmaryrd}
\usepackage{amssymb}
\usepackage{bbm}
\usepackage{graphicx}
\usepackage{floatflt}
\usepackage{float} 		
\usepackage{tikz}
\usetikzlibrary{decorations.pathreplacing,arrows.meta,shapes}

\usepackage{amsthm}
\usepackage{amscd}
\usepackage{hyperref}

\newtheorem{theorem}{Theorem}
\newtheorem{proposition}[theorem]{Proposition}
\newtheorem{example}[theorem]{Example}

\newtheorem{conjecture}[theorem]{Conjecture}
\newtheorem{remark}[theorem]{Remark}
\newtheorem{principle}[theorem]{General Principle}
\newtheorem{definition}[theorem]{Definition}
\newtheorem{definitionlemma}[theorem]{Definition + Lemma}
\newtheorem{lemma}[theorem]{Lemma}
\newtheorem{question}[theorem]{Question}
\newtheorem{corollary}[theorem]{Corollary}

\def\XXint#1#2#3{{\setbox0=\hbox{$#1{#2#3}{\int}$}
		\vcenter{\hbox{$#2#3$}}\kern-.5\wd0}}

\makeatletter
\@addtoreset{theorem}{section}
\@addtoreset{equation}{section}
\makeatother

\newcommand{\cT}{\mathcal{T}}
\newcommand{\cW}{\mathcal{W}}
\newcommand{\cS}{\mathscr{S}}
\newcommand{\set}[1]{\llbracket {#1} \rrbracket}

\DeclareMathOperator{\Res}{Res}

\allowdisplaybreaks[4]

\title[Universal $x$-$y$ duality in eq-TR and QKM]{Universal $x$-$y$ duality in equivariant topological recursion and quartic Kontsevich model}

\author{Alexander Hock}

\address{Section of Mathematics, University of Geneva, Rue du Conseil-G\'{e}n\'{e}ral 7-9,
	1205 Geneva, Switzerland} 
\email{alexander.hock@unige.ch}

\author{Jakob Lindner}

\address{Mathematisches Institut, Universität Münster, Einsteinstr.\ 62, 48149 Münster, Germany} 
\email{jakob.lindner@uni-muenster.de}

\author{Raimar Wulkenhaar}

\address{Mathematisches Institut, Universität Münster, Einsteinstr.\ 62, 48149 Münster, Germany} 
\email{raimar@math.uni-muenster.de}

\begin{document}

\begin{abstract}
Topological recursion is a universal procedure that generates an
infinite family of multi-differentials with different interpretations
in various areas of mathematics. Recently, a new understanding
of topological recursion has emerged from $x$-$y$ duality, which
relates two such families through explicit formulae.
We study $x$-$y$ duality, its parametrised version, and symplectic
duality in the setting of equivariant topological recursion.
We explicitly analyse the pole structure of the resulting dual
families and provide recursive formulae for the different types
of polar parts. In contrast to classical topological recursion,
the dual family has additional poles at fixed points
of the group action, and at anti-diagonals determined by the group action.

We show that the partition function 
$Z_P(E)=\int_{H_N} e^{-N\,\mathrm{Tr}(E\Phi^2+P(\Phi))}d\Phi$
of the Kontsevich matrix model with arbitrary potential $P$,
where $d\Phi$ is the Lebesgue measure on Hermitian $N\times N$ matrices and
$E$ a positive matrix, admits a Fock space realisation related to the
character expansion of the IZHC integral. The Fock space framework provides
two mutually $X$-$Y$ dual families of multi-differentials relative to an implicitly defined spectral
curve $(X,Y)$. The solution $Y(X)$ succeeds for the quartic Kontsevich model with potential
$P(\Phi)=\frac{\lambda}{4}\Phi^4$.
We identify a network of duality transformations and symplectic shifts
which relate the quartic Kontsevich model with the
$\mathbb{Z}/2\mathbb{Z}$-equivariant spectral curve.
The proposed construction yields a conjectural all-genus solution to
the quartic Kontsevich model, which reproduces 
recursive structures previously established for genus $g\leq 1$.

\end{abstract}


\maketitle

\section{Introduction}

This article proposes a solution to a longstanding problem related to a quantum field theory model on noncommutative geometry originally studied more than twenty years ago by the last-named author together with Harald Grosse \cite{Grosse:2004yu}. In zero dimensions, this model can be interpreted as a quartic analogue of the celebrated Kontsevich model, which was used by Kontsevich \cite{Kontsevich:1992ti} to prove Witten's conjecture \cite{Witten:1990hr} on the integrability of $\psi$-class intersection numbers on the moduli space of stable complex curves. The quartic and original Kontsevich
models are the cases $P(\Phi)=\frac{\lambda}{4}\Phi^4$ and $P(\Phi)=\frac{\mathrm{i}}{3}\Phi^3$ of a general family
\begin{align}
Z_P(E_{\set{N}}):=
\int_{H_N} e^{-N\,\mathrm{Tr}(E\Phi^2+P(\Phi))} d\Phi
\label{ZKontsevich-intro}
\end{align}
of $N\times N$ hermitian matrix models, where $d\Phi$ is the Lebesgue measure and $E=\mathrm{diag}(E_1,...,E_N)$ an external matrix.
Many developments concerning the Kontsevich model have since been established, including its enumerative interpretation in terms of matrix models, its integrability, and its algebro-geometric interpretation. A generalisation of this model ($r$-spin Kontsevich model) was introduced in \cite{Kharchev:1991}, and its connection to intersection numbers was proposed by Witten \cite{Witten:1993} to satisfy the $r$-KdV equation. It took considerable time to obtain a rigorous proof of this conjecture \cite{Faber:2010}. Notably this $r$-spin Kontsevich model does not cover the quartic Kontsevich model mentioned above, those are in two different classes of generalisations, see \cite[Sec. 2.1]{Branahl:2020uxs} for a discussion.

In parallel, a universal structure was identified that repeatedly appeared in different matrix models and required only some initial data as input. These initial data are called a \emph{spectral curve} and essentially consist of two meromorphic functions $x,y$ (together with some additional data that will be detailed later). The universal structure built upon them is now known as \emph{Topological Recursion} (TR) \cite{Chekhov:2005rr,Eynard:2007kz}. The output of TR is an infinite system of multi-differentials denoted by $\omega_{g,n}$. It turned out that TR and the multi-differentials it generates have remarkably many applications far beyond matrix models. Without attempting to give an exhaustive list, these include Mirzakhani's Weil--Petersson volumes \cite{Mirzakhani:2007wp,Eynard:2007wp}, Gromov--Witten invariants of $\mathbb{P}^1$ \cite{Pandharipande:2000,Norbury:2014}, and toric Calabi--Yau threefolds \cite{Aganagic:2003db,Bouchard:2007ys}. The higher $r$-spin version of Witten's conjecture also admits a formulation in terms of TR \cite{Bouchard:2016,Belliard:2025}. However, throughout all these formulations and generalisations, the quartic analogue of the Kontsevich model, regarded as a Hermitian matrix model (\ref{ZKontsevich-intro}) with non-trivial covariance (i.e.\ dependent on $E_{\set{N}}$) and quartic interaction $P(\Phi)=\frac{\lambda}{4}\Phi^4$, could not be formulated in terms of either intersection theory or TR.

Although TR appeared to be a universal structure applicable in its original form to many different areas of mathematics and physics, recent years have brought a substantially revised understanding of the theory itself. This development started with a conjectural connection between TR and free probability theory \cite{Borot:2021thu}, which led to an extensive study of the so-called $x$-$y$ duality \cite{Alexandrov:2022ydc,Hock:2022wer,Hock:2022pbw}. The $x$-$y$ duality changes the projection of the spectral curve from $x$ to $y$, with fundamental implications for all differentials generated by TR. A universal formula was found relating the multi-differentials $\omega_{g,n}$ generated by TR for the spectral curve $(x,y)$ to the \emph{dual} multi-differentials $\omega_{g,n}^\vee$ generated by the spectral curve $(y,x)$, literally exchanging $x$ and $y$. This duality between two different systems of multi-differentials has already had considerable impact on known applications of TR \cite{Hock:2023qii,Alexandrov:2023oov} and has further led to proofs of previously open conjectures \cite{Alexandrov:2023jcj,Alexandrov:2024zku}.

It turned out that the $x$-$y$ duality is, in fact, even more universal than TR itself. A new understanding of TR has emerged, leading to a global formulation of the recursive structure that is nevertheless equivalent to ordinary TR in the basic setting \cite{Alexandrov:2024tjo}. This new formulation will be called \emph{Generalised Topological Recursion} (Gen-TR) throughout this article and should not be confused with the higher-order ramification formulation of TR \cite{Bouchard:2012yg}. Gen-TR has, for instance, already provided a general understanding of Theta classes from the perspective of TR \cite{Bouchard:2025rid}.

An a priori different direction concerns additional symmetries of the spectral curve, originally motivated by connections to Hurwitz theory and now referred to as \emph{equivariant Topological Recursion} (eq-TR) \cite{Giacchetto:2021atk}. This framework has subsequently been used to prove new relations among kappa classes \cite{Alexandrov:2025kzn} and to study Gromov--Witten invariants with orbifold structure on Calabi--Yau threefolds \cite{fang2026}.

In parallel with these developments in the theory of TR, the first- and last-named authors made considerable progress towards a solution of the quartic Kon\-tsevich model \cite{Grosse:2019jnv,Branahl:2020yru,Hock:2021tbl,Hock:2023nki, hock2025symmetry}, showing, in particular, that ordinary TR does not solve the quartic Kontsevich model.

The present article proposes an all-genus/all-order solution of the quartic Kontsevich model (QKM) by combining equivariant topological recursion with the universal $x$-$y$ duality and further symplectic shifts of the spectral curve. \\

To establish this proposal, we derive and prove several properties describing how the universal $x$-$y$ duality extends to the setting of equivariant topological recursion. In doing so, we analyse the additional poles arising from the equivariant group action. The starting point is a system of multi-differentials generated by equivariant topological recursion, where an additional group acts on the spectral curve, in particular with a prescribed group action on the coverings $x$ and $y$ entering the spectral curve. Let $\omega^{(x,y)}_{g,n}$ (excluding $\omega_{0,1}^{(x,y)}$) denote the system of multi-differentials generated by eq-TR. In Sec.~\ref{sec:xydualofequivariant}, we prove structural formulae for the poles of the associated $x$-$y$ dual multi-differentials $\omega^{\vee(x,y)}_{g,n}$. Importantly, this dual system has poles not only at the ramification points of $y$, but also at the fixed points of the group action. All these poles are described precisely. We show that the dual system $\omega^{\vee(x,y)}_{g,n}$ satisfies Generalised Topological Recursion (as defined in \cite{Alexandrov:2024tjo}), with the dual key points chosen to be the ramification points of $y$. Furthermore, we prove that at the ramification points of $y$ where the equivariant group $G$ acts freely, the pole structure can be equivalently computed by a topological recursion-like formula. In summary, this means that the dual system $\omega^{\vee(x,y)}_{g,n}$ is in general given by a \emph{blobbed topological recursion formula} (as defined in \cite{Borot:2015hna}), where the ``blob'' of $\omega_{0,2}$ arises from the equivariant extension and other blobs arise from the fixed points of the group action and have a precise recursive structure given by Gen-TR.

We then further allow a shift of $y$ in the equivariant spectral curve, which leaves the original equivariant system $\omega_{g,n}^{(x,y)}=\omega_{g,n}^{(x,y+\psi(x))}$ invariant. This, however, takes us outside the realm of eq-TR, since $y+\psi(x)$ in general  does not satisfy the required group-action property. Considering the $x$-$y$ dual $\omega_{g,n}^{\vee(x,y+\psi(x))}$ of the system $\omega_{g,n}^{(x,y+\psi(x))}$ leads to a completely different pole structure, which we study in Sec.~\ref{sec:xtydualofequivariant}. In particular, the system $\omega_{g,n}^{\vee(x,y+\psi(x))}$ has, in general, poles at the ``anti-diagonal'', i.e.\ at the points where $B$ already has poles. Note, however, that the system $\omega_{g,n}^{\vee(x,y)}$ does not have poles at these points, except for $\omega_{0,2}^{\vee(x,y)}=\omega_{0,2}^{(x,y)}=B$. Here, a clear distinction can be observed between the two different dual systems.

\begin{figure}[ht]
\centering
\begin{tikzpicture}[
    scale=0.9,
    transform shape,
    >=Latex,
    every node/.style={font=\small},
    corr/.style={
        draw,
        rounded corners=12pt,
        minimum width=2.4cm,
        minimum height=1.2cm,
        align=center,
        inner sep=5pt
    },
    upper/.style={
        corr,
        draw=orange!80!black,
        fill=orange!8
    },
    lower/.style={
        corr,
        draw=blue!70!black,
        fill=blue!3
    },
    calloutorange/.style={
        rectangle callout,
        draw=orange!80!black,
        fill=orange!3,
        rounded corners=4pt,
        align=center,
        font=\small,
        inner sep=4pt
    },
    arrowlabel/.style={
        font=\scriptsize,
        align=center,
        fill=white,
        inner sep=1pt
    },
    vertlabel/.style={
        font=\scriptsize,
        align=center,
        inner sep=0pt,
        rotate=90
    }
]


\node[upper] (UL) at (-3.8,3.2)
    {$\omega^{(x,y)}_{g,n}$};

\node[upper] (UR) at (3.8,3.2)
    {$\omega^{(x,y+\psi(x))}_{g,n}$};

\node[lower] (LL) at (-3.8,-2.5)
    {$\omega^{\vee(x,y)}_{g,n}$};

\node[lower] (LR) at (3.8,-2.5)
    {$\omega^{\vee(x,y+\psi(x))}_{g,n}$};


\node[
    calloutorange,
    callout absolute pointer={(UL.north)},
    above=8mm of UL
] (eqtr)
    {eq-TR};


\draw[->]
    (UL.east) -- (UR.west)
    node[midway, above=1mm, arrowlabel]
        {trivial shift}
    node[midway, below=3mm, arrowlabel]
        {$(x,y)\mapsto(x,y+\psi(x))$};


\draw[->]
    (UL.south) -- (LL.north)
    node[midway, xshift=-3mm, vertlabel]
        {$x$-$y$ dual}
    node[midway, xshift=5mm, vertlabel]
        {$(x,y)$};


\draw[->]
    (UR.south) -- (LR.north)
    node[midway, xshift=-3mm, vertlabel]
        {$x$-$y$ dual}
    node[midway, xshift=5mm, vertlabel]
        {$(x,y+\psi(x))$};


\draw[<-]
    (LR.west) -- (LL.east)
    node[midway, above=1mm, arrowlabel]
        {sympl.\ dual}
    node[midway, below=3mm, arrowlabel]
        {$(y,x)\mapsto(y+\psi(x),x)$};


\node[below=5mm of LL, font=\small]
    {Sec.~\ref{sec:xydualofequivariant}};

\node[below=5mm of LR, font=\small]
    {Sec.~\ref{sec:xtydualofequivariant}};

\end{tikzpicture}

\caption{Relations between the different systems of differentials.}
\label{fig:four-families}
\end{figure}
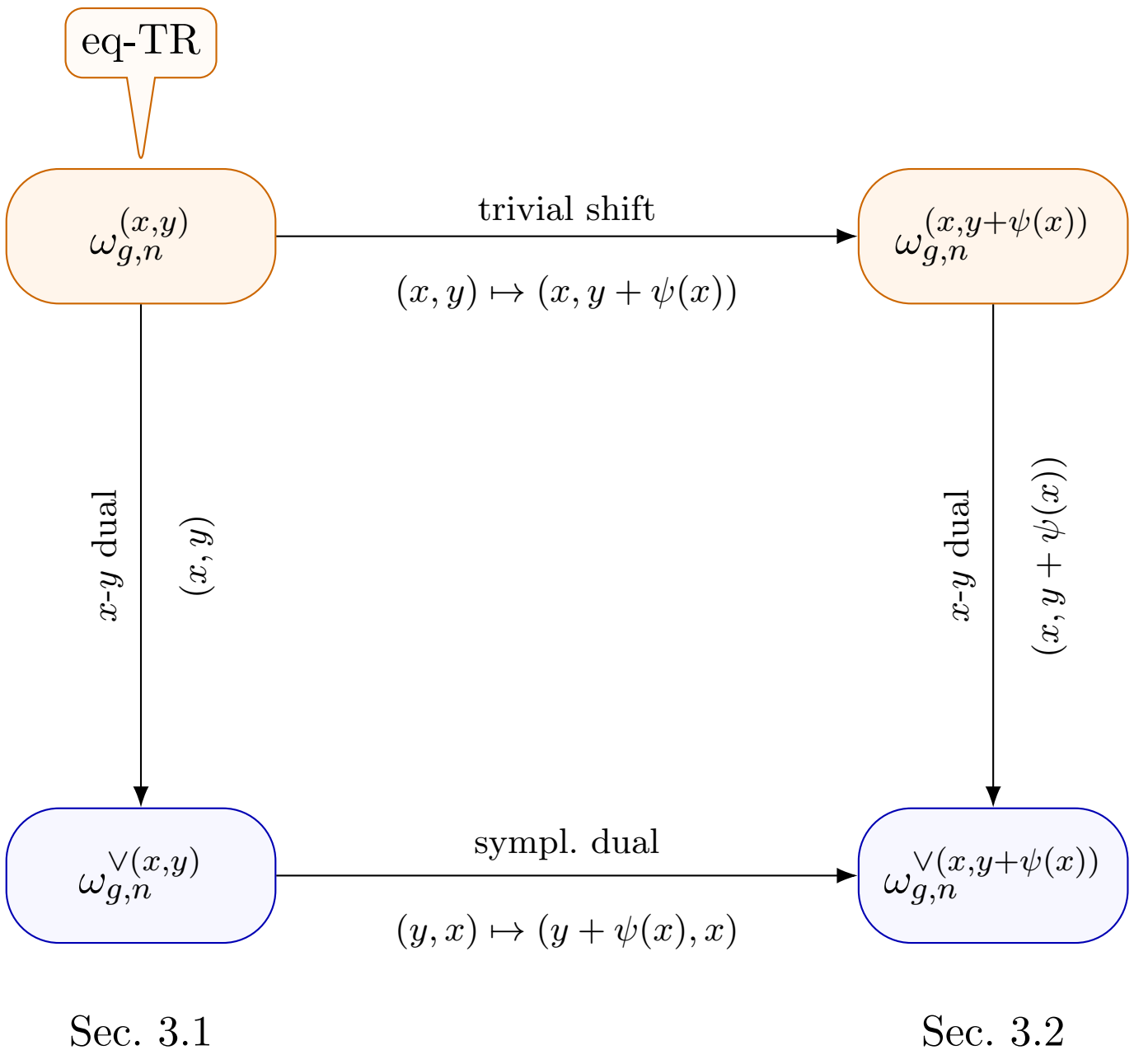

The relation between the two different systems $\omega_{g,n}^{\vee(x,y+\psi(x))}$ and $\omega_{g,n}^{\vee(x,y)}$ is called symplectic duality and agrees with the already known symplectic duality observed in ordinary topological recursion \cite{Bychkov:2022wgw,Alexandrov:2024ajj}. Fig.~\ref{fig:four-families} graphically summarises the chain of transformations starting from eq-TR between the different systems of multi-differentials. The diagram is commutative.

\medskip

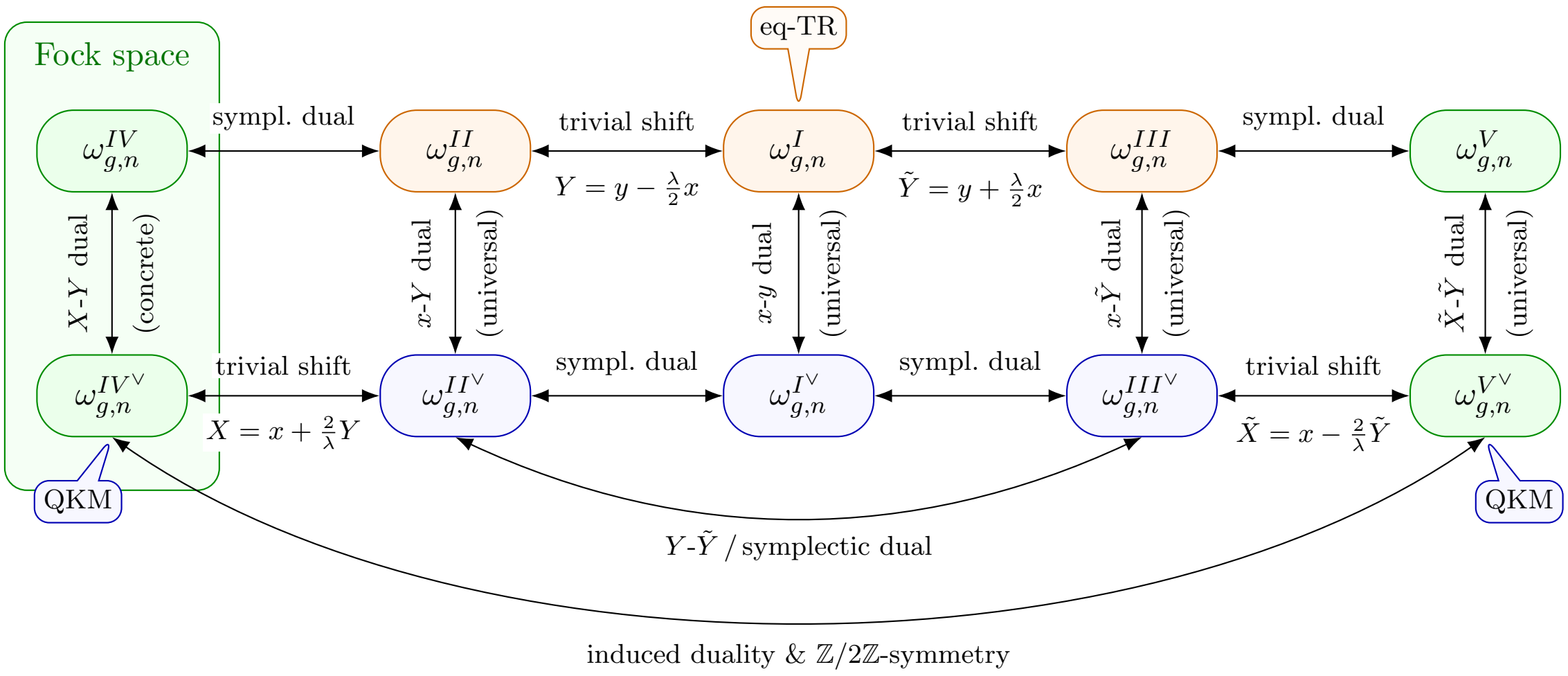
\begin{figure}[ht]
\centering

\begin{tikzpicture}[
    scale=0.82,
    transform shape,
    >=Latex,
    every node/.style={font=\small},
    corr/.style={
        draw,
        rounded corners=10pt,
        minimum width=1.75cm,
        minimum height=0.95cm,
        align=center,
        inner sep=4pt
    },
    upper/.style={
        corr,
        draw=orange!80!black,
        fill=orange!8
    },
    lower/.style={
        corr,
        draw=blue!70!black,
        fill=blue!3
    },
    greenbox/.style={
        corr,
        draw=green!55!black,
        fill=green!8
    },
    calloutorange/.style={
        rectangle callout,
        draw=orange!80!black,
        fill=orange!3,
        rounded corners=4pt,
        align=center,
        font=\scriptsize,
        inner sep=3pt
    },
    calloutblue/.style={
        rectangle callout,
        draw=blue!70!black,
        fill=blue!3,
        rounded corners=4pt,
        align=center,
        font=\scriptsize,
        inner sep=3pt
    },
    arrowlabel/.style={
        font=\scriptsize,
        align=center,
        fill=white,
        inner sep=1pt
    },
    clearlabel/.style={
        font=\scriptsize,
        align=center,
        inner sep=0pt
    },
    vertlabel/.style={
        font=\scriptsize,
        align=center,
        rotate=90,
        inner sep=0pt
    }
]




\draw[
    draw=green!55!black,
    fill=green!4,
    rounded corners=6pt
]
    (-9.25,-2.25)
    rectangle
    (-6.75,3.20);

\node[
    text=green!45!black,
    font=\small
] at (-8,2.80)
    {Fock space};


\node[greenbox] (IV) at (-8,1.70)
    {$\omega^{IV}_{g,n}$};

\node[upper] (II) at (-4,1.70)
    {$\omega^{II}_{g,n}$};

\node[upper] (I) at (0,1.70)
    {$\omega^{I}_{g,n}$};

\node[upper] (III) at (4,1.70)
    {$\omega^{III}_{g,n}$};

\node[greenbox] (V) at (8,1.70)
    {$\omega^{V}_{g,n}$};


\node[greenbox] (IVvee) at (-8,-1.15)
    {$\omega^{IV^\vee}_{g,n}$};

\node[lower] (IIvee) at (-4,-1.15)
    {$\omega^{II^\vee}_{g,n}$};

\node[lower] (Ivee) at (0,-1.15)
    {$\omega^{I^\vee}_{g,n}$};

\node[lower] (IIIvee) at (4,-1.15)
    {$\omega^{III^\vee}_{g,n}$};

\node[greenbox] (Vvee) at (8,-1.15)
    {$\omega^{V^\vee}_{g,n}$};


\node[
    calloutorange,
    callout absolute pointer={(I.north)},
    above=7mm of I
] (eqtr)
    {eq-TR};


\node[
    calloutblue,
    callout absolute pointer={(IVvee.south)},
    below left=5mm and -10mm of IVvee
] (qkmIV)
    {QKM};

\node[
    calloutblue,
    callout absolute pointer={(Vvee.south)},
    below right=5mm and -10mm of Vvee
] (qkmV)
    {QKM};


\draw[<->]
    (IV.east) -- (II.west)
    node[midway, above=2mm, arrowlabel]
        {sympl.\ dual};

\draw[<->]
    (II.east) -- (I.west)
    node[midway, above=2mm, arrowlabel]
        {trivial shift}
    node[midway, below=2mm, arrowlabel]
        {$Y=y-\tfrac{\lambda}{2}x$};
    
\draw[<->]
    (I.east) -- (III.west)
    node[midway, above=2mm, arrowlabel]
        {trivial shift}
        node[midway, below=2mm, arrowlabel]
        {$\tilde{Y}=y+\tfrac{\lambda}{2}x$};

\draw[<->]
    (III.east) -- (V.west)
    node[midway, above=2mm, arrowlabel]
        {sympl.\ dual};


\draw[<->]
    (IV.south) -- (IVvee.north)
    node[midway, xshift=-4mm, vertlabel]
    {$X$-$Y$ dual}
    node[midway, xshift=4mm, vertlabel]
    {(concrete)};

\draw[<->]
    (II.south) -- (IIvee.north)
    node[midway, xshift=-4mm, vertlabel]
        {$x$-$Y$ dual}
    node[midway, xshift=4mm, vertlabel]
    {(universal)};    

\draw[<->]
    (I.south) -- (Ivee.north)
    node[midway, xshift=-4mm, vertlabel]
        {$x$-$y$ dual}
    node[midway, xshift=4mm, vertlabel]
    {(universal)};    

\draw[<->]
    (III.south) -- (IIIvee.north)
    node[midway, xshift=-4mm, vertlabel]
    {$x$-$\tilde{Y}$ dual}
    node[midway, xshift=4mm, vertlabel]
    {(universal)};

\draw[<->]
    (V.south) -- (Vvee.north)
    node[midway, xshift=-4mm, vertlabel]
    {$\tilde{X}$-$\tilde{Y}$ dual}
        node[midway, xshift=4mm, vertlabel]
    {(universal)};


\draw[<->]
    (IVvee.east) -- (IIvee.west)
    node[midway, above=2mm, arrowlabel]
        {trivial shift}
        node[midway, below=2mm, arrowlabel]
        {$X=x+\tfrac{2}{\lambda}Y$};

\draw[<->]
    (IIvee.east) -- (Ivee.west)
    node[midway, above=2mm, arrowlabel]
        {sympl.\ dual};

\draw[<->]
    (Ivee.east) -- (IIIvee.west)
    node[midway, above=2mm, arrowlabel]
        {sympl.\ dual};

\draw[<->]
    (IIIvee.east) -- (Vvee.west)
    node[midway, above=2mm, arrowlabel]
        {trivial shift}
        node[midway, below=2mm, arrowlabel]
        {$\tilde{X}= x-\tfrac{2}{\lambda}\tilde{Y}$};


\draw[
    <->,
    preaction={
        draw=white,
        line width=5pt
    }
]
    (IIvee.south)
    to[out=-25,in=-155,looseness=0.95]
    node[
        pos=.50,
        below=1mm,
        arrowlabel
    ]
    {$Y$-$\tilde{Y}$\,/\,symplectic dual}
    (IIIvee.south);


\draw[<->]
    (IVvee.south)
    to[out=-35,in=-145,looseness=0.8]
    node[
        pos=.50,
        below=2mm,
        arrowlabel
    ]
        {induced duality \& $\mathbb{Z}/2\mathbb{Z}$-symmetry}
    (Vvee.south);

\end{tikzpicture}

\caption{Relations between the different systems of multi-differentials.}
\label{fig:duality-network}
\end{figure}

A main application of these structures is the construction of multi-differentials in the quartic Kontsevich model, which we now describe. Let us consider the following choice of $x$ and $y$:
\begin{align*}
    x(z)&=\frac{R(z)-R(-z)}{(-\lambda)},  &X(z)&=\frac{R(z)+R(-z)}{\lambda}, &\tilde{X}(z)&=\frac{R(z)+R(-z)}{(-\lambda)},\\
    y(z)&=\tfrac{1}{2}\big(R(z)+R(-z)\big), &Y(z)&=R(z),&\tilde{Y}(z)&=R(-z)
\end{align*}
for some rational function $R$ with simple poles and simple zeros of $dR$. For the quartic Kontsevich model, this function is implicitly defined \cite{Grosse:2019jnv, Schurmann:2019mzu}, but the general construction below is independent of the precise form of the rational function $R$.
We introduce the following ``spectral curves'', organised into two groups related by $x$-$y$ duality, which exchanges the r\^oles of the two functions:
\begin{align*}
I&=(x(z),y(z)),\qquad &II&=(x(z),Y(z)),\qquad &III&=(x(z),\tilde{Y}(z)),\\&&
IV&=(X(z),Y(z)),\qquad &V&=(\tilde{X}(z),\tilde{Y}(z)).
\end{align*}
They have associated $x$-$y$ duals $(I^\vee,II^\vee,III^\vee,IV^\vee,V^\vee)$.
To each ``curve'' $\star\in\{I,II,III,IV,V,I^\vee,II^\vee,III^\vee,IV^\vee,V^\vee\}$, we want to associate a system of multi-differentials denoted by $\omega_{g,n}^\star$, where the actual recursive procedure for deriving these differentials is not known \emph{a priori}. We impose only that a recursive formula is provided for the system $\omega^I_{g,n}$ by eq-TR, while the other systems are derived from it by certain transformations. Thus, the aim is to find recursive formulae for the different systems, although all of them can be computed from $\omega^I_{g,n}$ via the chain of transformations described in Fig.~\ref{fig:duality-network}.

As mentioned above, for $I$ we define $\omega^I_{g,n}$ by eq-TR with group $G=\mathbb{Z}/2\mathbb{Z}$, and perform several transformations starting from this family. The transformations to $\omega^{II}_{g,n}$ and $\omega^{III}_{g,n}$ are trivial, meaning that all $\omega^\star_{g,n}$ are identical for $\star\in\{I,II,III\}$. However, eq-TR is defined only for $I$, due to the required equivariant $\mathbb{Z}/2\mathbb{Z}$-action on $y$. Having these three identical systems $\omega^\star_{g,n}$, but with three different associated spectral curves, we apply the universal $x$-$y$ duality, which gives a precise combinatorial definition of a dual system and depends crucially on the choice of $(x,\tilde{y})$, where $\tilde{y}\in\{y,y\pm \tfrac{\lambda}{2}x\}$. In this way, we obtain three different dual systems of multi-differentials $\omega_{g,n}^{\star^\vee}$ with $\star^\vee\in\{I^\vee,II^\vee,III^\vee\}$, which are themselves related by symplectic duality as defined in \cite{Bychkov:2022wgw}.

From the system $\omega^{II^\vee}_{g,n}$ a further trivial shift connects it to $\omega^{IV^\vee}_{g,n}$, which is $x$-$y$ dual to $\omega^{IV}_{g,n}$ which by itself is further symplectic dual to $\omega^{II}_{g,n}$.

On the other hand, we prove that processing the general Kontsevich matrix model (\ref{ZKontsevich-intro}) with the character expansion of the Itzykson--Zuber--Harish-Chandra formula (see e.g.\ \cite{Alexandrov:2010th}) leads to a Fock space realisation and an induced duality between two families of multi-differentials. This is true for any potential $P$ in (\ref{ZKontsevich-intro}). To fully exploit this duality a Lagrange inversion $Y(X)$ is necessary. We control this inversion in the case $P(\Phi)=\frac{\lambda}{4}\Phi^4$ 
of the quartic Kontsevich model, thanks to results from \cite{Grosse:2019jnv, Schurmann:2019mzu, Branahl:2020yru}, but not in general. It turns 
out that for $P(\Phi)=\frac{\lambda}{4}\Phi^4$ the mutually dual families of multi-differentials, which are concrete Fock space expectation values, are precisely $\omega^{IV}_{g,n}$ and $\omega^{IV^{\vee}}_{g,n}$ of the above picture.

In a previous work \cite{Hock:2021tbl} 
on the genus-0 sector of the quartic Kontsevich model we observed that the $\omega_{0,n}$ can be obtained from a $\mathbb{Z}/2\mathbb{Z}$-reflection formula, extending the $\mathbb{Z}/2\mathbb{Z}$-symmetry of $\omega_{0,2}$ discovered in \cite{Branahl:2020yru}. In this paper we prove that the general symplectic duality between the 
two blue-coloured systems $\omega^{III^\vee}_{g,n}$ and $\omega^{II^\vee}_{g,n}$ takes the form of an $x$-$y$ duality (Thm.\ \ref{thm:xydualtyforeqivariantduals}) and provides a 
$\mathbb{Z}/2\mathbb{Z}$-reflection formula
(Thm.~\ref{thm:reflection}) valid to any genus $g$.

In other words, we can interpret this duality of the quartic Kontsevich model in three different ways: as a symplectic duality, an $x$-$y$ duality, and a $\mathbb{Z}/2\mathbb{Z}$-symmetry. With the help of Thm.\ \ref{thm:xydualtyforeqivariantduals}, we are finally able to derive a recursive formula for $\omega^{II^\vee}_{g,n}=\omega^{IV^\vee}_{g,n}$ which, upon identifying $R(z)$ with the spectral curve of the quartic Kontsevich model \cite{Grosse:2019jnv, Schurmann:2019mzu}, coincides with the one found in \cite{Hock:2021tbl,Hock:2023nki}.

It is highly plausible that the network of dualities around the eq-TR family $\omega^I_{g,n}$, summarised in Fig.~\ref{fig:duality-network}, 
constructs the complete 
solution of the quartic Kontsevich model. 
The universal $x$-$y$-duality expresses the multi-differentials $\omega^{II^\vee}_{g,n}=\omega^{IV^\vee}_{g,n}$ of the 
quartic Kontsevich model in terms of
$\omega^{II}_{g,n}=\omega^{I}_{g,n}$. 
To use this relation we have to split the poles
in the $x$-$y$-duality relation. Eq-TR tells us how to split. It remains to prove that the Fock space formulation gives the same splitting rule. For $g\leq 1$ this is the case.


\subsection*{Declaration of Generative AI and AI-assisted Technologies in the Writing Process}
Generative AI was used only to assist with proofreading, identifying typographical errors and helping to create diagrams; all mathematical content and scientific results are the authors' own work where no generative AI was used at all.

\subsection*{Acknowledgment}
A.H.\ was funded by the Swiss National Science Foundation (SNSF) through the
Ambizione project ``TRuality: Topological Recursion, Duality and Applications''
under the grant agreement PZ00P2\_223297 and partially by the DFG -- project ID 551478549.
J.L.\ and R.W.\ were funded by
  the Deutsche Forschungsgemeinschaft (DFG, German Research
  Foundation) -- Project-ID 427320536 -- SFB 1442 ``Geometry: Deformations and Rigidity'', as well as under
  Germany's Excellence Strategy EXC 2044 390685587/2, ``Mathematics
  M\"unster: Dynamics -- Geometry -- Structure''.

\subsection*{Parallel related work by Alexandrov, Bychkov,
Dunin-Barkowski, Kazarian, and Shadrin}

While this article was in the final stages of preparation,
we were informed by S.~Shadrin about their forthcoming work
on spin enumeration of hypermaps using $\mathbb{Z}/2\mathbb{Z}$-equivariant
topological recursion, together with $x$-$y$ duality and
symplectic duality. This work has since appeared
\cite{ABDKS2026SpinHypermaps}. Thus, some of the results
established here overlap with those of \emph{op. cit.}
We note, however, that our approach avoids logarithmic
topological recursion \cite{Hock:2023dno,Alexandrov:2023tgl,Alexandrov:2024ajj},
making the construction more explicit, while also allowing
for more general groups than $\mathbb{Z}/2\mathbb{Z}$ in
Sec.~\ref{sec:xydualofequivariantgeneral}.
We thank S.~Shadrin for informing us about this parallel work.

\section{Review of Universal $x$--$y$ Duality and Equivariant Topological Recursion}\label{sec:xydualtiyandequivariantTRgeneral}
Before reviewing the recent definitions and constructions in the realm of the universal $x$-$y$ duality, we want to mention that its origin (and still the most important connection and application) comes from topological recursion. This means that one is naturally interested in this duality for multi-differentials defined by topological recursion, however as we will review below and actually apply later, the universal $x$-$y$ duality can be defined, constructed and applied completely independently of topological recursion. Thus, we make no assumption that the multi-differentials are defined by, or related to, topological recursion if not explicitly specified. Thus, we need to define the necessary conditions which we impose on the multi-differentials.

\subsection{Universal $x$-$y$ Duality}\label{sec:xyduality}
We denote multi-differentials by $\omega_{g,n}$ with $(g,n)\in \mathbb{Z}_{\geq 0}\times\mathbb{Z}_{>0}$. They  are $n$-differentials on $\Sigma^n$, where $\Sigma$ is some Riemann surface. In other words, $\omega_{g,n}$ can be viewed as a 1-form on each copy $\Sigma$. The assumptions on $\Sigma$ (for instance whether it is connected, compact, etc.) depend on the context. 

\begin{definition}[System of multi-differentials]\label{def:systemofcorrelators}
    We call a collection $\{\omega_{g,n}\}_{g,n}$ indexed by $(g,n)\in \mathbb{Z}_{\geq 0}\times\mathbb{Z}_{>0}$  with $2g-2+n\geq 0$ a system of multi-differentials if the following conditions are satisfied
    \begin{itemize}
        \item $\omega_{g,n}$ are symmetric meromorphic $n$-differentials;
        \item $\omega_{0,2}$ has a double pole on the diagonal with bi-residue 1, that is, it behaves locally in any coordinate as $\omega_{0,2}(z_1,z_2)\overset{z_1\to z_2}{\sim} \frac{dz_1\,dz_2}{(z_1-z_2)^2}+\mathcal{O}\big((z_1-z_2)^0\big) dz_1dz_2$;
        \item for $2g-2+n>0$, $\omega_{g,n}$ is regular on the diagonal.
    \end{itemize}
\end{definition}
\begin{remark}
    In Def.~\ref{def:systemofcorrelators}, the 1-form $\omega_{0,1}$ is excluded by the condition of $2g-2+n\geq 0$. In our definition, $\omega_{0,1}$ is not part of a system of multi-differentials and, as seen later, should actually be viewed as some additional input. 
\end{remark}
\begin{example}
    A quite trivial example of a system of multi-differentials is a bi-differential $\omega_{0,2}$ satisfying the conditions of Def.~\ref{def:systemofcorrelators} and $\omega_{g,n}=0$ for $2g-2+n>0$.
\end{example}
\begin{example}
    A very important example of a system of multi-differentials arises from topological recursion (see Ex.~\ref{ex:G1topologicalrecursion}). For two given meromorphic 1-forms $dx,dy$ on $\Sigma$, topological recursion generates a system of multi-differentials defined locally around the critical points of $dx$. If, moreover, $\Sigma$ is compact and $\omega_{0,2}$ is chosen to be the Bergman kernel normalised with respect to a fixed symplectic basis $(\mathcal{A}_i,\mathcal{B}_i)_i$ of $H_1(\Sigma)$, then topological recursion determines uniquely a system of multi-differentials globally defined on $\Sigma$.
\end{example}
Def.~\ref{def:systemofcorrelators} is an abstract definition for a set of differentials $\{\omega_{g,n}\}_{g,n}$, and of course this system of multi-differentials is of particular interest if it has some specific geometric origin, combinatorial interpretation or is defined uniquely from some recursive structure. 

However, independent of the origin of a system of multi-differentials, one can uniquely define dual multi-differentials associated with given 1-forms $dx$ and $dy$. This duality between two sets of multi-differentials is known as the $x$-$y$ duality and appeared already in several recent works (see for instance \cite{Borot:2021thu,Hock:2022wer,Hock:2022pbw,Alexandrov:2022ydc,Alexandrov:2023jcj}). We will follow the presentation of \cite{Alexandrov:2024tjo} which gave rise to a profound generalisation of topological recursion. However, we prefer to present the formula in terms of bipartite graphs rather than in terms of
multiedges; both combinatorial structures are obviously equivalent.

The construction of the dual multi-differentials is quite involved. The definition proceeds in two steps, by first defining the extended multi-differentials denoted by $\mathbb{W}^{(x,y)}_{g,n}$ from bipartite graphs. Those extended multi-differentials already depend on two given 1-forms $dx$ and $dy$. The second step consists of defining the dual multi-differentials denoted by $\omega^{\vee(x,y)}_{g,n}$ from $\mathbb{W}^{(x,y)}_{g,n}$.

\begin{definition}[Bipartite graphs]\label{def:bipartite}
  A bipartite graph $\Gamma$ consists of a labelled set of `white' vertices
  $\mathcal{V}(\Gamma)$ and an unlabelled set of `black' vertices
  $\mathcal{B}(\Gamma)$. Edges connect only vertices in
  $\mathcal{V}(\Gamma)$ to vertices in $\mathcal{B}(\Gamma)$, and
  multiple edges between two vertices are allowed. Every vertex has
  valency at least one, except for the graph consisting of one single
  labelled vertex. A graph is called connected if every pair of
  vertices can be joined by a sequence of edges.

For a vertex $\flat\in \mathcal{B}(\Gamma)$, we denote by
$I(\flat)$ the subset of edges of $\Gamma$ incident to $\flat$. Then an edge 
$e\in I(\flat)$ is incident to $\flat$ and a unique
$\tilde{v}(e)\in \mathcal{V}(\Gamma)$. We let $v(e)$ be the label of
$\tilde{v}(e)$. In general $v(I(\flat))$ is a multiset.
The degree of $\flat$ is $|\flat|=|I(\flat)|$.

If $L$ is a set of labels,
we denote by $G_L$ the set bipartite connected graphs whose
$\mathcal{V}$-vertices are labelled by $L$.
\end{definition}
The set $G_{\set{n}}$ is infinite. It remains infinite even if one
restricts to a finite number of unlabelled vertices, since multiple
edges are allowed, or to a fixed first Betti number, since unlabelled
$1$-valent vertices are allowed. However, for a fixed Betti number and
a fixed upper bound on the number of unlabelled vertices, the
corresponding set is finite.

\begin{figure}[h]
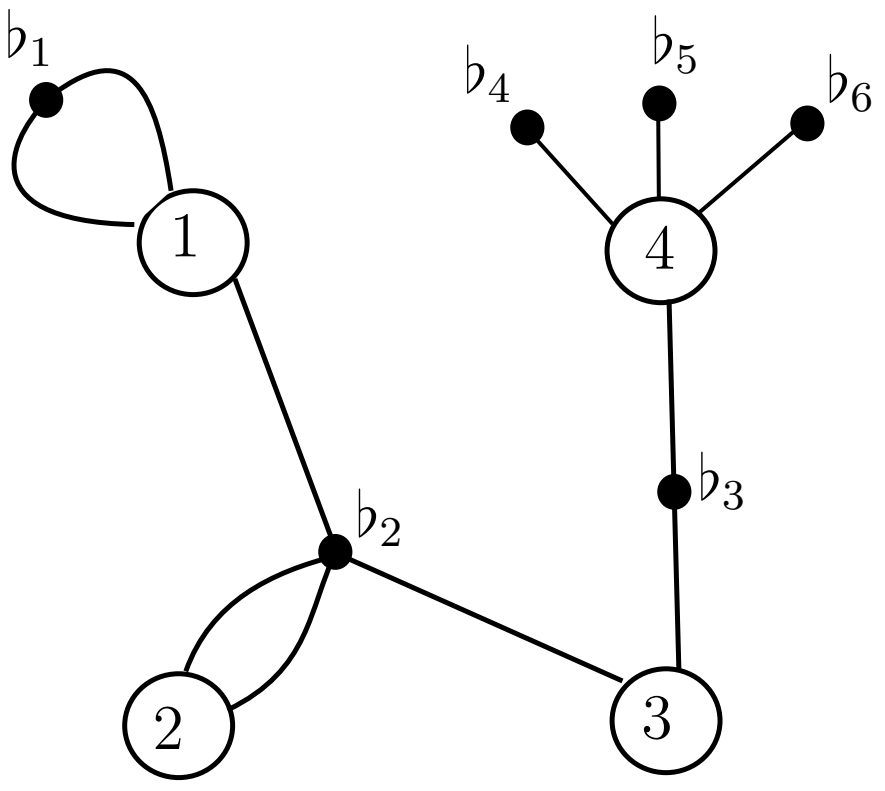
\caption{This is an example of a bipartite graph
  $\Gamma\in G_{\set{4}}$ with four labelled `white' vertices and six unlabelled
  `black' vertices $\flat_i\in \mathcal{B}(\Gamma)$. The multisets 
  associated to the vertices $\mathcal{B}(\Gamma)$ are
  $v(I(\flat_1)) =\{1,1\}, v(I(\flat_2))=\{1,2,2,3\}$,
  $v(I(\flat_3))=\{3,4\}$ and
  $v(I(\flat_4))=v(I(\flat_5))=v(I(\flat_6))=\{4\}$ with
  $|\flat_4|=|\flat_5|=|\flat_6|=1$, $|\flat_1|=|\flat_3|=2$,
  $|\flat_2|=4$. The automorphism group has order
  $|\operatorname{Aut}(\Gamma)|=2!\,2!\,3!=24,$ where the factors
  $2!$ and $2!$ come from permuting the two edges attached to the
  same labelled vertex in $\flat_1$ and $\flat_2$, respectively, and the
  factor $3!$ comes from permuting the three indistinguishable
  vertices $\flat_4,\flat_5,\flat_6$.}
\label{Fig:multiedge}
\end{figure}

With the set of bipartite graphs at hand, we are ready to define the
extended $n$-differentials as weighted sums over connected bipartite
graphs. This definition is understood as a formal power series in
$\hbar$ and additional formal parameters $u_i$. Since the set of
graphs in $G_{\set{n}}$ with fixed first Betti number and a fixed upper bound
on the number of unlabelled vertices is finite, the following
definition consists, at each order in $\hbar$, of only finitely many
contributing graphs.

\begin{definition}[Extended multi-differentials]\label{def:extendemulti}
  Let
  \begin{equation}\label{def:S}
  \cS(t)=\frac{\sinh(t/2)}{t/2} =\sum_{n\geq0}\frac{t^{2n}}{2^{2n}
    (2n+1)!}, 
\end{equation}
and define from a system of multi-differentials
  $\{\omega_{g,n}\}_{g,n}$ the formal series
\begin{equation}
  \omega_n=\sum_{\substack{g\geq0,\\(g,n)\neq (0,1)}} \hbar^{2g-2+n}\omega_{g,n}.
\end{equation}
Let $dx, dy$ be two 1-forms on $\Sigma$. Then, the extended
$n$-differentials associated to $\omega_{g,n},dx,dy$ are defined by
$\mathbb{W}^{(x,y)}_{g,n}:= [\hbar^{2g-2+n}]\mathbb{W}^{(x,y)}_n$ with
\begin{align} \label{eq:def-bW}
  &      \mathbb{W}^{(x,y)}_n(u_{\set{n}};z_{\set{n}})
  \\
  &:=
      \prod_{i=1}^n
      \frac{e^{\frac{u_i}{dx_i}\big(\cS(u_i\hbar \,d_{z_i}
          \frac{1}{dx_i})
          -1\big) y_idx_i}}{u_i\hbar}
      \nonumber
    \\
        &\times \sum_{\Gamma\in G_{\set{n}}} 
      \frac{1}{|\mathrm{Aut}(\Gamma)|}\bigg(
     \prod_{\flat\in \mathcal{B}(\Gamma)}
     \prod_{e\in I(\flat)} \!\frac{u_e\hbar}{dx_e}
     \cS\Big( u_e \hbar  \, d_{z_e} \frac{1}{dx_e}\Big)
     {:}\omega_{|\flat|}^{(x,y)}{:}( z_{I(\flat)})
     \bigg)\Big|_{\substack{z_e\mapsto z_{v(e)}\\
         u_e\mapsto u_{v(e)}}}.
     \nonumber
   \end{align}
   Here $d_{z_i}$ is the exterior differential in $z_i$ and
   $x_i = x(z_i)$, $x_e = x(z_e)$, $y_i=y(z_i)$. The symbol ${:}\omega_{|\flat|}{:}$ stands for
   the normal-ordered multi-differential
\begin{align}
  {:}\omega_{|\flat|}^{(x,y)}{:}(z_{e_1},...,z_{e_{|\flat|}})
  =\omega_{|\flat|}(z_{e_1},...,z_{e_{|\flat|}})
  -\delta_{|\flat|,2}\delta_{v(e_1),v(e_2)}
\frac{dx_{e_1}dx_{e_2}}{(x_{e_1}-x_{e_2})^2}.
\label{normal-ordering-2}
\end{align}
The differential operator $d_{z_j}\frac{1}{dx_j}$  first divides by $dx_j$ and then takes the exterior differential, so it has a 1-form as input and generates a 1-form as output.
\end{definition}
Let us mention some very important properties about this quite heavy combinatorial definition which are everything else than obvious from the first glance: The extended $n$-differential $\mathbb{W}^{(x,y)}_{g,n}(u_{\set{n}};z_{\set{n}})$ is
\begin{itemize}
\item a function of $z_1,...,z_n$. Other conventions exist in the literature.
  
\item a finite combination of terms involving $dx_i$, $dy_i$ and $\omega_{g',n'}$ with $2g'-2+n'\leq 2g-2+n$ and derivatives thereof. The dependence on $x_i$ or $y_i$ disappears by the action the differential operator or taking limits on the diagonal.
    \item a polynomial in $u_i$, with exception of
      $\mathbb{W}^{(x,y)}_{0,1}(u;z)=\frac{1}{u}$ from the graph without any edge.

    \item symmetric by simultaneously exchanging $(z_i,u_i)\leftrightarrow (z_j,u_j)$.
    \item at leading order $[u_1^0\cdots u_n^0]$ given by
      $\omega_{g,n}(z_{\set{n}})/\prod_{i=1}^n dx_i$,
      unless $(g,n)=(0,1)$. All $[u_1^{k_1}\cdots u_n^{k_n}]$-contributions 
      with $\sum_ik_i>0$ involve only
      $\omega_{g',n'}$  with $2g'-2+n'\leq 2g-2+n$.

\end{itemize}
The normal-ordering (\ref{normal-ordering-2}) leads to a well-defined
diagonal substitution $z_{e_1},z_{e_2}\mapsto z_i$.

\begin{remark}
  In the representation \eqref{eq:def-bW} of $\mathbb{W}_n^{(x,y)}$, we have
  chosen to work with bipartite graphs, as in \cite{Hock:2022pbw},
  rather than with multigraphs, as in
  \cite{Alexandrov:2022ydc,Alexandrov:2024tjo}; the two descriptions
  are, of course, equivalent. Nevertheless, the combinatorial
  structure underlying \eqref{eq:def-bW} can be refined further by
  decorating the unlabelled vertices and edges according to the
  expansions of $\omega_{|\flat|}$ and the function $\cS$ providing a
  grading on the set of graphs. The resulting expression can then be
  represented as a sum over connected decorated bipartite graphs of
  fixed Euler characteristic; see, for instance, \cite[Def.~2.9 and
  Prop.~2.11]{Hock:2022pbw}.

\end{remark}

Now we are ready to define the dual differentials
$\omega_{g,n}^{\vee(x,y)}$ associated to the two 1-forms $dx,dy$,
which again form a system of multi-differentials in the
sense of Def.~\ref{def:systemofcorrelators}. The last step in the
definition of $\omega_{g,n}^{\vee(x,y)}$ from the extended
$n$-differentials $\mathbb{W}_{g,n}^{(x,y)}$ of
Def.~\ref{def:extendemulti} can be interpreted as replacing the formal
variable $u_i$ by differentials operators $-d_{z_i}\frac{1}{dy_i}$
\begin{align*}
  \omega_{g,n}(z_{\set{n}})
  \xrightarrow{\text{Def.~\ref{def:extendemulti}}}
  \mathbb{W}_{g,n}^{(x,y)}(u_{\set{n}};z_{\set{n}})
  {\textstyle \prod_{i=1}^n (-dx_i)}  
  \xrightarrow{u_i\to -d_{z_i}\frac{1}{dy_i}}\omega_{g,n}^{\vee(x,y)}
  (z_{\set{n}}).
\end{align*}

\begin{definition}[$x$-$y$ dual multi-differentials]\label{def:xydualmultidiff}
  For a given system of multi-differentials $\{\omega_{g,n}\}_{g,n}$
  and two meromorphic 1-forms $dx,dy$, the associated $x$-$y$ dual
  system of multi-differentials denoted by
  $\{\omega_{g,n}^{\vee(x,y)}\}_{g,n}$ is defined by
  \begin{align}
    \label{eq:def-omvee}
    \omega_{g,n}^{\vee(x,y)}(z_{\set{n}})
    =(-1)^n \!\!\!\! \sum_{r_1,...,r_n\geq 0}
      \prod_{i=1}^n\Big({-}d_i\frac{1}{dy_i}\Big)^{r_i}[u_i^{r_i}]
      \Big(
      \mathbb{W}_{g,n}^{(x,y)}(u_{\set{n}};z_{\set{n}})
      \prod_{i=1}^n dx_i
      \Big).
 \end{align}
 Again, we emphasize that the case $(g,n)=(0,1)$ is excluded from our notion of a system of multi-differentials.
\end{definition}
From the definition, it is clear that $\omega_{g,n}^{\vee(x,y)}$ is
symmetric, since it is constructed from a system of symmetric
multi-differentials $\{\omega_{g,n}\}_{g,n}$ through a symmetric
combinatorial formula.
\begin{remark}
\label{remark:signs}
  Various other sign conventions are possible and equally justifiable.
  We could omit the leading $(-1)^n$ and still get a perfectly consistent
  dual family $\{\omega_{g,n}^{\vee(x,y)}\}$. Also various
  combinations of $u_i\mapsto -u_i$, $dx_i\mapsto -dx_i$, 
  $dy_i\mapsto -dy_i$ could be chosen, and such choices appear
  in the literature. The equations \eqref{eq:def-bW}, \eqref{eq:def-omvee}
  and their inverses in Thm.~\ref{thm:xyduality} below
  have their origin in identities
  for operators on (bosonic) Fock space. This connection was pioneered
  in \cite{Bychkov:2020ujd,BDKS24} and then more and more refined in
  \cite{Bychkov:2022wgw,Alexandrov:2022ydc}.
  There is still some freedom with conventions. For instance, there is
  an `old' choice of the parametrisation \cite{Bychkov:2022wgw}
  which is conceptionally simpler but leads to a more complicated
  $\mathbb{W}_{g,n}^{(x,y)}$, and a `new' parametrisation
  \cite{Alexandrov:2022ydc}. We follow the new parametrisation with 
  systematic sign changes. The reason is that we connect these dualities
  for 10 families of multi-differentials, see Fig.~\ref{fig:duality-network}.
  The above relation between $\omega_{g,n}$ and $\omega_{g,n}^{\vee(x,y)}$
  addresses the middle vertical pair. The duality between the 
  leftmost vertical pair is a concrete Fock space analysis related to
  the character expansion (see e.g.\ \cite{Alexandrov:2010th}) of the 
  Itzykson--Zuber--Harish-Chandra (IZHC) integral 
  \cite{Harish-Chandra:1957, Itzykson:1992ya} in a matrix model.
  We provide two longer Appendices \ref{app:functionalrelations}+\ref{app:XYFock} where we
  rederive the Fock space dualities in the IZHC setting
  and thereby determine the signs
  uniquely. That any $x$-$y$ duality can be universally traced back to
  an abstract  Fock space is discussed in  \cite{Alexandrov:2022ydc}.
\end{remark}

Furthermore, since the dual multi-differentials are built from the
$\omega_{g,n}$, possible poles of $\omega_{g,n}^{\vee(x,y)}$ can only
occur at poles of $\omega_{g,n}$ with $2g-2+n>0$, or at critical
points of $dx$ and $dy$ (if one takes the standard topological
recursion Bergman kernel $B=\omega_{0,2}$). Note further that since
the definition of extended $n$-differentials involves evaluating
several $\omega_{g,n}$ on diagonals, additional poles may arise in
$\omega_{g,n}^{\vee(x,y)}$ from a non-standard $\omega_{0,2}$, even
when they are not present in the original system
$\{\omega_{g,n}\}_{g,n}$.

As said before, the bi-differential $\omega_{0,2}^{\vee(x,y)}$ is
special: it satisfies $\omega_{0,2}^{\vee(x,y)}=\omega_{0,2}$
independently of any choice of $dx$ and $dy$. In particular, it
retains a double pole on the diagonal with bi-residue $1$, and its
other poles if present.

\begin{example}\label{ex:extendedom02}
    An new type of example consists of the compact genus-zero Riemann surface $\mathbb{P}^1$ together with the non-standard symmetric bi-differential
    \[
    \omega_{0,2}(z_1,z_2)=\frac{dz_1\,dz_2}{(z_1-z_2)^2}+\frac{dz_1\,dz_2}{(z_1+z_2)^2}.
    \]
    This bi-differential is usually not used in topological recursion
    itself, but appears in equivariant topological recursion
    \cite{Giacchetto:2021atk} and also in the quartic Kontsevich model
    \cite{Branahl:2020yru}. Consider the system of multi-differentials
    consisting of this $\omega_{0,2}$ together with the trivial
    multi-differentials $\omega_{g,n}=0$ for $2g-2+n>0$. For two
    meromorphic $1$-forms $dx,dy$, the corresponding $x$-$y$ dual
    multi-differentials $\omega_{g,n}^{\vee(x,y)}(z_{\set{n}})$
    will in general have poles on the anti-diagonal coming from
    $\omega_{0,2}$, that is, at the points where
    $z_i+z_j=0$. Furthermore, since $\omega_{0,2}$ is evaluated on the
    diagonal in the definitions of the extended $n$-differentials and
    the $x$-$y$ dual multi-differentials, all
    $\omega_{g,n}^{\vee(x,y)}$ with $g\geq 1$ also have poles at
    $z_i=0$ in general. Thus, allowing $\omega_{0,2}$ to have a
    singular anti-diagonal term indeed affects the pole structure of
    the $x$-$y$ dual
    multi-differentials. 
\end{example}

The reason why the $x$-$y$ dual multi-differentials are called
``dual'' is the following duality theorem, which relates the system
$\{\omega_{g,n}^{\vee(x,y)}\}_{g,n}$ back to the original system of
multi-differentials $\{\omega_{g,n}\}_{g,n}$ from which it was
constructed. The idea is to apply the same procedure once again, but
with the r\^oles of the two $1$-forms $dx$ and $dy$ exchanged: 
\begin{align*}
  \omega^{\vee(x,y)}_{g,n}(z_{\set{n}})
  \xrightarrow{\text{Def.~\ref{def:extendemulti}}}
  \mathbb{W}_{g,n}^{\vee(x,y)}(w_{\set{n}};z_{\set{n}})
  \xrightarrow{\text{Def.~\ref{def:xydualmultidiff}}}
  \omega_{g,n}(z_{\set{n}}).
\end{align*}
Here, $\mathbb{W}_{g,n}^{\vee(x,y)}$ is defined by the same formula as in Def.~\ref{def:extendemulti}, with $x$ and $y$ interchanged and the system of multi-differentials $\{\omega_{g,n}^{\vee(x,y)}\}_{g,n}$ used as input. The following theorem was proved in \cite{Alexandrov:2022ydc}.

\begin{theorem}[Universal $x$-$y$ Duality Theorem]\label{thm:xyduality}
  Let $\{\omega_{g,n}\}_{g,n}$ be a system of multi-differentials,
  $dx,dy$ two 1-forms and $\{\omega_{g,n}^{\vee(x,y)}\}$ the
  corresponding $x$-$y$ dual system of multi-differentials. Then,
  applying the same construction with the r\^ole of $dx,dy$ exchanged,
  one gets back the original system of multi-differentials
    \begin{align*}
        \big(\omega_{g,n}^{\vee(x,y)}\big)^{\vee(y,x)}=\omega_{g,n}.
      \end{align*}
Concretely, the dual formulae read
  \begin{align}
    \label{identity:om}
      \omega_{g,n}(z_{\llbracket n \rrbracket})=(-1)^n
      \!\!\!\! \sum_{r_1,...,r_n\geq 0}
      \prod_{i=1}^n\Big({-}d_{z_i}\frac{1}{dx_i}\Big)^{r_i}[w_i^{r_i}]
      \Big(
      \mathbb{W}_{g,n}^{\vee(x,y)}(w_{\set{n}};z_{\set{n}})
      \prod_{i=1}^n dy_i      \Big)
    \end{align}
for $2g+n\geq 2$, 
where $\mathbb{W}_{g,n}^{\vee(x,y)}(w_{\set{n}};z_{\set{n}})
= [\hbar^{2g+n-2}] \mathbb{W}_{n}^{\vee(x,y)}(w_{\set{n}};z_{\set{n}})$ and 
\begin{align} \label{eq:def-bWvee}
  &      \mathbb{W}^{\vee(x,y)}_n(w_{\set{n}};z_{\set{n}})
  \\
  &:=
      \prod_{i=1}^n
      \frac{e^{\frac{w_i}{dy_i}\big(\cS(w_i\hbar \,d_{z_i}
          \frac{1}{dy_i})
          -1\big) x_idy_i}}{w_i\hbar}
      \nonumber
    \\
        &\times \sum_{\Gamma\in G_{\set{n}}} 
      \frac{1}{|\mathrm{Aut}(\Gamma)|}\bigg(
     \prod_{\flat\in \mathcal{B}(\Gamma)}
     \prod_{e\in I(\flat)} \!\frac{w_e\hbar}{dy_e}
     \cS\Big( w_e \hbar  \, d_{z_e} \frac{1}{dy_e}\Big)
     {:}\omega_{|\flat|}^{\vee(x,y)}{:}( z_{I(\flat)})
     \bigg)\Big|_{\substack{z_e\mapsto z_{v(e)}\\
         w_e\mapsto w_{v(e)}}}.
     \nonumber
   \end{align}
   Here $\omega_{|\flat|}^{\vee(x,y)}=\sum_{g\geq 0, \,(g,|\flat|)\neq (0,1)} \hbar^{2g+|\flat|-2}
   \omega_{g,|\flat|}^{\vee(x,y)}$ and $x_i = x(z_i)$, $y_i = y(z_i)$, $y_e=y(z_e)$. The normal-ordering is defined by
\begin{align*}
  {:}\omega_{|\flat|}^{\vee(x,y)}{:}(z_{e_1},...,z_{e_{|\flat|}})
  =\omega^{\vee(x,y)}_{|\flat|}(z_{e_1},...,z_{e_{|\flat|}})
  -\delta_{|\flat|,2}\delta_{v(e_1),v(e_2)}
\frac{dy_{e_1}dy_{e_2}}{(y_{e_1}-y_{e_2})^2}.
\end{align*}
\end{theorem}
We reprove this theorem in App.~\ref{app:functionalrelations}+\ref{app:XYFock} by Fock space techniques in the 
sign conventions dictated by the IZHC integral (see Rem.~\ref{remark:signs}). The universal validity
would be argued as in \cite{Alexandrov:2022ydc}.

\begin{remark}
For Thm.~\ref{thm:xyduality} to hold, it is not necessary to have a compact Riemann surface or for $\omega_{0,2}$ to have poles only on the diagonal. In particular, the bi-differential of Ex.~\ref{ex:extendedom02} is perfectly fine.
\end{remark}
The Universal $x$-$y$ Duality Theorem is a very strong statement and
highly relies on the fact that the system of multi-differentials is
symmetric and that $\omega_{0,2}$ has a double pole on the diagonal
with bi-residue one. Furthermore, the function $\cS(t)$ in \eqref{def:S} is of paramount
importance. It arises from the action of the matrix units
$E_{ij}$ on Fock space and, through several operator identities,
relates perfectly to the
combinatorics of bipartite graphs and their automorphisms. Let us
illustrate this on an example:
\begin{example}
  By definition, we have $\omega_{0,2}
  =\omega_{0,2}^{\vee(x,y)}$ and
    \begin{align*}
        \omega_{1,1}^{\vee(x,y)}(z)
        &=
        -\omega_{1,1}(z)
        +\frac{1}{2}d_z\bigg(\frac{{:}\omega_{0,2}^{(x,y)}{:}(z,z)}{
          dx(z)\,dy(z)}\bigg)
        -\frac{1}{24}
        d_z\bigg(\frac{d_z\frac{d_z\frac{dy(z)}{dx(z)}}{dy(z)}}{dy(z)}\bigg),\\
        \omega_{1,1}(z)
        &=
        -\omega_{1,1}^{\vee(x,y)}(z)
        +\frac{1}{2}d_z\bigg(\frac{{:}\omega_{0,2}^{\vee(x,y)}{:}(z,z)}{
          dx(z)\,dy(z)}\bigg)
        -\frac{1}{24}
        d_z\bigg(\frac{d_z\frac{d_z\frac{dx(z)}{dy(z)}}{dx(z)}}{dx(z)}\bigg),
    \end{align*}
    where ${:}\omega_{0,2}^{(x,y)}{:}(z,z) =
    \big(\omega_{0,2}(z,z')-\frac{dx(z)dx(z')}{(x(z)-x(z'))^2}
    \big)|_{z'\to z}$ and
${:}\omega_{0,2}^{\vee(x,y)}{:}(z,z) =
    \big(\omega_{0,2}(z,z')-\frac{dy(z)dy(z')}{(y(z)-y(z'))^2}
    \big)|_{z'\to z}$.
    Thus, inserting the first expression into the second one, the equality, or equivalently the Universal $x$-$y$ Duality Theorem for $(g,n)=(1,1)$, holds if
    \begin{align*}
      &\frac{1}{2}d_z\bigg( \frac{ \lim_{z'\to z} \big(
        \frac{dy(z)dy(z')}{(y(z)-y(z'))^2} 
        -\frac{dx(z)dx(z')}{(x(z)-x(z'))^2}\big)
        }{dx(z)\,dy(z)}\bigg)
        -\frac{1}{24}d_z\bigg(\frac{d_z
          \frac{d_z\frac{dx(z)}{dy(z)}}{dx(z)}}{dx(z)}
        -\frac{d_z\frac{d_z
            \frac{dy(z)}{dx(z)}}{dy(z)}}{dy(z)}\bigg)=0,
    \end{align*}
    which indeed vanishes identically. In this computation, we have
    not used the assumption that $\omega_{0,2}$ has poles only on the
    diagonal. In particular, the $\omega_{0,2}$ of
    Ex.~\ref{ex:extendedom02} is perfectly fine here as well. The
    factor $\frac{1}{2}$ comes from the non-trivial automorphism group
    $\mathrm{Aut}(\Gamma)$, and the factor $\frac{1}{24}$ comes from
    the expansion $\cS(t)=1+\frac{t^2}{24}+\mathcal{O}(t^4)$. Thus,
    the cancellation relies on a non-trivial interplay between these
    two contributions.
\end{example}

The Universal $x$-$y$ Duality Theorem can actually be viewed as
particular case of a more general parametrised version relating the
two extended $n$-differentials \cite{Alexandrov:2022ydc}:
\begin{theorem}[Parametrised Universal $x$-$y$ Duality Theorem]\label{thm:parametrizedxyduality}
    For a system $\{\omega_{g,n}\}_{g,n}$, two meromorphic forms $dx,dy$ and its corresponding $x$-$y$ dual multi-differentials $\{\omega_{g,n}^{\vee(x,y)}\}_{g,n}$, the corresponding extended $n$-differentials are related by
    \begin{align}
      &\mathbb{W}_n^{\vee(x,y)}(w_{\set{n}};z_{\set{n}})
      \label{eq:parametric-vee}
      \\
      &= \frac{\delta_{n,1}}{\hbar w_1}
      +
      (-1)^n\!\!\! \sum_{r_1,...,r_n\geq 0}
     \prod_{i=1}^n \bigg(\frac{e^{-w_ix_i}}{dy_i}
      \Big({-}d_{z_i}\frac{1}{dy_i}\Big)^{r_i}[u_i^{r_i}]e^{w_ix_i}dx_i
      \bigg)
      \mathbb{W}_{n}^{(x,y)}(u_{\set{n}};z_{\set{n}}),
      \nonumber
      \\
      &\mathbb{W}_n^{(x,y)}(u_{\set{n}};z_{\set{n}})
      \label{eq:parametric}
      \\
      &= \frac{\delta_{n,1}}{\hbar u_1}
      +
      (-1)^n\!\!\!\sum_{r_1,...,r_n\geq 0}
     \prod_{i=1}^n \bigg(\frac{e^{-u_iy_i}}{dx_i}
      \Big({-}d_{z_i}\frac{1}{dx_i}\Big)^{r_i}[w_i^{r_i}]e^{u_iy_i}dy_i
      \bigg)
      \mathbb{W}_{n}^{\vee(x,y)}(w_{\set{n}};z_{\set{n}}),
      \nonumber
    \end{align}
    where both identities are understood at order $\hbar^{2g-2+n}$
    and $dx_i=dx(z_i),dy_i=dy(z_i)$.
\end{theorem}
As far as we can see, the proof of this theorem requires some
additional steps when generalising the universal $x$-$y$ duality
arguments to the parametrised version in \cite{Alexandrov:2022ydc}. We provide these steps later in this paper.
In Appendix \ref{app:XYFock} we give a complete
proof  by Fock space techniques in the 
sign conventions dictated by the IZHC integral. It involves some
non-trivial combinatorial identities, in particular a
statement in Lemma~\ref{lemma-deformation} about an $\hbar$-deformation.

\subsection{Mixed multi-differentials}\label{sec.mixedcorrelators}
For a given system of multi-differentials $\omega_{g,n}$, one can
naturally interpolate between this system of multi-differentials and
the dual family $\omega_{g,n}^{\vee(x,y)}$. Those interpolating
multi-differentials will be denoted by $\omega^{(g)(x,y)}_{n,m}$ where
$g,n,m\in \mathbb{Z}_{\geq 0}$ with $2g-2+n+m\geq 0$. The
identification with the two families is as follows
\begin{align*}
  \omega^{(g)(x,y)}_{n,0}=\omega_{g,n},\qquad
  \omega^{(g)(x,y)}_{0,m}=\omega_{g,m}^{\vee(x,y)}.
\end{align*}
The interpolation works like 
\begin{align*}
  \omega_{g,n}=\omega^{(g)(x,y)}_{n,0} \,
  \rightarrow\,\omega^{(g)(x,y)}_{n-1,1}\,\rightarrow\,\omega^{(g)(x,y)}_{n-2,2}\,\rightarrow...\rightarrow\,\omega^{(g)(x,y)}_{1,n-1}\,\rightarrow\,\omega^{(g)(x,y)}_{0,n}=\omega_{g,n}^{\vee(x,y)}.
\end{align*}
Each interpolated $\omega^{(g)(x,y)}_{n,m}$ depends explicitly on the
choice of the 1-forms $dx,dy$ except for
$\omega^{(g)(x,y)}_{n,0}=\omega_{g,n}$ corresponding to the
original system of multi-differentials (depending on the origin of
$\omega_{g,n}$). The formulae to determine all
$\omega^{(g)(x,y)}_{n,m}$ are such that they can be determined from
$\omega_{g',n'+m'}=\omega^{(g)(x,y)}_{n'+m',0}$ with
$2g'-2+n'+m'\leq 2g-2+n+m$ \cite[Prop. 4.2]{Alexandrov:2022ydc}.

For later presentation, there is no necessity to give all those
formulae going step by step from $\omega^{(g)(x,y)}_{n+1,m}$ to
$\omega^{(g)(x,y)}_{n,m+1}$. However, there is one step of great importance
which recently got some reformulation in terms of topological
recursion. For this let us define:
\begin{definition}[1-variable extended $n+1$-differentials]\label{def:1variableextended}
  From a system of multi-differentials $\{\omega_{g,n}\}_{g,n}$
  and two 1-forms $dx, dy$  on $\Sigma$, build the
  extended multi-differentials $\mathbb{W}^{(x,y)}_{g,n}(u_{\set{n}};z_{\set{n}})$
  according to Def.~\ref{def:extendemulti}.
  Then, the 1-variable extended $n+1$-differentials associated to
  $\omega_{g,n},dx,dy$ are defined by $\mathcal{W}^{(x,y)}_{g,n+1}
  :=[\hbar^{2g-2+n+1}]\mathcal{W}^{(x,y)}_{n+1}$ and
	\begin{align}
          \mathcal{W}^{(x,y)}_{n+1}(u;z;z_{\set{n}})
          :=
          [u_1^0\cdots u_n^0] 
          \mathbb{W}^{(x,y)}_{n+1}(u,u_{\set{n}};z,z_{\set{n}})
          \prod_{i=1}^n dx_i.
	\end{align}
\end{definition}
The 1-variable extended $n+1$-differentials
$\mathcal{W}^{(x,y)}_{n+1}(u;z;z_{\set{n}})$ are
symmetric with respect to the variables
$z_i\in z_{\set{n}}$, but not permuting $z$ with any
$z_i\in  z_{\set{n}}$. Since
$ \mathcal{W}^{(x,y)}_{n+1}(u;z;z_{\set{n}})$ is
defined from the extended $n+1$-differentials, it
inherits its graphical representation of
Def.~\ref{def:extendemulti}.
Recall that 
$\mathbb{W}^{(x,y)}_{g,n+1}(u,u_{\set{n}};z,z_{\set{n}})$ is for 
$2g+n\geq 2$ a polynomial in all $u,u_i$. Extracting $[u_i^0]$ then amounts to
restricting to graphs where all $\set{n}$-labelled vertices are one-valent
(there can only be one edge contributing $u_i$) and moreover, all
$\mathscr{S}$-operators restrict to $1$.
In contrast, the distinguished labelled vertex (whose label corresponds
to $u,z$) can have any valency and full $\mathscr{S}$-operators
act on the edges.
One can convince oneself (see for instance
\cite{Bychkov:2022wgw}) that the analytic expression corresponding to
this smaller class of graphs simplifies as follows: 
\begin{corollary}\label{cor:1variableextended}
  The 1-variable extended $n+1$-differentials are equivalently given
  by
 \begin{align}
	\label{eq:cW1}
	\cW_{n+1}^{(x,y)}(u;z;z_{\set n})&=
               \frac{e^{\cT_1(u;z)}}{u\hbar}
	\sum_{\substack{\set{n}=\sqcup_{\alpha} J_\alpha
			\\J_\alpha\ne\emptyset}}
	\prod_{\alpha}\cT_{|J_\alpha|+1}(u;z;z_{J_\alpha}),
\end{align}
where 
\begin{align}
	\label{eq:cT1x}
\cT_{n+1}&(u;z;z_{\set n})
\\
&:=\sum_{k=1}^\infty\frac1{k!}
\bigg(\Bigl(\prod_{i=1}^k\frac{u\hbar}{dx}
\cS\Big(u\hbar d_{\tilde{z}_i}\frac{1}{d\tilde{x}_i}\Big)\Bigr)
	\bigl(\omega_{k+n}(\tilde z_{\set{k}},z_{\set{n}})-
	\delta_{n,0}\delta_{k,2}\tfrac{d\tilde x_{1}d\tilde x_{2}}{(\tilde x_{1}-\tilde x_{2})^2}\bigr) \bigg)\Big|_{\tilde{z}_i\mapsto z}
       \nonumber
	\\\notag
        &
        +\delta_{n,0}\frac{u}{dx}\Big(\cS\Big(u\hbar d_z
        \frac{1}{dx }\Big)-1\Big) ydx.
	\end{align}
	Here $x=x(z),y=y(z)$ and $\tilde{x}_i=x(\tilde{z}_i)$. The summation is over ordered set partitions of $\set n$.
\end{corollary}
Note that $\cW_{n+1}^{(x,y)}(u;z;z_{\set n})$ is a function in $z$ and a differential in the variables $z_{\set{n}}$. From the 1-variable extended $n+1$-differential, the mixed multi-differential $\omega^{(g)(x,y)}_{n,1}$ is then given by the similar formula as in Def.~\ref{def:xydualmultidiff} but just for the variable $z$ taking the powers of $u$
\begin{align}\label{mixed1omegagn}
  \omega^{(g)(x,y)}_{n,1}(z_{\set{n}};z) &= -\sum_{r\geq 0}
  \Big({-}d_z\frac{1}{dy}\Big)^{r}[u^{r}]
  \big(\mathcal{W}_{g,n+1}^{(x,y)}(u;z;z_{\set{n}}) dx(z)\big)
  \\
    \nonumber
    &=-\omega_{g,n+1}(z,z_{\set{n}})
    -\sum_{r\geq 1}
  \Big({-}d_z\frac{1}{dy}\Big)^{r}[u^{r}]
  \big(\mathcal{W}_{g,n+1}^{(x,y)}(u;z;z_{\set{n}}) dx(z)\big),
\end{align}
where the leading order in $u$ was written in the second line
explicitly, $[u^0]\mathcal{W}_{g,n+1}^{(x,y)}(u;z;z_{\set{n}})dx(z)
=\omega^{(g)(x,y)}_{n+1,0}(z,z_{\set{n}})
=\omega_{g,n+1}(z,z_{\set{n}})$.
The next-to-leading order in $u$ reads (for $n\geq 1$)
\begin{align}\label{Wfirstexpansion}
  [u^1]\mathcal{W}_{g,n+1}^{(x,y)}(u;z;z_{\set{n}})
  & =\frac{1}{2} \frac{\omega_{g-1,n+2}(z,z,z_{\llbracket n \rrbracket})
}{dx(z) dx(z)}
\\
&+\frac{1}{2} \sum_{\substack{J_1\sqcup J_2=\llbracket n \rrbracket\\g_1+g_2=g\\(g_i,J_i)\neq (0,\emptyset)}}^{\text{ordered}}
\frac{\omega_{g_1,|J_1|+1}(z,z_{J_1})}{dx(z)}
\frac{\omega_{g_2,|J_2|+1}(z,z_{J_2})}{dx(z)}.
\nonumber
\end{align}
The factor $\frac{1}{2}$ in the first line comes from the order of the
automorphism
group of the graph or, equivalently, the factor $\frac{1}{k!}$ with $k=2$
in (\ref{eq:cT1x}). The factor $\frac{1}{2}$ in the second line corrects
the overcount from the convention that $J_1\sqcup J_2=\set{n}$
is the ordered set partition.

Let us mention the pole structure of the mixed multi-differentials
$\omega^{(g)(x,y)}_{n,m}(z_{\llbracket n \rrbracket};z_{\llbracket n+m
  \rrbracket\setminus \llbracket n \rrbracket})$. It has 
potential poles in the variables
$z_{\llbracket n+m \rrbracket\setminus \llbracket n \rrbracket}$ at
the poles of $\omega_{g',k'}$ where $2g'-2+k'\leq 2g-2+n+m$, at the
critical points of $dx,dy$ and the diagonal if a variable of the set
$z_{\llbracket n \rrbracket}$ meets a variable of the set
$z_{\llbracket n+m \rrbracket\setminus \llbracket n \rrbracket}$. In
the variables $z_{\llbracket n \rrbracket}$, it has potential poles 
at the poles of $\omega_{g',k'}$ where $2g'-2+k'\leq
2g-2+n+m$. This pole structure is easily seen by recursion.

\subsection{(Equivariant) Topological Recursion}
Let us review the definition of a $G$-equivariant topological recursion, as it was defined in \cite{Giacchetto:2021atk}. This can be seen as an extension of topological recursion with symmetries coming from a finite group $G$.
\begin{definition}[$G$-equivariant topological recursion]\label{defn:equiv:SC}
	Let $G$ be a finite group. A \emph{$G$-equivariant spectral curve} $\mathcal{S} = (\Sigma,\phi, x,y,B, \chi, \upsilon, \beta)$ consists of
	\begin{itemize}
		\item a Riemann surface $\Sigma$ (not necessarily compact or connected) with an action $ \phi \colon G \times \Sigma \to \Sigma $, which we will often write $\phi (\gamma,z) = \phi_\gamma z = \gamma z$;
		\item a function $x \colon \Sigma \to \mathbb{C}$ such that its differential $dx$ is meromorphic and has finitely many zeros $a_1,\dots,a_{r|G|}$ that are simple;
        \item the group action is free on some neighbourhood of the ramification locus;
		\item a meromorphic function $y \colon \Sigma \to \mathbb{C}$ such that $d y$ does not vanish at the zeros of $d x$;
		\item a symmetric bi-differential $B$ on $\Sigma \times \Sigma$;
		\item three one-dimensional representations $ \chi, \upsilon, \beta \colon G \to \mathbb{C}^\times $,
	\end{itemize}
	such that for any $ \gamma \in G$
	\begin{equation}\label{conditionequivariantTR}
		dx (\gamma z) = \chi_\gamma \, dx (z) \,, \qquad y (\gamma z) = \upsilon_\gamma \, y(z)\,, \qquad B(\gamma z_1,z_2) = \beta_\gamma \, B(z_1,z_2)\,,
	\end{equation}
	and $ B(z_1,z_2) - B^{G,\beta}(z_1,z_2) $ is regular as $ z_1 \to \gamma z_2$, where
	\begin{equation}
		B^{G,\beta}(z_1,z_2) :=  \sum_{\eta \in G} \beta_\eta^{-1} \frac{d (\eta z_1) dz_2}{(\eta z_1 - z_2)^2}\,.
	\end{equation}

    Let $\omega_{0,1}=\frac{1}{|G|}ydx$ and $\omega_{0,2}=B$. Then, $\omega_{g,n}$ is defined recursively on $2g-2+n > 0$ as
	\begin{equation}\label{eqn:TR}
		\begin{split}
			\omega_{g,n}(z_1,\dots,z_n) := \sum_{i=1}^{r|G|}& \Res\displaylimits_{z = a_i} K_i(z_1,z) \bigg( 
			\omega_{g-1,n+1}(z,\sigma_i(z),z_2,\dots,z_n) \\
			& +
			\sum_{\substack{g_1+g_2 = g \\ J_1 \sqcup J_2 = \{2,\dots,n\}\\(g_i,J_i)\neq (0,\emptyset)}}^{\text{ordered}}
			\omega_{g_1,1+|J_1|}(z,z_{J_1})
			\omega_{g_2,1+|J_2|}(\sigma_i(z),z_{J_2})
			\bigg),
		\end{split}
	\end{equation}
	where $K_i$, called the topological recursion kernels, are locally defined in a neighbourhood $U_i$ of $a_i$ as
	\begin{equation}\label{eqn:TR:kernel}
		K_i(z_1,z) := \frac{\frac{1}{2} \int_{w = \sigma_i(z)}^z B(z_1,w)}{\bigl( y(z) - y(\sigma_i(z)) \bigr) d x(z)},
	\end{equation}
	and $\sigma_i \colon U_i \to U_i$ is the Galois involution near the ramification point $a_i \in U_i$. In particular, if $a_i$ is a ramification point of $x$, then $\phi_\gamma(a_i)=\gamma a_i$ is also a ramification point for every $\gamma\in G$.
\end{definition}
\begin{remark}
    The definition used here differs slightly from the original one by the absence of a factor of $1/|G|$ in $B$ but included in $\omega_{0,1}$. This, however, amounts merely to a rescaling by $|G|$.
\end{remark}

\begin{remark}\label{rmk:symplecticshiftofequivariantTR}
    The condition on $y$ in \eqref{conditionequivariantTR} is important but can be relaxed in a certain sense. For instance, shifting $y$ by a function of $x$, that is, $y\to\tilde{y}=y+\psi(x)$, leaves the recursion kernel $K$ in \eqref{eqn:TR:kernel}, and hence the generated multi-differentials $\omega_{g,n}$, invariant. However, the new function $\tilde{y}$ does not carry a natural action of the group $G$. As the definition states, it is nevertheless important that there exists a function $y$ (possibly after such a shift by a function of $x$) satisfying the assumptions of the definition. That is, there exists a representation $\upsilon$ such that $y(\gamma z)=\upsilon_\gamma\,y(z)$ for all $\gamma \in G$. The difference between $y$ and $\tilde{y}$ has a great impact in considering the $x$-$y$ dual side which is separately discussed in Sec.~\ref{sec:xydualofequivariant} and Sec.~\ref{sec:xtydualofequivariant}.
\end{remark}
It follows from the construction and as shown in \cite{Giacchetto:2021atk}, that the multi-differentials generated by $G$-equivariant topological recursion form a system of multi-differentials in the sense of Def.~\ref{def:systemofcorrelators}, thus they are for instance symmetric. We want, however, to emphasise that parts of the spectral curve $\mathcal{S} = (\Sigma,\phi, x,y,B, \chi, \upsilon, \beta)$ are already the two forms $dx,dy$, thus the $\omega_{g,n}$'s constructed here have for instance just poles at the points $a_i$. Note also that even though the function $y$ is used, just its derivative and thus the 1-form $dy$ is needed and appearing in $\omega_{g,n}$ since $\omega_{0,1}$ is excluded in a system of multi-differentials. Topological recursion itself \cite{Eynard:2007kz} can be seen as an example with trivial group.
\begin{example}[Topological Recursion]\label{ex:G1topologicalrecursion}
	Let $G=1$ the trivial group; then any representation is trivially 1. Let further $\Sigma$ be a compact Riemann surface with  $B$ the Bergman kernel normalised with respect to a fixed symplectic basis $(\mathcal{A}_i,\mathcal{B}_i)_i$ of $H_1(\Sigma)$ around the $\mathcal{A}$-cycles. Then, the $\omega_{g,n}$ are by definition the same as the original topological recursion \cite{Eynard:2007kz}. The spectral curve reduces to $ (\Sigma,\phi, x,y,B, \chi, \upsilon, \beta)= (\Sigma, x,y,B)$ which is the classical spectral curve in topological recursion. Applying the universal $x$-$y$ duality to this system of multi-differentials with $dx,dy$ chosen to be the one which generates the family itself, then the $x$-$y$ dual family $\omega^{\vee,(x,y)}_{g,n}$ coincides with the family generated by topological recursion of the spectral curve  $(\Sigma,y,x,B)$ where $x$ and $y$ are swapped \cite{Alexandrov:2022ydc}. Note that this can only be true if $dy$ has just simple zeros; otherwise the $x$-$y$ dual recursion of  \eqref{eqn:TR} is not defined. 
\end{example}

\begin{remark}
It was shown in \cite{Giacchetto:2021atk} that the representation $\beta$ satisfies
$\beta_\gamma^2=1$
for every $\gamma\in G$. Moreover, in order for $G$-equivariant topological recursion to produce non-trivial multi-differentials, one imposes the additional condition
$\chi_\gamma\upsilon_\gamma=\beta_\gamma$
for every $\gamma\in G$.
\end{remark}

\begin{example}
  \label{ex:eqTRZ2}
    Let $G=\mathbb{Z}/2\mathbb{Z}=\{0,1\}$, with non-trivial element acting by
    $\phi_1(z)=-z.$ Let $\upsilon$ be the trivial representation and let $\chi=\beta$ be the sign representation. For the Riemann sphere $\Sigma=\mathbb{P}^1$, the associated $\mathbb{Z}/2\mathbb{Z}$-equivariant topological recursion has
    \begin{align}\label{extendedBergman}
        B=\frac{dz_1\,dz_2}{(z_1-z_2)^2}
        +\frac{dz_1\,dz_2}{(z_1+z_2)^2}.
    \end{align}

    Every ramification point $a_i$ of $x$ is accompanied with a $\mathbb{Z}/2\mathbb{Z}$-image $-a_i$. The fixed point of the involution $\phi_1$ at $z=0$ is not necessarily a ramification point, since $\chi$ is the sign representation and not the trivial representation, where this would necessarily be the case. Moreover, the multi-differentials satisfy
    \[
        \omega_{g,n}(-z,z_{\llbracket n-1 \rrbracket})=-\omega_{g,n}(z,z_{\llbracket n-1 \rrbracket})
    \]
    in each variable $z_i$.
\end{example}

Even though a general definition for $G$-equivariant TR was provided in \cite{Giacchetto:2021atk}, there are no enumerative interpretations known beyond  the trivial group (which is just ordinary TR) and the group $G=\mathbb{Z}/2\mathbb{Z}$, see for instance also \cite{Alexandrov:2025kzn,fang2026}. Interestingly, in the study of a specific matrix model (quartic Kontsevich model \cite{Branahl:2020yru,Hock:2021tbl,Hock:2023nki}) a bi-differential $\omega_{0,2}$ of the form \eqref{extendedBergman} appeared, but the recursive structure of all multi-differentials is different from Def.~\ref{defn:equiv:SC}. Although the $dx,y$ chosen in  \cite{Branahl:2020yru,Hock:2021tbl,Hock:2023nki} do not satisfy the equivariant assumptions of Def.~\ref{defn:equiv:SC}, the authors were able to interpret the multi-differentials of interest in terms of "blobbed topological recursion" \cite{Borot:2015hna} for the $g=0,1$ sectors.

The multi-differentials generated by $G$-equivariant topological recursion can also be computed in two other ways, one of which was already proved in \cite{Giacchetto:2021atk}:

\begin{lemma}\label{lemma:equivalentomegagn}
    Let $(\Sigma,\phi, x,y,B, \chi, \upsilon, \beta)$ be a spectral curve satisfying Def.~\ref{defn:equiv:SC}. Let $K^{\tilde{B}}$ be the kernel defined from some symmetric bi-differential $\tilde{B}$ by Definition~\eqref{eqn:TR:kernel}, with $B$ replaced by $\tilde{B}$ (note that $\tilde{B}$ does not necessarily satisfy condition~\eqref{conditionequivariantTR}). Let $\omega_{g,n}(K^{\tilde{B}},r|G|)$ be the multi-differentials defined by \eqref{eqn:TR} with $K$ replaced by $K^{\tilde{B}}$, using all $r|G|$ ramification points, and let $\omega_{g,n}(K^{\tilde{B}},r)$ be the multi-differentials defined by \eqref{eqn:TR} with $K$ replaced by $K^{\tilde{B}}$, using one representative from each of the $r$ $G$-orbits of ramification points. Then,
    \begin{align*}
        |G|^{2-2g-n}\omega_{g,n}(K^{B},r|G|)
        =
        \omega_{g,n}(K^{B},r)
        =
        \omega_{g,n}(K^{B^{TR}},r|G|),
    \end{align*}
    where $B^{TR}$ is the usual Bergman kernel used in classical topological recursion. In particular, it has no poles other than those on the diagonal. For the second equality assume additionally $B(z_1,z_2)= B^{G,\beta}(z_1,z_2)$.

    \begin{proof}
        Since the spectral curve is given by  $(\Sigma,\phi, x,y,B, \chi, \upsilon, \beta)$ of Def.~\ref{defn:equiv:SC}, we still have $\omega_{0,2}=B$ in the recursion procedure, where $B$ satisfies condition~\eqref{conditionequivariantTR}. We interested in just changing the recursion kernel. The first equality is essentially the statement of \cite[Prop.~8.8]{Giacchetto:2021atk}.

        The second equality follows straightforwardly from $B(z_1,z_2)= B^{G,\beta}(z_1,z_2)$ and the fact that
        \begin{align*}
             \frac{1}{2}\int_{w=\sigma_i(z)}^z B^{TR}(z_1,w)
        \end{align*}
        together with taking the residue at the ramification point $a_i$, defines a projection operator onto the principal part of the Laurent expansion. Since, for $\omega_{g,n}(K^{B^{TR}},r|G|)$, we sum over all $r|G|$ ramification points, the resulting multi-differentials have poles at all of them. Replacing $K^{B^{TR}}$ by $K^{B}$ amounts to multiplying by a factor of $|G|$ at each recursive step, because for the residue at each ramification point $(a_i)_{i=1,\ldots,r|G|}$, the residue operator defined by
        \begin{align*}
            \frac{1}{2}\int_{w=\sigma_i(z)}^z B(z_1,w)
        \end{align*}
        projects simultaneously onto the principal parts at all points $\gamma a_i$ with $\gamma\in G$.
    \end{proof}
\end{lemma}

Even though it is not obvious, the linear and quadratic loop equations (as formalised abstractly in \cite{Borot:2013lpa}) are satisfied in $G$-equivariant topological recursion. Furthermore, the projection property is also satisfied:
\begin{lemma}\label{lemma:linearquadraticloop}
    The multi-differentials generated by $G$-equivariant TR satisfy the linear and quadratic loop equations and the projection property, that is:
    \begin{itemize}
        \item Linear loop equation: for all $g,n\geq 0$
        \begin{align}
        \omega_{g,n+1}(z,z_{\llbracket n \rrbracket})+\omega_{g,n+1}(\sigma_i(z),z_{\llbracket n \rrbracket})
        \end{align}
        is holomorphic as $z\to a_i$ and has at least a simple zero in $z$ at $a_i$.

        \item Quadratic loop equation: for all $g,n\geq 0$
        \begin{align}
        \omega_{g-1,n+2}(z,\sigma_i(z),z_{\llbracket n \rrbracket})+\sum_{\substack{g_1+g_2=g\\ J_1\sqcup J_2=\llbracket n \rrbracket}}^{\text{ordered}}\omega_{g_1,|J_1|+1}(\sigma_i(z),z_{J_1})\omega_{g_2,|J_2|+1}(z,z_{J_2})
        \end{align}
        is holomorphic as $z\to a_i$ and has at least a double zero in $z$ at $a_i$.

        \item Projection property: for $2g-2+n>0$, $\omega_{g,n}$ can be reconstructed from their principal parts
        \begin{align}
			\omega_{g,n}(z_1,\dots,z_n) =  \Big(\frac{1}{|G|^n} \prod_{i=1}^n \sum_{a_j}\Res_{z'_i = a_j}	\int_{a_j}^{z'_i} B(\bullet,z_i) \Big) \omega_{g,n}(z'_1,\dots,z'_n).	
		\end{align}
    \end{itemize}

\begin{proof}
    These three properties are equivalent to Def.~\ref{defn:equiv:SC}. The proof is literally the same as in ordinary TR, using the fact that if $a_i$ is a ramification point, then so is $\gamma a_i$, which ensures the projection property and for each coordinate the normalisation $\frac{1}{|G|}$, since the action of the group $G$ is free on the ramification points. Thus, the proof can be recycled from several articles; see \cite{Eynard:2007kz,Borot:2013lpa,Borot:2015hna}. However, note that $\omega_{0,1}=\frac{1}{|G|}ydx$, which is consistent with the rescaling identified in Lemma~\ref{lemma:equivalentomegagn} coming from the free group action. If one assumes $B(z_1,z_2)= B^{G,\beta}(z_1,z_2)$, one can use the multi-differentials $\omega_{g,n}(K^{B},r)$ or $\omega_{g,n}(K^{B^{TR}},r|G|)$, the 1-form $\omega_{0,1}=ydx$ must be used in the quadratic loop equation. In that case, the projection can also be written without the symmetry factor, but with the TR Bergman kernel,
    \begin{align}
			\omega_{g,n}(z_1,\dots,z_n) =  \Big(\prod_{i=1}^n \sum_{a_j}\Res_{z'_i = a_j}	\int_{a_j}^{z'_i} B^{TR}(\bullet,z_i) \Big) \omega_{g,n}(z'_1,\dots,z'_n).	
		\end{align}
\end{proof}
\end{lemma}
\begin{definition}\label{def:blobb}
    If a system of multi-differentials (including $\omega_{0,1}=\frac{1}{|G|}ydx$ or $\omega_{0,1}=ydx$, respectively) satisfies the linear and quadratic loop equations but not the projection property, that is, if the $\omega_{g,n}$ with $2g-2+n>0$ have additional poles that are not located at the ramification points of $x$, then the system is said to satisfy \emph{blobbed topological recursion} \cite{Borot:2015hna}.
\end{definition}

\section{Universal $x$-$y$ duality and equivariant topological recursion }\label{sec:xydualofequivariantgeneral}
\subsection{$x$-$y$ dual side of equivariant topological recursion multi-differentials}\label{sec:xydualofequivariant}

We now study the $x$-$y$ dual side (in Fig.~\ref{fig:duality-network}
in the middle of the lower row) of the multi-differentials
generated by equivariant topological recursion.  Their definition is
straightforward. The interesting question is what structure these
multi-differentials possess and how they can be characterised.  We can
already observe that the dual side is considerably more complicated
than in the classical case, where the group is trivial, $G=1$, and one
simply works with ordinary topological recursion; see
Ex.~\ref{ex:G1topologicalrecursion}. In that case, the dual side is
again given by topological recursion (provided that $y$ has only
simple ramification points distinct from those of $dx$), obtained
simply by interchanging $x$ and $y$ \cite{Alexandrov:2022ydc}.

Let now $\omega_{g,n}$ be generated by $G$-equivariant topological
recursion as in Def.~\ref{defn:equiv:SC}, but rescaled by
$|G|$. Assume that $B(z_1,z_2)= B^{G,\beta}(z_1,z_2)$, equivalently,
one can use the definition given in
Lemma~\ref{lemma:equivalentomegagn} for $\omega_{g,n}(K^{B},r)$ or
$\omega_{g,n}(K^{B^{TR}},r|G|)$, which ensures that $\omega_{0,1}=ydx$
 in the quadratic loop equation, as
provided by Lemma~\ref{lemma:linearquadraticloop}. We assume this
setting for the rest of the section.
%
%
Just looking at the definition of the $x$-$y$ dual
$\omega_{g,n}^{\vee(x,y)}$, we observe automatically that
$\omega_{g,n}^{\vee(x,y)}$ will have possible poles at the ramification
points of $y$ (the zeros of $dy_i$), at the (``anti-diagonal'') points
satisfying $\gamma z_i=z_j$ for all $\gamma\in G$ with $\gamma\neq 1$,
coming from the bi-differential $B$, and at the fixed points of the
group action satisfying $\gamma z_i=z_i$, arising from sending the
regularised $B$ (i.e.\ the normal-ordered ${:}\omega_{0,2}^{(x,y)}{:}$)
to the diagonal. In principle, the dual
multi-differentials $\omega_{g,n}^{\vee(x,y)}$ could also have poles at the
ramification points of $x$ (the zeros of $dx_i$) and at the diagonal
$z_i=z_j$; these were however already absent in the original topological
recursion setting. Using the $x$ and $y$ from the equivariant spectral
curve $\mathcal{S} = (\Sigma,\phi, x,y,B, \chi, \upsilon, \beta)$ to
define the $x$-$y$ dual family $\omega_{g,n}^{\vee(x,y)}$, we will show
that the poles at the anti-diagonal ($\gamma z_i=z_j$) are absent as
well.

Let us first give an equivalent formulation to the definition of linear and quadratic loop equations mentioned in Lemma~\ref{lemma:linearquadraticloop}.
\begin{definitionlemma} \label{deflemma:linearquadraticnew}
     Let $\Xi^x$ be the span of meromorphic differentials of the form $\bigl(d\frac{1}{dx}\bigr)^k\alpha$, where $k\geq 0$ and $\alpha$ is holomorphic at the ramification points of $x$. 
     Then the linear and quadratic loop equations of multi-differentials from equivariant topological recursion of Lemma~\ref{lemma:linearquadraticloop} are equivalent to
     \begin{gather*}
         \omega_{g,n+1}(z,z_{\llbracket n \rrbracket})\in \Xi^x
         \qquad \text{and}
         \\
         \frac{1}{dx}\bigg(\omega_{g-1,n+2}(z,z,z_{\llbracket n \rrbracket})+\sum_{\substack{J_1\sqcup J_2=\llbracket n \rrbracket\\g_1+g_2=g}}^{\text{ordered}}\omega_{g_1,|J_1|+1}(z,z_{J_1})\omega_{g_2,|J_2|+1}(z,z_{J_2}) \bigg)\in \Xi^x.
       \end{gather*}
\begin{proof}
       This is a straightforward generalisation from topological
       recursion to equivariant topological recursion.  First, note
       that the fixed points of the group action do not coincide with
       the ramification points of $x$ since the group action is free
       on the ramification points of $x$. Then, this definition and
       lemma is by Lemma~\ref{lemma:linearquadraticloop} equivalent 
       to what is proved in
       \cite{Alexandrov:2022ydc,Alexandrov:2024tjo}. The linear and
       quadratic loop equations are local conditions at the
       ramification points of $x$. Since $dx$ is globally defined,
       these local conditions are equivalent to the corresponding
       global conditions that the above multi-differentials belong to
       $\Xi^x$.
     \end{proof}
\end{definitionlemma}
The $x$-$y$ dual multi-differentials are also $G$-equivariant in the following sense:
\begin{lemma}\label{lem:xydualomegatransformation}
    Let $\omega_{g,n}$ be generated by $G$-equivariant topological recursion on the spectral curve $(\Sigma,\phi, x,y,B, \chi, \upsilon, \beta)$. Then the $x$-$y$ dual multi-differentials $\omega_{g,n}^{\vee(x,y)}$, defined by Def.~\ref{def:xydualmultidiff}, satisfy in each variable
    \begin{align*}
      \omega_{g,n}^{\vee(x,y)}(z_1,...,\gamma z_i,...,z_n)
      =\beta_\gamma \omega_{g,n}^{\vee(x,y)}(z_1,..., z_i,...,z_n).
    \end{align*}
    \begin{proof}
      The one-dimensional representations $\chi,\upsilon,\beta$
      satisfy the identity
    \begin{align*}
        \chi_\gamma \upsilon_\gamma=\beta_\gamma\qquad \forall\gamma\in G
    \end{align*}
    and $\beta^2=1$. We send $z_1\to \gamma z_1$ in the definition of
    the extended multi-differential (for all the other $z_i$ it is the
    same argument by symmetry), that is, we consider
    $\mathbb{W}_{n}^{(x,y)}(u_{\set{n}};\gamma z_1,z_2,...,z_n)$.
    According to Def.~\ref{def:xydualmultidiff} the dual
    $\omega^{\vee(x,y)}_{g,n}$ are obtained from
    $(-1)^n\mathbb{W}_{n}^{(x,y)}(u_{\set{n}};\gamma z_1,z_2,...,z_n)
    dx(\gamma z_1)\prod_{i=2}^n dx_i$ by turning $u_i$ into the
    differential operator $-d_{z_i}\frac{1}{dy_i}$. We see from
    (\ref{eq:def-bW}) that each power of $u_1$ in 
$\mathbb{W}_{n}^{(x,y)}(u_{\set{n}};\gamma z_1,z_2,...,z_n)
    dx(\gamma z_1)\prod_{i=2}^n dx_i$
    comes with a
    factor $\frac{1}{\chi_\gamma}$ from the transformation of
    $x_1^\gamma=x(\gamma z_1)=\chi_\gamma x(z_1)=\chi_\gamma x_1$. Let
    us now collect all contributions in the definition
    \eqref{eq:def-bW} of the extended multi-differential with
    $z_1\to \gamma z_1$: From each edge $e \in I(\flat)$ with $v(e)=1$ 
    in
    ${:}\omega_{|\flat|}^{(x,y)}{:}(z_{I(\flat)})$
    (also for $y_1dx_1$ in the exponent), we get
    a factor of $\beta_\gamma$, which comes additionally with
    $(\frac{u_1}{\chi_\gamma})^{1+2h}$, where $2h$ comes from the
    expansion of $\cS$. Collecting all those, we get
    in $\mathbb{W}_{n}^{(x,y)}(u_{\set{n}};\gamma z_1,z_2,...,z_n)
    dx(\gamma z_1)\prod_{i=2}^n dx_i$
    a product of the
    form
    \begin{align*}
      \frac{\chi_\gamma}{u_1}\prod
      \beta_\gamma \big(\frac{u_1}{\chi_\gamma}\big)^{1+2h_i}
    \end{align*}
    for each graph $\Gamma\in G_{\set{n}}$, where the product runs over edges $e$
    of $\Gamma$ with $v(e)=1$ and powers of $y_1dx_1$ from the exponent.
    The factor $\frac{\chi_\gamma}{u_1}$ in front comes from the very first $\frac{dx_1}{u_1}$ in \eqref{eq:def-bW} times $dx_1$.
    Finally, according to Def.~\ref{def:xydualmultidiff},
    to get the dual multi-differentials $\omega_{g,n}^{\vee(x,y)}$ 
    we have to substitute each $u_1$ by $\frac{1}{\upsilon_\gamma}$
    from the operator $d_{\gamma z_1} \frac{1}{dy(\gamma z_1)}$.
This  ultimately gives a contribution
    \begin{align*}
        \chi_\gamma\upsilon_\gamma\prod \beta_\gamma \big(\frac{1}{\chi_\gamma \upsilon_\gamma}\big)^{1+2h_i}=\chi_\gamma\upsilon_\gamma\prod \beta_\gamma \big(\frac{1}{\beta_\gamma}\big)^{1+2h_i}=\chi_\gamma\upsilon_\gamma\prod1=\beta_\gamma,
    \end{align*}
    where we have used the imposed conditions $\beta^2=1$ and $\chi_\gamma \upsilon_\gamma=\beta_\gamma$. This finishes the proof.
    \end{proof}
\end{lemma}

Using the linear and quadratic loop equation of the form of Def.+Lemma~\ref{deflemma:linearquadraticnew}, the regularity of $\omega_{g,n}^{\vee(x,y)}$ at the ramification points of $x$, the diagonal and anti-diagonal can be proved:
\begin{proposition}\label{prop:regularityofxydual}
    Let $\omega_{g,n}$ be generated by $G$-equivariant topological recursion on the spectral curve $(\Sigma,\phi, x,y,B, \chi, \upsilon, \beta)$. Then for $2g-2+n>0$, the $x$-$y$ dual multi-differentials $\omega_{g,n}^{\vee(x,y)}$, defined by Def.~\ref{def:xydualmultidiff}, are regular at the ramification points of $x$, the diagonal and the anti-diagonals $z_i=\gamma z_j$ for all $\gamma \in G$. If $y$ has only simple ramification points and $G$ acts freely on their locus, then $\omega_{g,n}^{\vee(x,y)}$ satisfies the linear and quadratic loop equations with respect to $y$.
    \begin{proof}
The cancellation of the poles at the ramification points of $x$ follows one-to-one from \cite[Prop.~5.8]{Alexandrov:2022ydc}, using the fact that the multi-differentials generated by equivariant topological recursion also satisfy the linear and quadratic loop equations with respect to $x$ by Lemma~\ref{lemma:linearquadraticloop}. This is a local property, and replacing the Bergman kernel used in \textit{op.~cit.} by the equivariant bi-differential $B$ does not affect the argument. It uses furthermore the parametrised universal $x$-$y$ duality (Thm.~\ref{thm:parametrizedxyduality}) and further intermediate steps of it.  We therefore omit the technical details of the proof. For the cancellation of the poles at the ramification points of $x$ and at the diagonal $z_i=z_j$, the group $G$ does not necessarily have to act freely on the ramification locus of $y$.

As $\omega_{g,n}^{\vee(x,y)}$ is for $2g-2+n>0$ regular at the
diagonal $z_i=z_j$ and transforms by
Lemma~\ref{lem:xydualomegatransformation}, we conclude that
\begin{align}
  \omega_{g,n}^{\vee(x,y)}(\gamma z_1,z_2,...)
  =\beta_\gamma \omega_{g,n}^{\vee(x,y)}( z_1,z_2,...)
\end{align}
is regular at $\gamma z_1=z_j$ for all $\gamma \in G$ and $j=2,...,n$
due to the regularity of $\omega_{g,n}^{\vee(x,y)}(z_1,z_2,...)$ at
the diagonal.  By symmetry, this proves regularity at any
anti-diagonal $z_i=\gamma z_j$ of $\omega_{g,n}^{\vee(x,y)}$ for
$2g-2+n>0$.

Next we prove that the $x$-$y$ dual multi-differentials $\omega_{g,n}^{\vee(x,y)}$ satisfy the linear and quadratic loop equations with respect to $y$
if the group acts additionally freely on the ramification locus of $y$.
Iterating inductively through
 \begin{align*}
          \omega_{g,n}=\omega^{(g)(x,y)}_{n,0}\rightarrow
          \omega^{(g)(x,y)}_{n-1,1}\rightarrow\dots\rightarrow
          \omega^{(g)(x,y)}_{1,n-1}\rightarrow \omega^{(g)(x,y)}_{0,n}
          =\omega_{g,n}^{\vee(x,y)}
   \end{align*}
as described in Sec.~\ref{sec.mixedcorrelators}, iteratively cancels the poles at the ramification points of $x$ and produces poles at the ramification points of $y$. This means that $\omega^{(g)(x,y)}_{1,n-1}(z;z_{\llbracket n-1 \rrbracket})$ is given by the $x$-$y$ dual version of \eqref{mixed1omegagn} by the universal $x$-$y$ duality, Theorem~\ref{thm:xyduality}. Thus, $\omega^{(g)(x,y)}_{1,n-1}(z;z_{\llbracket n-1 \rrbracket})$ still has poles in $z$ at the ramification points of $x$, but is regular at the ramification points of $y$. Furthermore, from the representation \eqref{mixed1omegagn}, together with the ($x$-$y$ dual) expansion \eqref{Wfirstexpansion} and Definition and Lemma~\ref{deflemma:linearquadraticnew}, we conclude that
        \begin{gather*}
        \omega_{g,n}^{\vee(x,y)}\in \Xi^y
        \qquad \text{and}
        \\
          \frac{1}{dy}
          \bigg(\omega^{\vee(x,y)}_{g-1,n+2}(z,z,z_{\llbracket n \rrbracket})
          +\sum_{\substack{J_1\sqcup J_2=\llbracket n \rrbracket\\g_1+g_2=g}}^{\text{ordered}}
          \omega^{\vee(x,y)}_{g_1,|J_1|+1}(z,z_{J_1})
          \omega^{\vee(x,y)}_{g_2,|J_2|+1}(z,z_{J_2}) \bigg)\in \Xi^y,
        \end{gather*}
by induction. In the definition of the space $\Xi^y$, it is important that the span consists of differentials of the form $\bigl(d\frac{1}{dy}\bigr)^k\alpha$, where $\alpha$ is holomorphic at the ramification points of $y$. However, if $G$ does not act freely on the ramification locus of $y$, then a ramification point may coincide with a pole of $B$ on the diagonal (that is, with a fixed point satisfying $\gamma z=z$). By the definition of $\omega_{g,n}^{\vee(x,y)}$, this would produce multi-differentials of the above form for which an $\alpha'$ is no longer holomorphic. Therefore, if we assume that the group action of $G$ is free on the ramification locus of $y$, then this problem does not occur, and $\alpha$ remains holomorphic in a neighbourhood of the ramification locus of $y$.
    \end{proof}
\end{proposition}

It follows from the fact that the $x$-$y$ dual multi-differentials $\omega_{g,n}^{\vee(x,y)}$ satisfy the linear and quadratic loop equations with respect to $y$ that their pole structure at the ramification points of $y$ (under the assumption that $G$ acts freely on the ramification locus of $y$) is given by the usual TR-like recursive formula. More precisely:
 
\begin{corollary}\label{cor:polarpartxydualside}
Let $\omega_{g,n}$ be generated by $G$-equivariant topological recursion on the spectral curve $(\Sigma,\phi, x,y,B, \chi, \upsilon, \beta)$ with a free action of $G$ on the ramification loci of $x$ and $y$. Let $B^{TR}$ be the standard Bergman kernel used in ordinary topological recursion. Then the polar structure of the $x$-$y$ dual multi-differentials $\omega_{g,n}^{\vee(x,y)}$ at the ramification points of $y$, denoted by $(a_i^\vee)_i$, is given by the $x$-$y$ dual topological recursion, that is,
\begin{align*}
      &\Res\displaylimits_{z = a^\vee_i}
      \int^z_{a_i^\vee}B^{TR}(z_1,\bullet)
      \omega_{g,n}^{\vee(x,y)}(z,z_{\llbracket n-1 \rrbracket})\\
      &=\Res\displaylimits_{z = a^\vee_i}\frac{\frac{1}{2}
        \int_{ \sigma_i^\vee(z)}^z B^{TR}(z_1,\bullet)}{
        \bigl( x(z) - x(\sigma^\vee_i(z)) \bigr) d y(z)}\bigg(
      \omega^{\vee(x,y)}_{g-1,n+1}
      (z,\sigma^\vee_i(z),z_2,\dots,z_n)
      \\
& \qquad \qquad \qquad \qquad \qquad \qquad  +
\sum_{\substack{g_1+g_2 = g \\ J_1 \sqcup J_2 = z_{\set {n-1}}\\
  (g_i,J_i)\neq (0,\emptyset)}}^{\text{ordered}}
	\omega^{\vee(x,y)}_{g_1,1+|J_1|}(z,z_{J_1})
\omega^{\vee(x,y)}_{g_2,1+|J_2|}(\sigma^\vee_i(z),z_{J_2})\bigg).
\end{align*}
Since $\omega_{g,n}^{\vee(x,y)}(z,z_{\set{n-1}})$ does not necessarily satisfy the projection property for non-trivial $G$, due to a possible presence of additional poles, it naturally falls into the framework of blobbed topological recursion; see Def.~\ref{def:blobb}.
\end{corollary}

The corollary provides the recursive structure  at the ramification points of $y$ of the $x$-$y$ dual multi-differentials $\omega_{g,n}^{\vee(x,y)}$ associated with a system of multi-differentials generated by equivariant topological recursion for a given spectral curve $(\Sigma,\phi, x,y,B, \chi, \upsilon, \beta)$. However, the corollary does not provide a recursive procedure to compute $\omega_{g,n}^{\vee(x,y)}$ solely from the knowledge of $\omega_{g',n'}^{\vee(x,y)}$ with $2g-2+n>2g'-2+n'$. The pole structure at the remaining poles at the group action fixed points has not yet been characterised. 

Thus, we want to finish this subsection by finally giving a full recursive procedure to compute the dual family $\omega_{g,n}^{\vee(x,y)}$ just in terms of $\omega_{g',n'}^{\vee(x,y)}$ with $2g-2+n>2g'-2+n'$. This formula is of the form of Gen-TR \cite{Alexandrov:2024tjo} at all poles (ramification points of $y$ and fixed points of the group action). We want to emphasise that it also includes the case where a ramification point of $y$ coincides with a fixed point (the action of $G$ is not free on the ramification points of $y$). However, if this happens, the formula of Prop.~\ref{prop:GenTRxydual} is not equivalent to Cor.~\ref{cor:polarpartxydualside} at the ramification points of $y$, since the linear and quadratic loop equations are not satisfied with a similar argument as in the proof of Prop.~\ref{prop:regularityofxydual}. The precise recursive formula is the following:

\begin{proposition}\label{prop:GenTRxydual}
    Let $\omega_{g,n}$ be generated by $G$-equivariant topological recursion on the spectral curve $(\Sigma,\phi, x,y,B, \chi, \upsilon, \beta)$ (this includes a free action of $G$ on the ramification locus of $x$ but not necessarily on the ramification locus of $y$). Let $B^{TR}$ be the standard Bergman kernel used in ordinary topological recursion. Then, the $x$-$y$ dual multi-differentials $\omega_{g,n}^{\vee(x,y)}$ are derived recursively by 
    \begin{align}\label{eq:formulaGenTRxydual}
      &\omega_{g,n}^{\vee(x,y)}(z_{\llbracket n \rrbracket})
      \\
      &=- \!\!\!\!\! \sum_{q\in \{a_i^\vee\}\cup \{z^\star\}} \!\! \Res\displaylimits_{z = q}
      \int^z_{q} \!\!\! B^{TR}(z_1,\bullet)
      \sum_{r\geq 1}
      \Big({-}d_z\frac{1}{dx(z)}\Big)^{r}[w^{r}]
      \mathcal{W}_{g,n}^{\vee(x,y)}(w;z;z_{2},...,z_n)dy(z),
      \nonumber
    \end{align}
where the 1-variable extended $n$-differential
\[
\mathcal{W}_{g,n}^{\vee(x,y)}(w;z;z_{2},...,z_n)
:=[w_2^0\cdots w_n^0]\mathbb{W}^{\vee(x,y)}_{g,n}(w,w_2,...,w_n;z,z_2,...,z_n)
\prod_{i=2}^n dy_i
\]
is defined analogously to Def.~\ref{def:1variableextended}
by just replacing
$\omega_{g,n}$ by $\omega_{g,n}^{\vee(x,y)}$, and swapping $x$ and $y$.
Equivalently, it is obtained from Cor.~\ref{cor:1variableextended} by the same
replacements. 
The set $\{a_i^\vee\}$ is the set of ramification points of $y$ and $\{z^\star\}$ is the set of fixed points of the group action, that is, $z^\star=\gamma z^\star$ for some $\gamma \in G$ with $\gamma \neq 1$. Note that the sets $\{a_i^\vee\}$ and $\{z^\star\}$ may have a non-trivial intersection.
    
    \begin{proof}
        The idea of the proof essentially follows the original idea of the definition of Gen-TR in \cite{Alexandrov:2024tjo} with some additional small adaptations. Let us first note that by Prop.~\ref{prop:regularityofxydual} the only poles of $\omega_{g,n}^{\vee(x,y)}$ lie in $\{a_i^\vee\}\cup \{z^\star\}$ by the construction of the $x$-$y$ dual family. Furthermore, from the definition of mixed  multi-differentials in Sec.~\ref{sec.mixedcorrelators} together with Prop.~\ref{prop:regularityofxydual}, we conclude that $\omega^{(g)(x,y)}_{n-1,1}(z_1,...,z_{n-1};z)$ can be expressed in terms of $\omega_{g',n'}(z_I)$ such that one can conclude that $\omega^{(g)(x,y)}_{n-1,1}(z_1,...,z_{n-1};z)$ has no poles in $z_1,...,z_{n-1}$ at $\{a_i^\vee\}\cup \{z^\star\}$. This argument can be iterated to conclude that actually any mixed multi-differential $\omega^{(g)(x,y)}_{n,m}(z_1,...,z_n;z_{n+1},...,z_{n+m})$ has no poles in the first $n$ variables $z_1,...,z_n$ in $\{a_i^\vee\}\cup \{z^\star\}$, in particular, the multi-differential $\omega^{(g)(x,y)}_{1,n-1}(z_1;z_{2},...,z_{n})$ in $z_1$. Using, thus, the $x$-$y$ dual formula \eqref{mixed1omegagn} between $\omega^{(g)(x,y)}_{1,n-1}(z_1;z_{2},...,z_{n})$ and $\omega^{(g)(x,y)}_{0,n}(z_1,z_{2},...,z_{n})=\omega_{g,n}^{\vee(x,y)}(z_1,z_{2},...,z_{n})$ together with the projection operator to the principal part of the Laurent series at the poles $\{a_i^\vee\}\cup \{z^\star\}$, we finish the proof.
    \end{proof}
\end{proposition}
\begin{remark}
    If $G$ acts freely on the ramification locus of $y$, the ramification points of $y$ are distinct from the fixed points of the group action. Thus, Cor.~\ref{cor:polarpartxydualside} applies at the ramification points of $y$, while Prop.~\ref{prop:GenTRxydual} provides the polar part at the fixed points $z^\star$ via \eqref{eq:formulaGenTRxydual}. Together, these two formulae provide precisely the structure of blobbed topological recursion in the sense of Def.~\ref{def:blobb}, where the blobs are given by the explicit Gen-TR recursive formula projected onto the fixed points of the group action.
\end{remark}

\subsection{Families of $x$-$y$ dual equivariant topological recursion multi-differentials}\label{sec:xtydualofequivariant}

In the previous subsection, we defined the $x$-$y$ dual
multi-differentials $\omega_{g,n}^{\vee(x,y)}$ associated with the
multi-differentials generated by equivariant topological recursion for
the spectral curve $(\Sigma,\phi,x,y,B,\chi,\upsilon,\beta)$, using
the distinguished pair $(x,y)$ from the spectral curve. The definition
of the dual, however, allows the freedom to choose a pair
$(\tilde{x},\tilde{y})$, as discussed in Sec.~\ref{sec:xyduality}.
We now relax the canonical choice and consider
dual multi-differentials $\omega^{\vee(\tilde{x},\tilde{y})}_{g,n}$ for
other choices  than $(\tilde{x},\tilde{y})=(x,y)$.

As mentioned in Rem.~\ref{rmk:symplecticshiftofequivariantTR}, the function $y$ may be shifted by a rational function of $x$, that is,
\begin{align*}
    \tilde{y}=y+\psi(x).
\end{align*}
Replacing $y$ by $\tilde{y}$ in the recursion formula \eqref{eqn:TR} generates exactly the same multi-differentials $\omega_{g,n}$. However, now $\tilde{y}$ does no longer necessarily satisfy condition~\eqref{conditionequivariantTR}. We may therefore ask what the corresponding $x$-$\tilde{y}$ dual family
$\omega_{g,n}^{\vee(x,\tilde{y})}$ of multi-differentials is,
constructed from the same family $\omega_{g,n}$ together with
$(x,\tilde{y}=y+\psi(x))$.

Let us first observe that the linear and quadratic loop equations of Lemma~\ref{lemma:linearquadraticloop} are not affected by replacing $y$ with $\tilde{y}$, since the multi-differentials $\omega_{g,n}$ remain unchanged. Indeed, the only place where $\tilde{y}$ enters is through the quadratic loop equation,
\begin{align*}
    \tilde{y}(z)dx(z)\omega_{g,n+1}(\sigma_i(z),z_{\llbracket n \rrbracket})
    +
    \tilde{y}(\sigma_i(z))dx(\sigma_i(z))
    \omega_{g,n+1}(z,z_{\llbracket n \rrbracket}),
\end{align*}
whose behaviour near the ramification point $z\to a_i$ is unchanged because of the linear loop equation. Consequently, the linear and quadratic loop equations remain valid for the family $\omega_{g,n}$ with respect to $(x,\tilde{y})$. It follows immediately that Definition and Lemma~\ref{deflemma:linearquadraticnew} also remains valid. Hence, Prop.~\ref{prop:regularityofxydual} extends to the $x$-$\tilde{y}$ dual family except for the regularity at the anti-diagonal, where the equivariant transformation of $y$ was used, which now fails.

\begin{corollary}\label{cor:symplecticxyregularity}
    Let $\omega_{g,n}$ be generated by $G$-equivariant topological recursion on the spectral curve $(\Sigma,\phi,x,y,B,\chi,\upsilon,\beta)$. Let $\tilde{y}=y+\psi(x)$, and assume that the ramification points of $\tilde{y}$ do not coincide with the fixed points of the group action $\gamma z=z$ for any $\gamma\in G$, $\gamma\neq 1$, or with the ramification points of $x$. Then the $x$-$\tilde{y}$ dual multi-differentials $\omega_{g,n}^{\vee(x,\tilde{y})}$, defined by Def.~\ref{def:xydualmultidiff}, are regular at the ramification points of $x$ and on the diagonal. Furthermore, $\omega_{g,n}^{\vee(x,\tilde{y})}$ satisfy the linear and quadratic loop equations with respect to $\tilde{y}$.
\end{corollary}

Such families of dual multi-differentials $\omega_{g,n}^{\vee(x,y)}\mapsto\omega_{g,n}^{\vee(x,\tilde{y})}$ have already been studied in the context of ordinary topological recursion. The two families $\omega_{g,n}^{\vee(x,y)}$ and $\omega_{g,n}^{\vee(x,\tilde{y})}$ are related by the so-called symplectic duality \cite{Bychkov:2022wgw}. Consequently, under the assumptions of Cor.~\ref{cor:symplecticxyregularity}, the polar part of the $x$-$\tilde{y}$ dual multi-differentials is again given by the topological recursion-like formula of Cor.~\ref{cor:polarpartxydualside}, namely
\begin{align}\label{xtydualpolarpart}
  &\Res\displaylimits_{z = \tilde{a}^\vee_i}
  \int^z_{\tilde{a}_i^\vee}B^{TR}(z_1,\bullet)
  \omega_{g,n}^{\vee(x,\tilde{y})}(z,z_{\llbracket n \rrbracket\setminus \{1\}} )
  \\\nonumber
  &=\Res\displaylimits_{z = \tilde{a}^\vee_i}\frac{\frac{1}{2}
    \int_{ \tilde{\sigma}_i^\vee(z)}^z B^{TR}(z_1,\bullet)}{\bigl( x(z)
    - x(\tilde{\sigma}_i^\vee(z)) \bigr) d \tilde{y}(z)}\bigg(
\omega^{\vee(x,\tilde{y})}_{g-1,n+1}(z,\tilde{\sigma}_i^\vee(z),z_2,\dots,z_n) \\\nonumber
& \qquad \qquad \qquad \qquad \qquad  +
      \sum_{\substack{g_1+g_2 = g \\ J_1 \sqcup J_2 = z_{\set{n} \setminus \{1\}}
           \\(g_i,J_i)\neq (0,\emptyset)}}^{\text{ordered}}
\omega^{\vee(x,\tilde{y})}_{g_1,1+|J_1|}(z,z_{J_1})
	\omega^{\vee(x,\tilde{y})}_{g_2,1+|J_2|}(\tilde{\sigma}_i^\vee(z),z_{J_2})\bigg).
\end{align}
Here, $\tilde{a}_i^\vee$ denotes a ramification point of $\tilde{y}$, and $\tilde{\sigma}_i^\vee$ is the corresponding local Galois involution.

At first sight, this construction of whole families of $x$-$\tilde{y}$ dual multi-differentials may appear to be merely a mathematical curiosity. However, it is expected to have interesting applications. Indeed, in the original definition of equivariant topological recursion (Def.~\ref{defn:equiv:SC}), the representation $\upsilon$ is allowed to be the trivial representation, implying that $y(\gamma z)=y(z)$. Consequently, every fixed point of the group action satisfying $\gamma z=z$ is automatically a ramification point of $y$. In this situation, for instance parts of Prop.~\ref{prop:regularityofxydual} and Cor.~\ref{cor:polarpartxydualside} no longer apply. However, on the other hand Prop.~\ref{prop:GenTRxydual} remains valid.  Nevertheless, if the shifted function $\tilde{y}=y+\psi(x)$ does not have ramification points that do coincide with the fixed points of the group action, Cor.~\ref{cor:symplecticxyregularity} and equation~\eqref{xtydualpolarpart} remain valid, even though parts of Prop.~\ref{prop:regularityofxydual} and Cor.~\ref{cor:polarpartxydualside} fail for the original function $y$.

Now, we want to mention an important difference between the two $x$-$y$ dual families $\omega_{g,n}^{\vee(x,y)}$ and $\omega_{g,n}^{\vee(x,\tilde{y})}$ regarding their transformation and pole structure. First, Lemma~\ref{lem:xydualomegatransformation} does not apply to $\omega_{g,n}^{\vee(x,\tilde{y})}$, that is, the multi-differentials $\omega_{g,n}^{\vee(x,\tilde{y})}$ do not transform in any precise way under the group action. This clearly follows from the fact that $\tilde{y}$ transforms arbitrarily. From this observation, the pole structure of $\omega_{g,n}^{\vee(x,\tilde{y})}$ becomes more complicated:

\begin{corollary}\label{cor:polesxtydual}
    Let $\omega_{g,n}$ be generated by $G$-equivariant topological recursion on the spectral curve $(\Sigma,\phi,x,y,B,\chi,\upsilon,\beta)$. Let $\tilde{y}=y+\psi(x)$, and assume that the ramification points of $\tilde{y}$ do not coincide with the fixed points of the group action $\gamma z=z$ for any $\gamma\in G$, $\gamma\neq 1$, or with the ramification points of $x$. Then the $x$-$\tilde{y}$ dual multi-differentials $\omega_{g,n}^{\vee(x,\tilde{y})}$, defined by Def.~\ref{def:xydualmultidiff}, admit poles only at the ramification points of $\tilde y$, at the fixed points $\gamma z^\star=z^\star$ with $\gamma\neq 1$, and at the anti-diagonal $\gamma z_i=z_j$.
    \begin{proof}
        The only difference consists of the poles at the anti-diagonal, which directly follows from the fact that there is no canonical transformation of $\omega_{g,n}^{\vee(x,\tilde{y})}$ under the group action, which in Prop.~\ref{prop:regularityofxydual} actually cancelled the poles on the anti-diagonal. However, due to Cor.~\ref{cor:symplecticxyregularity}, the pole on the diagonal is still absent, though.
    \end{proof}
\end{corollary}

Let us finish the subsection on the family of $x$-$y$ dual multi-differentials $\omega_{g,n}^{\vee(x,\tilde{y})}$ with a kind of negative structural result concerning the poles at the anti-diagonal $\gamma z_i=z_j$. We have seen in Prop.~\ref{prop:GenTRxydual} that for $\omega_{g,n}^{\vee(x,y)}$ a recursive definition exists which is of the form of Gen-TR \cite{Alexandrov:2024tjo}, even at the poles at the fixed points of the group action. However, a natural question is whether this is also true for the newly arising poles at the anti-diagonal $\gamma z_i=z_j$ for $\gamma \neq 1$ of the multi-differentials $\omega_{g,n}^{\vee(x,\tilde{y})}$. It turns out that this is not the case:

\begin{proposition}\label{prop:notGenTratantidiagonal}
    The polar structure of $\omega_{g,n}^{\vee(x,\tilde{y})}$ at the anti-diagonal, that is, for $\gamma z_i=z_j$, can in general not be computed by the recursive formula of Gen-TR.
    \begin{proof}
      The idea is very simple. The definition of Gen-TR relies on the
      $x$-$y$ duality formula in one variable, as described in
      Sec.~\ref{sec.mixedcorrelators}. However, let us consider this
      formula in the following form. Let
      $\omega^{(g)(x,\tilde{y})}_{n,0}=\omega_{g,n}$ be generated by
      equivariant TR, and let
      $\omega^{(g)(x,\tilde{y})}_{0,n}=\omega_{g,n}^{\vee(x,\tilde{y})}$ be
      its $x$-$y$ dual associated to $(x,\tilde{y})$. The Gen-TR
      formula is derived from the fact that
      $\omega^{(g)(x,\tilde{y})}_{1,n-1}(z_1;z_2,...,z_n)$, as described in
      Sec.~\ref{sec.mixedcorrelators}, does not have a pole in $z_1$
      at the poles of $\omega_{g,n}^{\vee(x,\tilde{y})}(z_1,...,z_n)$
      in $z_1$. By Cor.~\ref{cor:polesxtydual}, the poles of
      $\omega_{g,n}^{\vee(x,\tilde{y})}(z_1,...,z_n)$ in $z_1$ now
      also occur at the anti-diagonal $z_1=\gamma z_i$. However, it
      turns out that $\omega^{(g)(x,\tilde{y})}_{1,n-1}(z_1;z_2,...,z_n)$ also has
      poles at the anti-diagonal. An explicit computation of
      $\omega^{(0)(x,\tilde{y})}_{1,2}(z_1;z_2,z_3)$ already shows that. Thus, the
      projection onto those poles is not trivial and the argument
      fails. The non-zero contribution from $\omega^{(g)(x,\tilde{y})}_{1,n-1}$ must
      be
      retained. 
    \end{proof}
\end{proposition}
\begin{remark}
    Prop.~\ref{prop:notGenTratantidiagonal} does not state that it is
    impossible to derive the recursive pole structure. It just states
    that, in general, it is not given by the Gen-TR
    formula. Since $\omega_{g,n}=\omega^{(g)(x,\tilde{y})}_{n,0}$ does not have
    poles along the diagonal or anti-diagonal for $2g-2+n>0$, one
    can still observe from the mixed correlators that, for $n\geq 2$,
    $\omega^{(g)(x,\tilde{y})}_{2,n-2}(z_1,z_2;z_3,\ldots,z_n)$ does not have
    poles at $z_1=\gamma z_2$, with
    $\gamma\in G\setminus\{1\}$. Therefore, another recursive formula
    can in principle be derived by swapping two variables and then
    projecting, which is generally simpler than swapping all variables
    but more complicated than just swapping one variable to the other
    side, as in Gen-TR. In Sec.~\ref{sec:spectralcurveZ2}, another
    trick will be
    used to derive a recursive formula at the anti-diagonal,
    especially constructed for the symmetry group $G=\mathbb{Z}/2\mathbb{Z}$.
\end{remark}

\section{Fock space formulation of the Kontsevich matrix model with arbitrary potential}

\label{sec:Kontsevich}

\subsection{Matrix models of Kontsevich type}

Let $H_N$ be the space of self-adjoint $N\times N$-matrices
and $E=\mathrm{diag}(E_{\set{N}})$ a diagonal positive matrix with pairwise
different eigenvalues $\{E_q\}$. We consider the partition function
\begin{align}
Z=Z_P(E_{\set{N}})=\int_{H_N} d\Phi  \; e^{-N\,\mathrm{Tr}(E\Phi^2+P(\Phi))} ,
\label{Z-KontsevichV}
\end{align}
where $d\Phi$ is the Lebesgue measure and $P(\Phi)$ a polynomial with at
least third-order zero at $\Phi=0$. For $P(\Phi)=\frac{\mathrm{i}}{3}\Phi^3$,
(\ref{Z-KontsevichV}) is the partition function of the
Kontsevich matrix model. Kontsevich proved in \cite{Kontsevich:1992ti}
that 
\begin{align}
  \tau_K(\mathbf{t}):= \log\Big(\frac{Z_{\mathrm{i}\Phi^3/3}(E_{\set{N}})}{Z_0(E_{\set{N}})}\Big)\Big|_{t_{2k+1}=-(2k-1)!! \sum_{q=1}^NE_q^{-2k-1}}
\end{align}
is the generating series
for intersection numbers of $\psi$-classes on the moduli space of
stable complex curves. Moreover, $\tau_K(\mathbf{t})$ is a tau function
of the KdV hierarchy, which proved a conjecture due to Witten
\cite{Witten:1990hr}.
In the last decade considerable progress
for the case $P(\Phi)=\frac{\lambda}{4}\Phi^4$, the
quartic Kontsevich model, was achieved
\cite{Panzer:2018tvy,
Grosse:2019jnv, Schurmann:2019mzu, Branahl:2020yru, Hock:2021tbl,
  Hock:2023nki, hock2025symmetry}, but not a complete
understanding. Up to an open Question~\ref{mainquestion},
this paper completes the analysis.
In \cite{Borot:2023thu} it was shown that for an additional deformation
\[
  P_{\mathbf{t}}(\Phi)=P(\Phi)+\sum_{k=0}^\infty t_{2k+1}  \Phi^{2k+1}
\]
by an odd formal power series, 
$\tau_B(\mathbf{t})=\log(\frac{Z_{P_{\mathbf{t}}}(E_{\set{N}})}{Z_0(E_{\set{N}})})$
is a tau function of the BKP integrable hierarchy. Otherwise, to the best or our knowledge, Kontsevich matrix models with a potential $P$ of degree
$\geq 5$ have not been studied so far.

The following definition was given in 
\cite{Branahl:2020yru} for a quartic potential 
$P(\Phi)=\frac{\lambda}{4}\Phi^4$. It generalises to any $P$:
\begin{definition}
  \label{def:Omega}
  For the partition function \eqref{Z-KontsevichV} of the Kontsevich
  matrix model with arbitrary potential $P$, consider the derivatives
\begin{align}
  W_n(E_{q_1},...E_{q_n}):=(-N)^{n-2} \frac{\partial^n \log Z_P(E_{\set{N}})}{
    \partial E_{q_1}\cdots
    \partial E_{q_n}} +\frac{\delta_{n,2}}{(E_{q_1}-E_{q_2})^2}
\label{Omega-def}
\end{align}
with respect to pairwise different eigenvalues of  $E$. 
Extend $W_n(E_{q_1},...E_{q_m})$ to functions
$W_n(Y_1,...,Y_n)$ defined in a neighbourhood of $E_{\set{N}}$.
Then by
\begin{align}
\varpi^\vee_n(Y_1,...,Y_n)
&=W_n(Y_1,...,Y_n)dY_1\cdots dY_n\\
&=
\sum_{g=0}^\infty N^{-2g}\varpi^\vee_{g,n}(Y_1,...,Y_n)\nonumber
\end{align}
a family of multi-differentials $\{\varpi^\vee_{g,n}\}$ is defined.
\end{definition}
\begin{example}
\label{ex:Gaussian}
  For $P=0$ the Gau\ss{}ian integral evaluates to
\begin{align}
  Z_0(E_{\set{N}})=\int_{H_N} d\Phi  \; e^{-N\,\mathrm{Tr}(E\Phi^2))}
  =\frac{2^{N/2}(\pi/N)^{N^2/2}}{\prod_{1\leq k,l\leq N} \sqrt{E_k+E_l}} .
\end{align}
This gives 
\[
  W_1(E_{q_1})
  = \frac{1}{N} \sum_{k=1}^N \frac{1}{E_{q_1}+E_k},\qquad
  W_2(E_{q_1},E_{q_2})
  = \frac{1}{(E_{q_1}+E_{q_2})^2}
  +\frac{1}{(E_{q_1}-E_{q_2})^2},
\]
which directly extends to
\begin{align}
\varpi_1^\vee(Y_1)
  &= \frac{1}{N} \sum_{k=1}^N \frac{dY_1}{Y_1+E_k},  &
  \varpi_2^\vee(Y_1,Y_2)
  = \frac{dY_1dY_2}{(Y_1+Y_2)^2}
  +\frac{dY_1dY_2}{(Y_1-Y_2)^2}.
\end{align}
All $\varpi_n^\vee$ with $n\geq 3$ vanish identically.
Note that for $Y_i=z_i$, $\varpi_2^\vee(z_1,z_2)
=B(z_1,z_2)$ is the symmetric bi-differential \eqref{extendedBergman} of
$\mathbb{Z}/2\mathbb{Z}$ equivariant TR.
\end{example}
\noindent The extension from
$W_n(E_{q_1},...E_{q_n})$ to $W_n(Y_1,...,Y_n)$ is always possible
as formal power series in $P$ about the Gau\ss{}ian 
Example~\ref{ex:Gaussian}.

\bigskip


Our goal is to show that the $\varpi^\vee_n$ fit naturally into the
framework of Alexandrov, Bychkov, Dunin-Barkowski, Kazarian and
Shadrin which is built around Fock space techniques, $x$-$y$ and
symplectic dualities and generalised topological recursion.  Notations
and conventions have been evolved over the years. Moreover, the evaluation
of the matrix integral (\ref{Z-KontsevichV})
passes via the Itzykson--Zuber--Harish-Chandra integral \cite{Harish-Chandra:1957, Itzykson:1992ya}. This
integral has a character expansion (see e.g.\
\cite [Sec.~1]{Alexandrov:2010th}) which introduces a certain operator
$\mathcal{D}$ on Fock space.  When translating the IZHC integral to
Fock space, the inverse $\mathcal{D}^{-1}$ appears, whereas the
standard formulation works with $\mathcal{D}$. The only effect is a
systematic change of signs. These relative signs and also the scale
$\lambda$ of the potential $P$ are important for us, in particular in
view of the application to the quartic Kontsevich model
\cite{Branahl:2020yru}. We therefore provide in two longer
Appendices~\ref{app:functionalrelations}+\ref{app:XYFock} a
detailed proof of the $x$-$y$ dualities in the conventions dictated by
the matrix integral (\ref{Z-KontsevichV}).

\subsection{Bosonic Fock space and Itzykson--Zuber--Harish-Chandra integral}

In the bosonic Fock space $\mathbb{C}[[p_1,p_2,p_3,...]]$ we introduce:
\begin{itemize}
\item the vacuum $|0\rangle=1$,
\item the covacuum $\langle 0|$ as the functional 
which  sets all $p_i=0$,
\item multiplication operators $J_{-k}:=p_k$
for $k\in \mathbb{Z}_{>0}$,
  
\item differentiation operators $J_k:=k\frac{\partial}{\partial p_k}$
  for $k\in \mathbb{Z}_{>0}$,

\item $J_0:= 0$. 
\end{itemize}
Consequently, the commutation relation $[J_k,J_l]=k \delta_{k+l,0}$ holds.

A basis in $\mathbb{C}[[p_1,p_2,p_3,...]]$ is given by the power-sum Schur
polynomials $\hat{s}_\lambda(\mathbf{p})$ labelled by partitions
$\lambda\vdash n$, i.e.\ weakly decreasing sequences
$\lambda=(\lambda_1\geq \lambda_2\geq ...\geq \lambda_\ell>0,0,0,...)$
with $\lambda_1+...+\lambda_\ell=n$. Then 
\begin{align}
\hat{s}_\lambda(\mathbf{p})= \det_{1\leq i,j\leq \ell(\lambda)}
\big(  h_{\lambda_i-i+j}   (\mathbf{p})\big)
\end{align}
(Jacobi-Trudi formula), where 
$h_k(\mathbf{p})=[z^k] \exp(\sum_{i=1}^\infty \frac{1}{i} z^i p_i)$.
Of particular importance is the Cauchy-Littlewood identity
\begin{align}
  e^{\sum_{k=1}^\infty \frac{1}{k} p_k q_k}
  =\sum_\lambda \hat{s}_\lambda(\mathbf{p})\hat{s}_\lambda(\mathbf{q}).
\label{Cauchy-Littlewood}
\end{align}
We have the following identity
\begin{align}
\langle 0|  \hat{s}_\nu(J_1,J_2,J_3,\dots)
  \hat{s}_\mu(J_{-1},J_{-2},J_{-3},\dots) |0\rangle 
  =\delta_{\mu,\nu} ,
\label{Schur-orthogonal}
\end{align}
which directly follows from (\ref{Cauchy-Littlewood}) after setting $\mathbf{p}=(J_{-1},J_{-2},J_{-3},\dots)=0$:
\begin{align*}
  &\sum_\nu \hat{s}_\nu(\mathbf{q})\hat{s}_\nu(J_1,J_2,J_3,\dots) \hat{s}_{\mu}(J_{-1},J_{-2},J_{-3},\dots)
  \\
    &\qquad = e^{\sum_{k\geq 1}\frac{q_k}{k}J_k} \hat{s}_\mu(J_{-1},J_{-2},J_{-3},\dots) = \hat{s}_\mu (\mathbf{p}+\mathbf{q}).
\end{align*}

A key ingredient of the $x$-$y$ duality is an operator $\mathcal{D}$ on
bosonic Fock space by the
following\footnote{We follow the sign convention of
\cite[eq. (13)]{Borot:2021thu}. In 
\cite[eq.\ (1.2)]{Alexandrov:2022ydc} and
\cite[eq. (2.11)]{Alexandrov:2023jcj}, 
a different sign $i-j$ compared with $j-i$ is chosen!}
diagonal action on the Schur basis:
\begin{align}
  \mathcal{D} \hat{s}_\lambda(\mathbf{p})
  =\prod_{(i,j)\in\lambda}(1+\hbar(j-i)) \hat{s}_\lambda(\mathbf{p}) ,
\label{D-Borot}
\end{align}
where the product is over the boxes of the Young diagram of the partition
$\lambda$, and $j-i$ is the content of the box in the $i$-th row
and $j$-th column.

\medskip


Let $dU$ be the Haar measure on $\mathcal{U}(N)$ normalised to
$\int_{\mathcal{U}(N)} dU=1$. Let $A=\mathrm{diag}(A_1,A_2,...,A_N)$ and
$B=\mathrm{diag}(B_1,B_2,...,B_N)$ be diagonal real matrices. Then
\begin{align}
  \int_{\mathcal{U}(N)} dU \;e^{N\, \mathrm{Tr} (UAU^\dag B)}
\label{IZHC-def}
\end{align}
is known as the Itzykson--Zuber--Harish-Chandra (IZHC) integral
\cite{Harish-Chandra:1957, Itzykson:1992ya}.
It is the special case $p_k=N\delta_{k,1}$ of 
\begin{align}
&  \int_{\mathcal{U}(N)} dU \;
  e^{\sum_{k=1}^\infty \frac{1}{k} p_k  \mathrm{Tr} (UAU^\dag B)^k}
  \nonumber
  \\
  &= \sum_\lambda \hat{s}_\lambda(\mathbf{p})
  \int_{\mathcal{U}(N)} dU \;
  \hat{s}_\lambda(\mathrm{Tr} (UAU^\dag B)^1,
  \mathrm{Tr} (UAU^\dag B)^2,\mathrm{Tr} (UAU^\dag B)^3,\dots)
\label{IZHC-moregeneral}
\end{align}
when taking the Cauchy-Littlewood identity (\ref{Cauchy-Littlewood}) into account. The
integral in the last line can be found in the literature
(e.g.\ \cite{Alexandrov:2010th}): 
\begin{align}
  \int_{\mathcal{U}(N)} dU \;
  \hat{s}_\lambda(\mathrm{Tr} (UAU^\dag B)^1,
  \mathrm{Tr} (UAU^\dag B)^2,\mathrm{Tr} (UAU^\dag B)^3,\dots)
  = \frac{\hat{s}_\lambda(\mathbf{a})\hat{s}_\lambda(\mathbf{b})}{
    \mathrm{dim}_\lambda}\,,
\end{align}
where $\mathbf{a}=(\mathrm{Tr}(A),\mathrm{Tr}(A^2),\mathrm{Tr}(A^3),\dots)$ and
similarly for $\mathbf{b}$, and $\mathrm{dim}_\lambda=\prod_{1\leq i<j\leq N}
\frac{\lambda_i-\lambda_j+j-i}{j-i}$. 
For the IZHC integral (\ref{IZHC-def}) we need to multiply with
$\hat{s}_\lambda(N\delta_{k,1})=N^{|\lambda|} \hat{s}_\lambda(\delta_{k,1})$,
so that (see e.g.\ \cite[eq.\ (6.1)+(6.2)]{Alexandrov:2015tau})
\begin{align}
  \int_{\mathcal{U}(N)} dU \;e^{N\, \mathrm{Tr} (UAU^\dag B)}
  &=\sum_\lambda N^{|\lambda|}\hat{s}_\lambda(\mathbf{a})
  \hat{s}_\lambda(\mathbf{b})
    \prod_i \frac{\Gamma(N-i+1)}{\Gamma(\lambda_i +N-i+1)}
    \nonumber
    \\
    &=
    \sum_\lambda \hat{s}_\lambda(\mathbf{a})
  \hat{s}_\lambda(\mathbf{b})
    \prod_{(i,j)\in \lambda} \frac{1}{1+ \frac{(j-i)}{N}} .
\label{IZHC}
  \end{align}
  Here the sum is over all partitions $\lambda$, and the product is
  over the positions $(i,j)$ of the boxes of the Young diagram of the
  partition $\lambda$. As before, $A,B$ are diagonal
  $N\times N$-matrices with
  $\mathbf{a}=(\mathrm{Tr}(A),\mathrm{Tr}(A^2),\mathrm{Tr}(A^3),\dots)$
  and
  $\mathbf{b}=(\mathrm{Tr}(B),\mathrm{Tr}(B^2),\mathrm{Tr}(B^3),\dots)$.

  \subsection{Fock space formulae for partition function
  and multi-differentials}

\label{sec:partitionfunction}

We return to (\ref{Z-KontsevichV}) and its derivatives (\ref{Omega-def}).
The change of variables $\Phi=U M U^\dag$ with
$M=\mathrm{diag}(m_1,...,m_N)$ and $U\in \mathcal{U}(N)$ has Jacobian 
\begin{align}
  d\Phi= C_N (\Delta(M))^2 dU \prod_{i=1}^N dm_i ,\qquad 
C_N= \frac{\pi^{\frac{N(N-1)}{2}}}{\prod_{k=1}^N k!} ,
\label{Jacobian}
\end{align}
where $\Delta(M)=\prod_{1\leq j<i\leq N} (m_i-m_j)$ is the
Vandermonde determinant. Inserting this transformation and the 
IZHC integral
(\ref{IZHC}) brings 
 (\ref{Z-KontsevichV})  into the form
\begin{align}
  &Z_P(E_{\set{N}})
  \label{Z-IZHC}
\\
  &=C_N  \int_{\mathbb{R}^N} \prod_i \Big(dm_i e^{-N \,P(m_i)} \Big)\;
(\Delta(M))^2  \int_{\mathcal{U}(N)} dU\;e^{-N\,\mathrm{Tr}(UM^2U^\dag E)}
\nonumber
\\
&= \sum_\lambda
\int_{\mathbb{R}^N} \prod_i \Big(dm_i e^{-N\,P(m_i)} \Big)
C_N (\Delta(M))^2
\hat{s}_\lambda (\mathbf{-\tilde{m}^2})
 \prod_{(i,j)\in \lambda} \frac{1}{1+ \frac{(j-i)}{N}}
\hat{s}_\lambda (\mathbf{\tilde{e}}),
\nonumber
\end{align}
where 
$\mathbf{\tilde{e}}=(\mathrm{Tr}(E),\mathrm{Tr}(E^2),\dots)$ and
$(\mathbf{-\tilde{m}^2})=(-\mathrm{Tr}(M^2),\mathrm{Tr}(M^4),-\mathrm{Tr}(M^6),\dots)$. We will later prefer normalised traces 
\begin{align}
    \mathbf{e}_k&:=\tfrac{1}{N}\mathrm{Tr}(E^k) , &
    \mathbf{m}_{-k}&:=\tfrac{(-1)^k}{N}\mathrm{Tr}(M^{2k}) .
     \label{eq:em}
\end{align}

In order to evaluate the derivatives (\ref{Omega-def}) with respect to
the diagonal entries $E_q$ of $E$ we pass to the bosonic Fock space
$\mathbb{C}[[p_1,p_2,..]]$. Let $\hbar=1/N$.
The Cauchy-Littlewood identity (\ref{Cauchy-Littlewood}) gives
\begin{align}
  e^{\sum_{k=1}^\infty \frac{1}{k \hbar} J_{-k}  
    \mathbf{e}_{k}}
  &=\sum_\lambda \hat{s}_\lambda (\mathbf{\tilde{e}})\hat{s}_\lambda
  (J_{-1},J_{-2},J_{-3},
  \dots)
   ,
  \nonumber
  \\
  e^{\sum_{k=1}^\infty \frac{1}{k\hbar } J_{k}  \mathbf{m}_{-k}}
  &=\sum_\nu \hat{s}_\nu(\mathbf{-\tilde{m}^2})
  \hat{s}_\nu(J_1,J_2,J_3,\dots) .
  \label{Cauchy-Fock}
\end{align}
We let the first line act on the vacuum $ |0\rangle$ and then
apply the inverse of $\mathcal{D}$ (given in (\ref{D-Borot});  understood
to act only
on $\mathbb{C}[[p_1,p_2,..]]$, not on $s_\mu(\mathbf{\tilde{e}})$):
\begin{align}
  \mathcal{D}^{-1}
  e^{\sum_{k=1}^\infty   \frac{1}{k\hbar } J_{-k}     \mathbf{e}_{k} }|0\rangle
  \Big|_{\hbar=1/N}
  =\sum_\lambda
  \hat{s}_\lambda(\mathbf{\tilde{e}})
\prod_{(i,j)\in \lambda} \frac{1}{1+ \frac{(j-i)}{N}}
  \hat{s}_\lambda(\mathbf{p}) .
\label{D-Cauchy}
\end{align}
This recovers the factor we need in (\ref{Z-IZHC}), traced back to
(\ref{IZHC}). On the rhs of (\ref{D-Cauchy}) we can also write
$  \hat{s}_\mu(\mathbf{p})=  \hat{s}_\mu(J_{-1},J_{-2},J_{-3},\dots)|0\rangle
$. We multiply 
with the second line of
(\ref{Cauchy-Fock}) and apply the functional $\langle0|$. 
Then the identity (\ref{Schur-orthogonal}) leads to
\begin{align}
Z_P(E_{\set{N}}) &=\Big\langle 0\Big|
  \int d_P\vec{m}\;
  e^{\sum_{l=1}^\infty \frac{1}{l\hbar} J_{l} \mathbf{m}_{-l}}
  \mathcal{D}^{-1}
e^{\sum_{k=1}^\infty \frac{1}{k\hbar} J_{-k} \mathbf{e}_{k}} \Big|0\Big\rangle ,
\label{Z-Fock}
\end{align}
with integration over $\mathbb{R}^n$, the given dependence
of $\mathbf{m}_{-k}$ on $\vec{m}$, and 
measure 
\begin{align}
  d_P \vec{m}:= \frac{\pi^{\frac{N(N-1)}{2}}}{\prod_{k=1}^N k!}
  \Big(\prod_{1\leq i\leq j \leq N} (m_j-m_i)^2\Big)
\Big(  \prod_{i=1}^N e^{-N\,P( m_i)} dm_i\Big) .
\label{eq:dPm}
\end{align}

\begin{remark}
\label{note-psi}
In our conventions, $\mathcal{D}^{-1}$ must appear in \eqref{Z-Fock} and
subsequent formulae. In the literature always $\mathcal{D}$ is placed
in the vacuum expectation value. Sometimes (e.g.\ \cite{Bychkov:2022wgw})
a more flexible
$\mathcal{D}_{\hat{\psi}}$ is considered which specifies to
$\mathcal{D}$ for $\hat{\psi}(\theta)=\log(1+\theta)$ and
$\mathcal{D}^{-1}$ for $\hat{\psi}(\theta)=-\log(1+\theta)$.
This changed sign in front of $\log(1+\theta)$ will later be relevant, see
Remark~\ref{remark:DXY}.
\end{remark}

\begin{remark}
 In vacuum expectation values as in \eqref{Z-Fock} it is always understood that 
 $e^{\sum_{l=1}^\infty \frac{1}{l\hbar} J_{l} \mathbf{m}_{-l}}$ is expanded by the Cauchy-Littlewood identity \eqref{Cauchy-Fock}. The integral 
 $\int d_P\vec{m} \,\hat{s}_{\lambda}(\mathbf{-\tilde{m}^2})$ is evaluated before the sum over partitions, see
 \eqref{Z-IZHC}. It converges e.g.\
 for $P$ a polynomial of even degree with positive leading coefficient. The convergence of the sum over partitions should follow from the bound 
 $e^{N \,\mathrm{Tr}(M^2) \max_i E_i}$
 of the IZHC integral \eqref{IZHC} and asymptotic estimates in \cite{Goulden:2011qn}.
\end{remark}

Using $\frac{\partial \mathbf{e}_l}{\partial E_q}=
\frac{1}{N} l E_q^{l-1}$ it is now easy to evaluate the differentiations
(\ref{Omega-def}):
\begin{align}
  &  W_n(E_{q_1},...,E_{q_n})
\label{Omega-Fock}
  \\
  &=(-N)^{n-2}
\Big\langle 0\Big|
\int d_P\vec{m}\;
e^{\sum_{l=1}^\infty \frac{1}{l\hbar } J_{l} \mathbf{m}_{-l}}
\mathcal{D}^{-1}
\prod_{i=1}^n\Big(
  \sum_{l_i=1}^\infty E_{q_i}^{l_i-1} J_{-l_i}
  \Big)
e^{\sum_{k=1}^\infty \frac{1}{k\hbar } J_{-k} \mathbf{e}_{k}}
 \Big|0\Big\rangle^\circ
\nonumber
\\
&+\frac{\delta_{n,2}}{(E_{q_1}-E_{q_2})^2}.
\nonumber
\end{align}
Here $\langle 0|\dots|0\rangle^\circ$ means to take the connected part,
defined as
\begin{align}
  &  \Big\langle 0\Big| \mathcal{O}_1 \mathcal{D}^{-1}
  \prod_{i=1}^n \mathcal{J}_i \mathcal{O}_2
  \Big| 0\Big\rangle^\circ
  \nonumber
  \\[-2ex]
  & =\sum_{\pi \in \Pi(\set{n})}
 (-1)^{|\pi|-1} (|\pi|-1)! \prod_{B\in \pi}
 \frac{
   \Big\langle 0\Big| \mathcal{O}_1 \mathcal{D}^{-1}
   \prod_{i\in B}\mathcal{J}_i \mathcal{O}_2
   \Big| 0\Big\rangle}{
   \Big\langle 0\Big| \mathcal{O}_1 \mathcal{D}^{-1}\mathcal{O}_2
   \Big| 0\Big\rangle},
\label{moment-cumulant}
\end{align}
where $\Pi(\set{n})$ is the set of set partitions of
$\set{n}$ and $|\pi|$ the number of blocks $B$ of a set partition $\pi$.

Following Def.~\ref{def:Omega} we substitute $E_{q_i}\mapsto Y_i$
in (\ref{Omega-Fock}) to produce functions $W_n(Y_1,...,Y_n)$
and multiply by $\prod_{i=1}^n dY_i$ to get the  multi-differentials
$\varpi_n^\vee(Y_1,...,Y_n)$. Our aim is to prove an $x$-$y$ duality
with another family of multi-differentials. This makes two changes necessary:
\begin{itemize}
\item We rescale $\varpi_n^\vee(Y_1,...,Y_n)=
  (-N)^{n-2}\omega_n^\vee(Y_1,...,Y_n)$. The factor  $(-N)^{n-2}$ is
  introduced in (\ref{Omega-def}) for historic reasons; it now turned out
  inconvenient.

\item $\omega_{0,1}$ is not part of the family of multi-differentials.
  We therefore introduce a projection $\mathcal{P}_n=
  \mathrm{id}-\delta_{n,1}\hbar^{-1}[\hbar^{-1}]$ which removes $\omega_{0,1}$.
\end{itemize}
In addition use 
the following convention of the genus expansion:
$\omega^\vee_n(Y_1,...,Y_n)=\sum_{g=0}^\infty \hbar^{2g+n-2}
\omega^{\vee}_{g,n}(Y_1,...,Y_n)$ (excluding $(g,n)=(0,1)$)
and
$\varpi^\vee_n(Y_1,...,Y_n)=\sum_{g=0}^\infty \hbar^{2g}
\varpi^{\vee}_{g,n}(Y_1,...,Y_n)$, giving
$\varpi^{\vee}_{g,n}(Y_1,...,Y_n)=
(-1)^m \omega^{\vee}_{g,n}(Y_1,...,Y_n)$.
The above adaptations bring (\ref{Omega-Fock}) with Def.~\ref{def:Omega} into
the following form: 
\begin{proposition}
 The  multi-differentials 
 of the Kontsevich matrix model with arbitrary potential $P$
 \eqref{Z-KontsevichV} are 
$\varpi_n^\vee=(-N)^{n-2}\omega_n^\vee$ with
\begin{align}
&\omega_n^\vee(Y_1,...,Y_n)
  \label{omega-complex}
  \\*[-1ex]
  &=\mathcal{P}_n
\Big\langle 0\Big|
\int d_P\vec{m}\;
e^{\sum_{l=1}^\infty \frac{1}{l\hbar } J_{l} \mathbf{m}_{-l}}
\mathcal{D}^{-1}
\Big(\prod_{i=1}^n
\sum_{k_i=1}^\infty Y^{k_i-1}J_{-k_i}\Big)
e^{\sum_{k=1}^\infty \frac{1}{k\hbar } J_{-k} \mathbf{e}_{k}}
 \Big|0\Big\rangle^\circ \prod_{i=1}^n dY_i
\nonumber
\\*[-1ex]
&+\delta_{n,2}\frac{dY_1dY_2}{(Y_1-Y_2)^2},
\nonumber
\end{align}
where the measure $d_P\vec{m}$ on $\mathbb{R}^N$ is given by \eqref{eq:dPm}
and $\mathbf{e}_k,\mathbf{m}_{-l}$ are defined in \eqref{eq:em}.
\end{proposition}

The result is very close to the definition of
multi-differentials associated with Orlov-Scherbin tau functions
\cite{Orlov:2001ped, Bychkov:2020ujd, Bychkov:2022wgw}. We identify the parameters 
\begin{itemize}
\item 
  $s_i= \mathbf{e}_i$,

\item $t_i= \mathbf{m}_{-i}$ with an additional integration acting on
  the $t_i$. This integration commutes with the Fock space operations,
  but has drastic consequences for the primary
  multi-differentials $\omega_n$. See Remark~\ref{remark-negativemodes}.

\item $\hat{\psi}(\theta)=-\log(1+\theta)$. See Remark~\ref{note-psi}.

\item The correction terms $\delta_{m,2} \frac{dY_1 dY_2}{(Y_1-Y_2)}$
  are introduced in a later step in
  \cite{Bychkov:2022wgw}.

\end{itemize}

\subsection{Spectral curve and primary multi-differentials}

Generalised Topological Recursion proposes a powerful strategy to compute the expectation 
values (\ref{omega-complex}): derive functional 
relations with a $x$-$y$-dual family and separate the poles (see \cite[Section\,5.4]{Alexandrov:2022ydc} and \cite{Alexandrov:2024tjo}). 
Accordingly, we develop in Appendix~\ref{app:functionalrelations}+\ref{app:XYFock} the duality framework
adapted to the $\mathcal{D}^{-1}$-convention (\ref{omega-complex}). A key ingredient will be the \emph{spectral curve}
$(X,Y)$ which is governed by a function $\Theta^\vee$ defined as
\begin{align}
  \label{Theta-prelim}
  \Theta^\vee(Y)
&=[\hbar^{-1}]\frac{\displaystyle \Big\langle 0\Big|
\int d_P\vec{m} 
e^{\sum_{l=1}^\infty \frac{1}{l\hbar } J_{l} \mathbf{m}_{-l}}
\mathcal{D}^{-1}
\sum_{j=1}^\infty Y^{j}J_{-j}
e^{\sum_{k=1}^\infty \frac{1}{k\hbar } J_{-k} \mathbf{e}_{k}}
\Big|0\Big\rangle
}{\displaystyle \Big\langle 0\Big|
\int d_P\vec{m}\;
e^{\sum_{l=1}^\infty \frac{1}{l\hbar } J_{l} \mathbf{m}_{-l}}
\mathcal{D}^{-1}
e^{\sum_{k=1}^\infty \frac{1}{k\hbar } J_{-k} \mathbf{e}_{k}}
\Big|0\Big\rangle }
+\sum_{k=1}^\infty \frac{\mathbf{e}_k}{Y^k}.
\end{align}
The function $\Theta^\vee(Y)$ will first be differently defined via a
prominent r\^ole in the functional relations. That it reduces to
(\ref{Theta-prelim}) will be proved in Lemma~\ref{lemma:WW-u0}. The
concrete recursive evaluation of the $\omega_n,\omega_n^\vee$ via the
$X$-$Y$-functional relations needs as input the prior determination of
$\Theta^\vee(Y)$.  To the best of our knowledge this is not possible
via Fock space techniques. Instead techniques specific to matrix
models or to enumerative geometry are necessary, in particular loop
equations and Dyson--Schwinger equations. We have one example
where $\Theta^\vee$ has been found: the Kontsevich type
matrix model with potential $P(\Phi)=\frac{\lambda}{4}\Phi^4$
\cite{Branahl:2020yru}. With the knowledge of this particular solution and the
insight from the present paper we can universally identify a
part of the spectral curve:
\begin{principle}
  \label{principle}
  The spectral curve $(X,Y)$ of Kontsevich matrix
  models with arbitrary potential P \eqref{Z-KontsevichV} is
  \begin{align}
    YX= 1+ \Theta^\vee(Y)
\label{ThetaX}
  \end{align}
  with $\Theta^\vee$ given by \eqref{Theta-prelim}, or equivalently 
  \begin{align}
 X(Y)
&:=[\hbar^{-1}]\frac{\displaystyle \Big\langle 0\Big|
\int d_P\vec{m}\;
e^{\sum_{l=1}^\infty \frac{1}{l\hbar } J_{l} \mathbf{m}_{-l}}
\mathcal{D}^{-1}
\sum_{j=1}^\infty Y^{j-1}J_{-j}
e^{\sum_{k=1}^\infty \frac{1}{k\hbar } J_{-k} \mathbf{e}_{k}}
\Big|0\Big\rangle
}{\displaystyle \Big\langle 0\Big|
\int d_P\vec{m}\;
e^{\sum_{l=1}^\infty \frac{1}{l\hbar } J_{l} \mathbf{m}_{-l}}
\mathcal{D}^{-1}
e^{\sum_{k=1}^\infty \frac{1}{k\hbar } J_{-k} \mathbf{e}_{k}}
\Big|0\Big\rangle }
\nonumber
\\[-1ex]
&+\frac{1}{N} \sum_{a=1}^N \frac{1}{Y -E_a} .
    \label{spectralcurve-Y}
  \end{align}
  The ramified covering $Y:\mathbb{P}^1\to \mathbb{P}^1$ has to be identified
  such that $X(z):=X(Y(z))$ induced by \eqref{spectralcurve-Y}
  is of the form
\begin{align}
  X(z)=
  \frac{1}{N}\sum_{a=1}^N \frac{1}{Y'(\varepsilon_a)(z-\varepsilon_a)}
  + \sum_{i,r} \frac{c_{ir}}{(z-\xi_i)^{r}} ,
\label{Y-Kontsevich}
\end{align}
where $\xi_i$ are poles of $Y$ and $Y(\varepsilon_a)=E_a$.
These $\varepsilon_a \in Y^{-1}(E_a)$ are the principal preimages with
$\lim_{P\to 0} \varepsilon_a=E_a$.  
If $\infty$ is a pole of $Y$, the corresponding term
is understood as $\sum c_r z^r$.
\end{principle}
This principle implies that for a contour $\mathcal{C}$ which encircles
all those poles $\varepsilon_a$ of $X$ which are not poles of $Y$, one has
with $\omega^{\vee}_{0,1}(z):=X(z)dY(z)$
\begin{align}
  \int_{\mathcal{C}} (Y(z))^k \omega^{\vee}_{0,1}(z)
  =  
  \frac{1}{N}\sum_{a=1}^N (Y(\varepsilon_a))^k= \mathbf{e}_k .
\end{align}
This is the general relation for KP time variables.
It is insensitive to symplectic transformations
$X\mapsto X+\text{polynomial}(Y)$ discussed in
Sec.~\ref{sec:xtydualofequivariant}.

We remark that for the original Kontsevich model
$P(\Phi)=\frac{\mathrm{i}}{3} \Phi^3$, although the $\omega^\vee_{g,n}$ are
differently defined, in the spectral curve $(x_K,y_K)$ we have
$y_K(z)=z^2
$ and $x_K(z)$  of the same form (\ref{Y-Kontsevich})
(when swapping $x_K \leftrightarrow y_K$ to view the 
$\omega^\vee_{g,n}$ as dual), see \cite{Eynard:2007kz}.

In the longer appendices~\ref{app:functionalrelations}+\ref{app:XYFock} we prove the
following main result:
\begin{theorem}
\label{thm:Kontsevich-duality}
  Define a primary family $\{\omega_{g,n}(X_{\set{m}})\}$ of multi-differentials
  by Fock space expectation values
\begin{align}
&\omega_n(X_1,...,X_n)
  \label{omega-complex-1}
  \\*[-1ex]
  &=\mathcal{P}_n
\Big\langle 0\Big|
\int d_P\vec{m}\;
e^{\sum_{l=1}^\infty \frac{J_{l} \mathbf{m}_{-l}}{l\hbar } }
\Big(\prod_{i \in \set{n}}^{R.O.}
\sum_{k_i \in \mathbb{Z}\setminus \{0\}} X^{k_i-1}J_{k_i}\Big)
\mathcal{D}^{-1}
e^{\sum_{k=1}^\infty \frac{J_{-k} \mathbf{e}_{k}}{k\hbar } }
 \Big|0\Big\rangle^\circ \prod_{i=1}^n dX_i,
 \nonumber
\end{align}
where the radially-ordered product is defined as
\begin{align}
  \prod_{i \in B}^{R.O.}\mathcal{J}(X_i)
  &:= \mathcal{J}(X_{\pi(1)})
  \mathcal{J}(X_{\pi(2)})\cdots
  \mathcal{J}(X_{\pi(|B|)})
  \\
  &\text{ if } |X_{\pi(1)}|<|X_{\pi(2)}|<...<|X_{\pi(B)}|
\text{ for a permutation $\pi$ of  $B$}.
\nonumber
\end{align}
The connected expectation value is defined analogous to
\eqref{moment-cumulant}; only with operators $\mathcal{J}_i$ placed
to the left of $\mathcal{D}^{-1}$.

Assume that for $\Theta^\vee$ defined by
\eqref{Theta-prelim} the spectral curve $XY=1+\Theta^\vee(Y)$
is identically solved by  two ramified coverings
$X,Y:\mathbb{P} \to \mathbb{P}^1$, and define
\begin{align}
\omega^{\vee(X,Y)}_n(z_{\set{n}}):=
\omega^{\vee}_n(Y(z_1),...,Y(z_n)),\quad
\omega^{(X,Y)}_n(z_{\set{n}}):=
  \omega_n(X(z_1),...,X(z_n))
\end{align}
for  $\omega^{\vee}_n(Y_1,...,Y_n),\omega_n(X_1,...,X_n)$
given by \eqref{omega-complex} and \eqref{omega-complex-1}, respectively.
Define from these $\{\omega^{(X,Y)}_n(z_{\set{n}}),\omega^{\vee(X,Y)}_n(z_{\set{n}}),
X(z),Y(z)\}$ 
the extended multi-differentials
$\mathbb{W}^{(X,Y)}_n(u_{\set{n}};z_{\set{n}})$ and 
$\mathbb{W}^{\vee(X,Y)}_n(w_{\set{n}};z_{\set{n}})$ 
according to \eqref{eq:def-bW} and \eqref{eq:def-bWvee}
(with $(x_i,y_i)$ replaced by $(X(z_i),Y(z_i))$).
Then $\{\omega^{\vee(X,Y)}_n\}$ and  $\{\omega^{(X,Y)}_n\}$ are related
by the duality formulae \eqref{eq:def-omvee} and \eqref{identity:om}
(with $(x_i,y_i)$ replaced by $(X(z_i),Y(z_i))$).
\end{theorem}

\begin{remark}
  The universal $x$-$y$-duality discussed in Sec.~\ref{sec:xyduality}
  states that to a family $\{\omega_{g,n}\}$ of multi-differentials
  and two 1-forms $dx,dy$ there is always some dual family
  $\{\omega^{\vee(x,y)}_{g,n}\}$.  Thm.~\ref{thm:Kontsevich-duality}
  states far more: it gives a concrete formula \eqref{omega-complex-1}
  for the primary family associated with the dual \eqref{omega-complex}
  and the spectral curve \eqref{spectralcurve-Y}.

  In fact, the
  proof of the universal $x$-$y$ duality proceed via an abstract Fock
  space calculation.  It is discussed in \cite[Proof of Proposition 4.8]{Alexandrov:2022ydc} that to any family of multi-differentials
  together with a generic point one can always associate a Fock space
  theory. The proof of the universal duality then proceeds in this
  abstract Fock space in the same manner as done in our proof of
  Thm.~\ref{thm:Kontsevich-duality} in
  Appendix~\ref{app:functionalrelations}+\ref{app:XYFock}. The only difference is that
  usually $\mathcal{D}$ is used instead of $\mathcal{D}^{-1}$ in our
  proof. 
\end{remark}

\begin{remark}
  With the exception of the special term
  $\delta_{n,2}\frac{dY_1dY_2}{(Y_1-Y_2)^2}$, only non-negative powers
  of $Y_i$ contribute to the multi-differentials 
  $\omega_n^\vee(Y_{\set{n}})$ in \eqref{omega-complex}. In contrast,
  in the primary $\omega_n(X_{\set{n}})$ in \eqref{omega-complex-1}
  the $X_i$ appear with positive and negative power. When commuting
  the $J_{<0}$ through
  $e^{\sum_{l=1}^\infty \frac{1}{l\hbar } J_{l} \mathbf{m}_{-l}}$ towards the
  co\-vacuum $\langle 0|$, a complicated dependence on the
  matrix traces $\mathbf{m}_{-l}$ arises.
\end{remark}

\begin{example}
\label{ex:om02Fock}
 We give the explicit form in the case of $\omega^{\vee(X,Y)}_{0,2}(z_1,z_2)=
\omega^{(X,Y)}_{0,2}(z_1,z_2)$ where only the single 
graph
$\Gamma=~~\begin{picture}(53,5)\put(0,0){$1$}\put(3,3){\circle{15}}
  \put(40,0){$2$}\put(43,3){\circle{15}}
  \put(10,3){\line(1,0){25}}  \put(20,0){$\bullet$}
\end{picture}$ contributes to the extended multi-differentials:
\begin{align}
&\omega^{(0)}_2(z_1,z_2)
\label{omega02-Fock}
\\
&=\frac{dX(z_1)dX(z_2)}{
  X(z_1)X(z_2)}
[\hbar^0] \Bigg(\frac{X(z_1)X(z_2)}{(X(z_1)-X(z_2))^2} 
\nonumber
\\
&
+\frac{1}{Z}  
\Big\langle 0\Big|
\!\! \int \!\! d_P\vec{m}\,
e^{\sum_{l=1}^\infty \!\! \frac{J_{l} \mathbf{m}_{-l}}{l\hbar }}
\prod_{i=1}^2\!\Big(\!\sum_{k_i=1}^\infty \!\Big(\!(X(z_i))^{k_1}J_{k_i}+
\frac{\mathbf{m}_{-k_i}}{\hbar (X(z_i))^{k_i}}\Big)\!\Big)
\mathcal{D}^{-1}
e^{\sum_{k=1}^\infty \!\!\frac{J_{-k} \mathbf{e}_{k}}{k\hbar } }
\Big|0\Big\rangle
\nonumber
\\
&-\prod_{i=1}^2\frac{1}{Z}  
\Big\langle 0\Big|
\!\! \int \!\! d_P\vec{m}\,
e^{\sum_{l=1}^\infty \!\! \frac{J_{l} \mathbf{m}_{-l}}{l\hbar }}
\sum_{k_i=1}^\infty \Big((X(z_i))^{k_1}J_{k_i}+
\frac{\mathbf{m}_{-k_i}}{\hbar (X(z_i))^{k_i}}\Big)
\mathcal{D}^{-1}
e^{\sum_{k=1}^\infty \!\!\frac{J_{-k} \mathbf{e}_{k}}{k\hbar } }
\Big|0\Big\rangle
\!\! \Bigg)
\nonumber
\\
&\equiv\frac{dY(z_1)dY(z_2)}{
  Y(z_1)Y(z_2)}
[\hbar^0] \Bigg(\frac{Y(z_1)Y(z_2)}{(Y(z_1)-Y(z_2))^2} 
\nonumber
\\
&
+\frac{1}{Z}  
\Big\langle 0\Big|
\!\! \int \!\! d_P\vec{m}\,
e^{\sum_{l=1}^\infty \!\! \frac{J_{l} \mathbf{m}_{-l}}{l\hbar }}
\mathcal{D}^{-1}
\prod_{i=1}^2 \sum_{k_i=1}^\infty (Y(z_i))^{k_1}J_{-k_i}
e^{\sum_{k=1}^\infty \!\!\frac{J_{-k} \mathbf{e}_{k}}{k\hbar } }
\Big|0\Big\rangle
\nonumber
\\
&-\prod_{i=1}^2\frac{1}{Z}  
\Big\langle 0\Big|
\!\! \int \!\! d_P\vec{m}\,
e^{\sum_{l=1}^\infty \!\! \frac{J_{l} \mathbf{m}_{-l}}{l\hbar }}
\mathcal{D}^{-1}
\sum_{k_i=1}^\infty (Y(z_i))^{k_1}J_{-k_i}
e^{\sum_{k=1}^\infty \!\!\frac{J_{-k} \mathbf{e}_{k}}{k\hbar } }
\Big|0\Big\rangle
\!\! \Bigg),
\nonumber
\end{align}
where $Z\equiv Z_P(E_{\set{N}})$. This identity holds for any
potential $P$ in \eqref{Z-KontsevichV}. It requires, of course, to
determine $X(z),Y(z)$ via \eqref{spectralcurve-Y}.

For quartic potential $P(\Phi)=\frac{\lambda}{4} \Phi^4$, the coverings
$X,Y$ can be found (see \eqref{X0Y0-4} in the next subsection). Moreover,
it was proved in \cite{Branahl:2020yru} that in this case
\eqref{omega02-Fock} equals the bi-differential
$B(z_1,z_2)=\frac{dz_1dz_2}{(z_1-z_2)^2}+  \frac{dz_1dz_2}{(z_1+z_2)^2}$
of $\mathbb{Z}/2\mathbb{Z}$ equivariant TR (which is remarkable when looking at
the complicated integrals \eqref{omega02-Fock}).
\end{example}

\subsection{Spectral curve of the quartic Kontsevich model}

The two conditions (\ref{spectralcurve-Y}) and (\ref{Y-Kontsevich}) 
in General Principle~\ref{principle} are difficult to satisfy
simultaneously. Typically the ramified covering $Y$ has some degree $d+1$ with $d>0$ and hence $d+1$ preimages
$Y^{-1}(E_a)=\{\varepsilon_a,\widehat{\varepsilon_a}^1,
...,\widehat{\varepsilon_a}^d\}$. Then
\begin{align*}
  \frac{1}{Y(z)-E_a}=  \frac{1}{Y'(\varepsilon_a)(z-\varepsilon_a)}+\sum_{j=1}^d
  \frac{1}{Y'(\widehat{\varepsilon_a}^j)(z-\widehat{\varepsilon_a}^j)}.
\end{align*}
Inserted into the last line of (\ref{spectralcurve-Y}), the poles
$\widehat{\varepsilon}_a$ must cancel against poles from the second
line of (\ref{spectralcurve-Y}), because $\widehat{\varepsilon_a}^j$
cannot be a pole of $Y$. On the other hand, it can be seen from a
perturbative expansion about the Gau\ss{}ian Example~\ref{ex:Gaussian}
that the ratio in the second line of (\ref{spectralcurve-Y}) will have
many poles different from $Y_i=\infty$. They must be arranged into
$\{\widehat{\varepsilon_a}^j\}$; so the cheap guess $Y(z)=z$ does not
work. This tension between (\ref{spectralcurve-Y}) and
(\ref{Y-Kontsevich}) can be solved for the quartic Kontsevich model
with potential $P(\Phi)=\frac{\lambda}{4} \Phi^4$ in
(\ref{Z-KontsevichV}), thanks to results established in
\cite{Grosse:2019jnv,Schurmann:2019mzu,Branahl:2020yru}:
\begin{theorem}
\label{Thm:BHW}
  For the quartic analogue of the Kontsevich model with potential
$P(\Phi)=\frac{\lambda}{4} \Phi^4$ in \eqref{Z-KontsevichV},
the spectral curve according to the General Principle \ref{principle}
is 
\begin{align}
  Y(z)&= R(z) ,\qquad X(z)=\frac{R(z)+R(-z)}{\lambda} \qquad \text{with}
  \label{X0Y0-4}
  \\
  R(z)&:= 
  z-\frac{\lambda}{N} \sum_{a=1}^N \frac{1}{R'(\varepsilon_a)
    (z+\varepsilon_a)}  .
  \nonumber
\end{align}
\begin{proof}
  In \cite{Grosse:2019jnv,Schurmann:2019mzu} it was found that the
  quartic Kontsevich model is governed by a relation
\begin{align}
  -R(-z)=R(z)+\frac{\lambda}{N} \sum_{a=1}^N \frac{1}{E_a-R(z)}
  +\frac{\lambda}{N} \sum_{a=1}^N \mathcal{G}^{(0)}(\varepsilon_a,z) ,
\label{curve-quartic}
\end{align}
where
\begin{align}
  \sum_{a=1}^N \mathcal{G}^{(0)}(\varepsilon_a,z)
  &:= \sum_{a=1}^N  G_{|aq|} \Big|_{\substack {E_q \mapsto R(z) \\
      \text{leading $\hbar$-order}}}  ,
  \nonumber
  \\
  G_{|pq|}  
  &:= \frac{N}{Z_{\lambda\Phi^4/4}(E_{\set{N}})} \int_{H_N} d\Phi \;\Phi_{qp}\Phi_{pq}
  \;e^{-N\;\mathrm{Tr}(E\Phi^2+\frac{\lambda}{4} \Phi^4)}
  \text{ if } p\neq q .
\label{2pt-function}
\end{align}
Making (\ref{curve-quartic}) as an ansatz, $R(z)$ is necessarily
given by (\ref{X0Y0-4}), see  \cite{Schurmann:2019mzu}, 
and the then algebraically resulting $\sum_{a=1}^N  [\hbar^0]G_{|aq|}$
and $[\hbar^0]G_{|pq|}$ solve the Dyson-Schwinger equation
for the 2-point function (\ref{2pt-function}).
See \cite{Grosse:2019jnv,Schurmann:2019mzu} for details.

According to (\ref{Z-KontsevichV}) we have
$\frac{1}{N}\sum_{a=1}^N  G_{|aq|}  
=-\hbar \frac{\partial \log Z_{\lambda\Phi^4/4}(E_{\set{N}})}{\partial E_q}$.
We have proved in sec.~\ref{sec:partitionfunction} the representation
(\ref{Z-Fock}) for the partition function $Z_{\lambda\Phi^4/4}(E_{\set{N}})$, which depends on
the matrix entries $E_q$ via
$\mathbf{e}_k=\frac{1}{N}\mathrm{Tr}(E^k)$.
With $\frac{\partial \mathbf{e}_k}{\partial E_q}
= \hbar k E_q^{k-1}$ we get 
\begin{align}
\frac{1}{N}  \sum_{a=1}^N  G_{|aq|}
  &= -\hbar^2 \sum_{k=1}^\infty k E_q^{k-1} 
  \frac{\partial (\log Z_{\lambda\Phi^4/4}(E_{\set{N}}))}{\partial \mathbf{e}_k}
\label{Omega01-4}
  \\
  &=-\hbar
\frac{\displaystyle  \Big\langle 0\Big|
  \int \!\!d_P\vec{m}\;
  e^{\sum_{k=1}^\infty \frac{J_{k} \mathbf{m}_{-k}}{k \hbar}}
\mathcal{D}^{-1} 
\Big(\sum_{k=1}^\infty E_q^{k-1} J_{-k}\Big)
  e^{\sum_{k=1}^\infty \frac{J_{-k}\mathbf{e}_k}{k \hbar}}
\Big|0\Big\rangle}{
\displaystyle  \Big\langle 0\Big|
  \int \!\!d_P\vec{m}\;
  e^{\sum_{k=1}^\infty \frac{J_{k} \mathbf{m}_{-k}}{k \hbar}}
\mathcal{D}^{-1} 
  e^{\sum_{k=1}^\infty \frac{J_{-k}\mathbf{e}_k}{k \hbar}}
\Big|0\Big\rangle}.
  \nonumber
\end{align}
With $Y(z)=R(z)$, we substitute $E_q\mapsto R(z)$ in the first line of
(\ref{2pt-function}). Leading $\hbar$-order
means to take the $[\hbar^{-1}]$-coefficient of the ratio in the last line of 
(\ref{Omega01-4}), so that
\[
\frac{1}{N} \sum_{a=1}^N \mathcal{G}^{(0)}(\varepsilon_a,z)
= -X(z)+\frac{1}{N}\sum_{a=1}^N \frac{1}{R(z)-E_a}
\]
by comparison with (\ref{spectralcurve-Y}). Comparing with 
(\ref{curve-quartic}) we confirm $X(z)$ in (\ref{X0Y0-4}).
\end{proof}
\end{theorem}

\section{The $\mathbb{Z}/2\mathbb{Z}$-equivariant spectral curve related to the quartic Kontsevich model}\label{sec:spectralcurveZ2}

\subsection{Duality network}

In Sec.~\ref{sec:Kontsevich} we established a Fock space formulation of the Kontsevich matrix model (\ref{Z-KontsevichV}) with arbitrary potential. This framework permitted us to prove a duality theorem \ref{thm:Kontsevich-duality} between two families $\{\omega_{g,n}^{\vee(X,Y)}\}$ and 
$\{\omega_{g,n}^{(X,Y)}\}$ of Fock space expectation values associated with (\ref{Z-KontsevichV}). A key step is a Lagrange inversion between $Y$ and $X$ induced by $XY=1+\Theta^\vee(Y)$, where $\Theta^\vee(Y)$ is a concrete vacuum expectation value (\ref{Theta-prelim})
in Fock space. We do not control $\Theta^\vee(Y)$ for a
general potential $P$ in (\ref{Z-KontsevichV}). For the special case of the
quartic potential $P( \Phi)=\frac{\lambda}{4}\Phi^4$, results of \cite{Grosse:2019jnv,Schurmann:2019mzu,Branahl:2020yru})
permitted to identify $X,Y$, see (\ref{X0Y0-4}). Of central importance for this section is the proof in \cite{Branahl:2020yru} that
for $P(\Phi)=\frac{\lambda}{4}\Phi^4$ one has
\begin{align}
  \omega^\vee_{0,2}(z_1,z_2)=\frac{dz_1\,dz_2}{(z_1-z_2)^2}
  +\frac{dz_1\,dz_2}{(z_1+z_2)^2}.  
\label{om02-BHW}
\end{align}
It is a general fact that $\omega_{0,2}$ is invariant
under $x$-$y$ and symplectic transformations. There is thus a chain
of dualities which produces from
$\omega_{g,n}^{\vee}=:\omega_{g,n}^{\vee(X,Y)}$ a family
$\omega_{g,n}^{(x,y)}$ where $x(z),y(z)$ transform equivariantly under the
$\mathbb{Z}/2\mathbb{Z}$-symmetry of (\ref{om02-BHW}), namely
\begin{align}
  x(z)&=X(z)-\frac{2}{\lambda} Y(z)=
  \frac{R(z)-R(-z)}{(-\lambda)},
\label{def:xy}
  \\
  y(z)&=Y(z)+\frac{\lambda}{2} x(z)=
  \frac{R(z)+R(-z)}{2}.\nonumber  
  \end{align}
The meromorphic function $R$ is given in (\ref{X0Y0-4}), but its precise form will not play a r\^ole.
The data (\ref{def:xy}) and (\ref{om02-BHW})
give rise to the 
$\mathbb{Z}/2\mathbb{Z}$-equivariant spectral curve 
\begin{align}
  \label{eqGWspectralcurve}
    \mathcal{S} = (\mathbb{P}^1,\phi,x,y,B,\chi,\upsilon,\beta),
\end{align}
with $\gamma z=-z$ for $\gamma\neq 1$,
$\chi=\beta$ the sign representation,  $\upsilon$ the trivial
representation and $B=\omega^\vee_{0,2}$. Equivariant TR associates to the curve 
(\ref{eqGWspectralcurve}) a unique family $\{\omega_{g,n}\}$ of
multi-differentials. The central question is:
\begin{question}
\label{mainquestion}
  Is it true that
  \begin{align}
    \omega_{g,n}=\omega_{g,n}^{(x,y)} ?
    \label{eq:question}
  \end{align}
  Equivalently, do the $\omega_{g,n}^{(x,y)}$ constructed from
  $\omega_{g,n}^{\vee}=\omega_{g,n}^{\vee(X,Y)}$ via the duality chain
  of Fig.~\ref{fig:duality-network}
  \begin{align}
    \omega_{g,n}^{\vee}=:\omega_{g,n}^{\vee(X,Y)}
    \xrightarrow{\text{trivial shift}}
  \omega_{g,n}^{\vee(x,Y)}
    \xrightarrow{\text{$x$-$y$ duality}}
  \omega_{g,n}^{(x,Y)}
    \xrightarrow{\text{trivial shift}}
  \omega_{g,n}^{(x,y)}
    \label{matrix-to-eq}
\end{align}
satisfy the projection property, i.e.\ have poles only at the
ramification points of $x$?
\end{question}
If the answer is affirmative, then reverting the arrows to
  \begin{align}
    \omega_{g,n}=:\omega_{g,n}^{(x,y)}
    \xrightarrow{\text{trivial shift}}
  \omega_{g,n}^{(x,Y)}
    \xrightarrow{\text{$x$-$y$ duality}}
  \omega_{g,n}^{\vee(x,Y)}
    \xrightarrow{\text{trivial shift}}
    \omega_{g,n}^{\vee(X,Y)},
    \label{eq-to-matrix}
\end{align}
we get from the recursively computable equivariant $\omega_{g,n}$
the multi-differentials of the quartic Kontsevich model
\cite{Branahl:2020yru}. These were only partially known so far.

We prove in this section that the 
$\omega_{g,n}^{\vee(X,Y)}$ defined by the chain (\ref{eq-to-matrix})
have exactly the properties
established previously for the  multi-differentials of the
quartic Kontsevich model:
\begin{itemize}
\item   The $\omega_{0,n}^{\vee(X,Y)}$ have the reflection property found in
  \cite{Hock:2021tbl} and the symmetry proved in \cite{hock2025symmetry}.
  Moreover, we identify the generalisation of the reflection formula
  to all $(g,n)$.

\item   The $\omega_{0,n}^{\vee(X,Y)}$ and $\omega_{1,n}^{\vee(X,Y)}$
  are recursively computable via the formula found in
  \cite{Hock:2023nki}.
  Moreover, we prove a recursion formula for the pole at the anti-diagonal
  for any $(g,n)$.

\end{itemize} 
These results and the elegance of the duality network 
of Fig.~\ref{fig:duality-network} make it highly plausible that the
answer to Question~\ref{mainquestion} is `yes'. The longstanding
open problem to understand and to construct the
multi-differentials of the quartic Kontsevich model
\cite{Branahl:2020yru} is then solved.

\subsection{Symplectic duality and reflection formula}

Let us now start from the spectral curve (\ref{eqGWspectralcurve}).
Then Def.~\ref{defn:equiv:SC} for 
equivariant topological recursion 
constructs a family $\{\omega_{g,n}\}$ of multi-differentials. We
slightly deviate and use $B^{TR}(z_1,z_2)=\frac{dz_1\,dz_2}{(z_1-z_2)^2}$
in the recursion kernel instead,
as described in Lemma~\ref{lemma:equivalentomegagn}, to obtain the
correct scaling by $2$ such that $\omega_{0,1}=ydx$ can be taken.
Let us first collect some properties:
\begin{itemize}
    \item the set of fixed points of the group action is $\{z^\star\}=\{0,\infty\}$;
    \item $y$ has a ramification point at the fixed point $z^\star=0$.
\end{itemize}
From these two observations, we can define the $x$-$y$ dual family $\omega_{g,n}^{\vee(x,y)}$ associated with $\omega_{g,n}$ of the spectral curve \eqref{eqGWspectralcurve}. We have to be careful with the polar structure of this dual family because the ramification locus intersects non-trivially with the set of fixed points of the group action. Thus, Cor.~\ref{cor:polarpartxydualside}, for instance, does not hold, but Prop.~\ref{prop:GenTRxydual} is still valid.

However, we want to make another observation and are actually more interested in other families of $x$-$y$ dual multi-differentials in the sense of Sec.~\ref{sec:xtydualofequivariant}. Thus, we build the dual multi-differentials with
two pairs $(x,Y)$ and $(x,\tilde{Y})$. To make contact with the quartic Kontsevich model we choose $Y=y-\frac{\lambda}{2}x$ as in (\ref{def:xy}) and
$\tilde{Y}=y+\frac{\lambda}{2}x$. Note that $Y(-z)=\tilde{Y}(z)$,
which will be relevant below. The corresponding construction of 
Def.~\ref{def:xydualmultidiff} yields two families
    $\omega_{g,n}^{\vee(x,Y)}$ and $\omega_{g,n}^{\vee(x,\tilde{Y})}$, see Fig.~\ref{fig:yyy-dualities}.
\begin{figure}[ht]
    \centering
    \begin{tikzpicture}[>=stealth]

    \node (omega) at (0,2.5) {$\omega_{g,n}$};

    \node (omegavee) at (0,0) {$\omega^{\vee(x,y)}_{g,n}$};

    \node (omegaleft) at (-4,2.5)
    {$\omega_{g,n}^{(x,Y)}$};

    \node (omegaright) at (4,2.5)
    {$\omega_{g,n}^{(x,\tilde{Y})}$};

    \node (omegaplus) at (-4,0)
    {$\omega_{g,n}^{\vee(x,Y)}$};
    
    \node (omegaminus) at (4,0)
    {$\omega_{g,n}^{\vee(x,\tilde{Y})}$};

    \draw[<->]
    (omegaleft) -- 
    node[pos=0.5, left, font=\tiny] {$(x,Y)$}
    node[pos=0.5, right, font=\tiny] {duality}
    (omegaplus);

    \draw[<->]
    (omegaright) -- 
    node[pos=0.5, left, font=\tiny] {$(x,\tilde{Y})$}
    node[pos=0.5, right, font=\tiny] {duality}
    (omegaminus);

    \draw[<->]
    (omega) -- 
    node[pos=0.5, left, font=\tiny] {$(x,y)$}
    node[pos=0.5, right, font=\tiny] {duality}
    (omegavee);

    \draw[double]
    (omega) -- 
    node[pos=0.5, below, font=\tiny] {trivial shift}
    node[pos=0.5, above, font=\tiny] {$Y=y-\frac{\lambda}{2}x$}
    (omegaleft);
    
    \draw[double]
    (omega) -- 
    node[pos=0.5, below, font=\tiny] {trivial shift}
    node[pos=0.5, above, font=\tiny] {$\tilde{Y}=y+\frac{\lambda}{2}x$}
    (omegaright);
    
    \end{tikzpicture}
    \caption{\label{fig:yyy-dualities}}
\end{figure}

We can conclude from the results of Sec.~\ref{sec:xtydualofequivariant} and the assumptions that $\omega_{g,n}^{\vee,(x,Y)}$
\begin{itemize}
    \item has no poles at the ramification points of $x$ or (for $(g,n)\neq (0,2)$) at the diagonal;
    \item is symmetric;
    \item has poles at the ramification points of $Y$, that is, at $Y'(z)=0$, at the anti-diagonal $z_i=-z_j$, and, for $g>0$, at $z_i=0$;
    \item satisfies the linear and quadratic loop equations with respect to
      $(Y,x)$, and thus falls into the setting of blobbed topological recursion.
\end{itemize}
Now, the interesting observation is that the other dual family
$\omega_{g,n}^{\vee(x,\tilde{Y})}$ has the same properties by simply reflecting all $z_i\to-z_i$ simultaneously for all $i=1,...,n$.

The construction of the dual multi-differentials starts from
the extended multi-differentials in Def.~\ref{def:extendemulti}
and thus depends on the choice of $\{dy,dY,d\tilde{Y}\}$ via the
exponential in the second line of (\ref{eq:def-bW}).
Note that every order of $\mathscr{S}-1$ is a least a quadratic differential operator. We have for linear shifts 
\[
  \Big(u \hbar d_z \frac{1}{dx}\Big)^2 (y+\alpha x) dx
  =u\hbar   \Big(u \hbar d_z \frac{1}{dx}\Big) (dy+\alpha dx)
  =(u\hbar)^2 d_z  \Big(\frac{dy}{dx}\Big)
\]
independently of $\alpha$, which remains true at higher order.
This means that the corresponding extended multi-differentials are all the same:
\begin{align}  
\mathbb{W}^{(x,y)}_n(u_{\set{n}};  z_{\set{n}})
&=\mathbb{W}^{(x,Y)}_n(u_{\set{n}};  z_{\set{n}})
=\mathbb{W}^{(x,\tilde{Y})}_n(u_{\set{n}};  z_{\set{n}}).
\label{trivial-shift-top}
\end{align}
For the same reason, the linear shifts $X=x+\frac{2}{\lambda}Y$ and
$\tilde{X}=x-\frac{2}{\lambda}\tilde{Y}$ in the bottom row
of  Fig.~\ref{fig:duality-network} do not change the corresponding dual
extended multi-differentials (\ref{eq:def-bWvee}):
\begin{align}
  \mathbb{W}^{\vee(X,Y)}_n(w_{\set{n}};  z_{\set{n}})
  &= \mathbb{W}^{\vee(x,Y)}_n(w_{\set{n}};  z_{\set{n}})  ,
  \nonumber
  \\
  \mathbb{W}^{\vee(\tilde{X},\tilde{Y})}_n(w_{\set{n}};  z_{\set{n}})
  &= \mathbb{W}^{\vee(x,\tilde{Y})}_n(w_{\set{n}};  z_{\set{n}})  .
\label{trivial-shift-bottom}
\end{align}  
This implies:
\begin{lemma}
  The following reflection identity holds:
  \begin{align}
 \omega^{\vee(x,\tilde{Y})}_{g,n}(z_1,...,z_n)=
 (-1)^m\omega^{\vee(x,Y)}_{g,n}(-z_1,...,-z_n).
 \label{reflectionproperty-new}
\end{align}
\begin{proof}
 According to (\ref{eq:def-omvee}), the dual multi-differentials are given by
\begin{align}
&\omega^{\vee(x,Y)}_{g,n}(z_{\set{n}})
\label{varpiIV}
\\
&=
(-1)^n \!\!\!\!\!\! \sum_{r_1,...,r_n=0}^\infty
\prod_{i=1}^n
\Big({-}d_{z_i}\frac{1}{d Y(z_i)}\Big)^{r_i} [u_i]
\Big(\mathbb{W}_{g,n}^{(x,y)}\big(u_{\set{n}},z_{\set{n}}\big)
\prod_{i=1}^n dx(z_i)
\Big),
\nonumber
\\
&\omega^{\vee(x,\tilde{Y})}_{g,n}(z_{\set{n}})
\label{varpiV}
\\
&=
(-1)^n \!\!\!\!\! \sum_{r_1,...,r_n=0}^\infty
\prod_{i=1}^n
\Big({-}d_{z_i}\frac{1}{d \tilde{Y}(z_i)}\Big)^{r_i} [u_i]
\Big(\mathbb{W}_{g,n}^{(x,y)}\big(u_{\set{n}},z_{\set{n}}\big)
\prod_{i=1}^n dx(z_i)
\Big),
\nonumber
\end{align}
where (\ref{trivial-shift-top}) was used to map to the common
$\mathbb{W}_{g,n}^{(x,y)}$.  We flip the signs $z_i\mapsto -z_i$ in
(\ref{varpiIV}) and note $Y(-z_i)=\tilde{Y}(z_i)$ and
$x(-z_i)=-x(z_i)$. This gives 
$({-}d_{z_i}\frac{1}{d Y(z_i)})_{z_i \mapsto -z_i}=
({-}d_{z_i}\frac{1}{d \tilde{Y}(z_i)})$.
The function $\mathbb{W}_n^{(x,y)}$ is of the form
(\ref{eq:def-bW}).  We have
${:}\omega^{(x,y)}_{|\flat|}{:}(-z_1,...,-z_{|\flat|}) =(-1)^{|\flat|}
{:}\omega^{(x,y)}_{|\flat|}{:}(z_1,...,z_{|\flat|})$ by
Example~\ref{ex:eqTRZ2}; the factor $(-1)^{|\flat|}$ is cancelled by
the $|\flat|$ sign changes of $\frac{1}{dx(-z_e)}=-\frac{1}{dx(z_e)}$
for every edge. Moreover, $\mathscr{S}$ is an even function. Thus,
$\mathbb{W}_{g,n}^{(x,y)}$ is invariant under $z_i\mapsto -z_i$,
leaving a $(-1)^n$ from
$\prod_{i=1}^n dx(-z_i)=(-1)^n \prod_{i=1}^n dx(z_i)$.  In summary,
the rhs of (\ref{varpiIV})$_{z_i\mapsto -z_i}$ equals $(-1)^n$ times
the rhs of (\ref{varpiV}).
\end{proof}
\end{lemma}

From (\ref{trivial-shift-bottom}) we conclude that 
  \begin{align}
 \omega^{\vee(\tilde{X},\tilde{Y})}_{g,n}(z_1,...,z_n)=
 (-1)^n\omega^{\vee(X,Y)}_{g,n}(-z_1,...,-z_n)
 \label{reflectionproperty-new1}
\end{align}
in the bottom row of Fig.~\ref{fig:duality-network}.


The extended multi-differential $\mathbb{W}^{\vee(x,\tilde{Y})}_n$
is by the parametric formula (\ref{eq:parametric-vee}) expressed in terms of
$\mathbb{W}^{(x,\tilde{Y})}_n$, which by (\ref{trivial-shift-top})
equals $\mathbb{W}^{(x,Y)}_n$. Then,
the parametric formula (\ref{eq:parametric}) expresses
$\mathbb{W}^{(x,Y)}_n$ 
in terms of
$\mathbb{W}^{\vee(x,X)}_n$. The composition of these operations reads
\begin{align}
  &\mathbb{W}^{\vee(x,\tilde{Y})}_n(v_{\set{n}};z_{\set{n}})
  -\frac{\delta_{n,1}}{\hbar v_1}
\label{bbW-double-1}
\\
&= \!\!\!\! \sum_{\substack{r_1,...,r_n,~\\
    s_1,...,s_n=0}}^\infty \!\!\!
\Big(\prod_{i=1}^n e^{-v_ix_i}\Big({-}\frac{\partial}{
  \partial \tilde{Y}_i}\Big)^{r_i}
\Big)[u_1^r\cdots u_n^r]
\Bigg(
\Big(\prod_{i=1}^n e^{v_ix_i-u_iY_i}
\frac{dx_i}{d\tilde{Y}_i}\Big)
\nonumber
\\
&\qquad \times 
\prod_{i=1}^n\Big({-}\frac{\partial}{\partial x_i}\Big)^{s_i}
\bigg(\Big(\prod_{i=1}^n \frac{dY_i}{dx_i}e^{u_iY_i}\Big)
[w_1^{s_1}\cdots w_n^{s_n}] \mathbb{W}^{\vee(x,Y)}_n(w_{\set{n}};
z_{\set{n}})\bigg)\Bigg),
\nonumber
\end{align}
where we have used $(d_{z}\frac{1}{dx(z)})^r (f(z)dY(z))=
dx(z) (\frac{\partial}{\partial x(z)})^r (f(z)\frac{dY(z)}{dx(z)})$
and
$(d_{z}\frac{1}{d\tilde{Y}(z)})^r (f(z)dx(z))=
d\tilde{Y}(z) (\frac{\partial}{\partial \tilde{Y}(z)})^r (f(z)\frac{dx(z)}{d\tilde{Y}(z)})$.
This equation simplifies:
\begin{theorem}
\label{thm:xydualtyforeqivariantduals}
  The  extended multi-differentials
$\mathbb{W}^{\vee(x,\tilde{Y})}_n(v_{\set{n}};z_{\set{n}})$
are related to   $\mathbb{W}^{\vee(x,Y)}_n(w_{\set{n}};  z_{\set{n}})$
by a $Y$-$\tilde{Y}$-duality
\begin{align}
  &\mathbb{W}^{\vee(x,\tilde{Y})}_n(v_{\set{n}};z_{\set{n}})
  -\frac{\delta_{n,1}}{\hbar v_1}
\label{bbW-double}
\\
&=\!\!\!\!\! \sum_{ s_1,...,s_n=0}^\infty \!\!\!
\Big(\prod_{i=1}^n \!\frac{e^{\frac{v_i}{\lambda} Y_i}}{d\tilde{Y}_i}
\Big({-}\lambda d_{z_i} \frac{1}{d \tilde{Y}_i}\Big)^{s_i} \Big)\!
\bigg(\!
\Big(\prod_{i=1}^n \! e^{-\frac{v_i}{\lambda} Y_i}
dY_i [w_i^{s_i}]\Big)
 \mathbb{W}^{\vee(x,Y)}_n(w_{\set{n}};
z_{\set{n}})\!\bigg),
\nonumber
\end{align}
where $Y_i=Y(z_i)$ and $\tilde{Y}_i=\tilde{Y}(z_i)$.
Consequently,
\begin{align}
  \label{varpi-double}
  \omega^{\vee(\tilde{X},\tilde{Y})}_{g,n}(z_{\set{n}})
  &=\!\!\!\! \sum_{ s_1,...,s_n=0}^\infty 
\prod_{i=1}^n \Big({-}\lambda d_{z_i}\frac{1}{d  \tilde{Y}_i}\Big)^{s_i} 
\bigg(\!
\Big(\prod_{i=1}^n dY_i [w_i^{s_i}]\Big)
\mathbb{W}^{\vee(X,Y)}_{g,n}(w_{\set{n}};
z_{\set{n}})\!\bigg).
\end{align}
\begin{proof}
    It suffices to prove the identity in a single set of variables/parameters $(v_i,u_i,w_i,z_i)$ while keeping the others with $j\neq i$ fixed. We omit the index $i$.  In (\ref{bbW-double-1}) we write
  $e^{\pm uY}=e^{\pm u(\tilde{Y}-\lambda x)}$.  Set
  $A_s(u,z)=e^{-\lambda u x}
  \frac{dY}{dx}[w^s]\mathbb{W}_n^{\vee(x,Y)}(w;z)$ and
  $B(u,v,z)=e^{vx+u\lambda x}$. Then the rhs of (\ref{bbW-double-1})
  in one family $\{r,s,v,u,w,z\}$ has the following structure:
\begin{align*}
  W(v;z)=
  e^{-vx}
\sum_{r,s=0}^\infty 
\Big({-}\frac{\partial}{
  \partial \tilde{Y}}\Big)^{r}
\bigg(
\frac{dx}{d\tilde{Y}} [u^{r}]\Big( B(u,v,z) e^{-u\tilde{Y}}
\Big({-}\frac{\partial}{\partial x}\Big)^{s}
\Big(e^{u\tilde{Y}}
A_{s}(u,z)\Big)\Big)\bigg).
\end{align*}
For the contribution with $s=0$ this agrees with (\ref{bbW-double}).
In general, \cite[Lemma~5.2]{Alexandrov:2024ajj} states 
(in our variables)
\begin{align*}
&\sum_{r=0}^\infty 
\Big({-}\frac{\partial}{
  \partial \tilde{Y}}\Big)^{r}
\bigg(
\frac{dx}{d\tilde{Y}} [u^{r}]\Big( B(u,v,z) e^{-u\tilde{Y}}
\Big(-\frac{\partial}{\partial x}\Big)^{s}
\Big(e^{u\tilde{Y}}
A_{s}(u,z)\Big)\Big)\bigg)
\\
&=
\sum_{r=0}^\infty 
\Big({-}\frac{\partial}{
  \partial \tilde{Y}}\Big)^{r}
\bigg(
\frac{dx}{d\tilde{Y}} [u^{r}]\Big( 
A_{s}(u,z)
\Big(\frac{\partial}{\partial x}\Big)^{s}  B(u,v,z)\Big)
\bigg).
\tag{*}
\end{align*}
The proof in \cite{Alexandrov:2024ajj} is given in the situation
of a single derivative acting on
$e^{u\tilde{Y}}A_{s}(u,z)$ and $B(u,v,z)$, and is 
based on
\begin{align*}
0 \equiv \sum_{r=0}^\infty \Big({-}\frac{\partial}{
  \partial \tilde{Y}}\Big)^{r}
[u^r]\Big(\Big(\frac{\partial}{\partial \tilde{Y}}+u\Big)
A(u,z)B(u,v,z)\Big).
\tag{**}
\end{align*}
Then (*) for any $s$ follows by induction on the number of derivatives on
$e^{u\tilde{Y}}A_{s}(u,z)$ and $B(u,v,z)$, where the decisive step is (**)
for adjusted $A,B$.

With 
$(\frac{\partial}{\partial x})^{s}  B(u,v,z)
=(v+\lambda u)^s B(u,v,z)$ we get
\begin{align*}
  W(v;z) &=e^{-vx}
\sum_{s=0}^\infty \sum_{r=0}^s \binom{s}{r} v^{s-r} \lambda^r
\Big({-}\frac{\partial}{
  \partial \tilde{Y}}\Big)^{r}
\bigg(
\frac{dY}{d\tilde{Y}} e^{vx}
[w^s]\mathbb{W}^{\vee(x,Y)}(w;z)
\Big)\bigg)
\\
&=e^{-vx}
\sum_{s=0}^\infty 
\Big(\frac{\partial}{\partial t}{-}\lambda \frac{\partial}{
  \partial \tilde{Y}}\Big)^{s}
\bigg(
\frac{dY}{d\tilde{Y}} e^{v(t+x)}
[w^s]\mathbb{W}^{\vee(x,Y)}(w;z)
\Big)\bigg)\Big|_{t=0}. \tag{***}
\end{align*}
From $\lambda x=\tilde{Y}-Y$ we get $(\frac{\partial}{\partial t}
-\lambda \frac{\partial}{
  \partial \tilde{Y}})^{r}e^{v(t+x)}|_{t=0}
=e^{\frac{v}{\lambda}\tilde{Y}}
  (-\lambda \frac{\partial} {\partial \tilde{Y}})^r e^{-\frac{v}{\lambda}Y}$.
We insert this into (***) and 
convert $d\tilde{Y}(\frac{\partial}{\partial \tilde{Y}})^r
\frac{1}{d\tilde{Y}}
= (d_{z_i}\frac{1}{d\tilde{Y}_i})^r$. Repeating the discussion for 
all $i\in \set{m}$ we confirm (\ref{bbW-double}).

Its $[v_1^0\cdots v_m^0]$-coefficient is
$\frac{\omega^{\vee(x,\tilde{Y})}_m(z_{\set{m}})}{\prod_{i=1}^m d\tilde{Y}(z_i)}$,
from  which (\ref{varpi-double}) follows with 
(\ref{trivial-shift-bottom}).
\end{proof}  
\end{theorem}

The functional relation (\ref{bbW-double}) strongly resembles a
$Y$-$\tilde{Y}$ duality. In addition, the variable
$x=\frac{1}{\lambda}(\tilde{Y}-Y)$ appears in
$\mathbb{W}^{\vee(x,Y)}_m$ only in the exponent
$w(\mathscr{S}(\hbar w \frac{\partial}{\partial Y})-1)x \equiv
w(\mathscr{S}(\hbar w \frac{\partial}{\partial Y})-1)
\frac{\tilde{Y}}{\lambda}$, and in $\mathbb{W}^{\vee(x,\tilde{Y})}_m$
via
$v(\mathscr{S}(\hbar v \frac{\partial}{\partial \tilde{Y}})-1)x
\equiv -v(\mathscr{S}(\hbar v \frac{\partial}{\partial \tilde{Y}})-1)
\frac{Y}{\lambda}$. As indicated, $x$ can be entirely expressed in
terms of the $Y$-$\tilde{Y}$ opposite variable, making the
$Y$-$\tilde{Y}$ duality (\ref{bbW-double}) even more striking.  The
only difference to a standard $x$-$y$ duality is that, as consequence
of $x=\frac{1}{\lambda}(\tilde{Y}-Y)$, the inverse transformation
comes with reflected $\lambda\mapsto -\lambda$.
\begin{align}
   &\mathbb{W}^{\vee(x,Y)}_n(w_{\set{n}};z_{\set{n}})
   -\frac{\delta_{n,1}}{\hbar w_1}
 \label{bbW-double-inverse}
 \\
 &=\!\!\!\!\! \sum_{ s_1,...,s_n=0}^\infty \!\!\!
 \Big(\prod_{i=1}^n \! \frac{e^{-\frac{w_i}{\lambda} \tilde{Y}_i}}{dY_i}
 \Big(\lambda d_{z_1}\frac{1}{d Y_i}\Big)^{s_i} \Big)\!
 \bigg(\!
 \Big(\prod_{i=1}^n \! e^{\frac{w_i}{\lambda} \tilde{Y}_i}
d\tilde{Y}_i [v_i^{s_i}]
\mathbb{W}^{\vee(x,\tilde{Y})}_n(v_{\set{n}};
 z_{\set{n}})\!\bigg).
 \nonumber
\end{align}

\begin{remark}
  The duality \eqref{bbW-double}$\leftrightarrow$\eqref{bbW-double-inverse}
  is a special case of a   {\sl symplectic duality}
  \cite{Bychkov:2022wgw,Alexandrov:2023oov,Alexandrov:2024ajj}, which can
  be seen as a composition of a symplectic shift and $x$-$y$ dualities. 

  For a system of multi-differentials $\omega_{g,n}$, we define the $x$-$y$ dual by $\omega_{g,n}^{\vee(x,y)}$. For this dual family, we now shift $x\to x+\psi(y)$ while leaving the actual system of multi-differentials invariant (in a sense, only $\omega_{0,1}$ is changed, which is actually not included in the system of multi-differentials in the sense of Def.~\ref{def:systemofcorrelators})\footnote{Usually, symplectic duality also involves a $\hbar^2$-deformation $\tilde{\psi}$ of $\psi$, which results in shifts of the $\omega_{1,n}^{\vee(x,y)}$. See \cite{Alexandrov:2024ajj} for details.}. Then, we consider for the $x$-$y$ dual $\omega_{g,n}^{\vee(x,y)}$ the transformation back using the newly chosen $x,y$, thus taking $(y,x+\psi(y))$ for this transformation and defining the symplectic dual $\omega_{g,n}^{\dagger(x,y,\psi)}$. Formally, this means
\begin{align*}
    \omega_{g,n}^{\dagger(x,y,\psi)}
    =(\omega_{g,n}^{\vee(x,y)})^{\vee(y,x+\psi(y))}.
\end{align*}
Fig.~\ref{fig:symplecticdual} gives the schematic idea behind the symplectic duality.
\begin{figure}[ht]
    \centering
    \begin{tikzpicture}[
        >=stealth,
        node distance=3.5cm and 6cm,
        every node/.style={align=center}
    ]

    \node (omega) 
    {$\omega_{g,n}$};

    \node[below=of omega] (omegavee)
    {$\omega_{g,n}^{\vee(x,y)}$};

    \node[right=of omega, xshift=1.5cm] (omegadagger)
    {$\omega_{g,n}^{\dagger(x,y,\psi)}$};

    \node[below=of omegadagger, yshift=1.5mm] (omegadaggervee)
    {$\omega_{g,n}^{\vee(x,y)}
    +\delta_{g,0}\delta_{n,1}\psi(y)dy$};

    \draw[<->]
    (omega) --
    node[above, font=\small] {symplectic duality}
    node[below, font=\small] {$(x,y)\mapsto(x+\psi(y),y)$}
    (omegadagger);

    \draw[<->]
    (omega) --
    node[left, font=\small] {$x$-$y$ duality}
    node[right, font=\small] {$(x,y)$}
    (omegavee);

    \draw[<->]
    (omegadagger) --
    node[right, font=\small] {$x$-$y$ duality}
    node[left, font=\small] {$(x+\psi(y),y)$}
    (omegadaggervee);

    \draw[<->]
    (omegavee) --
    node[above, font=\small]
    {$\omega_{g,n}^{\vee}$ invariant for $(g,n)\neq(0,1)$}
    node[below, font=\small]
    {$(y,x)\mapsto(y,x+\psi(y))$}
    (omegadaggervee);

    \end{tikzpicture}
    \caption{    \label{fig:symplecticdual}}
\end{figure}
It is important to emphasise that symplectic duality is actually independent of topological recursion and is a formal construction ultimately consisting of an explicit formula relating the two systems of multi-differentials $\omega_{g,n}$ and $\omega_{g,n}^\dagger$, depending on $x,y,\psi$. 
\end{remark}

The universal property of such dualities guarantees that 
also the partial dualities where only one variable is
moved to the other side of the (abstract Fock space)
$\mathcal{D}^{-1}$-operator
hold:
\begin{align}
&\mathcal{W}^{\vee(x,\tilde{Y})}_{1,n}(v;z;z_{\set{n}})
\label{calWVvee}
\\
&=\frac{\delta_{n,0}}{\hbar v}+\sum_{s=0}^\infty
\frac{e^{\frac{v}{\lambda} Y}}{d\tilde{Y}} \Big({-}\lambda d_{z}
\frac{1}{d  \tilde{Y}}\Big)^{s} 
\Big(e^{-\frac{v}{\lambda} Y}dY
[w^{s}] \mathcal{W}^{\vee(x,Y)}_{n+1}(w;z;z_{\set{n}})\Big).
\nonumber
\end{align}
Here the subscript $_{1,n}$ is understood in the sense of a
$Y$-$\tilde{Y}$ duality, and not with respect to $(x,\tilde{Y})$.
It induces
\begin{align}
\omega^{\vee(x,\tilde{Y})}_{1,n}(z;z_{\set{n}})
&:=d\tilde{Y} [v^0]\mathcal{W}^{\vee(x,\tilde{Y})}_{1,n}(v;z;z_{\set{n}})
\label{reflectionappendix-new}
\\
&=\sum_{s=0}^\infty
\Big({-}\lambda d_z \frac{1}{d \tilde{Y}}\Big)^{s} 
\Big(dY
[w^{s}] \mathcal{W}^{\vee(x,Y)}_{n+1}(w;z;z_{\set{n}})\Big).
\nonumber
\end{align}
On the other hand, repeating the proof of 
(\ref{reflectionproperty-new}) in the situation where only one
loop insertion is moved to the other side we get
$\omega^{\vee(x,\tilde{Y})}_{1,n}(z;z_{\set{n}})=
-\omega^{\vee(x,Y)}_{1+n}(-z,z_{\set{n}})$.
We have thus
established:
\begin{theorem}
\label{thm:reflection}
  The multi-differentials 
$\omega^{\vee(x,Y)}_{g,m}$ satisfy the reflection formula
\begin{align}
&\omega^{\vee(x,Y)}_{g,1+n}(-z_0,z_{\set{n}})
\label{reflection-formula}
\\
&=-\sum_{s=0}^\infty
\Big({-}\lambda d_{z_0}\frac{1}{d \tilde{Y}(z_0)}\Big)^{s} 
\Big(dY(z_0) 
[w^{s}\hbar^{2g+n-1}] \bigg\{
\frac{e^{\frac{w}{\lambda dY(z_0)} (\mathscr{S}(\hbar w
    d_{z_0}\frac{1}{d Y(z_0)})-1) \tilde{Y}(z_0) dY(z_0)}}{\hbar w}
\nonumber
\\
&\times
\sum_{\substack{\Gamma\in G_{\{0,1,...,n\}}\\
    \text{white vertices }\set{n} \\
    \text{one-valent}}} \!\!\!\!\!\!
 \frac{1}{|\mathrm{Aut}(\Gamma)|} 
\nonumber
\\[-3ex]
&\qquad\qquad\quad \times 
\prod_{\flat \in \mathcal{B}(\Gamma)} \!\!
\bigg(\!
\Big(\!\! \prod_{\substack{e \in I(\flat) \\ v(e)=0}} 
\frac{\hbar w}{dY(z_e)} \mathscr{S}\Big(\!\hbar w d_{z_e}
\frac{1}{d Y(z_e)}\Big)\Big) 
{:}\omega^{\vee(x,Y)}_{|I(\flat)|}{:}(z_{I(\flat)})
\!\bigg)_{\!z_e\mapsto z_{v(e)}}\bigg\}.
\nonumber
\end{align}
The graphical notation is explained in Def.~\ref{def:bipartite}.
The normal-ordered ${:}\omega^{\vee(x,Y)}_{|I(\flat)|}{:}$ is
defined in \eqref{varpi-normal}.
\end{theorem}

Converting to $\varpi^{\vee}_{g,n}(z_{\set{n}})=(-1)^{n}
\omega^{\vee(x,Y)}_{g,n}(z_{\set{n}})$,
we have thus proved a universal reflection formula for the multi-differentials
of the quartic Kontsevich model.

For $g=0$, (\ref{reflection-formula}) reduces to the involution identity
found in \cite{Hock:2021tbl}. Indeed, for $g=0$ only $\mathscr{S}=1$
contributes, and every black vertex $\flat$ can be incident to only one edge
$e$ with $v(e)=0$ (making the automorphism group trivial and
normal ordering absent):
\begin{align}
&\omega^{\vee(x,Y)}_{0,1+n}(-z_0,z_{\set{n}})
\label{reflection-formula-0}
\\
&=-\sum_{s=1}^{n} \frac{1}{s!}
\sum_{I_1\sqcup ...\sqcup I_s=\set{n}}^{\text{ordered}}
\Big({-}\lambda d_{z_0}\frac{1}{d \tilde{Y}(z_0)}\Big)^{s-1} 
\bigg(dY(z_0) 
\prod_{i=1}^s 
\frac{\omega^{\vee(x,Y)}_{0,1+|I_i|}(z_0,z_{I_i})}{dY(z_0)}
\bigg)\nonumber
\\
&=-d\tilde{Y}(z_0) \sum_{s=1}^{n} \frac{({-}\lambda)^{s-1}}{s}\!\!\!\!\!\!
\sum_{I_1\sqcup ...\sqcup I_s=\set{n}}^{\text{ordered}}
\Res\displaylimits_{q=z} \frac{d\tilde{Y}(q)}{(\tilde{Y}(q){-} \tilde{Y}(z_0))^s}
\bigg(\frac{dY(q)}{d\tilde{Y}(q)}
\prod_{i=1}^s 
\frac{\omega^{\vee(x,Y)}_{0,1+|I_i|}(q,z_{I_i})}{dY(q)}
\bigg).
\nonumber
\end{align}
The factor $\frac{1}{s!}$ in the second line corrects the overcounting resulting from the convention that $I_1\sqcup ...\sqcup I_s=\set{n}$ is the ordered set partition. With the following dictionary we confirm
\cite[Def.~1.1]{Hock:2021tbl}:
\begin{align}
  \label{dictionary}
\begin{array}{|c||c|c|c|} \hline 
  \text{this paper}          & Y(z) & \tilde{Y}(z) &
         \omega^{\vee(x,Y)}_{g,k+1}(z,z_{\set{k}}) \\ \hline
  \text{\cite{Hock:2021tbl}} & x(z) & -y(z) &
  (-\lambda)^{2g+k-1} \omega_{g,k+1}(u_1,...,u_k,z) \\ \hline
\end{array}
\end{align}

\subsection{Recursive formula at the ramification points}

The $x$-$Y$ duality between $\omega_{g,n}\equiv \omega_{g,n}^{(x,y)}$
and $\omega_{g,n}^{\vee(x,Y)}\equiv\omega_{g,n}^{\vee(X,Y)}$
permits, in principle, to compute all
$\omega_{g,n+1}^{\vee(X,Y)}$ by first computing the $\omega_{g,n}$ via
equivariant TR and then applying the universal $x$-$y$ duality formula as
described above. However, one can also ask what kind of recursive formula is
satisfied by $\omega_{g,n+1}^{\vee(x,Y)}$ itself at the different types of
poles. As follows from the construction and definition, these multi-differentials
have poles at the ramification points of $Y$, at the anti-diagonal, and starting from $g=1$ also at
the fixed points of the group action, namely $z=0$ (at $z=\infty$ there will be no pole). These points are of
different nature and satisfy different recursive formulae. In the following,
we derive the recursive formula at each type of point. There are no poles at 
the ramification points of $x$. This is exactly the statement of 
Cor.~\ref{cor:symplecticxyregularity}.

\medskip

Let us start with the simplest poles, namely those at the ramification points of
$Y(z)=R(z)$. The recursive formula governing these poles was already derived in
Sec.~\ref{sec:xtydualofequivariant}, see equation~\eqref{xtydualpolarpart}.
We simply recall the result:

\begin{corollary}\label{cor:recursivepoleatramR}
  Let $a_i^\vee$ be the zeros of $dY$ and let $\sigma^\vee_i(z)$ be the
  corresponding local Galois involution, that is, $
Y(z)=Y(\sigma^\vee_i(z)),  \sigma^\vee_i(a_i^\vee)=a_i^\vee$.
Then, the polar part of
$\omega_{g,n}^{\vee(x,Y)}(z_1,\ldots,z_n)$ at $z_1=a_i^\vee$ is recursively
determined by
\begin{align}
    &\Res\displaylimits_{z = a_i^\vee}
    \frac{\frac{\lambda}{2}\int_{\sigma_i^\vee(z)}^z B^{TR}(z_1,\bullet)}
    {\bigl(\tilde{Y}(z)-\tilde{Y}(\sigma_i^\vee(z))\bigr)\,dY(z)}
    \bigg(
        \omega^{\vee(x,Y)}_{g-1,n+1}
        (z,\sigma_i^\vee(z),z_2,\ldots,z_n)
        \label{eq:pole-avee}
        \\
    &\qquad\qquad\qquad\qquad\qquad
        +
        \sum_{\substack{
            g_1+g_2=g\\
            J_1\sqcup J_2=z_{\llbracket n\rrbracket\setminus\{1\}}\\
            (g_i,J_i)\neq(0,\emptyset)
        }}^{\text{ordered}}
        \omega^{\vee(x,Y)}_{g_1,1+|J_1|}(z,z_{J_1})
        \omega^{\vee(x,Y)}_{g_2,1+|J_2|}
        (\sigma_i^\vee(z),z_{J_2})
        \bigg).
        \nonumber
\end{align}
Thus, at the ramification points of $Y(z)=R(z)$, the
$\omega_{g,n}^{\vee(x,Y)}$ satisfy a TR-like recursion for the spectral
curve $(Y,\tilde{Y})$.

\begin{proof}
  This follows immediately from Cor.~\ref{cor:symplecticxyregularity}
  and formula~\eqref{xtydualpolarpart}
  with $\tilde{y}:=Y$ and $\tilde{a}_i^\vee=:a_i^\vee$.
We have 
$x=\frac{\tilde{Y}-Y}{\lambda}$.
Shifting $x \mapsto x+\frac{Y}{\lambda}$, which is invariant under the Galois involution, yields the denominator
\[
\frac{1}{x(z)-x(\sigma_i^\vee(z))}=\frac{\lambda}{\tilde{Y}(z)-\tilde{Y}(\sigma_i^\vee(z))}
\]
appearing in the recursion kernel.
\end{proof}
\end{corollary}

The dictionary (\ref{dictionary}) converts 
(\ref{eq:pole-avee}) into the corresponding part of 
\cite[eq.\ (1.5)]{Hock:2021tbl} for $g=0$ and 
\cite[eq.\ (1.3)]{Hock:2023nki} for $g \leq 1$ (when rewriting with the
linear and quadratic loop equation).

\subsection{Recursive formula at the anti-diagonal}
One can find different types of recursive formulae at the
anti-diagonal. One can just impose the universal $x$-$y$ duality
formula from the $\omega_{g,n}=:\omega_{g,n}^{(x,y)} \equiv
\omega_{g,n}^{(x,Y)}$ to
$\omega_{g,n}^{\vee(x,Y)}(z_1,\ldots,z_n)$ and get a formula which
will be of the complexity the $x$-$y$ duality formula itself by just
projecting to the pole at the anti-diagonal. However, the aim is to
derive a ``simple'' formula at the anti-diagonal which might be of
similar form as the ordinary TR-like formula. This however can not be
the case since at a ramification point $a_i$, there is the local
Galois involution $\sigma_i$ which is fundamental for the recursion at
a ramification point. Thus, since the TR-like formula at a
ramification point turns out to be equivalent to the global formula of
Gen-TR \cite{Alexandrov:2024tjo} (formula of
Prop.~\ref{prop:GenTRxydual}), one can ask if the polar part at the
anti-diagonal $\omega_{g,n}^{\vee(x,Y)}(z_1,\ldots,z_n)$ follows
the Gen-TR formula. This was proved to be not the case in
Prop.~\ref{prop:notGenTratantidiagonal}. So, is there a nice recursive
formula at the anti-diagonal and how does it look like? The following
proposition provides an answer to this question:
\begin{proposition}\label{prop:poleantidiagonal}
    The polar part of
$\omega_{g,n}^{\vee(x,Y)}(z_1,\ldots,z_n)$ at $z_1=-z_i$ is recursively
determined by
\begin{align*}
&  d_{z_i}\bigg[\Res\displaylimits _{q = -z_i}
  \frac{\lambda \int_{-z_i}^qB^{TR}(z_1,\bullet)}{2
    (\tilde{Y}(q)-\tilde{Y}(-z_i))dY(q)} \bigg\{
 -d^{-1}_{z_i}{:}\omega^{\vee(x,Y)}_{g-1,n+2}{:}(q,q,z_{\set n})
  \\
&  -
  \sum_{\substack{g_1+g_2+g\\ \llbracket n \rrbracket  \setminus \{1\} =J_1 \sqcup J_2\\ (g_i,J_i)\neq (0,\emptyset)}}^{\text{ordered}}
  d^{-1}_{z_i}\omega^{\vee(x,Y)}_{g_1,|J_1|+1}(q,z_{J_1})
  \omega^{\vee,(x,Y)}_{g_2,|J_2|+1}(q,z_{J_2})\\
    &+(dY(q))^2\sum_{h=0}^{g-1}\lambda^{1+2h}\frac{2B_{2h+2}}{(2h{+}2)!}
    \Big(\frac{\partial}{\partial Y(q)}\Big)^{2+2h}
    \Big(\frac{\partial}{\partial Y(z_i)}\Big)^{2h}
    \frac{\omega^{\vee(x,Y)}_{g-h-1,n+1}(q,z_{\set n})}{dY(q)dY(z_i)}
    \bigg\}\bigg],
  \end{align*}
  where $B_k$ are the Bernoulli numbers.
  In the first line we have
  ${:}\omega^{\vee(x,Y)}_{0,2}{:}(q,q):=
  \lim_{q'\to q}(\omega^{\vee(x,Y)}_{0,2}(q,q')-\frac{dY(q)dY(q')}{
    (Y(q)-Y(q'))^2})$ for $(g,n)=(1,0)$ and
${:}\omega^{\vee(x,Y)}_{g-1,n+2}{:}(q,q,z_{\set n})=
\omega^{\vee(x,Y)}_{g-1,n+2}(q,q,z_{\set n})$ otherwise.  
\begin{proof}
    The proof is outsourced to the App.~\ref{app:proofpoleantidiagonal}, since it is technical and lengthy.
\end{proof}
\end{proposition}

Note that $\omega_{g,n}^{\vee(x,Y)}\equiv \omega_{g,n}^{\vee(X,Y)}
=(-1)^n \varpi^\vee_{g,n}$, where $\varpi^\vee_{g,n}$ are the
multi-differentials associated with the quartic Kontsevich model.
With the dictionary (\ref{dictionary}) the formula in Prop.~\ref{prop:poleantidiagonal} specifies for
$g=0$ and $g=1$ to \cite[eq.~(1.3)]{Hock:2023nki}.
For $g=0$, the last line
in Prop.~\ref{prop:poleantidiagonal} drops out. For $g=1$, we only need
$h=0$, which gives an overall rational factor of
$\frac{2B_2}{2!}=\frac{1}{6}$, in agreement with
\cite[eq.~(1.3)]{Hock:2023nki}.

Let us emphasise the structure of the recursion. Interestingly, in contrast to
what usually appears in topological recursion, we have an exterior derivative
acting after the residue has been taken, namely an additional $d_{z_i}$. This
derivative does not commute with the recursion kernel
\[
\frac{\lambda \int_{-z_i}^q B^{TR}(z_1,\bullet)}
{2\bigl(\tilde{Y}(q)-\tilde{Y}(-z_i)\bigr)dY(q)},
\]
and therefore cannot cancel the integrations $d^{-1}_{z_i}$. Furthermore, the
last line exhibits a new type of contribution related to the $\cS$-function,
which appears throughout the universal $x$-$y$ duality.

At this point, the original motivation for the project comes to the surface.
For $g=0$, the same recursion was found in \cite{Hock:2021tbl}, based on
\cite{Branahl:2020yru}, for the quartic Kontsevich model with a very specific
choice of $R(z)$. The symmetry of the multi-differentials at $g=0$ defined by
the reflection property of Thm.~\ref{thm:reflection} was proved independently
in \cite{hock2025symmetry}. The recursive formula for the multi-differentials
of the quartic Kontsevich model was extended to $g=1$ in \cite{Hock:2023nki}.
Importantly, the pole at the anti-diagonal in
\cite[Thm.~1.1]{Hock:2023nki} is exactly of the form appearing in
Prop.~\ref{prop:poleantidiagonal}. Thus, we expect and conjecture that 
(\ref{eq:question}) in Question~\ref{mainquestion} is true, which means that 
poles of the $\omega_{g,n}$ of the quartic Kontsevich model at the
anti-diagonal follow the structure of Prop.~\ref{prop:poleantidiagonal} for
arbitrary genus.

The derivations in the articles on the quartic Kontsevich model were very
different and relied on Dyson--Schwinger equations; in particular, neither
$x$-$y$ duality nor symplectic transformations were used. It is therefore
quite striking that the recursive structure seems to be encoded in a
combination of equivariant topological recursion and the universal
$x$-$y$ duality.

\subsection{Recursive formula at zero}
There are several possible recursive formulae for the derivation of $\omega_{g,n}(z_1,\ldots,z_n)$ at $z_i=0$. Since we are interested in the principal part of the Laurent series at this point derived by a residue formula, one can ask for the \emph{simplest} recursive formula. As a comment, in some cases of ordinary TR, the Gen-TR recursive formula is equivalent to TR. Thus, we regard the ordinary TR recursion as simple \cite{Alexandrov:2024tjo}, whereas Gen-TR is not really simple. We have in principle already established a rather complicated formula for the principal part of the Laurent series at the fixed points of the group action in Prop.~\ref{prop:GenTRxydual}. One can easily see that the proof is adaptable to this setting. Thus, one can show that the following recursive formula holds.
\begin{corollary}\label{cor:GenTRatz=0}
The polar part of
$\omega_{g,n}^{\vee(x,Y)}(z_1,\ldots,z_n)$ at $z_1=0$ is recursively
determined by
\begin{align*}
&\Res\displaylimits_{q = 0}\int^q_{0}B^{TR}(z_1,\bullet)
\sum_{r=1}^\infty  \Big({-}d_q\frac{1}{dx(q)}\Big)^{r}[w^{r}]
\mathcal{W}_{g,n}^{\vee(x,Y)}(w;q;z_{2},\ldots,z_n)dY(q),
\end{align*}
where the 1-variable extended $n$-differential $\mathcal{W}_{g,n}^{\vee(x,Y)}$
is defined in Def.~\ref{def:1variableextended}, or equivalently in
Cor.~\ref{cor:1variableextended}, by just replacing $\omega_{g,n}$ by
$\omega_{g,n}^{\vee(x,Y)}$ and swapping $x$ and $Y$.
\begin{proof}
This literally follows the same argument as used in Prop.~\ref{prop:GenTRxydual} for the spectral curve defined by \eqref{eqGWspectralcurve}. Taking the system $\omega_{g,n}^{\vee(x,Y)}$ and performing the duality with
\begin{align*}
(\omega_{g,n}^{\vee(x,Y)})^{\vee(Y,x)}
\end{align*}
generates a system with no pole at $z=0$, which is why the argument of
Prop.~\ref{prop:GenTRxydual} applies. However, if one takes the system
$(\omega_{g,n}^{\vee(x,Y))})^{\vee(Y,\tilde{Y})}$, Gen-TR would not
hold (with $x$ replaced by $\tilde{Y}$), because the dual family has a
pole at $z=0$. Thus, one has to be careful about which functions
$(\tilde{x},\tilde{y})$ are actually used in this $x$-$y$ duality
formula.
\end{proof}
\end{corollary}

As mentioned before, we will not regard a Gen-TR recursive formula as \emph{simple}, because it involves a heavy combinatorial definition which is hidden in the 1-variable extended $n$-differential. For $g=1$, we found a simpler though equivalent formula for the polar part of $\omega_{1,n}^{\vee(x,Y)}$ at $z=0$ which is exactly of the form as of \cite{Hock:2023nki} found by so-called extended loop equations. 
\begin{proposition}\label{prop:poleatzero}
    The polar part of
$\omega_{1,n}^{\vee(x,Y)}(z_1,\ldots,z_n)$ at $z_1=0$ is recursively
determined by
\begin{align*}
&\Res\displaylimits_{z = 0}K_0(z;q)	 	\Big\{{:}\omega^{\vee(x,Y)}_{0,n+1}{:}(q,q,z_2,...,z_n)+
\sum_{\substack{ I_1\sqcup I_2=\set{n}\setminus\{1\}\\ g_1+g_2=1\\
(g_i,I_i)\neq (0,\emptyset)   }}^{\text{ordered}}
\omega^{\vee(x,Y)}_{g_1,|I_1|+1}(q,z_{I_1})
\omega^{\vee(x,Y)}_{g_2,|I_2|+1}(q,z_{I_2})
	 	\nonumber\\
&\qquad\qquad +\frac{(d Y(q))^2}{2}
\frac{\partial}{\partial Y(q)}
\Big( \frac{d_{q'}^{-1} 
{:}\omega^{\vee(x,Y)}_{0,n+1}{:}(q,q',z_2,...,z_n)}{dY(q)}\Big|_{q'=q}
\Big)\Big\} ,
\end{align*}
where
${:}\omega^{\vee(x,Y)}_{0,2}{:}(q,q)
=\lim_{q'\to q}(\omega_{0,2}(q,q')-\frac{dY(q)dY(q')}{(Y(q)-Y(q'))^2})$
for $n=1$ and
${:}\omega^{\vee(x,Y)}_{0,|I|+2}{:}(q,q,z_2,...,z_n)=
\omega^{\vee(x,Y)}_{0,|I|+2}(q,q,z_2,...,z_n)$ otherwise. The 
recursion kernel is given by
$K_0(z;q)=-\frac{ \int^q_0B^{TR}(\bullet,z) }{2x(q)dY(q)}
=\frac{-\lambda \int^q_0B^{TR}(\bullet,z) }{2(\tilde{Y}(q)-Y(q))dY(q)}
$.
\begin{proof}
    The proof can be found in App.~\ref{app:proofpoleatzero}.
\end{proof}
\end{proposition}

\section{Summary and outlook}
\subsection{Dualities in equivariant topological recursion}

After independently reviewing equivariant topological recursion and
$x$-$y$ duality in the theory of topological recursion in
Sec.~\ref{sec:xydualtiyandequivariantTRgeneral}, we formulated the
corresponding dualities ($x$-$y$ and symplectic) in the theory of
equivariant topological recursion in
Sec.~\ref{sec:xydualofequivariantgeneral}. These results apply to
global equivariant spectral curves satisfying the stated assumptions.
We explicitly computed the pole structure of the corresponding dual
families, starting from a family $\omega_{g,n}$ coming from equivariant
topological recursion.

In doing so, it was important to distinguish the actual $x$-$y$ dual
family $\omega_{g,n}^{\vee(x,y)}$ from its symplectic dual
families. The former has not only poles at the ramification points of $y$
but also at the fixed points of the group action.
We proved that the principal parts of the Laurent expansions at those
points admit a recursive structure. In the case of simple ramification
points of $y$ that are distinct from the fixed points of the group
action, these polar parts satisfy a topological recursion-like
formula; see Cor.~\ref{cor:polarpartxydualside}. At the fixed points,
including those possibly coinciding with ramification points of $y$, another
recursive formula for the polar parts can be formulated in terms of
Gen-TR; see Prop.~\ref{prop:GenTRxydual}.

Instead of studying the $x$-$y$ dual multi-differentials, one can study
a more general family related to them by symplectic duality. We studied
the pole structure of these symplectic duals
$\omega_{g,n}^{\vee(x,\tilde{y})}$, where $\tilde{y}$ is not necessarily
the function used in equivariant topological recursion, but is related
to it by a shift involving $x$. It turned out that this dual family
$\omega_{g,n}^{\vee(x,\tilde{y})}$ may have additional poles in comparison to the family $\omega_{g,n}^{\vee (x,y)}$.
Notably, we found that poles may occur along the anti-diagonals,
corresponding to the poles already appearing in the equivariant
$\omega_{0,2}$. Surprisingly, for the $x$-$y$ dual family
$\omega_{g,n}^{\vee(x,y)}$, these anti-diagonal poles are absent for $2g-2+n>0$. We explained how the  different polar parts can be computed recursively purely in terms of
previously known $\omega_{g',n'}^{\vee(x,\tilde{y})}$.

These different recursive polar structures are then related explicitly
to the so-called quartic Kontsevich model in
Sec.~\ref{sec:spectralcurveZ2}, after we revisited the Fock space
formulation of Kontsevich-type matrix models in
Sec.~\ref{sec:Kontsevich} (even more general for arbitrary potentials). The proposed
correlators of the quartic Kontsevich model are obtained from an
equivariant topological recursion spectral curve by a chain of
$x$-$y$ and symplectic duality transformations, see Fig.~\ref{fig:duality-network}. The final differentials of interest
are denoted by $\omega_{g,n}^{\vee(x,Y)}$, and their identification
with the matrix-model correlators remains conjectural in higher genera.

\subsection{Recursive derivation of $\omega^{\vee(x,Y)}_{g,n}$}

We summarise the different approaches
to compute $\omega^{\vee(x,Y)}_{g,n}$. The starting point was a spectral curve of the form of $\mathbb{Z}/2\mathbb{Z}$-equivariant topological recursion
\begin{align}\label{eqGWspectralcurve1}
    \mathcal{S} = (\mathbb{P}^1,\phi,x,y,B,\chi,\upsilon,\beta),
\end{align}
where
\begin{align*}
    x(z)&=\frac{R(z)-R(-z)}{(-\lambda)},\qquad
    y(z)=\frac{R(z)+R(-z)}{2},
    \\
    \phi_\gamma(z)&=\gamma z=-z,\quad \text{for }\gamma\neq 1,
\end{align*}
$\chi=\beta$ is the sign representation, $\upsilon$ is the trivial
representation, and $R(z)=Y(z)$ is a rational function with simple
poles given by (\ref{X0Y0-4}). This means, in particular, that
\begin{align*}
   B(z_1,z_2)=\frac{dz_1\,dz_2}{(z_1-z_2)^2}+\frac{dz_1\,dz_2}{(z_1+z_2)^2}.
\end{align*}
For this curve, equivariant topological recursion (Def.~\ref{defn:equiv:SC}) defines recursively the system of multi-differentials denoted by $\omega_{g,n}$. Then, we \emph{define} a second system of differentials derived from $\omega_{g,n}$ by the $x$-$Y$ duality. This system is denoted by $\omega^{\vee(x,Y)}_{g,n}$ and is, by definition, obtained by using
\begin{align*}
    (x,Y)
\end{align*}
in the universal $x$-$y$ duality formula of Def.~\ref{def:xydualmultidiff}. Importantly, $\omega^{\vee(x,Y)}_{0,2}=\omega_{0,2}=B$ of \eqref{extendedBergman} needs to be used. Another important remark is that in the recursive definition of equivariant TR \eqref{eqn:TR}, one can use $Y
=y-\frac{\lambda}{2}x$ instead of $y$. The point is that $Y(z)=R(z)$ does
not satisfy any group action transformation, but the obtained $\omega_{g,n}$ are still the same.

This gives the first recursive way to compute all $\omega^{\vee(x,Y)}_{g,n}$, just by their definition, which is the $x$-$Y$ dual of $\omega_{g,n}$:
\begin{align}
&  \omega_{g,n}^{\vee(x,Y)}(z_{\set{n}})
\label{omegavee-final}
\\
&=(-1)^n [\hbar^{2g+n-2}]\!\!\!\! 
\sum_{r_1,...,r_n\geq 0} \!\!
\bigg(      \prod_{i=1}^n\Big({-}d_{z_i}\frac{1}{dY_i}\Big)^{r_i}[u_i^{r_i}]
\bigg)\bigg\{      \prod_{i=1}^n
\frac{dx_i}{u_i\hbar} 
  e^{\frac{u_i}{dx_i}\big(\cS(u_i\hbar \,d_{z_i}
          \frac{1}{dx_i})-1\big) Y_idx_i}
      \nonumber
    \\
        &\times \sum_{\Gamma\in G_{\set{n}}} 
      \frac{1}{|\mathrm{Aut}(\Gamma)|}\bigg(
     \prod_{\flat\in \mathcal{B}(\Gamma)}
     \prod_{e\in I(\flat)} \!\frac{u_e\hbar}{dx_e}
     \mathscr{S}\Big( u_e \hbar  \, d_{z_e} \frac{1}{dx_e}\Big)
\nonumber
\\
& \qquad\qquad\times
\Big( \!\!\!\! \sum_{\substack{g'=0 \\ (g',|\flat|)\neq (0,1)}}^\infty \!\!\!\!\!\!\!
\hbar^{2g'+|\flat|-2}
\omega_{g',|\flat|}(z_{I(\flat)})-\delta_{|\flat|,2}\delta_{v(e_1),v(e_2)}
\frac{dx_{e_1}dx_{e_2}}{(x_{e_1}{-}x_{e_2})^2}\Big)
     \bigg)\Big|_{\substack{z_e\mapsto z_{v(e)}\\
         u_e\mapsto u_{v(e)}}}\bigg\},
     \nonumber
   \end{align}
   where $x_i=x(z_i)$, $x_e=x(z_e)$, $Y_i=Y(z_i)$ and $\mathscr{S}(t)
=\frac{\sinh (t/2)}{t/2}$.  
The graphical language is explained in Def.~\ref{def:bipartite}.

From this, we have established several facts about the system of multi-differentials $\omega^{\vee(x,Y)}_{g,n}$. We have shown that a further $x$-$y$ duality transformation of those multi-differentials coming from a symplectic duality
gives rise to the reflection formula
(\ref{reflection-formula}) which expresses
$\omega^{\vee(x,Y)}_{g,n+1}(-z_0,z_{\set{n}})$ in terms of
$\omega^{\vee(x,Y)}_{g',m}(\underbrace{z_0,...,z_0}_k,z_{I_{m-k}})$
with $2g'+m\leq 2g+n+1$ and $I_{m-k}$ subsets of $z_{\set{n}}$ of cardinality
$m-k$.
This provides a highly non-trivial symmetry structure on this system of multi-differentials.

Furthermore, we concluded from the general theory of $x$-$y$ duality in equivariant topological recursion, which we developed in Sec.~\ref{sec:xydualofequivariantgeneral}, that the $\omega^{\vee(x,Y)}_{g,n}$ have poles only at the ramification points of $Y$, at the anti-diagonal $z_i=-z_j$, and, for $g>0$, also at the fixed point $z_i=0$. These three different sets of poles were further characterised. We found recursive formulae for all three sets of poles:
\begin{itemize}
    \item at the ramification points of $Y$: the pole structure is of TR-like form, see Cor.~\ref{cor:recursivepoleatramR};
    \item at the anti-diagonal $z_i=-z_j$: the pole structure is recursively given by Prop.~\ref{prop:poleantidiagonal};
    \item at $z_i=0$: the pole structure can be computed by Gen-TR, see
      Cor.~\ref{cor:GenTRatz=0}. For $g=1$ we have the simpler formula of Prop.~\ref{prop:poleatzero}.
\end{itemize}
Let us also mention that for the poles at the ramification points of $Y$, the Gen-TR recursive formula also holds, but not for the poles at the anti-diagonal, as follows from Prop.~\ref{prop:notGenTratantidiagonal}!

Last but not least, we have shown that up to $g=1$, the recursive formulae described above for $\omega_{g,n}^{\vee(x,Y)}$ agree exactly with
the recursive formulae for multi-differentials of the quartic Kontsevich model of \cite{Hock:2023nki} and the specific choice
(\ref{X0Y0-4}) of $R(z)=Y(z)$. Note that in \emph{op.cit.} only a formula up to $g=1$ was derived. Thus, the natural conjecture is:
\begin{conjecture}\label{conj:mainconjecture}
  The multi-differentials of the quartic Kontsevich model as defined in
  \cite{Branahl:2020yru} agree up to a conventional factor $(-\lambda)^{2g-2+n}$ with
  the $\omega^{\vee(x,Y)}_{g,n}$ given by \eqref{omegavee-final}, or with
  the equivalent recursive form provided in
  Cor.~\ref{cor:recursivepoleatramR}, Prop.~\ref{prop:poleantidiagonal}
  and Cor.~\ref{cor:GenTRatz=0}. Equivalently, the answer to Question
  \ref{mainquestion} is `yes'.
\end{conjecture}

\subsection{Outlook}

Our interest in the quartic Kontsevich model originally arose from a different
subject, the study of a scalar quantum field theory with quartic interaction
on noncommutative Moyal space with harmonic propagation
\cite{Grosse:2004yu,Grosse:2012uv}. At a critical parameter value (self-duality point), this theory admits a
matrix formulation whose finite-matrix counterpart is the quartic Kontsevich
model, which thus has independent interest in theoretical physics. The
corresponding higher-dimensional quantum field theory limits require renormalisation
to provide finite results. This quantum field theory was introduced by the
last-named author together with Harald Grosse and is known as the
Grosse--Wulkenhaar model. The highly structured relations among its
correlation functions have long motivated the search for an underlying
integrable hierarchy.

Thanks to results established in this paper, we can now outline two possible paths to integrability. First, 
the Fock space formulation (\ref{Z-Fock}) of the partition function (\ref{Z-KontsevichV}) 
is, for any potential $P$, \emph{formally} an Orlov-Scherbin tau function \cite{Orlov:2001ped, Bychkov:2020ujd}. One family of time variables is integrated out, so it remains a KP tau function in the times
$s_k=\mathbf{e}_k=\frac{1}{N}\mathrm{Tr}(E^k)$, where $E$ is the external matrix that defines the matrix model (\ref{Z-KontsevichV}). The proof that it is a true KP tau functions requires more work. One should define (\ref{Z-KontsevichV}) 
not as integral over $H_N$ but over $\mathcal{U}(N)\times [-T,T]^N$ and measure (\ref{Jacobian}). The IZHC character expansion (\ref{IZHC}) should be viewed as convergent sum over partitions $\lambda\vdash d$, followed by a sum over $d$ which can only be expected to be asymptotic (see \cite{Goulden:2011qn}). 
Moreover, one should establish the Hirota bilinear relations directly in the Schur basis $\hat{s}_\lambda(\mathbf{e})$ instead of the Fock space representation. The latter is based on the Cauchy-Littlewood identity (\ref{Cauchy-Littlewood}) which does not converge in our situation. A particular  Hirota bilinear relation involves Schur functions $\hat{s}_\lambda(\mathbf{e})$ with $\lambda \vdash d$
for $d\leq d_0$, so that asymptoticity of the IZHC character expansion should suffice for a proof.

Our conjectural all-genus recursive construction
(Conj.~\ref{conj:mainconjecture}) suggests a possible second path towards identifying
an integrable hierarchy for the matrix model and, potentially, for the
corresponding higher dimensional quantum field theory. The idea comes from the conjectural
connection of the quartic Kontsevich differentials
$\omega_{g,n}^{\vee(x,Y)}$ to $\mathbb{Z}/2\mathbb{Z}$-equivariant topological recursion
via $x$-$y$ and symplectic duality. Motivated by known examples
\cite{Giacchetto:2021atk,Alexandrov:2021izt,ABDKS2026SpinHypermaps}, we expect that the multi-differentials
$\omega_{g,n}$ in Fig.~\ref{fig:duality-network}, generated by eq-TR, give rise
to a BKP tau-function. If our Conj.~\ref{conj:mainconjecture} holds, these differentials would be
related to the quartic Kontsevich model by the described chain of dualities.
This raises the question of whether BKP integrability is preserved under
$x$-$y$ and symplectic duality transformations. This question has recently
been partially answered in \cite[Thm.~B.16]{ABDKS2026SpinHypermaps}, which
establishes preservation under $x$-$y$ duality. The theorem also extends to certain symplectic
dualities, with additional conditions on $\tilde{y}=y+\psi(x)$ with the involution. However, the duality chain used here is
not directly covered by their assumptions. In particular, the shift
$\widetilde{y}=y\pm x$ does not in general preserve the required transformation under
the $\mathbb{Z}/2\mathbb{Z}$-involution. Thus, the BKP property of the quartic Kontsevich model does
not follow immediately from this theorem. Establishing the BKP property of
the initial equivariant system $\omega_{g,n}$, and understanding its behaviour under the
full chain of dualities, remain interesting questions.

On the other hand, in \cite{Borot:2023thu} it was shown that a
Kontsevich-like matrix model with an arbitrary
potential gives rise to a BKP tau-function in formal odd deformation
parameters of the potential. In particular, the even part can be kept
fixed. From the viewpoint of the original quantum field theory,
however, our main interest lies in the fixed quartic interaction and in a
hierarchy governed by the natural external matrix $E$. The distinction
relevant here is therefore the choice of integrable times:
\cite{Borot:2023thu} establishes BKP relations in the odd deformation
parameters, whereas the natural variables in our approach, as explained in
Sec.~\ref{sec:Kontsevich}, are associated with traces $s_k=\mathbf{e}_k=\frac{1}{N}\mathrm{Tr}(E^k)$ 
of powers of the external matrix $E$. These are precisely the KP time variables in the first path
outlined above. Note that all these time variables differ from the KdV times 
$t_{2k+1}=-(2k-1)!!\mathrm{Tr}(E^{-2k-1})$ 
of the original Kontsevich model \cite{Kontsevich:1992ti}.
Connecting these descriptions in the quartic case
remains an interesting problem. Furthermore, the analytic interpretation
of the formal expansions studied here and their relation to the
renormalised quantum field theory require further investigation.

Another direction is the general question of whether the correlators
studied in the Kontsevich-like matrix model of
Sec.~\ref{sec:Kontsevich} are related to eq-TR for more general even
potentials. In particular, one may ask whether the symmetry induced by the
evenness of the potential can be realised as a group action on spectral
data required for eq-TR. The quartic potential is the first
even polynomial example in this class of matrix models. A first step would
be to identify the disk and annulus amplitudes from the Dyson--Schwinger
equations (in the quartic case identified in \cite{Grosse:2019jnv, Branahl:2020yru}). For potentials of higher degree, these equations will clearly be of a different more involved form. The
disk amplitude might not satisfy a quadratic equation anymore, which makes the solution far more complicated.

\begin{appendix}

\section{Multi-differentials and functional relations}

\label{app:functionalrelations}

Our goal is to prove functional relations between the
multi-differentials $\omega^\vee_n$ defined in (\ref{omega-complex})
and Fock space expectation values where operators are inserted
to the other side (left) of $\mathcal{D}^{-1}$.
We follow closely the strategy of
\cite{Bychkov:2022wgw, Alexandrov:2022ydc}, which themselves rely
on \cite{Bychkov:2020ujd}. Because of
different conventions regarding $\mathcal{D}^{-1}$ and KP time variables
we cannot simply copy the formulae.

\subsection{The operator $\mathcal{E}$}

The strategy is based on the action of the Lie algebra
$\widehat{\mathfrak{gl}}(\infty)=\mathfrak{gl}(\infty)+\mathbb{C}\mathbf{1}$,
a one-dimensional central extension of the Lie algebra
$\mathfrak{gl}(\infty)$ of infinite band matrices, whose rows and columns
are labelled by half-integers and have only finitely many non-zero diagonals. Let
$\{\hat{E}_{i,j}\}_{i,j \in \mathbb{Z}+\frac{1}{2}}$ be its (standard)
matrix basis. Its representation on $\mathbb{C}\set{p_1,p_2,...}$ is
defined by the generating series
\begin{align}    
  \sum_{k,\ell\in \mathbb{Z}+\frac{1}{2}}
  x^\ell y^{-k} \hat{E}_{k,\ell}
  =x^{1/2}y^{1/2} 
  \frac{e^{\sum_{i=1}^\infty (y^{-i}-x^{-i}) p_i/i }
e^{\sum_{i=1}^\infty (x^{i}-y^{i}) \partial_{p_i}  }-1}{x-y}.
\label{hatE}
\end{align}
The operators $\hat{E}_{k,l}$ satisfy the commutation
relations 
\begin{align}
[\hat{E}_{a,b},\hat{E}_{c,d}]  
= \delta_{b,c} \hat{E}_{a,d}-\delta_{a,d} \hat{E}_{c,b}
+ \delta_{b,c} \delta_{a,d} (\delta_{b>0}-\delta_{d>0}) \mathbf{1}.
\end{align}
Setting $x=z e^{u/2}$ and $y=z e^{-u/2}$ and abbreviating
$\mathscr{S}(u)= \frac{e^{u/2}-e^{-u/2}}{u}$ as in (\ref{def:S}), the series (\ref{hatE})
is turned into 
\begin{align}    
\mathcal{E}(u,z)&:=  \sum_{\ell\in \mathbb{Z}+\frac{1}{2}}\sum_{m\in \mathbb{Z}}
 e^{u(\ell-m/2)} z^m \hat{E}_{\ell-m,\ell}
 \nonumber
 \\
 &=
  \frac{e^{\sum_{i=1}^\infty z^{-i} u \mathscr{S}(ui) J_{-i}}
e^{\sum_{i=1}^\infty z^i u\mathscr{S}(ui) J_i}-1}{u \mathscr{S}(u)}.
\label{calEuz}
\end{align}
The $[u^0][z^m]$-coefficient is 
$[u^0][z^m]\mathcal{E}(u,z)=  \sum_{\ell\in \mathbb{Z}+\frac{1}{2}}
\hat{E}_{\ell-m,\ell} = J_m$.
In particular, the identity matrix 
$\sum_{\ell\in \mathbb{Z}+\frac{1}{2}}\hat{E}_{\ell,\ell}$ acts as zero.
Its r\^ole is taken by the central extension $\mathbb{C}\mathbf{1}$.
The central extension also rescues the commutation relation
$[J_k,J_l]=k\delta_{k+l,0}$.

Another input is the fact that diagonal
matrices in $\mathfrak{gl}(\infty)$ act diagonally on the Schur basis:
\begin{align}
  \sum_{k\in \mathbb{Z}+\frac{1}{2}} w_k \hat{E}_{k,k}
  \hat{s}_\lambda(\mathbf{p})=  
  \sum_{(i,j)\in \lambda} \big(w_{j-i+\frac{1}{2}}
  -w_{j-i-\frac{1}{2}}\big)
  \hat{s}_\lambda(\mathbf{p}).
\end{align}
Here, we sum over the boxes of the Young diagram associated with the partition $\lambda$. Comparison with (\ref{D-Borot}) identifies
\begin{align}
  \label{calD}
  \mathcal{D}
  = \exp\Big(   \sum_{k\in \mathbb{Z}+\frac{1}{2}} d_k \hat{E}_{k,k}\Big)
  \quad\text{with}\quad
d_{n+1/2}-d_{n-1/2}=\log(1+\hbar n),
\end{align}
for $n \in \mathbb{Z}$. Note that the $d_k$'s are only defined up to a
common constant. It can be neglected since $e^{J_0}$ acts as the
identity on vectors that belong to the fermionic Fock space of charge
0.

We will later insert operators
$\mathcal{E}(\hbar u,X)$ and $\mathcal{E}^\vee(\hbar w,1/Y)$ into
the Fock space expectation value (\ref{Z-Fock}) for
$\log Z_P(E_{\set{N}})$.  Inserting $\mathcal{E}(\hbar u,X)$ to the
left of $\mathcal{D}^{-1}$ is the same as inserting
$\mathcal{D} \mathcal{E}(\hbar u,X) \mathcal{D}^{-1}$ to the right of
$\mathcal{D}^{-1}$. For that we need a result which is a reformulation
of \cite[eq.\ (59)]{Bychkov:2022wgw}:
\begin{proposition}
  \begin{subequations}
    \begin{align}
\mathcal{D}
  \mathcal{E}(\hbar u,X)  \mathcal{D}^{-1} 
  &= \sum_{m\in \mathbb{Z}}
X^{-m}
\sum_{r=0}^\infty
\partial_\theta^r \tilde{\phi}_m(\theta,u)\Big|_{\theta=0}
[w^r] [Y^{m}] \mathcal{E}(\hbar w,\tfrac{1}{Y}),
\label{adED}
\\
\mathcal{D}^{-1}
  \mathcal{E}(\hbar w,\tfrac{1}{Y})  \mathcal{D}
  &= \sum_{m\in \mathbb{Z}}
Y^{-m}
\sum_{r=0}^\infty
\partial_\theta^r \tilde{\phi}_m(\theta,w)\Big|_{\theta=0}
[u^r] [X^{m}] \mathcal{E}(\hbar u,X),
\label{adED-2}
\end{align}
\end{subequations}
where 
$\tilde{\phi}_m(\hbar p,u):=e^{d_{p+m/2}-d_{p-m/2}+p\hbar u}$.
The $d_k$ are related to $\mathcal{D}$ by \eqref{calD}.
\begin{proof}
Recall that we implicitly work in the $\widehat{\mathfrak{gl}}(\infty)$-representation
$\mathcal{D}=\exp\big(\sum_{l\in \mathbb{Z}+1/2} d_l \hat{E}_{l,l}\big)$ where
differences of adjacent diagonal entries are given by
$d_{k+\frac{1}{2}}-d_{k-\frac{1}{2}}=\log(1+\hbar k)$, for $k\in \mathbb{Z}$. 
This leads to
\begin{subequations}
 \begin{align}
  \mathcal{D} \mathcal{E}(\hbar u,X)  \mathcal{D}^{-1}
&=
\sum_{\ell \in \mathbb{Z}+\frac{1}{2}} \sum_{m\in \mathbb{Z}}
e^{d_{\ell+m}-d_\ell}
e^{u\hbar (\ell+m/2)} X^{-m} \hat{E}_{\ell+m,\ell}
\nonumber
\\
&=\sum_{m\in \mathbb{Z}}
\sum_{\ell \in \mathbb{Z}+\frac{1}{2}} 
\tilde{\phi}_m(\hbar(\ell+m/2),u)
X^{-m} \hat{E}_{\ell+m,\ell}
,
\label{adED-3a}
\\
  \mathcal{D}^{-1} \mathcal{E}(\hbar w,\tfrac{1}{Y})  \mathcal{D}
&=\sum_{\ell \in \mathbb{Z}+\frac{1}{2}} \sum_{m\in \mathbb{Z}}
e^{d_{\ell}-d_{\ell-m}}
e^{w\hbar (\ell-m/2)} Y^{-m} \hat{E}_{\ell-m,\ell}
\nonumber
\\
&=\sum_{m\in \mathbb{Z}}
\sum_{\ell \in \mathbb{Z}+\frac{1}{2}}
\tilde{\phi}_m(\hbar(\ell-m/2),w)
Y^{-m} \hat{E}_{\ell-m,\ell}.
\label{adED-3b}
\end{align}
\end{subequations}
We have 
\begin{subequations}
\label{DED}
\begin{align}
\tilde{\phi}_m(\hbar(\ell + m/2),u)\hat{E}_{\ell+ m,\ell}
&=
\sum_{r=0}^\infty \frac{(\hbar(\ell + m/2))^r}{r!} 
\partial_\theta^r \tilde{\phi}_m(\theta,u)\Big|_{\theta=0}
\hat{E}_{\ell + m,\ell}
\nonumber
\\
&=
\sum_{r=0}^\infty
\partial_\theta^r \tilde{\phi}_m(\theta,u)\Big|_{\theta=0}
[w^r] [Y^{m}] \mathcal{E}(\hbar w,1/Y),
\\
\tilde{\phi}_m(\hbar(\ell - m/2),w)\hat{E}_{\ell- m,\ell}
&=
\sum_{r=0}^\infty \frac{(\hbar(\ell - m/2))^r}{r!} 
\partial_\theta^r \tilde{\phi}_m(\theta,w)\Big|_{\theta=0}
\hat{E}_{\ell - m,\ell}
\nonumber
\\
&= \sum_{r=0}^\infty
\partial_\theta^r \tilde{\phi}_m(\theta,w)\Big|_{\theta=0}
[u^r] [X^{m}] \mathcal{E}(\hbar u,X)\,,
\end{align}
\end{subequations}
where (\ref{calEuz}) was used. Inserting this into
(\ref{adED-3a}) and (\ref{adED-3b}) gives the assertion.
\end{proof}
\end{proposition}

\begin{remark}
  \label{remark:DXY}
  Compared with \cite[eq.\ (59)]{Bychkov:2022wgw}, our $(X,Y)$ are
  $(1/Y,1/X)$ in \cite{Bychkov:2022wgw}, caused by the fact that
  $\mathcal{D}^{-1}$ in this paper is $\mathcal{D}$ in \cite{Bychkov:2022wgw}.
\end{remark}

\begin{lemma}[\cite{Bychkov:2020ujd}]
  For $\mathcal{D}=\exp\big(\sum_{l\in \mathbb{Z}+1/2} d_l \hat{E}_{l,l}\big)$
  with $d_{k+\frac{1}{2}}-d_{k-\frac{1}{2}}=\log(1+\hbar k)$
one has 
\begin{align}
 d_{p+m/2}-  d_{p-m/2}  &= m \frac{\mathscr{S}(m\hbar \partial_\theta)}{\
   \mathscr{S}(\hbar \partial_\theta)} \log (1+\theta)\Big|_{\theta=\hbar p},
 \nonumber
 \\
 \tilde{\phi}_m(\theta,u)&=e^{\theta u+
   m \frac{\mathscr{S}(m\hbar \partial_\theta)}{\
     \mathscr{S}(\hbar \partial_\theta)} \log (1+\theta) }.
\label{phi-Kazarian}
 \end{align}
 \begin{proof}
   We have
   $d_{p+|m|/2}- d_{p-|m|/2} = \sum_{i=-(|m|-1)/2}^{(|m|-1)/2}
   (d_{p+i+1/2}-d_{p+i-1/2})$ with
   $d_{p+i+1/2}-d_{p+i-1/2}= \log(1+\hbar p + \hbar i)$.  One writes
   $\log(1+\hbar p + \hbar i)= e^{\hbar i \partial_\theta}
   \log(1+\theta)\Big|_{\theta=\hbar p}$ and evaluates the geometric
   sum
   $ \sum_{i=-(|m|-1)/2}^{(|m|-1)/2} e^{\hbar i \partial_\theta} = |m|
   \frac{\mathscr{S}(m\hbar \partial_\theta)}{\ \mathscr{S}(\hbar
     \partial_\theta)}.$ Discussion of $m>0$ and $m<0$ gives the first
   equation (\ref{phi-Kazarian}); the second one follows when replacing
 $\hbar p\mapsto \theta$ in
 $\tilde{\phi}_m(\hbar p,u):=e^{d_{p+m/2}-d_{p-m/2}+p\hbar u}$.
\end{proof}
\end{lemma}

This turns (\ref{adED}) and (\ref{adED-2}) into
\begin{subequations}
  \begin{align}
&\mathcal{D}^{-1}
  \mathcal{E}(\hbar w,\tfrac{1}{Y})  \mathcal{D}
\label{adED-3}
\\
&=  \sum_{m\in \mathbb{Z}}
Y^{-m}
\sum_{r=0}^\infty
 \partial_\theta^r\Big(e^{\theta w+
   m \frac{\mathscr{S}(m\hbar \partial_\theta)}{
     \mathscr{S}(\hbar \partial_\theta)} \log (1+\theta) }
\Big)
\Big|_{\theta=0}
[u^r] [X^{m}] \mathcal{E}(\hbar u,X).
\nonumber
\\
&\mathcal{D} \mathcal{E}(\hbar u,X)  \mathcal{D}^{-1} 
\label{adED-4}
\\
&=  \sum_{m\in \mathbb{Z}}
X^{-m}
\sum_{r=0}^\infty
 \partial_\theta^r\Big(e^{\theta u+
   m \frac{\mathscr{S}(m\hbar \partial_\theta)}{
     \mathscr{S}(\hbar \partial_\theta)} \log (1+\theta) }
\Big)
\Big|_{\theta=0}
[w^r] [Y^{m}] \mathcal{E}(\hbar w,\tfrac{1}{Y}).
\nonumber
\end{align}
\end{subequations}

\subsection{Reparametrisation}

We introduce the following functions:
\begin{subequations}
  \begin{align}
  \mathscr{A}(u) &:=2\mathrm{artanh}(u/2)=\log\frac{1+u/2}{1-u/2},
  \\
  \mathscr{C}(u) &:=\cosh \big(\tfrac{1}{2}\mathscr{A}(u)) =
\frac{1}{\sqrt{1-u^2/4}}.
\end{align}
\end{subequations}
Following \cite{Alexandrov:2022ydc}
we introduce a reparametrisation of (\ref{calEuz}):
\begin{subequations}
  \begin{align}    
 \hat{\mathcal{E}}(u,X)&:=
\mathscr{C}(u)
  \mathcal{E}\big(\mathscr{A}(u),X/\mathscr{C}(u)\big) +\frac{1}{u}
  \nonumber
  \\
 &= 
 \frac{e^{\sum_{i=1}^\infty 
     \mathscr{S}(u,-i) X^{-i}J_{-i}}
    e^{\sum_{i=1}^\infty \mathscr{S}(u,i) X^i J_i}}{u},
\label{calEuz-new}
\\
 \hat{\mathcal{E}}^\vee(w,Y)&:=
\mathscr{C}(w)
  \mathcal{E}\big(\mathscr{A}(w),\mathscr{C}(w)/Y\big) +\frac{1}{w}
  \nonumber
  \\
  &= 
  \frac{e^{\sum_{i=1}^\infty \mathscr{S}(w,i) Y^i J_{-i}}
    e^{\sum_{i=1}^\infty \mathscr{S}(w,-i) Y^{-i} J_i}}{w},
\end{align}
\end{subequations}
where 
\[
  \mathscr{S}(w,i)
=\frac{(1+w/2)^i - (1-w/2)^{i}}{i}.
\]
We have
for any $i\in \mathbb{Z}\setminus \{0\}$
\begin{align}
  \mathscr{S}(\hbar w/x,i) x^i
  &=  \hbar w \sum_{k=0}^\infty \frac{1}{i}\binom{i}{2k+1}
  \Big(\frac{\hbar w}{2}\Big)^{2k}  x^{i-2k-1}
  \nonumber
  \\
  &=  \hbar w \sum_{k=0}^\infty \frac{1}{(2k+1)!} 
  \Big(\frac{\hbar w}{2}\Big)^{2k} \frac{\partial^{2k}}{\partial x^{2k}}
  \Big(\frac{x^{i}}{x}\Big) \nonumber \\
  &=\hbar w \mathscr{S}(\hbar w \partial_x)  \Big(\frac{x^{i}}{x}\Big)
  \label{scrS-x}
\end{align}
 and thus
\begin{subequations}
  \label{hatE-scrS}
\begin{align}
   \frac{1}{Y} \hat{\mathcal{E}}^\vee(\hbar w/Y,Y)
   &=\frac{1}{\hbar w}
   e^{\hbar w \mathscr{S}(\hbar u \frac{\partial}{\partial Y})
     \sum_{i=1}^\infty Y^{i-1} J_{-i}}
   e^{\hbar w \mathscr{S}(\hbar w \frac{\partial}{\partial Y})
     \sum_{i=1}^\infty Y^{-i-1} J_{i}},
\label{hatE-scrS-Y}
   \\
   \frac{1}{X} \hat{\mathcal{E}}(\hbar u/X,X)
   &=\frac{1}{\hbar u}
 e^{\hbar u \mathscr{S}(\hbar u \frac{\partial}{\partial X})
     \sum_{i=1}^\infty X^{-i-1} J_{-i}}
   e^{\hbar u \mathscr{S}(\hbar u \frac{\partial}{\partial X})
     \sum_{i=1}^\infty X^{i-1} J_{i}}.
 \end{align}
\end{subequations}
The limit $i \to 0$ of (\ref{scrS-x}) proves the identity 
\begin{align}
\mathscr{A}(\hbar w/x)= \hbar w \mathscr{S}(\hbar w \partial_x)\frac{1}{x}.
\label{identity-AS}
\end{align}

\subsection{Loop insertions}

To condense our equations we introduce formal \emph{loop
insertion operators} which insert to the left and right of the
distinguished operator $\mathcal{D}^{-1}$:
\begin{subequations}
\label{loopins-DZ}
   \begin{align}
  \overleftarrow{D}_{Y} \langle 0|\mathcal{O}_1
  \mathcal{D}^{-1} \mathcal{O}_2|0\rangle
  &:=  
  \Big\langle 0\Big|\mathcal{O}_1 
  \mathcal{D}^{-1} \sum_{k=1}^\infty \frac{Y^kJ_{-k}}{k}
  \mathcal{O}_2\Big|0\Big\rangle,
  \\
  \overleftarrow{D}_{Y}^{\mathbb{Z}} \langle 0|\mathcal{O}_1
  \mathcal{D}^{-1} \mathcal{O}_2|0\rangle
  &:=  
  \Big\langle 0\Big|\mathcal{O}_1
  \mathcal{D}^{-1} \Big(\sum_{k\in \mathbb{Z}\setminus \{0\}}
\frac{Y^k J_{-k}}{k}\Big)
  \mathcal{O}_2\Big|0\Big\rangle,
  \\
  \overrightarrow{D}_{X} \langle 0|\mathcal{O}_1
  \mathcal{D}^{-1} \mathcal{O}_2|0\rangle
  &:=  
  \Big\langle 0\Big|\mathcal{O}_1 
  \sum_{k=1}^\infty \frac{X^kJ_{k}}{k}
  \mathcal{D}^{-1}
  \mathcal{O}_2\Big|0\Big\rangle,
  \\
 \overrightarrow{D}_{X}^{\mathbb{Z}} \langle 0|\mathcal{O}_1
  \mathcal{D}^{-1} \mathcal{O}_2|0\rangle
  &:=  
  \Big\langle 0\Big|\mathcal{O}_1
  \Big(\sum_{k\in \mathbb{Z}\setminus \{0\}}
\frac{X^k J_{k}}{k}\Big)
  \mathcal{D}^{-1}
  \mathcal{O}_2\Big|0\Big\rangle.
\end{align}  
\end{subequations}
Loop insertions act as derivations, i.e.\ they respect linearity,
Leibniz rule and chain rule. They were
a central combinatorial tool e.g.\ in \cite{Hock:2022wer,Hock:2023nki}. 
For example, if
$Z=\langle 0|\mathcal{O}_1 \mathcal{D}^{-1} \mathcal{O}_2|0\rangle$,
then $\overleftarrow{D}_{Y} \log Z=\frac{1}{Z}\overleftarrow{D}_{Y} Z$
and
$\overleftarrow{D}_{Y_1}\overleftarrow{D}_{Y_1} \log Z
=\frac{1}{Z}\overleftarrow{D}_{Y_1}\overleftarrow{D}_{Y_2} Z
-(\frac{1}{Z}\overleftarrow{D}_{Y_1}Z)
(\frac{1}{Z}\overleftarrow{D}_{Y_2}Z) $. They allow us to write
(\ref{omega-complex}) short as
\begin{align}
  \omega^\vee_n(Y_{\set{n}})
  = \mathcal{P}_n d_{Y_1}\cdots d_{Y_n}
  \overleftarrow{D}_{Y_1}\cdots \overleftarrow{D}_{Y_n} \log Z_P(E_{\set{N}})
  +\delta_{n,2}\frac{dY_1dY_2}{(Y_1-Y_2)^2},
\label{omega-complex-loop}
\end{align}
where $d_{Y_i}$ is the exterior differential in $Y_i$ and 
$Z_P(E_{\set{N}})$ was given in (\ref{Z-Fock}).
The connected expectation values (\ref{moment-cumulant}) are
automatically produced by Leibniz rule.
We have identified $
\mathcal{O}_1=\int d_P\vec{m}\;
e^{\sum_{l=1}^\infty \frac{1}{l\hbar } J_{l} \mathbf{m}_{-l}}$
(which depends on the potential $P$)
and  $\mathcal{O}_2=e^{\sum_{k=1}^\infty \frac{1}{k\hbar } J_{-k} \mathbf{e}_{k}}$
(which depends on the eigenvalues $E_{\set{N}}$).

We will later see that loop insertions of (\ref{hatE-scrS}) produce,
among others, a dependence on
\begin{align}
  \Theta^\vee(Y)&:=Y\frac{\partial}{\partial Y}
  [\hbar^{-1}] \overleftarrow{D}^{\mathbb{Z}}_Y \log Z_P(E_{\set{N}})  ,
  &
\Theta(X)&:=X\frac{\partial}{\partial X}
[\hbar^{-1}] \overrightarrow{D}^{\mathbb{Z}}_X \log Z_P(E_{\set{N}})  .
\label{def-Theta}
\end{align}
We view $\Theta^\vee,\Theta$ as particular functions of
$Y$ and $X$, respectively.
It will later become clear that it is more convenient to rescale
(\ref{hatE-scrS}) by these functions:
\begin{definition}
\label{def:scrE}
  The following operators are introduced:
  \begin{subequations}
    \begin{align}
\mathbb{E}^\vee(w,Y)&:=
      \frac{e^{\frac{1}{\hbar}\mathscr{A}(\hbar w/Y) -w\frac{(1+\Theta^\vee(Y))}{Y}}}{Y}
    \hat{\mathcal{E}}^\vee(\hbar w/Y,Y),
\label{def:scrEY}
   \\
   \mathbb{E}(u,X)&:=
   \frac{e^{\frac{1}{\hbar}\mathscr{A}(\hbar u/X) -u\frac{(1+\Theta(X))}{X}}}{X}
 \hat{\mathcal{E}}(\hbar u/X,X).
\label{def:scrEX}
\end{align}
\end{subequations}
The following loop insertions of these operators are introduced:
 \begin{subequations}
\label{loopins-hatE}
   \begin{align}
 \overleftarrow{D}_{\mathbb{E}^\vee(w,Y)} \langle 0|\mathcal{O}_1
  \mathcal{D}^{-1} \mathcal{O}_2|0\rangle
  &:=  
  \langle 0|\mathcal{O}_1
  \mathcal{D}^{-1} \mathbb{E}^\vee(w,Y)
  \mathcal{O}_2|0\rangle,
  \\
  \overrightarrow{D}_{\mathbb{E}(u,X)} \langle 0|\mathcal{O}_1
  \mathcal{D}^{-1} \mathcal{O}_2|0\rangle
  &:=  
  \langle 0|\mathcal{O}_1
  \mathbb{E}(u,X)\mathcal{D}^{-1} \mathcal{O}_2|0\rangle.
\end{align}  
\end{subequations}
The following expectation values are introduced:
\begin{subequations}
  \label{calWW-m+1n}
  \begin{align}
  \mathbb{W}_m^\vee(w_{\set{m}};  Y_{\set{m}})
  &:=  
\bigg(  \prod_{\substack{i=1 \\
|Y_{\pi(1)}|>|Y_{\pi(2)}|>...>|Y_{\pi(m)}|}}^m \hspace*{-2em}
\overleftarrow{D}_{\mathbb{E}^\vee(w_ {\pi(i)},  Y_{\pi(i)})}\bigg)
  \log Z_P(E_{\set{N}}),
\label{def-bbWvee}
  \\
\mathbb{W}_m(u_{\set{m}};  X_{\set{m}})
&:= 
\bigg(  \prod_{\substack{i=1 \\
|X_{\pi(1)}|>|X_{\pi(2)}|>...>|X_{\pi(m)}|}}^m \hspace*{-2em}
\overrightarrow{D}_{\mathbb{E}(u_{\pi(i)},X_{\pi(i)})}
\bigg) \log Z_P(E_{\set{N}}),
\label{def-bbW}
\\
\mathcal{W}^\vee_{m+1}(w;Y;Y_{\set{m}})
&:=
d_{Y_1}\cdots d_{Y_m} \overleftarrow{D}_{\mathbb{E}^\vee(w ,Y)}
\overleftarrow{D}_{Y_{[[m]]}}
\log Z_P(E_{\set{N}}),
  \\
\mathcal{W}_{1,m}(u;X;Y_{\set{m}})
&:=d_{Y_1}\cdots d_{Y_m}\overleftarrow{D}_{Y_{[[m]]}}
\overrightarrow{D}_{\mathbb{E}(u,X)}
\log Z_P(E_{\set{N}}).
\end{align}
More generally, the following mixed expectation values are introduced:
\begin{align}
  &\mathbb{W}_{n,m}(u_{\set{n}}; w_{\set{m}};  X_{\set{n}}; Y_{\set{m}})
  \\
  &:=
\bigg(  \prod_{\substack{i=1 \\
|Y_{\pi(1)}|>...>|Y_{\pi(m)}|}}^m \hspace*{-1em}
\overleftarrow{D}_{\mathbb{E}^\vee(w_ {\pi(i)},  Y_{\pi(i)})}\bigg)
\bigg(  \prod_{\substack{i=1 \\
|X_{\pi(1)}|>...>|X_{\pi(n)}|}}^n \hspace*{-1em}
\overrightarrow{D}_{\mathbb{E}(u_{\pi(i)},X_{\pi(i)})}
\bigg) \log Z_P(E_{\set{N}}).
\nonumber
\end{align}
\end{subequations}
\end{definition}
\noindent
We have $\mathbb{W}_{0,m}(\emptyset; w_{\set{m}};  \emptyset; Y_{\set{m}})
=\mathbb{W}^\vee_{m}(w_{\set{m}};Y_{\set{m}})$ and
$\mathbb{W}_{n,0}(u_{\set{n}};  \emptyset; X_{\set{n}};\emptyset)
=\mathbb{W}_{n}(u_{\set{n}};X_{\set{n}})$.

From (\ref{scrS-x}) and (\ref{hatE-scrS}) we conclude
\begin{subequations}
\label{bbE}
  \begin{align}
&\mathbb{E}^\vee(w,Y)
\label{bbE-Y}
\\
&=\frac{e^{w \mathscr{S}(\hbar w \partial_Y)\frac{1}{Y}
    - \frac{w(1+\Theta^\vee(Y))}{Y}}}{\hbar w}
   e^{\hbar w \mathscr{S}(\hbar w \frac{\partial}{\partial Y})
     \sum_{i=1}^\infty Y^{i-1} J_{-i}}
   e^{\hbar w \mathscr{S}(\hbar w \frac{\partial}{\partial Y})
     \sum_{i=1}^\infty Y^{-i-1} J_{i}},
   \nonumber
   \\
&\mathbb{E}(u,X)\label{bbE-X}
\\
&=\frac{e^{u \mathscr{S}(\hbar u \partial_X)\frac{1}{X}
    - \frac{u(1+\Theta(X))}{X}}}{      \hbar u}
  e^{\hbar u \mathscr{S}(\hbar u \frac{\partial}{\partial X})
     \sum_{i=1}^\infty X^{-i-1} J_{-i}}
   e^{\hbar u \mathscr{S}(\hbar u \frac{\partial}{\partial X})
     \sum_{i=1}^\infty X^{i-1} J_{i}}.
   \nonumber
 \end{align}
\end{subequations}
In particular,
\begin{subequations}
  \label{hatE-u0}
  \begin{align}
\mathbb{E}^\vee(w,Y) 
&= 
\frac{1}{\hbar w}
-\frac{\Theta^\vee(Y)}{\hbar  Y}
+\sum_{i\in \mathbb{Z}\setminus \{0\}} Y^{i-1}J_{-i}
  +\mathcal{O}(w^1)  ,
  \label{scrEvee-w0}
  \\
\mathbb{E}(u,X) 
&=
\frac{1}{\hbar u}
-\frac{\Theta(X)}{\hbar  X}
+\sum_{i\in \mathbb{Z}\setminus \{0\}} X^{i-1}J_i
  +\mathcal{O}(u^1)  .
\end{align}
\end{subequations}

\begin{remark}
\label{rem:radial}
  The operators
$\mathbb{E}^\vee$ and 
$\mathbb{E}$ do not commute. The products in \eqref{calWW-m+1n}
are always understood in
standard order $\prod_{i=1}^m D_i= D_1D_2\cdots D_m$. Via 
\eqref{loopins-hatE}, the largest $|Y_i|$ or $|X_i|$ are inserted
next to $\mathcal{D}^{-1}$; then the next-largest $|Y_i|$ or $|X_i|$
is placed to the other side (seen from $\mathcal{D}^{-1}$)
of the first insertion, and so on in decreasing radial order away
from $\mathcal{D}^{-1}$.
We will later see that the insertions are only well-defined if the given
radial ordering of $|Y_i|$ and $|X_i|$ are respected. This radial
ordering makes
$\mathbb{W}_m^\vee$ and $\mathbb{W}_m$ symmetric in all their arguments. 
\end{remark}

\subsection{Functional relations I: from $\mathbb{W}^{(\vee)}_m$
  to $\omega^{(\vee)}_m$}

Our goal is to prove functional relations between the functions in
Def.~\ref{def:scrE}. We first study the 
$[w^0]$-coefficients.
Inside vacuum expectation values
involving $[w^0]\overleftarrow{D}_{\mathbb{E}^\vee(w,Y)}$
we commute the $J_{k}$ with $k>0$
to the right where they annihilate the vacuum $|0\rangle$.
The commutation relation $[J_k,J_l]=k\delta_{k+l,0}$ implies
\begin{subequations}
  \begin{align}
\Big[
\sum_{k=1}^\infty  Y^{-k-1} J_{k},
e^{\sum_{l=1}^\infty \frac{J_{-l}\mathbf{e}_l}{l\hbar}}
\Big]
=\frac{e^{\sum_{l=1}^\infty \frac{J_{-l}\mathbf{e}_l}{l\hbar}}}{\hbar Y}
\sum_{k=1}^\infty  Y^{-k} \mathbf{e}_k,
\label{Jneg-contrib-new-1}
\\
\Big[
\sum_{l_i=1}^\infty  Y_i^{-l_i-1} J_{l_i},
\sum_{k_j=1}^\infty  Y_j^{k_j-1} J_{-k_j}\Big]
= \frac{1}{Y_iY_j} \sum_{l=1}^\infty l \frac{Y_j^l}{Y_i^l}
=: \frac{1}{(Y_i-Y_j)^2}.
\label{Jneg-contrib-new-2}
\end{align}
\end{subequations}
In (\ref{Jneg-contrib-new-2}), according to Remark~\ref{rem:radial},
these operators
are placed with $|Y_i|>|Y_j|$, which makes the series well-defined.
The resulting extra terms 
$\frac{1}{Y}\sum_{k=1}^\infty  Y^{-k} \mathbf{e}_k$ and 
$\frac{1}{(Y_i-Y_j)^2}$ only contribute to $m=1$ and
$m=2$, respectively. They are independent of Fock space operators,
can be taken out of the vacuum expectation values and cancel 
in the decomposition (\ref{moment-cumulant}) unless $m=1$ or $m=2$.
Only the series $\sum_{l_i=1}^\infty  Y_i^{l_i-1} J_{-l_i}
= \frac{1}{dY_i} d_{Y_i}\big( \sum_{l_i=1}^\infty  \frac{1}{l_i}
Y_i^{l_i} J_{-l_i}\big)
$ with a negatively indexed $J$
remain. This gives:
\begin{lemma}
  \label{lemma:WW-u0}
The following relations hold:
\begin{subequations}
 \begin{align}
  \Theta^\vee(Y)
  =  Y\frac{\partial}{\partial Y} [\hbar^{-1}]
  (\overleftarrow{D}_Y \log Z_P(E_{\set{N}})) &+
  \sum_{k=1}^\infty \mathbf{e}_kY^{-k},
  \label{def:Theta1}
  \\
  [w_1^0\cdots w_m^0]
  \mathbb{W}^\vee_m(w_{\set{m}};Y_{\set{m}})&=
  \frac{\omega^\vee_m(Y_1,...,Y_m) }{\prod_{i=1}^m dY_i},
  \label{bbWvee-omega}
  \\
    [u_1^0\cdots u_m^0]
  \mathbb{W}^\vee_{m+1}(u_0,u_{\set{m}};Y_0,Y_{\set{m}})&=
  \frac{\mathcal{W}^\vee_{m+1}(u_{0};Y_0;Y_{\set{m}})}{\prod_{i=1}^m dY_i}
  .
  \label{bbWvee-calWvee}
\end{align}
\end{subequations}
\begin{proof}
Let us write $Z_P(E_{\set{N}})=\langle 0|\mathcal{O}_1
\mathcal{D}^{-1} 
\mathcal{O}_2|0\rangle$.
  For $m=1$ we have from (\ref{scrEvee-w0})
  \begin{align*}
[w^0]\overleftarrow{D}_{\mathbb{E}^\vee(w,Y)}
\log Z_P(E_{\set{N}})
&= \frac{
\langle 0|\mathcal{O}_1
\mathcal{D}^{-1} (-\frac{\Theta^{\vee}(Y)}{\hbar Y}
+ \sum_{i\in \mathbb{Z}\setminus\{0\}} Y^{i-1}J_{-i} )
\mathcal{O}_2|0\rangle}{
\langle 0|\mathcal{O}_1
\mathcal{D}^{-1} 
\mathcal{O}_2|0\rangle}
\\
&= -\frac{\Theta^{\vee}(Y)}{\hbar Y}
+\frac{\partial}{\partial Y}
\overleftarrow{D}^{\mathbb{Z}}_Y \log Z_P(E_{\set{N}}).
\tag{*}
\end{align*}
Comparing with (\ref{def-Theta}) we
confirm $[\hbar^{-1}]
  [w^0]
  \mathbb{W}^\vee_1(w;Y)=0$ as required by the definition of
  $\omega^\vee_1$. We have $\mathcal{O}_2
  =e^{\sum_{l=1}^\infty \frac{J_{-l}\mathbf{e}_l}{l\hbar}}$. By
  (\ref{Jneg-contrib-new-1}), the $J_{>0}$ of (\ref{scrEvee-w0}) contribute
  $\frac{1}{\hbar Y}\sum_{k=1}^\infty  Y^{-k} \mathbf{e}_k$ when
  commuting to the vacuum. This means that (*) evaluates further to
  \begin{align*}
[w^0]\overleftarrow{D}_{\mathbb{E}^\vee(w,Y)}
\log Z_P(E_{\set{N}})
&= -\frac{\Theta^{\vee}(Y)}{\hbar Y}
+\frac{1}{\hbar Y}\sum_{k=1}^\infty  Y^{-k} \mathbf{e}_k
+\frac{\partial}{\partial Y}
\overleftarrow{D}_Y \log Z_P(E_{\set{N}}).
\tag{**}
\end{align*}
The (vanishing) $[\hbar^{-1}]$-coefficient gives (\ref{def:Theta1}). At higher
$\hbar$-order we get in (**)
  \begin{align*}
    (1-\hbar^{-1}[\hbar^{-1}]) &
    [w^0]\overleftarrow{D}_{\mathbb{E}^\vee(w,Y)}
\log Z_P(E_{\set{N}})
\\
&=    (1-\hbar^{-1}[\hbar^{-1}])
\frac{\partial}{\partial Y}
\overleftarrow{D}_Y \log Z_P(E_{\set{N}})
\\
&=    (1-\hbar^{-1}[\hbar^{-1}])
\frac{1}{dY} d_Y
\overleftarrow{D}_Y \log Z_P(E_{\set{N}})
\equiv \frac{\omega_1^\vee(Y)}{dY}.
\end{align*}
This proves (\ref{bbWvee-omega}) for $m=1$. 

By (\ref{moment-cumulant}), a constant shift in a loop insertion into
$\log Z_P(E_{\set{N}})$ only contributes for a single such insertion
(i.e.\ $m=1$).
For $m=2$, the part $\sum_{k=1}^\infty Y_2^{-k-1}J_{k}$ of
(\ref{hatE-u0}) contributes $\frac{1}{(Y_1-Y_2)^2}$ from
(\ref{Jneg-contrib-new-2}). This means that for $|Y_1|>|Y_2|$ we have
\begin{align*}
  [w_1^0w_2^0]\overleftarrow{D}_{\mathbb{E}^\vee(w_1,Y_1)}
  \overleftarrow{D}_{\mathbb{E}^\vee(w_2,Y_2)}
\log Z_P(E_{\set{N}})
&=\frac{1}{(Y_1-Y_2)^2}
\\
&+\frac{\partial}{\partial Y_1}\frac{\partial}{\partial Y_2}
\overleftarrow{D}_{Y_1}\overleftarrow{D}_{Y_2} \log Z_P(E_{\set{N}}).
\end{align*}
Comparing with (\ref{omega-complex-loop}) for $\omega^\vee_2$ we
confirm (\ref{bbWvee-omega}) for $m=2$. Finally, by (\ref{moment-cumulant}) the constant terms
$\frac{1}{(Y_i-Y_j)^2}$ cancel in $\geq 3$ loop insertions into $\log Z$,
which proves (\ref{bbWvee-omega}) for $m\geq 3$. The discussion of (\ref{bbWvee-calWvee}) is similar.
\end{proof}  
\end{lemma}

Note that (\ref{def:Theta1}) agrees with (\ref{Theta-prelim}).
The identity (\ref{bbWvee-omega}) suggests:
\begin{definition}
\label{def:varpiX}
  Multi-differentials $\omega_m$ are defined by
  \begin{align}
  \omega_m(X_1,...,X_m):=
    [u_1^0\cdots u_m^0]
  \mathbb{W}_m(u_{\set{m}};u_{\set{m}}) \prod_{i=1}^m  dX_i.
  \label{bbW-omega}
\end{align}
\end{definition}
\noindent
Similar as in the proof of Lemma~\ref{lemma:WW-u0}, we have
$[\hbar^{-1}]  \omega_1(X)=0$. 
\begin{remark}
  The relations \eqref{bbWvee-omega} and \eqref{bbW-omega} 
  have a totally different nature. The $\omega_m^\vee$ are defined
  via derivatives of the partition function $Z_P(E_{\set{N}})$
  of the 
  Kontsevich matrix model with arbitrary
  potential (\ref{Z-KontsevichV}), see Def.~\ref{def:Omega}.
  By \eqref{omega-complex-loop} they are 
  given by multiple
  loop insertion of $\overleftarrow{D}_{Y_i}$-operators (which involve only
  $J_{<0}$) into $\log Z_P(E_{\set{N}})$. The identity  \eqref{bbWvee-omega} is a
  non-trivial statement. 
  In contrast, \eqref{bbW-omega} is a definition which provides
  a matrix integral formula for $\omega_m$. This formula is, however,
  very complicated as commutators of the negatively indexed part
  $J_{<0}$ of $\overrightarrow{D}^{\mathbb{Z}}_{X_i}$ through $\mathcal{O}_1$
  towards the covacuum $\langle 0|$ depend on the integration variables
  $m_i$ and cannot be taken out of the matrix integral.
  Whereas 
  $\overleftarrow{D}^{\mathbb{Z}}_{Y_i}-\overleftarrow{D}_{Y_i}$ only contributes to the lowest $\hbar$-order of $m=1$ and $m=2$, the difference
  $\overrightarrow{D}^{\mathbb{Z}}_{X_i}-\overrightarrow{D}_{X_i}$
  contributes to $\omega_m$ for any $m$ and any $\hbar$-order!
\end{remark}

\begin{remark}
\label{remark-negativemodes}
  In $\omega^\vee_m$, the modes $Y_i^{-k}$ with $k\in \mathbb{Z}_{>0}$
  are only present in the extra term $\frac{dY_1dY_2}{(Y_1-Y_2)^2}$.
  In contrast, $\omega_m(X_1,...,X_m)$ contains
  $X_i^{-k}$ with $k\in \mathbb{Z}_{>0}$ at any $m$ and
  any $\hbar$-order when commuting  
  $\sum_{k=1}^\infty X_i^{-k} J_{-k}$ through the factor 
  $e^{\sum_{l=1}^\infty \frac{J_{l} \mathbf{m}_{-l}}{l\hbar }}$ of the
  partition function. The commutator depends on the $\mathbf{m}_{-l}$,
  which are functions of the integration variables and cannot be taken
  out of the vacuum expectation value.  See Ex.~\ref{ex:om02Fock}.

\end{remark}

\subsection{Functional relations II: from $\omega_\bullet^{(\vee)}$
  to   $\mathbb{W}_m^{(\vee)}$}

Our next goal is to express 
$\mathbb{W}^\vee_m(w_{\set{m}};Y_{\set{m}})$ in terms of the $\omega_\bullet^\vee$
functions for any $w_i$. The most convenient approach goes via
normal ordering.
Let $p(J)$ be a polynomial or power series in
the operators $J_k$. Since $[J_k,J_l]=k\delta_{k+l,0}$, the order of factors
matters. The \emph{normal ordering}
${:} p(J_k) {:}$ is defined as the formal rearrangement of
monomials in $J_k$ such that all $J_{>0}$ are placed to the right of all
$J_{<0}$, without employing $[J_k,J_l]=k\delta_{k+l,0}$ to rearrange.
We can then write (\ref{bbE}) as
\begin{subequations}
\label{bbE-normal}
  \begin{align}
\mathbb{E}^\vee(w,Y) 
&=\frac{e^{w \mathscr{S}(\hbar w \frac{\partial}{\partial Y})\frac{1}{Y}
    - \frac{w(1+\Theta^\vee(Y))}{Y}}}{\hbar w}
\nonumber
\\
&\times
{:} \exp\Big( \hbar w \mathscr{S}\Big(\hbar w \frac{\partial}{\partial Y}\Big)
\sum_{k\in \mathbb{Z}\setminus \{0\}}  Y^{k-1} J_{-k} \Big){:}
,
\label{bbE-normal-Y}
\\
\mathbb{E}(u,X) 
&=\frac{e^{u \mathscr{S}(\hbar u \frac{\partial}{\partial X})\frac{1}{X}
    - \frac{u(1+\Theta(X))}{X}}}{\hbar u}
\nonumber
\\
&\times 
   {:} \exp\Big( \hbar u \mathscr{S}\Big(\hbar u \frac{\partial}{\partial X}\Big)
     \sum_{k\in \mathbb{Z}\setminus \{0\}}  X^{k-1} J_{k} \Big){:}
.
 \end{align}
\end{subequations}
In $\mathbb{W}^\vee_m$ there are insertions of several such factors.
We stress that normal ordering is only applied inside
(\ref{bbE-normal}), not between $J$'s of different loop insertions.
In general, $AB \neq {:}A{:}\, {:}B{:} \neq {:}AB{:} \neq AB$.

The loop insertions $\mathbb{W}^\vee_m$ into $\log Z_P(E_{\set{N}})$ in
(\ref{def-bbWvee}) amount to extract the connected part
$\langle ~\rangle^\circ$ of their insertion into $Z_P(E_{\set{N}})$.
According to
(\ref{moment-cumulant}), for that we need products of loop insertions
of (\ref{bbE-normal}) indexed by elements of a block into $Z_P(E_{\set{N}})$,
then
divided by $Z_P(E_{\set{N}})$.  We expand the exponentials in the second line of
(\ref{bbE-normal-Y}) into a power series and artificially distinguish
the factors in $Y_i^{n_i}$ in order to take out the operators
$ \hbar w_i \mathscr{S}(\hbar w_i \frac{\partial}{\partial Y_i})$:
\begin{align}
&\frac{1}{Z_P(E_{\set{N}})}\Big( \prod_{i\in B}^{R.O.}
\overleftarrow{D}_{\mathbb{E}^\vee(w_i,Y_i)}\Big)Z_P(E_{\set{N}})
\label{bbW-block}
\\
&= \frac{1}{Z_P(E_{\set{N}})} 
\Big(\prod_{i\in B}
\frac{e^{w_i \mathscr{S}(\hbar w_i \frac{\partial}{\partial Y_i})\frac{1}{Y_i}
    - \frac{w_i(1+\Theta^\vee(Y_i))}{Y_i}}}{\hbar w_i}
\Big)
\sum_{\substack{n_{i_1},...,n_{i_{|B|}}=0\\
    B=\{i_1,...,i_{|B|}\}}}^\infty \frac{1}{\prod_{i\in B } n_{i}!}\!
\nonumber
\\
&\times  
\Big(\prod_{i\in B}
\prod_{j_i=1}^{n_i} \!
\hbar u_i \mathscr{S}\Big(\hbar u_i 
\frac{\partial}{\partial Y_{i,j_i}}\Big)
\frac{\partial}{\partial Y_{i,j_i}} \Big)
\bigg(\Big(\prod_{i\in B}^{R.O.}
{:}\prod_{j_i=1}^{n_i}\!
\overleftarrow{D}^{\mathbb{Z}  }_{Y_{i,j_i}}{:}\Big)
Z_P(E_{\set{N}})\bigg)\!\Big|_{Y_{i,j_i}\mapsto Y_i} .
\nonumber
\end{align}
Here the symbol $\prod_{i\in B}^{R.O.}$ indicates the radial ordering
according to Remark~\ref{rem:radial}
where the leftmost factor has the largest $|Y_i|$.
We have introduced normal-ordered loop insertions
\begin{align}
  {:}\Big(\prod_ {j=1}^n  \overleftarrow{D}_{Y_j}^{\mathbb{Z}}
  \Big){:}\langle 0|\mathcal{O}_1
  \mathcal{D}^{-1} \mathcal{O}_2|0\rangle
  &=  
  \Big\langle 0\Big|\mathcal{O}_1 
  \mathcal{D}^{-1}
{:}\bigg(\prod_{j=1}^n
  \Big(\sum_{k\in \mathbb{Z}\setminus\{0\}} 
    \frac{Y_j^{k} J_{-k}}{k}\Big)\bigg){:}
\mathcal{O}_2\Big|0\Big\rangle
\label{loop-DZY}
\end{align}
of positively and negatively indexed $J$'s.  The last
$\prod_{i\in B}^{R.O.}$ takes products of these in radial
order. In the last line of (\ref{bbW-block}) we write
$Z_P(E_{\set{N}})=\sum_{n=0}^\infty \frac{1}{n!} (\log
Z_P(E_{\set{N}}))^n$ and distribute the normal-ordered and radially
ordered loop insertions to the various factors $\log Z_P(E_{\set{N}})$.  It is
well-known that the result has a graphical interpretation where the
combinatorics of prefactors $\frac{1}{n_i!},\frac{1}{n!}$ and
distribution of loop insertions is captured by the order of the
automorphism group of the graph:
\begin{lemma} \label{lemma-bbWW}
  Loop insertions $\overleftarrow{D}_{\mathbb{E}^\vee(w_i,Y_i)}$
  into the partition function evaluate to
  \begin{align}
&\frac{1}{Z_P(E_{\set{N}})}\Big( \prod_{i\in B}^{R.O.}
\overleftarrow{D}_{\mathbb{E}^\vee(w_i,Y_i)}\Big)Z_P(E_{\set{N}})
\label{bbWW-block-graph}
\\
&= 
\Big(\prod_{i\in B}
\frac{e^{w_i \mathscr{S}(\hbar w_i \frac{\partial}{\partial Y_i})\frac{1}{Y_i}
    - \frac{w_i(1+\Theta^\vee(Y_i))}{Y_i}}}{\hbar w_i}
\Big)
\sum_{\Gamma\in \mathbb{G}_{B}} \frac{1}{|\mathrm{Aut}(\Gamma)|}
\nonumber
\\
&\times 
\prod_{\flat \in \mathcal{B}(\Gamma)}
\bigg(\Big(\prod_{e \in I(\flat)} \!\!
\hbar w_e \mathscr{S}\Big(\hbar w_e 
\frac{\partial}{\partial Y_e}\Big) 
\frac{\partial}{\partial Y_e}\Big)
\Big( \Big(\prod_{i\in B}^{R.O}
{:}\!\!\! \prod_{\substack{e \in I(\flat) \\ v(e)=i}}  \!\!\!
\overleftarrow{D}^{\mathbb{Z}}_{Y_e}{:}
\Big) \log Z_P(E_{\set{N}}) \Big)
\bigg)\Big|_{\substack{Y_e\mapsto Y_{v(e)}\\
    w_e\mapsto w_{v(e)}}} .
\nonumber
\end{align}
Here $\mathbb{G}_{B}$ is the set of possibly
disconnected bipartite graphs with 
\begin{itemize}
\item  $|B|$ white vertices  of any valency
  $\geq 0$, labelled by the elements of $B$,

\item  an arbitrary number
  of unlabelled black vertices of valency $\geq 1$,

\item each edge connecting a black vertex with a white vertex.
\end{itemize}
Only the incidence relation between edges
and vertices matters, not a possible order or path
when drawing the graph concretely. Moreover: 
\begin{itemize}
\item   For
  $\Gamma \in \mathbb{G}_{B}$ we let $\mathcal{B}(\Gamma)$ be the
  set of black vertices of $\Gamma$.

\item For $\flat\in \mathcal{B}(\Gamma)$ we let $I(\flat)$ be the
  set   of edges incident to $\flat$, and we let
  $|\flat|=|I(\flat)|$ be the valency of $\flat$.

\item For $e\in I(\flat)$ we let $v(e)$ be the label of the
  white vertex incident to $e$. \hfill $\square$

\end{itemize}
\end{lemma}
By (\ref{Jneg-contrib-new-2}), the difference between ordinary and
normal-ordered
double loop insertions is 
\begin{align*}
\frac{\partial}{\partial Y_{e_1}}\frac{\partial}{\partial Y_{e_2}}
  \overleftarrow{D}^{\mathbb{Z}}_{Y_{e_1}}\overleftarrow{D}^{\mathbb{Z}}_{Y_{e_2}}
  \log  Z_P(E_{\set{N}})&-
  \frac{\partial}{\partial Y_{e_1}}\frac{\partial}{\partial Y_{e_2}}
  {:}\overleftarrow{D}^{\mathbb{Z}}_{Y_{e_1}}\overleftarrow{D}^{\mathbb{Z}}_{Y_{e_2}}
  {:} \log  Z_P(E_{\set{N}})
  \\
  &=  \frac{1}{(Y_{e_1}-Y_{e_2})^2}  .
\end{align*}
As discussed at the end of the proof of Lemma~\ref{lemma:WW-u0}, only
two loop insertions
into $\log  Z_P(E_{\set{N}})$ have a difference between ordinary and normal-ordered
loop insertion. This implies
\begin{align*}
\Big(\prod_{e\in I(\flat)} 
\frac{\partial}{\partial Y_e}\Big)
\Big(\prod_{i\in B}^{R.O.}
  \Big(  
{:}\!\!\! \prod_{\substack{e \in I(\flat) \\ v(e)=i}}  \!\!\!
\overleftarrow{D}^{\mathbb{Z}}_{Y_e}{:}
\Big)\Big) \log  Z_P(E_{\set{N}})
&= \Big(\prod_{e \in I(\flat)}
\frac{\partial}{\partial Y_e}
\overleftarrow{D}^{\mathbb{Z}}_{Y_e}
\Big) \log  Z_P(E_{\set{N}})
\\
&-\frac{\delta_{|\flat|, 2} \delta_{v(e_1),v(e_2)}}{(Y_{e_1}-Y_{e_2})^2}.
\end{align*}

Moreover, we consider the family of graphs which differ only by the
number $k_i$ of one-valent black vertices attached to the white vertex $i$:
\begin{align}
\sum_{k_{i}=0}^\infty \frac{1}{k_{i}!} \qquad \begin{picture}(60,20)
    \put(0,0){\circle{15}}
    \put(-1,-2){\small$i$}
    \qbezier(5,5)(25,9)(25,9)
    \qbezier(5,-5)(25,-9)(25,-9)
    \qbezier(4,7)(25,21)(25,21)
    \qbezier(4,-7)(25,-21)(25,-21)
    \put(25,-4.4){$\vdots$}
    \put(28,17){\small$1$}
    \put(28,-26){\small$k_{i}$}
    \put(23,6){$\bullet$}    
    \put(23,-13){$\bullet$}    
    \put(23,17){$\bullet$}    
    \put(23,-23){$\bullet$}
    \put(-5,-5){\line(-1,-1){10}}
    \put(-5,5){\line(-1,1){10}}
    \put(25,-4.4){$\vdots$}
  \end{picture} \rule[-8mm]{0mm}{10mm}
\label{one-valent-graphs}
\end{align}
When summing over $k_i$, a factor
\begin{align}
&  \exp\Big\{
u_i  \mathscr{S}\Big(\hbar u_i\frac{\partial}{\partial Y_i}\Big)  
\frac{\Theta^\vee(Y_i)}{Y_i}\Big\}
\label{one-point-hbar}
\\
&\times \exp\Big\{
\hbar u_i  \mathscr{S}\Big(\hbar u_i\frac{\partial}{\partial Y_i}\Big)  
\Big(\frac{\partial}{\partial Y_i}
\mathcal{P}_1 \overleftarrow{D}^{\mathbb{Z}}_{Y_i}\log Z_P(E_{\set{N}})
\Big)\Big\}
\nonumber
\end{align}
results in which we have separated the $[\hbar^{-1}]$-contribution
$\frac{\Theta^\vee(Y_i)}{Y_i}=
[\hbar^ {-1}]\big(\frac{\partial}{\partial Y_i}
\overleftarrow{D}^{\mathbb{Z}}_{Y_i}\log Z_P(E_{\set{N}})\big)$,
see (\ref{def-Theta}). We recall $\mathcal{P}_n=\mathrm{id}-\delta_{n,1}
\hbar^{-1}[\hbar^{-1}]$. 
The first line combines with the exponential factor of the second line of 
(\ref{prodDlogZ-2}). The second line of (\ref{one-point-hbar})
is reexpanded into
graphical contributions of $1$-valent black vertices attached to 
the white vertex $i$ whose weight has no $[\hbar^{-1}]$-contribution.

Noting
\begin{align}
\mathcal{P}_{|I|}\Big(\prod_{e \in I}
\frac{\partial}{\partial Y_e}
\overleftarrow{D}^{\mathbb{Z}}_{Y_e} \Big) \log Z_P(E_{\set{N}})
=\frac{\omega^\vee_{|I|}(Y_I)}{\prod_{e\in I} dY_e}
\label{varpi-logZ}
\end{align}
where $\omega^\vee_{|I|}(Y_I)\equiv 
\omega^\vee_{|I|}(Y_{e_1},...,Y_{e_{|I|}})$ 
if $I=\{e_1,...,e_{|I|}\}$,
we get the final result
  \begin{align}
&\frac{1}{Z_P(E_{\set{N}})}\Big( \prod_{i\in B}^{R.O.}
\overleftarrow{D}_{\mathbb{E}^\vee(w_i,Y_i)}\Big)Z_P(E_{\set{N}})
\label{prodDlogZ-2}
\\
&= 
\Big(\prod_{i\in B}
\frac{e^{w_i (\mathscr{S}(\hbar w_i \frac{\partial}{\partial Y_i})-1)
    \frac{1+\Theta^\vee(Y_i)}{Y_i}}}{\hbar w_i}
\Big)
\nonumber
\\
&\times \sum_{\Gamma\in \mathbb{G}_{B}} \frac{1}{|\mathrm{Aut}(\Gamma)|}
\prod_{\flat \in \mathcal{B}(\Gamma)}
\bigg(\Big(\prod_{e \in I(\flat)} \!\!
\hbar w_e \mathscr{S}\Big(\hbar w_e 
\frac{\partial}{\partial Y_e}\Big) \frac{1}{dY_e}\Big)
  {:}\omega^\vee_{|\flat|}{:}(Y_{I(\flat)})\bigg)
\Big|_{\substack{Y_e\mapsto Y_{v(e)}\\
    w_e\mapsto w_{v(e)}}} .
\nonumber
\end{align}
Here we
introduced 
normal-ordered (or renormalised) multi-differentials
\begin{align}
  {:}\omega^\vee_{|\flat|}{:}(Y_{e_1},...,Y_{e_{ |\flat|}})
= \omega^\vee_{|\flat|}(Y_{e_1},...,Y_{e_{|\flat|}})
- \delta_{|\flat|,2} \delta_{v(e_1),v(e_2)}
\frac{dY_{e_1} dY_{e_2}}{ (Y_{e_1}-Y_{e_2})^2}
\label{varpi-normal}
\end{align}
with ${:}\omega^\vee_{|\flat|}{:}(Y_{I(\flat)})
\equiv 
{:}\omega^\vee_{|\flat|}{:}(Y_{e_1},...,Y_{e_{|\flat|}})$
if $I(\flat)=\{e_1,...,e_{|\flat|}\}$.

We insert (\ref{prodDlogZ-2}) into
(\ref{moment-cumulant}). The product of prefactors
from the middle line (\ref{prodDlogZ-2}) over the blocks is the 
same in all terms
in (\ref{moment-cumulant}). For the sum over disconneted graphs, it is a
standard result that the combinations (\ref{moment-cumulant})
leave the sum over connected graphs. This proves:
\begin{proposition}
\label{prop:bbWvee}
  The following relations hold:
\begin{align}
  & \mathbb{W}^\vee_m(w_{\set{m}};Y_{\set{m}})
  \label{bbWvee}
  \\
&=
\prod_{i=1}^m \frac{e^{w_i (\mathscr{S}(\hbar w_i
    \frac{\partial}{\partial Y_i})-1)
    \frac{1+\Theta^\vee(Y_i)}{Y_i}}}{\hbar w_i}
\nonumber
\\
&\times \sum_{\Gamma\in G_{\set{m}}} \frac{1}{|\mathrm{Aut}(\Gamma)|}
\prod_{\flat \in \mathcal{B}(\Gamma)}
\bigg\{
\Big(\prod_{e \in I(\flat)} \!\!
\hbar w_e \mathscr{S}\Big(\hbar w_e 
\frac{\partial}{\partial Y_e}\Big)\frac{1}{dY_e}\Big) 
{:}\omega^\vee_{|\flat|}{:}(Y_{I(\flat)})
\bigg\}\Big|_{\substack{Y_e\mapsto Y_{v(e)}\\
    w_e\mapsto u_{v(e)}}}\,,
\nonumber
\end{align}
where $G_{\set{m}}$ is the set of connected bipartite graphs with
$m$ white vertices labelled by $\set{m}$ and an arbitrary number
of black vertices of any valency $\geq 1$.
The normal-ordered multi-differentials ${:}\omega^\vee_{|\flat|}{:}$
are defined in \eqref{varpi-normal} and \eqref{varpi-logZ}.
For $m=1$, the graph
consisting only of the white vertex $1$ and no edge is included.
Otherwise the notation is as in Lemma~\ref{lemma-bbWW}.
\end{proposition}
\noindent
Equation (\ref{bbWvee}) is the motivation for the definition
(\ref{eq:def-bWvee}) of dual extended multi-differentials, which arises
when inserting $Y_i=Y(z_i)$ from the spectral curve. One can then
use the identity $(\frac{\partial}{\partial Y(z)})^r\frac{1}{dY(z)}
=\frac{1}{dY(z)} (d_{z}\frac{1}{dY(z)})^r$.
The established graphical representation motivates Def.~\ref{def:bipartite}.

We can easily obtain a formula for
$\mathcal{W}^\vee_{m+1}(w_0;Y_0;Y_{\set{m}})$ by
prepending (\ref{bbWvee}) by $(w_0,Y_0)$ and taking
the $[w_1^0\cdots w_m^0]$-coefficient, see
(\ref{bbWvee-calWvee}). This 
sets $\mathscr{S} (\hbar w_i \frac{\partial}{\partial Y_i})\mapsto 1$
for $i\in \set{m}$, and only graphs with a single edge $e$ with $v(e) = i$
can contribute:
\begin{proposition}
\label{prop:calWvee}
  The following relations hold:
\begin{align}
  & \mathcal{W}^\vee_{m+1}(w;Y_0;Y_{\set{m}})
  \label{calWvee}
  \\
&=
\frac{e^{w (\mathscr{S}(\hbar w
    \frac{\partial}{\partial Y_0})-1)
    \frac{1+\Theta^\vee(Y_0)}{Y_0}}}{\hbar w}
\nonumber
\\
&\times \!\!\!\!\!\!
\sum_{\substack{\Gamma\in G_{\{0,1,...,m\}}\\
    \text{white vertices }\set{m} \\
    \text{one-valent}}} \!\!\!\!
 \frac{1}{|\mathrm{Aut}(\Gamma)|}
\prod_{\flat \in \mathcal{B}(\Gamma)}
\bigg\{
\Big(\!\! \prod_{\substack{e \in I(\flat) \\ v(e)=0}} \!\!
\hbar w \mathscr{S}\Big(\hbar w 
\frac{\partial}{\partial Y_e}\Big)\frac{1}{dY_e}\Big) 
{:}\omega^\vee_{|\flat|}{:}(Y_{I(\flat)})
\bigg\}\Big|_{Y_e\mapsto Y_{v(e)}}.
\nonumber
\end{align}
The notation is the same as in
Prop.~\ref{prop:bbWvee}
and Lemma~\ref{lemma-bbWW}.
\hspace*{\fill}$\square$%
\end{proposition}
\noindent
Equation (\ref{calWvee}) is the motivation for
the 1-variable extended $n+1$-differentials in
Def.~\ref{def:1variableextended}.

In complete analogy to Prop.~\ref{prop:bbWvee} we have:
\begin{proposition}
  \label{prop:bbW-2}
  The following relations hold:
\begin{align}
  & \mathbb{W}_m(u_{\set{m}};X_{\set{m}})
  \label{bbW-final}
 \\
&=
\prod_{i=1}^m \frac{e^{u_i (\mathscr{S}(\hbar u_i
    \frac{\partial}{\partial X_i})-1)
    \frac{1+\Theta(X_i)}{X_i}}}{\hbar u_i}
\nonumber
\\
&\times \sum_{\Gamma\in G_{\set{m}}} \frac{1}{|\mathrm{Aut}(\Gamma)|}
\prod_{\flat \in \mathcal{B}(\Gamma)}
\bigg\{
\Big(\prod_{e \in I(\flat)} \!\!
\hbar u_e \mathscr{S}\Big(\hbar u_e 
\frac{\partial}{\partial X_e}\Big)\frac{1}{dX_e}\Big) 
{:}\omega_{|\flat|}{:}(X_{I(\flat)})
\bigg\}\Big|_{\substack{X_e\mapsto X_{v(e)}\\
    w_e\mapsto w_{v(e)}}}\,,
\nonumber
\end{align}
where 
\begin{align}
  \frac{\omega_{|\flat|}(X_I)}{\prod_{e\in I(\flat)} dX_e}
  &=\mathcal{P}_{|\flat|}
\Big(\Big(\prod_{e \in I(\flat)}\frac{\partial}{\partial X_e}
\overrightarrow{D}^{\mathbb{Z}}_{X_e} \Big) \log Z_P(E_{\set{N}})\Big),
\nonumber
\\
  {:}\omega_{|\flat|}{:}(X_{I(\flat)})
&= \omega_{|\flat|}(X_{I(\flat)})
- \delta_{|\flat|,2} \delta_{v(e_1),v(e_2)}
\frac{dX_{e_1} dX_{e_2}}{ (X_{e_1}-X_{e_2})^2}.
\label{varpi-normal-2}
\end{align}
Otherwise the same notation as in Prop.~\ref{prop:bbWvee}
and Lemma~\ref{lemma-bbWW} is employed.
\end{proposition}
\noindent
Equation (\ref{bbW-final}) is the motivation for the definition
(\ref{eq:def-bW}) of extended multi-differentials, which arises
when inserting $X_i=X(z_i)$ from the spectral curve.

More generally, we have:
\begin{proposition}
  \label{prop:bbW}
  The following relations hold:
\begin{align}
  & \mathbb{W}_{n,m}(u_{\set{n}};w_{\set{m}};X_{\set{n}};Y_{\set{m}})    
  \label{bbW-mixed}
 \\
&=
\prod_{j=1}^n \frac{e^{u_j (\mathscr{S}(\hbar u_j
    \frac{\partial}{\partial X_j})-1)
    \frac{1+\Theta(X_j)}{X_j}}}{\hbar u_j}
\prod_{i=1}^m \frac{e^{w_i\mathscr{S}(\hbar w_i
    \frac{\partial}{\partial Y_i})-1)
    \frac{1+\Theta^\vee(Y_i)}{Y_i}}}{\hbar w_i}
\nonumber
\\
&\times \sum_{\Gamma\in G_{\set{n},\set{m}}} \frac{1}{|\mathrm{Aut}(\Gamma)|}
\prod_{\flat \in \mathcal{B}(\Gamma)}
\bigg\{
\Big(\prod_{e \in I(\flat)} \!\!
\hbar u_e \mathscr{S}\Big(\hbar u_e 
\frac{\partial}{\partial X_e}\Big)\frac{1}{dX_e}\Big)
\nonumber
\\
&\qquad \times
\Big(\prod_{f \in I^\vee(\flat)} \!\!
\hbar w_f \mathscr{S}\Big(\hbar w_e 
\frac{\partial}{\partial Y_f}\Big)\frac{1}{dY_f}\Big) 
{:}\omega_{|I(\flat)|,|I^\vee(\flat)|}{:}(X_{I(\flat)};Y_{I^\vee(\flat)})
\bigg\}\Big|_{
  \substack{X_e\mapsto X_{v(e)}\\
  Y_f\mapsto Y_{v(f)}\\
    w_e\mapsto w_{v(e)}\\
    u_f\mapsto u_{v(f)}}}\,,
\nonumber
\end{align}
where 
\begin{subequations}
  \begin{align}
 & \frac{\omega_{|I(\flat)|,|I^\vee(\flat)|}
    (X_{I(\flat)};Y_{I^\vee(\flat)})}{\prod_{e\in I(\flat)} dX_e
    \prod_{f\in I^\vee(\flat)} dY_f  }
  \\
  &:=\mathcal{P}_{|I(\flat)|+|I^\vee(\flat)|}\bigg(
  \Big(\prod_{e \in I(\flat)}
\frac{\partial}{\partial X_e}
\overrightarrow{D}^{\mathbb{Z}}_{X_e} \Big)
\Big(\prod_{f \in I^\vee(\flat)}
\frac{\partial}{\partial Y_f}
\overrightarrow{D}^{\mathbb{Z}}_{Y_f} \Big)
\log Z_P(E_{\set{N}})\bigg)
\nonumber
\end{align}
and
\begin{align}
{:}
\omega_{|I(\flat)|,|I^\vee(\flat)|}{:}    (X_{I(\flat)};Y_{I^\vee(\flat)})
&=
\omega_{|I(\flat)|,|I^\vee(\flat)|} (X_{I(\flat)};Y_{I^\vee(\flat)})
\\
&- \delta_{|I(\flat)|,2}\delta_{|I^\vee(\flat)|,0} \delta_{v(e_1),v(e_2)}
\frac{dX_{e_1} dX_{e_2}}{ (X_{e_1}-X_{e_2})^2}
\nonumber
\\
&
- \delta_{|I(\flat)|,0}\delta_{|I^\vee(\flat)|,2} \delta_{v(f_1),v(f_2)}
\frac{dY_{f_1} dY_{f_2}}{ (Y_{f_1}-Y_{f_2})^2}.
\nonumber
\end{align}
\end{subequations}
Here  $G_{\set{n},\set{m}}$ is the set of connected tripartite graphs with
\begin{itemize}
  \item $n$ white vertices labelled by $\set{n}$,
  \item $m$ dual white vertices labelled by $\set{m}$,
  \item an arbitrary number of unlabelled black vertices of valency
    $\geq 1$.
\end{itemize}    
We let $\mathcal{B}(\Gamma)$ be the set of black vertices of $\Gamma 
\in G_{\set{n},\set{m}}$. For $\flat \in \mathcal{B}(\Gamma)$, we let
\begin{itemize}

\item $I(\flat)$ be the set of edges attached to $\flat$ whose other end
  attaches to a white vertex;
\item $I^\vee(\flat)$ be the set of edges attached to $\flat$ whose other end
  attaches to a dual white vertex.  
\end{itemize}
For $e\in I(\flat)$ we let $v(e)$ be the label of the white vertex incident to $e$, and for $f\in I^\vee(\flat)$ we let $v(f)$ be the label of the dual
white vertex incident to $f$.
\end{proposition}

\section{Fock space proof of $X$-$Y$ duality}

\label{app:XYFock}

This appendix follows the strategy of \cite{Alexandrov:2022ydc}, 
which itself is based on \cite{Bychkov:2020ujd}.

\subsection{Conjugation by $\mathcal{D}$}

We conjugate (\ref{calEuz-new}) with $\mathcal{D}$
(ignoring the constant term $\frac{1}{u}$ for the moment)
and take (\ref{adED-4}) into account:
\begin{align*}
&\mathcal{D}\mathscr{C}(\hbar u)
  \mathcal{E}(\mathscr{A}(\hbar u),X/\mathscr{C}(\hbar u))  \mathcal{D}^{-1} 
  \\
  &= \sum_{m\in \mathbb{Z}}
X^{-m} (\mathscr{C}(\hbar u))^{m+1}
\sum_{r=0}^\infty
 \partial_\theta^r\Big(e^{\frac{\theta \mathscr{A}(\hbar u)}{\hbar}+
   m \frac{\mathscr{S}(m\hbar \partial_\theta)}{
     \mathscr{S}(\hbar \partial_\theta)} \log (1+\theta) }
\Big)
\Big|_{\theta=0}
[w^r] [Y^{m}] \mathcal{E}(\hbar w,\tfrac{1}{Y}).
\end{align*}
We use the following non-trivial identity
(where we replace $\theta\mapsto 1+\theta$):
\begin{lemma}[{\cite[Lemma 3.4]{Alexandrov:2022ydc}}]
  \begin{align}
    e^{m \frac{\mathscr{S}(m\hbar \partial_\theta)}{
     \mathscr{S}(\hbar \partial_\theta)} \log (1+\theta) }
 &= \sum_{s=0}^\infty \partial^s_\theta(1+\theta)^m [w^s]\Big(
 (1-w^2\hbar^2/4)^{\frac{m-1}{2}}e^{-w(1+\theta)}
 \Big(\frac{1+w\hbar/2}{1-w\hbar/2}\Big)^{\frac{1+\theta}{\hbar}}\Big)
 \nonumber
\\
&= \sum_{s=0}^\infty \partial^s_\theta(1+\theta)^m [w^s]\Big(
 (\mathscr{C}(\hbar w))^{1-m} e^{-w(1+\theta)}
 e^{\frac{1+\theta}{\hbar}
   \mathscr{A}(\hbar w)}\Big).
\label{Alexandrov-L}
\end{align}
\end{lemma}

\noindent
This gives
\begin{align*}
&\mathcal{D}\mathscr{C}(\hbar u)
  \mathcal{E}(\mathscr{A}(\hbar u),X/\mathscr{C}(\hbar u))  \mathcal{D}^{-1} 
  \\
  &= \sum_{m\in \mathbb{Z}}
X^{-m} (\mathscr{C}(\hbar u))^{m+1}
\sum_{r=0}^\infty
\partial_\theta^r\Big(
\sum_{s=0}^\infty \partial^s_\theta(1+\theta)^m
\\
&\qquad \times
[v^s]\Big(
(\mathscr{C}(\hbar v))^{1-m} e^{-v(1+\theta)}
e^{\frac{\theta \mathscr{A}(\hbar u)}{\hbar}}
 e^{\frac{1+\theta}{\hbar}
   \mathscr{A}(\hbar v)}\Big)
\Big)
\Big|_{\theta=0}
[w^r] [Y^{m}] \mathcal{E}(\hbar w,\tfrac{1}{Y})
  \\
  &= \sum_{m\in \mathbb{Z}}
X^{-m} (\mathscr{C}(\hbar u))^{m+1}
\sum_{s=0}^\infty \partial^s_\theta(1+\theta)^m [v^s]\Big(
(\mathscr{C}(\hbar v))^{1-m} e^{-v(1+\theta)}
e^{\frac{\theta \mathscr{A}(\hbar u)}{\hbar}}
 e^{\frac{1+\theta}{\hbar}
   \mathscr{A}(\hbar v)}
\\
&\qquad \times 
 \sum_{r=0}^\infty
 \Big(\frac{1}{\hbar}( \mathscr{A}(\hbar u)+\mathscr{A}(\hbar v))
\Big)^r
[w^r] [Y^{m}] \mathcal{E}(\hbar w,\tfrac{1}{Y})
 \Big)
\Big|_{\theta=0}.
\end{align*}
That $\partial^r_\theta$ just produces a factor
$(\frac{1}{\hbar}( \mathscr{A}(\hbar u)+\mathscr{A}(\hbar v))
)^r$ is explained in 
\cite[eq. (3.23)]{Alexandrov:2022ydc}.
We continue: 
\begin{align}
&\mathcal{D}\mathscr{C}(\hbar u)
  \mathcal{E}(\mathscr{A}(\hbar u),X/\mathscr{C}(\hbar u))  \mathcal{D}^{-1} 
 \label{DED-1}
  \\
  &= \sum_{m\in \mathbb{Z}}
X^{-m} (\mathscr{C}(\hbar u))^{m+1}
\sum_{s=0}^\infty \partial^s_\theta(1+\theta)^m
\Big|_{\theta=0}
[v^s]\Big(
(\mathscr{C}(\hbar v))^{1-m} e^{-v}
 e^{\frac{1}{\hbar}
   \mathscr{A}(\hbar v)}
 \nonumber
 \\
&\qquad \times 
[Y^{m}] \mathcal{E}(\mathscr{A}(\hbar u)+\mathscr{A}(\hbar v),
\tfrac{1}{Y})
 \Big)
 \nonumber
  \\
  &= \sum_{m\in \mathbb{Z}}
X^{-m}
\sum_{s=0}^\infty \partial^s_\theta(1+\theta)^m
\Big|_{\theta=0}
[w^s][Y^{m}]
\Big(
 e^{\frac{1}{\hbar} \mathscr{A}(\hbar w)-w}
 \mathscr{F}^\vee(\hbar u,\hbar w;Y)
 \Big),
 \nonumber
\end{align}
where we have used the Taylor formula and defined
\begin{align}
\mathscr{F}^\vee(\hbar u,\hbar w;Y)
&:=\mathscr{C}(\hbar u)\mathscr{C}(\hbar w)
\mathcal{E}(\mathscr{A}(\hbar u)+\mathscr{A}(\hbar w),
\tfrac{\mathscr{C}(\hbar w)}{Y\mathscr{C}(\hbar u)}).
\label{scrF}
\end{align}

Recall from (\ref{calEuz}) that 
$\mathcal{E}(\mathscr{A}(\hbar u)+\mathscr{A}(\hbar w),
\tfrac{\mathscr{C}(\hbar w)}{Y\mathscr{C}(\hbar u)})$
has the special constant term
$-1/((\mathscr{A}(\hbar u)+\mathscr{A}(\hbar w))
\mathscr{S}(\mathscr{A}(\hbar u)+\mathscr{A}(\hbar w)))
$. One finds
\[
  \frac{\mathscr{C}(\hbar u)\mathscr{C}(\hbar w)}{
  (\mathscr{A}(\hbar u)+\mathscr{A}(\hbar w))
\mathscr{S}(\mathscr{A}(\hbar u)+\mathscr{A}(\hbar w))}
= \frac{1}{\hbar(w+u)},
\]
which provides the constant term and more generally the numerator.
The operator in (\ref{DED-1}) applied to this constant term evaluates to
\begin{align*}
&\sum_{m\in \mathbb{Z}}
X^{-m} 
\sum_{s=0}^\infty \partial^s_\theta(1+\theta)^m
\Big|_{\theta=0}
[w^s][Y^{m}]
\Big(
 e^{\frac{1}{\hbar} \mathscr{A}(\hbar w)-w}
\frac{1}{\hbar(w+u)}
\Big)
\\
&= [w^0]\Big(
 e^{\frac{1}{\hbar} \mathscr{A}(\hbar w)-w}
\frac{1}{\hbar(w+u)}
\Big)=\frac{1}{\hbar u}.
\end{align*}
Adding this to (\ref{DED-1}), we get with
$\tilde{\mathscr{F}}^\vee(\hbar u,\hbar w;Y)
:= \mathscr{F}^\vee(\hbar u,\hbar w;Y)+\frac{1}{\hbar(u+w)}$
and (\ref{calEuz-new})
\begin{align}
  &
\mathcal{D}\Big(\mathscr{C}(\hbar u)
\mathcal{E}(\mathscr{A}(\hbar u),X/\mathscr{C}(\hbar u))
+\frac{1}{\hbar u}\Big)\mathcal{D}^{-1} 
=
\mathcal{D}\hat{\mathcal{E}}(\hbar u,X)\mathcal{D}^{-1} 
\label{DED-2}
  \\
  &= \sum_{m\in \mathbb{Z}}
X^{-m}
\sum_{s=0}^\infty \partial^s_\theta(1+\theta)^m
\Big|_{\theta=0}
[w^s][Y^{m}]
\Big(
 e^{\frac{1}{\hbar} \mathscr{A}(\hbar w)-w}
\tilde{\mathscr{F}}^\vee(\hbar u,\hbar w;Y)
 \Big)
\nonumber
     \\
  &= \sum_{m\in \mathbb{Z}}
X^{-m} 
\sum_{s=0}^\infty \partial^s_\theta(1+\theta)^m
\Big|_{\theta=0}
[w^s][Y^{m}]
\Big( Y^{s+1} 
 e^{\frac{1}{\hbar} \mathscr{A}(\hbar w/Y)-w/Y}
\frac{ \tilde{\mathscr{F}}^\vee(\hbar u,\hbar w/Y;Y)}{Y}
\Big).
\nonumber
\end{align}

\subsection{The $\mathcal{U}$-operation}

It will be convenient to define 
\begin{align}
&  \mathcal{U} f(u(X);\star;\bullet)(X)
\\
&:= \frac{1}{X}\sum_{m\in \mathbb{Z}}
X^{-m}
\sum_{s=0}^\infty \partial^s_\theta(1+\theta)^m
\Big|_{\theta=0}
[w^s][Y^{m}]
\Big( Y^{s+1} 
e^{(w/Y)\Theta^\vee(Y)} f(u(X), w;Y)\Big),
\nonumber
\end{align}
where $\Theta^\vee$ is defined by (\ref{def-Theta}). We include an 
additional parameter $u$ which might depend as $u(X)$ on $X$.
This turns (\ref{DED-2}) into
\begin{align}
\mathcal{D}\frac{1}{X}\hat{\mathcal{E}}(\hbar u,X)  \mathcal{D}^{-1}
\label{DED-3}
&= \mathcal{U}\Big(
    e^{\frac{1}{\hbar}\mathscr{A}(\hbar \star/\bullet)-
    \star(1+\Theta^\vee(\bullet))/\bullet}
    \frac{\tilde{\mathscr{F}}^\vee(\hbar u,\hbar \star/\bullet;\bullet)}{\bullet}
    \Big)(X),
  \end{align}
  where $u$ can be $X$-dependent.
We have the following crucial result that uses the \emph{spectral curve}:
\begin{lemma}
\label{Lemma:U}
  If $X,Y$ are related by the spectral curve
  $XY=1+\Theta^\vee(Y)$ and $Y(X)$ is the inverse solution, then
  \begin{align}
\mathcal{U} f(u(X);\star;\bullet)(X)    
= -\bigg(\sum_{s=0}^\infty 
\Big({-}\frac{\partial}{\partial X}\Big)^s
[w^s]\Big(
\frac{dY(X)}{dX} f(u(\hat{X}),w;Y(X))\Big)\bigg)\Big|_{\hat{X}=X}.
\end{align}
Here it is understood that $({-}\frac{\partial}{\partial X})^s$ does not act on
$u(\hat{X})$.
\begin{proof}
We evaluate the $\mathcal{U}$-operation by a sequence of
steps consisting of
\begin{enumerate}
\item rescaling $w\mapsto wY$ and back
\item Taylor formula \cite[Lemma 4.5]{Bychkov:2020ujd}
\item evaluation of derivative and reconstruction of another derivative
\item reexpression $[x^a](x^b f(x))=[x^{a-b}]f(x)$
\item shift $m\mapsto m+s$
\end{enumerate}

\begin{align*}
  &\mathcal{U} f(u(X);\star;\bullet)(X)
  \\
  &= \frac{1}{X}\sum_{m\in \mathbb{Z}}
X^{-m}
\sum_{s=0}^\infty \partial^s_\theta(1+\theta)^m
\Big|_{\theta=0} [w^s][Y^{m}]\Big( Y e^{w\Theta^\vee(Y)} f(u(X); wY;Y)\Big)
\tag{a}
\\
&= \frac{1}{X}\sum_{m\in \mathbb{Z}}
X^{-m}[Y^{m}]
\Big(
\sum_{s=0}^\infty \partial^s_\theta(1+\theta)^m
\Big|_{\theta=\Theta^\vee(Y)}
[w^s]
\Big( Y
 f(u(X);wY;Y)\Big)\Big)
\tag{b}
\\
&= \sum_{m\in \mathbb{Z}}
X^{-m-1}[Y^{m}]
\Big(
\sum_{s=0}^\infty
\prod_{i=0}^{s-1}(m{-}i)
(1{+}\Theta^\vee(Y))^{m-s}
[w^s]
\Big( Y  f(u(X); wY;Y)\Big)\Big)
\tag{c}
\\
&= \sum_{s=0}^\infty
(-1)^s\sum_{m\in \mathbb{Z}}
\frac{\partial^sX^{-m-1+s}}{\partial X^s}
[Y^{m}]\Big(\!
(1{+}\Theta^\vee(Y))^{m-s}
[w^s]
\big( Y f(u(X); wY;Y)\big)\Big)
\tag{c}
\\
&= \sum_{s=0}^\infty
(-1)^s\!\sum_{m\in \mathbb{Z}}
\frac{\partial^sX^{-m-1+s}}{\partial X^s}
[Y^{m}]\Big(\!
(1{+}\Theta^\vee(Y))^{m-s}
[w^s]
\big( Y^{s+1}  f(u(X); w;Y)\big)\!\Big)
 \tag{a}
 \\
&= \sum_{s=0}^\infty\!
(-1)^s[w^s]\Big( \sum_{m\in \mathbb{Z}}
\frac{\partial^sX^{-m-1+s}}{\partial X^s}
[Y^{m-1-s}]\big(
(1{+}\Theta^\vee(Y))^{m-s}
 f(u(X);w;Y)\big)\!\Big)
 \tag{d}
 \\
&= \sum_{s=0}^\infty
(-1)^s[w^s]\Big( \sum_{m\in \mathbb{Z}}
\Big(\frac{\partial^s}{\partial X^s}
X^{-m-1}\Big)
[Y^{m-1}]\Big(
(1+\Theta^\vee(Y))^{m}
 f(u(X); w;Y)\Big)\Big)
 \tag{e}
 \\
&= \bigg(\sum_{s=0}^\infty
\Big({-}\frac{\partial}{\partial X}\Big)^s
\big([w^s]
\hat{f}(u(\hat{X}); w;X)\big)\bigg)_{\hat{X}=X},
\end{align*}
where
\begin{align*}
\hat{f}(u(\hat{X}); w;X)
&:=\sum_{m\in \mathbb{Z}}
X^{-m-1}
[Y^{m-1}]\Big(
(1+\Theta^\vee(Y))^{m}
 f(u(\hat{X}); w;Y)\Big).
\end{align*}
Here it is important that 
$({-}\frac{\partial}{\partial X})^s$ only differentiates 
$X^{-m-1}$ in $\hat{f}(u(\hat{X}); w;X)$ but not a possible
$\hat{X}$-dependence of $u(\hat{X})$. Next,
 we apply step (d) and express $[Y^{-1}]f(Y)=-\Res_Y f(Y)dY$ as
 a residue to get
\begin{align*}
\hat{f}(u(\hat{X}); w;X)
 &=-\sum_{m\in \mathbb{Z}}
X^{-m-1}
\Res_Y dY\Big(
\frac{(1+\Theta^\vee(Y))^{m}}{Y^m}
 f(u(\hat{X}); w;Y)\Big).
\end{align*}
The sign arises from the convention that
$Y$ is a local variable at $z=\infty$ whereas we have
$[X^{-1}]f(X)=\Res_X f(X)dX$ for a local variable $X$ near $z=0$.
Now the spectral curve $1+\Theta^\vee(Y)=Y X(Y)$ is implemented.
It has an inverse $Y(X)$ so that we continue 
\begin{align*}
\hat{f}(u(\hat{X}); w;X)
&=-\sum_{m\in \mathbb{Z}}
X^{-m-1}
\Res_X dX\Big( \frac{dY(X)}{dX} X^m
 f(u(\hat{X}); w;Y(X))\Big)
\\
&=-\sum_{m\in \mathbb{Z}}
X^{-m-1} [X^{-1}]\Big( \frac{dY(X)}{dX} X^m
 f(u(\hat{X}); w;Y(X))\Big)
\\
&=-\sum_{m\in \mathbb{Z}}
X^{-m-1} [X^{-1-m}]\Big( \frac{dY(X)}{dX} 
 f(u(\hat{X});w;Y(X))\Big)
\\
&=-\frac{dY(X)}{dX} f(u(\hat{X}); w;Y(X)).
\end{align*}
This finishes the proof.
\end{proof}
\end{lemma}

\subsection{Invariance $\Theta(X)=\Theta^\vee(Y(X))$
under duality}

Before continuing, we derive with similar methods the following result:
\begin{proposition}
\label{prop:Theta}
  If $X$ and $Y$ are related by the spectral curve
  $XY=1+\Theta^\vee(Y)$, then
  \begin{align}
\Theta(X)=\Theta^\vee(Y(X))
\end{align}    
up to a constant shift (which with more effort
\cite[Sec.~3.4]{Bychkov:2022wgw} can be shown to vanish).
\begin{proof}
  We set $u=0$ in (\ref{DED-1}) and take
$\mathcal{E}(0,X)=\sum_{k\in \mathbb{Z}\setminus \{0\}}
X^k J_k$ as well as (\ref{calEuz-new}) and later (\ref{def:scrEY})
into account:
\begin{align*}
&\mathcal{D}
  \Big(\sum_{k\in \mathbb{Z}\setminus \{0\}}
  X^k J_k \Big)\mathcal{D}^{-1} 
  \\
  &=
\sum_{m\in \mathbb{Z}}
X^{-m}
\sum_{s=0}^\infty \partial^s_\theta(1+\theta)^m
\Big|_{\theta=0}
[w^s][Y^{m}]
\Big(
 e^{\frac{1}{\hbar} \mathscr{A}(\hbar w)-w}
\Big(\hat{\mathcal{E}}^\vee(\hbar w,Y)-\frac{1}{\hbar w}\Big)
\Big)
\\
&=
\sum_{m\in \mathbb{Z}}
X^{-m}
\sum_{s=0}^\infty \partial^s_\theta(1+\theta)^m
\Big|_{\theta=0}
[w^s][Y^{m}] 
\Big(Y^{s+1}
 e^{\frac{1}{\hbar} \mathscr{A}(\frac{\hbar w}{Y})-\frac{w}{Y}}
 \Big(\frac{\hat{\mathcal{E}}^\vee(\frac{\hbar w}{Y},Y)}{Y}
 -\frac{1}{\hbar w}\Big)
\Big)
\\
&=
\sum_{m\in \mathbb{Z}}
X^{-m}
\sum_{s=0}^\infty \partial^s_\theta(1+\theta)^m
\Big|_{\theta=0}
[w^s][Y^{m}] 
\Big(Y^{s+1}
 \Big(
e^{w\frac{\Theta^\vee(Y)}{Y}} \mathbb{E}^\vee(w,Y)
- \frac{
e^{\frac{1}{\hbar} \mathscr{A}(\frac{\hbar w}{Y})-\frac{w}{Y}}}{\hbar w}
\Big)\!\Big).
\end{align*}
We insert this identity to the right of $\mathcal{D}^{-1}$
into $\log Z_P(E_{\set{N}})$. For the lhs, this amounts to insert
$\sum_{k\in \mathbb{Z}\setminus \{0\}}
X^k J_k$ to the left of $\mathcal{D}^{-1}$ into $\log Z_P(E_{\set{N}})$. With
(\ref{def-bbWvee}) and (\ref{loopins-DZ}) and
we thus get
\begin{align}
&  X\frac{\partial}{\partial X}
  \overrightarrow{D}_X^{\mathbb {Z}}  \log Z_P(E_{\set{N}})
\\
&=
\sum_{m\in \mathbb{Z}}
X^{-m}
\sum_{s=0}^\infty \partial^s_\theta(1+\theta)^m
\Big|_{\theta=0}
[w^s][Y^{m}] 
\Big(Y^{s+1}
 \Big(
e^{w\frac{\Theta^\vee(Y)}{Y}} \mathbb{W}^\vee_1(w;Y)
- \frac{
e^{\frac{1}{\hbar} \mathscr{A}(\frac{\hbar w}{Y})-\frac{w}{Y}}}{\hbar w}
\Big)\!\Big).
\nonumber
\end{align}
Its $[\hbar^{-1}]$-coefficient is
\begin{align}
\Theta(X)&\equiv 
  X\frac{\partial}{\partial X}
 [\hbar^{-1}] \overrightarrow{D}_X^{\mathbb {Z}}  \log Z_P(E_{\set{N}})
\\
&=
\sum_{m\in \mathbb{Z}}
X^{-m}
\sum_{s=0}^\infty \partial^s_\theta(1+\theta)^m
\Big|_{\theta=0}
[w^s][Y^{m}] 
\Big(Y^{s+1}
 \Big(
 \frac{e^{w\frac{\Theta^\vee(Y)}{Y}} -1}{w}
\Big) \Big)
\nonumber
\\
&= 
\sum_{m\in \mathbb{Z}}
X^{-m}
\sum_{s=0}^\infty \partial^s_\theta(1+\theta)^m
\Big|_{\theta=0}
[w^s][Y^{m}] 
 \Big(
 \frac{e^{w \Theta^\vee(Y)}}{w}
\Big).
\nonumber
\end{align}
To evaluate this it is convenient to differentiate first:
\begin{align}
  X\frac{\partial \Theta(X)}{\partial X}
&= 
-\sum_{m\in \mathbb{Z}}
X^{-m}
\sum_{s=0}^\infty \partial^s_\theta(1+\theta)^m
\Big|_{\theta=0}
[w^s][Y^{m}] 
 \Big(
e^{w \Theta^\vee(Y)} Y\frac{\partial \Theta^\vee(Y)}{\partial Y}
\Big)
\nonumber
\\
&= 
-\sum_{m\in \mathbb{Z}}
X^{-m}
[Y^{m}]
\Big((1+\Theta^\vee(Y))^m
Y\frac{\partial \Theta^\vee(Y)}{\partial Y}
\Big)
\nonumber
\\
&= 
-\sum_{m\in \mathbb{Z}}
X^{-m}
[Y^{-1}]
\Big((1+\Theta^\vee(Y))^m Y^{-m}
\frac{\partial \Theta^\vee(Y)}{\partial Y}
\Big)
\nonumber
\\
&= 
\sum_{m\in \mathbb{Z}}
X^{-m} \Res_Y dY
\Big((1+\Theta^\vee(Y))^m Y^{-m}
\frac{\partial \Theta^\vee(Y)}{\partial Y}
\Big).
\nonumber
\end{align}
If $X$ and $Y$ are related by the spectral curve
$XY= 1+ \Theta^\vee(Y)$, then after a change of variables we have
\begin{align*}
  X\frac{\partial \Theta(X)}{\partial X}
&=
\sum_{m\in \mathbb{Z}}
X^{-m} \Res_X dX \Big(X^m 
\frac{\partial \Theta^\vee(Y(X))}{\partial X}
\Big)
\nonumber
\\
&=
\sum_{m\in \mathbb{Z}}
X^{-m} [X^ {-1}]\Big(X^m 
\frac{\partial \Theta^\vee(Y(X))}{\partial X}
\Big)
\nonumber
\\
&=
\sum_{m\in \mathbb{Z}}
X^{-m} [X^{-m}]\Big(X 
\frac{\partial \Theta^\vee(Y(X))}{\partial X}
\Big)
\equiv 
X\frac{\partial \Theta^\vee(Y(X))}{\partial X}.
\end{align*}
It integrates to the assertion, up to a constant.
In \cite[Sec.~3.4]{Bychkov:2022wgw} it is stated that vanishing of the
integration constant can be proved by taking residues with
$\frac{dX}{X}$.
\end{proof}
\end{proposition}

\subsection{Simplification of $\tilde{\mathscr{F}}$}

We return to (\ref{DED-3}) which takes with Lemma~\ref{Lemma:U} the form
\begin{align}
  &\mathcal{D}\frac{1}{X}\hat{\mathcal{E}}(\hbar u,X)
  \mathcal{D}^{-1}
  \label{DED-4}
  \\
&= -\sum_{s=0}^\infty
\Big({-}\frac{\partial}{\partial X}\Big)^s
[w^s]\bigg(\Big( \frac{dY}{dX} 
    e^{\frac{1}{\hbar}\mathscr{A}(\hbar w/Y)-
    w\frac{(1+\Theta^\vee(Y))}{Y}}
    \frac{\tilde{\mathscr{F}}^\vee(\hbar u,\hbar w/Y;Y)}{Y}
\Big)\Big|_{Y=Y(X)}\bigg).
\nonumber
\end{align}
In a later step we will rescale $u\mapsto u/X$. Then
according to Lemma~\ref{Lemma:U} 
the $({-}\frac{\partial}{\partial X})^s$ differentiation does not act on
this $u/X$, which we implement as rescaling $u\mapsto u/\hat{X}$
with substitution $\hat{X}\mapsto X$ after the differentiation.

The function $\frac{\tilde{\mathscr{F}}^\vee(\hbar u,\hbar w/Y;Y)}{Y}$
can be expressed as follows:
\begin{lemma} One has
\begin{align}
  \frac{ \tilde{\mathscr{F}}^\vee(\hbar u,\hbar w/Y;Y)}{Y}
  = \frac{1}{(1-\hbar^2u^2/4)\hat{Y}}
  \hat{\mathcal{E}}^\vee\Big(\tfrac{\hbar(w+uY)}{
1-\hbar^2u^2/4} /  \hat{Y},\hat{Y}\Big)
  \Big|_{\hat{Y} = \frac{Y+\hbar^2uw/4}{1-\hbar^2u^2/4}}.
\label{FE-shit}
\end{align}
where it is understood that the derivatives with respect to
$\hat{Y}$ in $\hat{\mathcal{E}}^\vee$, see equation \eqref{hatE-scrS-Y},
do not act on $Y$ in the first argument $\frac{\hbar(w+uY)}{1-\hbar^2u^2/4}$.
\begin{proof}
We have with (\ref{scrF}) and (\ref{calEuz-new})
\begin{align}
\tilde{\mathscr{F}}^\vee(\hbar u,\hbar w;Y)
 &:= \mathscr{F}^\vee(\hbar u,\hbar w;Y)+\frac{1}{\hbar(u+w)}
\label{hatscrF}
 \\
 &=\frac{1}{\hbar (w+u)}
e^{ \sum_{i=1}^\infty
\big(((1+\hbar w/2)(1+\hbar u/2))^{i}
-((1-\hbar w/2)(1-\hbar u/2))^{i}\big)
\frac{J_{-i} Y^i}{i(1-\hbar^2u^2/4)^i}}
\nonumber
\\
&\qquad \times
e^{ \sum_{i=1}^\infty
\big(((1+\hbar w/2)(1+\hbar u/2))^{-i}
-((1-\hbar w/2)(1-\hbar u/2))^{-i}\big)
\frac{J_{i} Y^{-i}}{-i(1-\hbar^2u^2/4)^{-i}}}
\nonumber
\\
&=\frac{1}{\hbar (w+u)}
e^{ \sum_{i=1}^\infty
\big((1+\frac{\hbar(u+w)}{2(1+\hbar^2 wu/4)})^{i}
-(1-\frac{\hbar(u+w)}{2(1+\hbar^2 wu/4)})^{i}
\big)
\frac{J_{-i} ((1+\hbar^2 wu/4)Y)^i}{i(1-\hbar^2u^2/4)^i}}
\nonumber
\\
&\qquad \times
e^{ \sum_{i=1}^\infty
\big((1+\frac{\hbar(u+w)}{2(1+\hbar^2 wu/4)})^{-i}
-(1-\frac{\hbar(u+w)}{2(1+\hbar^2 wu/4)})^{-i}
\big)
\frac{J_{i}  ((1+\hbar^2 wu/4)Y)^{-i}}{-i(1-\hbar^2u^2/4)^{-i}}}.
 \nonumber
\end{align}
After substitution $w\mapsto w/Y$ in (\ref{hatscrF}),
in the exponents of
$\tilde{\mathscr{F}}^\vee(\hbar u,\hbar w/Y;Y)$
we get the following combinations of $Y$, where
$i\in \mathbb{Z}\setminus \{0\}$:
\begin{align}
  &
\Big(\Big(1+\frac{\frac{1}{2}\hbar(w+uY)}{Y(1+\frac{\hbar^2 wu}{4Y})}\Big)^{i}
-\Big(1-\frac{\frac{1}{2}\hbar(w+uY)}{Y(1+\frac{\hbar^2 wu}{4Y})}\Big)^{i}
\Big)
\frac{ ((1+\frac{\hbar^2 wu}{4Y})Y)^i}{i
(1-\hbar^2u^2/4)^{i}}
\nonumber
\\
&=\sum_{k=0}^\infty \frac{2}{i}\binom{i}{2k+1}
\Big(\frac{1}{2} \frac{\hbar(w+uY)}{1-\hbar^2u^2/4}\Big)^{2k+1}
\Big(\frac{1+\hbar^2uw/(4Y)}{1-\hbar^2u^2/4}Y\Big)^{i-1-2k} 
\nonumber
\\
&=\bigg(
\frac{\hbar(w+uY)}{1-\hbar^2u^2/4}
\sum_{k=0}^\infty \frac{1}{(2k+1)!}
\Big(\frac{1}{2}
\frac{\hbar(w+uY)}{1-\hbar^2u^2/4}\Big)^{2k}
\frac{\partial^{2k}}{\partial \hat{Y}^{2k}} \frac{\hat{Y}^i}{\hat{Y}}
\bigg)\Big|_{\hat{Y} \mapsto \frac{Y+\hbar^2uw/4}{1-\hbar^2u^2/4}}
\nonumber
\\
&=\bigg(
\frac{\hbar(w+uY)}{1-\hbar^2u^2/4}
\mathscr{S}\Big(\frac{\hbar(w+uY)}{1-\hbar^2u^2/4}
\frac{\partial}{\partial \hat{Y}}\Big)
\frac{\hat{Y}^i}{\hat{Y}}
\bigg)\Big|_{\hat{Y} \mapsto \frac{Y+\hbar^2uw/4}{1-\hbar^2u^2/4}}
\label{hatF-exponent}
\end{align}
Comparison with (\ref{hatE-scrS-Y}) gives the assertion.
\end{proof}
\end{lemma}
We insert (\ref{FE-shit}) into (\ref{DED-4}) 
and multiply by $e^{\frac{1}{\hbar}\mathscr{A}(\hbar u)-u(1+\Theta(X))}$
so that the lhs becomes 
$\mathcal{D}\mathbb{E}(u X,X)\mathcal{D}^{-1}$,
see (\ref{def:scrEX}). On the resulting rhs 
we use
$\mathscr{A}(\hbar u)+\mathscr{A}(\hbar w/Y)
=\mathscr{A}(\frac{\hbar(w+ uY)}{1-\hbar^2u^2/4}/\hat{Y})$ to get
\begin{align}
  &
  \mathcal{D}\mathbb{E}(u X,X)\mathcal{D}^{-1} \equiv 
  e^{\frac{1}{\hbar}\mathscr{A}(\hbar u)-u(1+\Theta(X))}
  \mathcal{D}\frac{1}{X}\hat{\mathcal{E}}(\hbar u,X)
  \mathcal{D}^{-1}
\label{Dconj-general}
\\
&= -e^{-u(1+\Theta(X))} \sum_{s=0}^\infty
\Big({-}\frac{\partial}{\partial X}\Big)^s
\nonumber
\\
&\qquad 
\times [w^s]\bigg(\Big( \frac{dY}{dX}
e^{\frac{1}{\hbar}\mathscr{A}(\frac{\hbar(w+uY)}{1-\hbar^2u^2/4}
    /\hat{Y})-  w X}
    \frac{\hat{\mathcal{E}}^\vee(\frac{\hbar (w+uY)}{1-\hbar^2u^2/4}/
      \hat{Y},\hat{Y})}{(1-\hbar^2u^2/4)\hat{Y}}
    \Big)\Big|_{\substack{Y=Y(X)\\
        \hat{Y} \mapsto \frac{Y(X)+\hbar^2uw/4}{1-\hbar^2u^2/4}}}\bigg)
\nonumber
\\
&= -e^{-u (1+\Theta(X))} \sum_{s=0}^\infty
\Big({-}\frac{\partial}{\partial X}\Big)^s
\nonumber
\\
&\qquad 
\times [w^s]\bigg(\Big( \frac{dY}{dX}
\frac{e^{
\frac{w+uY}{1-\hbar^2u^2/4}\frac{1+\Theta^\vee(\hat{Y})}{\hat{Y}}
  -  w X}}{1-\hbar^2u^2/4}
  \mathbb{E}^\vee\Big(\frac{w+uY}{1-\hbar^2u^2/4},
      \hat{Y}\Big)
    \Big)\Big|_{\substack{Y=Y(X)\\
        \hat{Y} \mapsto \frac{Y(X)+\hbar^2uw/4}{1-\hbar^2u^2/4}}}\bigg).
\nonumber
\end{align}

\subsection{Loop insertion}

The identity (\ref{Dconj-general})
induces a corresponding identity between loop insertions
to the right of $\mathcal{D}^{-1}$. Inserting 
$\mathcal{D}\mathbb{E}(u X,X)\mathcal{D}^{-1}$ to the right of
$\mathcal{D}^{-1}$ is the same as inserting 
$\mathbb{E}(u X,X)$ to the left of $\mathcal{D}^{-1}$:
\begin{align}
  \overrightarrow{D}_{\mathbb{E}(uX,X)}
&= -e^{-u (1+\Theta(X))} \sum_{s=0}^\infty
\Big({-}\frac{\partial}{\partial X}\Big)^s
\nonumber
\\
&\qquad 
\times [w^s]\bigg(\Big( \frac{dY}{dX}
\frac{e^{
\frac{w+uY}{1-\hbar^2u^2/4}\frac{1+\Theta^\vee(\hat{Y})}{\hat{Y}}
  -  w X}}{1-\hbar^2u^2/4}
\overleftarrow{D}_{\mathbb{E}^\vee(\frac{w+uY}{1-\hbar^2u^2/4},
      \hat{Y})}
    \Big)\Big|_{\substack{Y=Y(X)\\
        \hat{Y} \mapsto \frac{Y(X)+\hbar^2uw/4}{1-\hbar^2u^2/4}}}\bigg).
\nonumber
\end{align}
This implies
\begin{align}
&\prod_{i\in \set{m}}  \overrightarrow{D}_{\mathbb{E}(uX_i,X_i)}\log Z_P(E_{\set{N}})
\label{XY-preliminary}
\\
&= (-1)^m e^{-\sum_{i=1}^m u_i (1+\Theta(X_i))}
\nonumber
\\
&
\times \!\!\!\!
\sum_{s_1,...,s_m=0}^\infty
\prod_{i=1}^m
\Big({-}\frac{\partial}{\partial X_i}\Big)^{s_i}\bigg(\!
[w^{s_1}_1\cdots w^{s_m}_m]\Big(\!
\Big(\prod_{i=1}^m
\frac{dY_i}{dX_i}\Big)
\mathbf{W}^\vee_m(u_{\set{m}};w_{\set{m}};Y_{\set{m}})\Big)\!\bigg),
\nonumber
\end{align}
where $Y_i=Y(X_i)$ is the inverse solution of the spectral curve
$X_iY_i=1+\Theta^\vee(Y_i)$ and
\begin{align}
 \mathbf{W}^\vee_m(u_{\set{m}};w_{\set{m}};Y_{\set{m}})
&=
  \prod_{i=1}^m
  \frac{e^{\frac{v_i(1+\Theta^\vee(\hat{Y}_i))}{\hat{Y}_i}
-w_iX_i}}{1-\hbar^2u_i^2/4 }
\mathbb{W}^\vee_m\big(v_{\set{m}};\hat{Y}_{\set{m}}\big)
\Big|_{\substack{v_i=\frac{w_i+u_iY_i}{1-\hbar^2u_i^2/4} \\
  \hat{Y}_i = Y_i+\hbar^2u_iv_i/4}}.
\end{align}
When later rescaling $u_i\mapsto u_i/\hat{X}_i$, the derivatives
$({-}\frac{\partial}{\partial X_i})^{s_i}$ do not act on $\hat{X}_i$.

From (\ref{bbWvee}) we know that, with the single exception of 
$[\hbar^{-1}]
\mathbb{W}^\vee_1(v_1;\hat{Y_1})
=\frac{1}{v_1}$, all other
$ \mathbb{W}^\vee_m(v_{\set{m}},\hat{Y}_{\set{m}})$ are polynomials in $v_i$.
So excluding the 
$[\hbar^{-1}]$-case of $m=1$ with the projection $\mathcal{P}_m
=\mathrm{id}-\delta_{m,1}\hbar^{-1}[\hbar^{-1}]$, we have
\begin{align*}
& [w^{s_1}_1\cdots w^{s_m}_m]\Big(\mathcal{P}_m  \prod_{i=1}^m
\frac{e^{
\frac{v_i(1+\Theta^\vee(\hat{Y}_i))}{\hat{Y}_i}
-w_iX_i}}{1-\hbar^2u_i^2/4 }
\mathbb{W}^\vee_m\big(v_{\set{m}};\hat{Y}_{\set{m}}\big)
\Big|_{v_i=\frac{w_i+u_iY_i}{1-\hbar^2u_i^2/4}}\Big)
\\
&=\mathcal{P}_m \sum_{r_1=s_1}^\infty\cdots\sum_{r_m=s_m}^\infty
\Big(\prod_{i=1}^m 
\binom{r_i}{s_i}
\frac{\partial^{r_i-s_i}}{\partial \hat{X}_i^{r_i-s_i}}
e^{u_iY_i(\hat{X}_i+X_i)}\Big)\Big|_{\hat{X}_i=0}
\\
& \times [v_1^{r_1}\cdots v_m^{r_m}]
\Big(\prod_{i=1}^m
\frac{e^{
    \frac{v_i(1+\Theta^\vee(\hat{Y}_i))}{\hat{Y}_i}
      -v_iX_i(1-\hbar^2 u_i^2/4)
    }}{(1-\hbar^2u_i^2/4)^{r_i+1} }
\mathbb{W}^\vee_m\big(v_{\set{m}};\hat{Y}_{\set{m}}\big)
\Big).
\end{align*}
We apply
$({-}\frac{\partial}{\partial X_i})^{s_i}$ and sum
over $s_i$. We have
\[
\sum_{s_i=0}^\infty \sum_{r_i=s_i}^\infty
\binom{r_i}{s_i}
\Big({-}\frac{\partial}{\partial X_i}\Big)^{s_i}
\frac{\partial^{r_i-s_i}}{\partial \hat{X}_i^{r_i-s_i}} F(r_i)
=\sum_{r_i=0}^\infty 
\Big(\frac{\partial}{\partial \hat{X}_i}{-}
\frac{\partial}{\partial X_i}\Big)^{r_i}F(r_i),
\]
because there is no further dependence on $s_i$.
The derivative
$\frac{\partial}{\partial \hat{X}}{-}\frac{\partial}{\partial X}$
kills $X+\hat{X}$ in 
$e^{u(X+\hat{X})Y}$. The same is achieved if we
replace  $e^{u(X+\hat{X})Y}\mapsto e^{u\hat{X}Y}$,
$\frac{\partial}{\partial \hat{X}}{-}\frac{\partial}{\partial X}
\mapsto {-}\frac{\partial}{\partial X}$ with substitution
$\hat{X}=X$ at the end, but understanding that
$\frac{\partial}{\partial X}$ does not act on
$\hat{X}$ in intermediate steps:
\begin{align}
&\mathcal{P}_m \!\!\!\!\!\sum_{s_1,...,s_m=0}^\infty
\prod_{i=1}^m
\Big({-}\frac{\partial}{\partial X_i}\Big)^{s_i}\!\bigg(\!
[w^{s_1}_1\cdots w^{s_m}_m]\Big(\!
\Big(\prod_{i=1}^m
\frac{dY_i}{dX_i}\Big)
\mathbf{W}^\vee_m(u_{\set{m}};w_{\set{m}};Y_{\set{m}})\Big)\!\bigg)
\nonumber
\\
&=\mathcal{P}_m  \!\!\!\!\!\sum_{r_1,...,r_m=0}^\infty
\bigg(\prod_{i=1}^m
\Big({-}\frac{\partial}{\partial X_i}\Big)^{r_i}\!\bigg(\!
[v^{r_1}_1\cdots v^{r_m}_m]\Big(\Big(\!
\prod_{i=1}^m
\frac{dY_i}{dX_i}
  \nonumber
  \\
  & \qquad \times 
\frac{e^{u_iY_i \hat{X}_i
    +    \frac{v_i(1+\Theta^\vee(\hat{Y}_i))}{\hat{Y}_i}
      -v_iX_i(1-\hbar^2 u_i^2/4)}}{(1-\hbar^2u_i^2/4)^{r_i+1} }
\mathbb{W}^\vee_m\big(v_{\set{m}};\hat{Y}_{\set{m}}\big)
  \Big)\Big|_{\substack{      \hat{Y}_i=Y_i+\hbar^2u_iv_i/4
      \\ Y_i=Y(X_i)}}    \bigg)\bigg)\Big|_{\hat{X}_i=X_i}
  \nonumber
  \\
&=\mathcal{P}_m  \!\!\!\!\!\sum_{r_1,...,r_m=0}^\infty
\bigg(\prod_{i=1}^m
\Big({-}\frac{\partial}{\partial X_i}\Big)^{r_i}\!\bigg(\!
[w^{r_1}_1\cdots w^{r_m}_m]\Big(\Big(\!
\prod_{i=1}^m
\frac{dY_i}{dX_i}
\frac{e^{u_iY_i \hat{X}_i -w_iX_i(1-\hbar^2 u_i^2/4)}}{(1-\hbar^2u_i^2/4)^{r_i+1} }
\nonumber
\\
& \times 
e^{\hbar^2u_iw_i/4\frac{\partial}{\partial Y_i}}\Big)
\Big(
e^{\sum_{i=1}^m w_i X_i}
\mathbb{W}^\vee_m\big(w_{\set{m}};Y_{\set{m}}\big)
  \Big)\Big|_{Y_i=Y(X_i)}    \bigg)\bigg)\Big|_{\hat{X}=X}.
\label{bfW-hbar}
\end{align}
In the last step we wrote the shift
$Y_i\mapsto \hat{Y}_i$ as Taylor operator
$e^{\hbar^2u_iv_i/4 \frac{\partial}{\partial Y_i}}$, and
we renamed $v_i\mapsto w_i$.

\subsection{$\hbar$-deformation}

We thus consider the following operator on functions
$f(w,Y)$ of $w$ and $Y$:
\begin{align}
&\mathcal{O}_{\hbar^2,\hat{X}} f(w,Y)
\\
&=\sum_{r=0}^\infty
\Big({-}\frac{\partial}{\partial X}\Big)^r [w^r]\Big(
\frac{dY}{dX}
\frac{e^{u \hat{X} Y-wX(1-\hbar^2 u^2/4)}}{(1-\hbar^2u^2/4)^{r+1}}
e^{\hbar^2uw/4 \frac{\partial}{\partial Y}}
\Big(e^{wX}f(w,Y)\Big)\Big)\,,
\nonumber
\end{align}
where $Y=Y(X)$ but $\hat{X}$ is for the time being
considered as independent of $X,Y$.
 The following is a non-trivial result:
 \begin{lemma}
 \label{lemma-deformation}
   Let $X(Y)$ and $Y(X)$ be functional inverses of each other
   and $f(w,Y)=\sum_{k=-1}^\infty
   w^kf_k(Y)$ for meromorphic functions $f_k$. Then
\begin{align}
\mathcal{O}_{\hbar^2,\hat{X}} f(w,Y)\Big|_{\hat{X}=X}
&=\mathcal{O}_{0,\hat{X}} f(w,Y)\Big)\Big|_{\hat{X}=X}
\\
&\equiv \bigg(
\sum_{r=0}^\infty
\Big({-}\frac{\partial}{\partial X}\Big)^r [w^r]\Big(
\frac{dY}{dX}
e^{u \hat{X} Y} f(w,Y)\Big)\bigg)\Big|_{\hat{X}=X}
. \nonumber
\end{align}
\begin{proof}
  By the fundamental theorem of calculus it is enough to show
  \[
\frac{\partial}{\partial s}
\Big(\mathcal{O}_{(t+s)\hbar^2,\hat{X}} f(w,Y)\Big)\Big|_{\substack{s=0\\\hat{X}=X}}=0
.
\]
We find
\begin{align*}
&\frac{\partial}{\partial s}
\Big(\mathcal{O}_{(t+s)\hbar^2} f(w,Y)\Big)\Big|_{s=0}
\\
&=
    \sum_{r=0}^\infty \Big({-}\frac{\partial}{\partial X}\Big)^r  [w^r]   
\bigg(\frac{\hbar^2}{4}\frac{dY}{dX}
e^{u \hat{X} Y}
\\
&\qquad \times \Big\{
\frac{u^2(r+1)}{(1-\hbar^2t u^2/4)^{r+2}}
\Big(e^{-wX(1-\hbar^2tu^2/4)} e^{\hbar^2 t u w/4   \frac{\partial}{\partial Y}}
\Big(e^{w X} f(w ,Y)\Big)\Big)
\\
&\qquad
+\frac{wX u^2}{(1-\hbar^2t u^2/4)^{r+1}}
\Big(e^{-wX(1-\hbar^2tu^2/4)} e^{\hbar^2 t u w/4   \frac{\partial}{\partial Y}}
\Big(e^{w X} f(w ,Y)\Big)\Big)
\\
&\qquad
+\frac{w u}{(1-\hbar^2t u^2/4)^{r+1}}
\Big(e^{-wX(1-\hbar^2tu^2/4)}
\frac{\partial}{\partial Y}\Big(
e^{\hbar^2 t u w/4   \frac{\partial}{\partial Y}}
\Big(e^{w X} f(w ,Y)\Big)\Big)\Big)\Big\}\bigg).
\end{align*}
Setting
\[
  \tilde{f}(w,Y)  
:=e^{-wX(1-\hbar^2tu^2/4)}
e^{\hbar^2 t u w/4   \frac{\partial}{\partial Y}}
(e^{w X} f(w ,Y)),
\]
we can rearrange the above formula to
\begin{align*}
&\frac{\partial}{\partial s}
\Big(\mathcal{O}_{(t+s)\hbar^2} f(w,Y)\Big)\Big|_{s=0}
\\
&=  \sum_{r=0}^\infty \Big({-}\frac{\partial}{\partial X}\Big)^r  [w^r]   
\bigg(\frac{\hbar^2}{4}\frac{dY}{dX}
e^{u \hat{X} Y}
\\
&\qquad \times \Big\{
\frac{u^2(r+1)}{(1-\hbar^2t u^2/4)^{r+2}}
\tilde{f}(w,Y)
+\frac{wX u^2}{(1-\hbar^2t u^2/4)^{r+1}}
\tilde{f}(w,Y)
\\
&\qquad
+\frac{w u}{(1-\hbar^2t u^2/4)^{r+1}}
\frac{\partial \tilde{f}(w,Y)}{\partial Y}
+\frac{w^2 u}{(1-\hbar^2t u^2/4)^{r}}
\frac{dX}{dY}
\tilde{f}(w,Y)
\Big\}\bigg)
\\
&=\sum_{\ell=-1}^\infty \frac{\hbar^2}{4(1-\hbar^2t u^2/4)^{\ell+2}}
\Big\{
(\ell+1) u^2
\Big({-}\frac{\partial}{\partial X}\Big)^{\ell}
\Big(\frac{dY}{dX}e^{u \hat{X} Y} \tilde{f}_\ell(Y)\Big)
\\
&+u^2\Big({-}\frac{\partial}{\partial X}\Big)^{\ell+1}
\Big(\frac{dY}{dX}e^{u \hat{X} Y} X  \tilde{f}_\ell(Y)\Big)
+u\Big({-}\frac{\partial}{\partial X}\Big)^{\ell+1}
\Big(\frac{dY}{dX}e^{u \hat{X} Y}
\frac{\partial \tilde{f}_\ell(Y)}{\partial Y}\Big)
\tag{*}
\\
&
+u\Big({-}\frac{\partial}{\partial X}\Big)^{\ell+2}
\Big( e^{u \hat{X} Y}\tilde{f}_\ell(Y)\Big)
\Big\},
\end{align*}
where $\tilde{f}_\ell(Y):=[w^\ell]\tilde{f}(w,Y)$.
The very last derivative is, by Leibniz and chain rule,
\begin{align*}
&u\Big({-}\frac{\partial}{\partial X}\Big)^{\ell+2}
\Big( e^{u \hat{X} Y}\tilde{f}_\ell(Y)\Big)
\\
&=- u^2 \hat{X} \Big({-}\frac{\partial}{\partial X}\Big)^{\ell+1}
\Big(\frac{dY}{dX} e^{u \hat{X} Y}\tilde{f}_\ell(Y)\Big)
-
u\Big({-}\frac{\partial}{\partial X}\Big)^{\ell+1}
\Big( \frac{dY}{dX} e^{u \hat{X} Y}\frac{\partial \tilde{f}_\ell(Y)}{\partial Y}
\Big).
\end{align*}
The Leibniz rule in the first term of the line (*)
with one factor $X$ then establishes
\begin{align*}
\frac{\partial}{\partial s}
\Big(\mathcal{O}_{(t+s)\hbar^2,\hat{X}} f(w,Y)\Big)
\Big|_{s=0}
&=\sum_{\ell=-1}^\infty \frac{\hbar^2u^2(X-\hat{X})}{4
  (1-\hbar^2t u^2/4)^{\ell+2}}
\Big({-}\frac{\partial}{\partial X}\Big)^{\ell+1}
\Big(\frac{dY}{dX} e^{u \hat{X} Y}
\\
&\quad \times [w^\ell]
\Big(e^{-wX(1-\hbar^2tu^2/4)}
e^{\hbar^2 t u w/4   \frac{\partial}{\partial Y}}
(e^{w X} f(w ,Y))\Big)\Big).
\end{align*}
Setting $\hat{X}=X$ confirms the assertion.
\end{proof}
\end{lemma}

\subsection{Completing the duality formulae}

Lemma~\ref{lemma-deformation} simplifies (\ref{bfW-hbar})
considerably.  We insert the result into (\ref{XY-preliminary}),
rescale $u_i\mapsto u_i/\hat{X}_i$ and take the following observations
into account:
\begin{itemize}
\item   $(1+\Theta(X))=XY(X)$  (see Prop.~\ref{prop:Theta});

\item $\hat{X}_i$ in  $u_i/\hat{X}_i$ is not differentiated by
  $\frac{\partial}{\partial X_i}$;
  
\item $\mathcal{P}_m$ can be omitted in   
  (\ref{bfW-hbar}) because the term 
  $\frac{\delta_{m_1}}{\hbar w_1}$ excluded by $\mathcal{P}_m$
  is anyway removed by $[w_1^{r_1}]$ for $r\geq 0$;
  
\item on the lhs, the missing term to complete
  $\mathcal{P}_m$ to the identity is easily added.
\end{itemize}
We have thus proved the following main result,
from which the universal parametrised $x$-$y$ duality
Thm.~\ref{thm:parametrizedxyduality} is abstractly deduced:
\begin{theorem}
  \label{Thm:XYpar}
The following parametrised $X$-$Y$-duality holds:
\begin{align}
\mathbb{W}_m(u_{\set{m}};X_{\set{m}})
\label{bbW-bbV}
&= \frac{\delta_ {m,1}}{\hbar u_1}+
(-1)^m \!\!\!\! \sum_{r_1,...,r_m=0}^\infty
\Big(\prod_{i=1}^m\! e^{-u_i Y_i}
\Big({-}\frac{\partial}{\partial X_i}\Big)^{r_i}\Big)
\\
&\qquad \times [w_1^{r_1}\cdots w_m^{r_m}]
 \Big(\Big(\prod_{i=1}^m \frac{dY_i}{dX_i}
  e^{u_i Y_i}\Big)
\mathbb{W}^\vee_m\big(w_{\set{m}},Y_{\set{m}}\big)\Big)
  \Big|_{Y_i=Y(X_i)},
  \nonumber
\end{align}
where $Y(X)$ is the inverse of the spectral curve
$X(Y)=\frac{1+\Theta^\vee(Y)}{Y}$.
\hfill $\square$%
\end{theorem}

From Def.~\ref{def:varpiX} we deduce immediately:
\begin{corollary}
  \label{Cor:XY}
  The following $X$-$Y$-duality holds:
\begin{align}
\omega_m(X_{\set{m}})
&=
(-1)^m \!\!\!\! \sum_{r_1,...,r_m=0}^\infty
\Big(\prod_{i=1}^m dX_i
\Big({-}\frac{\partial}{\partial X_i}\Big)^{r_i}\frac{1}{dX_i}\Big) 
\nonumber
\\
&\qquad \times [w_1^{r_1}\cdots w_m^{r_m}]\Big(
\mathbb{W}^\vee_m\big(w_{\set{m}},Y_{\set{m}}\big)
\prod_{i=1}^m dY_i\Big)
\Big|_{Y_i=Y(X_i)},
  \nonumber
\nonumber
\end{align}
where $Y(X)$ is the inverse of the spectral curve
$X(Y)=\frac{1+\Theta^\vee(Y)}{Y}$.
\hfill $\square$%
\end{corollary}
\noindent
If $X_i=X(z_i)$ and $Y_i=Y(z_i)$ solve identically the spectral curve, then
$dX(z_i)(\frac{\partial}{\partial X(z_i)})^r \frac{1}{dX(z_i)}
=d_{z_i}\frac{1}{dX(z_i)}$, and the universal formula
(\ref{identity:om}) follows.

Since the relations (\ref{adED-3}) and
(\ref{adED-4}) and the functions and operations
in Def.~\ref{def:scrE}
are mutually symmetric when exchanging
$(w,Y)\leftrightarrow (u,X)$, we have the following dual
$X$-$Y$-relations, which have their $x$-$y$ universal counterpart stated in
Sec.~\ref{sec:xyduality}:
\begin{theorem}
The following (parametrised) $X$-$Y$-duality holds:
\begin{align}
\mathbb{W}_m^\vee(w_{\set{m}};Y_{\set{m}})
\label{bbV-bbW}
&= \frac{\delta_ {m,1}}{\hbar w_1}+
(-1)^m \!\!\!\! \sum_{r_1,...,r_m=0}^\infty
\Big(\prod_{i=1}^m\! e^{-w_i X_i}
\Big({-}\frac{\partial}{\partial Y_i}\Big)^{r_i}\Big)
\\
&\qquad \times [u_1^{r_1}\cdots u_m^{r_m}]
 \Big(\Big(\prod_{i=1}^m \frac{dX_i}{dY_i}
  e^{w_i X_i}\Big)
\mathbb{W}_m\big(u_{\set{m}},X_{\set{m}}\big)\Big)
  \Big|_{X_i=X(Y_i)}
  \nonumber
\end{align}
as well as the $X$-$Y$ duality
\begin{align}
\omega^\vee_m(Y_{\set{m}})
\label{varpi-bbV}
&=
(-1)^m \!\!\!\! \sum_{r_1,...,r_m=0}^\infty
\Big(\prod_{i=1}^m dY_i
\Big({-}\frac{\partial}{\partial Y_i}\Big)^{r_i}\frac{1}{dY_i}\Big)
\\
&\qquad \times [u_1^{r_1}\cdots u_m^{r_m}]
 \Big(
\mathbb{W}_m\big(u_{\set{m}},X_{\set{m}}\big)\prod_{i=1}^m dX_i\Big)
  \Big|_{X_i=X(Y_i)}.
  \nonumber
\end{align}
Here $X(Y)=\frac{1+\Theta^\vee(Y)}{Y}$ is the spectral curve, or the inverse of
$Y(X)=\frac{1+\Theta(X)}{X}$.
\hfill $\square$%
\end{theorem}

We can also move just one loop insertion to the other side of
$\mathcal{D}^{-1}$ and take the $[u_i^0]$-coefficient of the others:
\begin{proposition}
The following relations holds:  
\begin{align}
\mathcal{W}_{1,m}(u;X;Y_{\set{m}})
&= \frac{\delta_ {m,0}}{\hbar u}-
\sum_{r=0}^\infty
e^{-u Y}
\Big({-}\frac{\partial}{\partial X}\Big)^{r}
[w^{r}] \Big(\frac{dY}{dX}
  e^{u Y} \mathcal{W}^\vee_{m+1}\big(w;Y;Y_{\set{m}}\big)\Big), 
  \nonumber
\\
\mathcal{W}^\vee_{m+1}\big(w;Y;Y_{\set{m}})
&= \frac{\delta_ {m,0}}{\hbar w}-
\sum_{r=0}^\infty
e^{-w X}
\Big({-}\frac{\partial}{\partial Y}\Big)^{r}
[u^{r}] \Big(\frac{dX}{dY}
e^{w X} \mathcal{W}_{1,m}(u;X;Y_{\set{m}})
\big)\Big),
\label{calW-calVnew}
\end{align}
where $Y=Y(X)$ is the inverse of the spectral curve.
\end{proposition}

The various functional relations established in Appendices
\ref{app:functionalrelations} and \ref{app:XYFock} complete the proof
of Thm.~\ref{thm:Kontsevich-duality}.

\section{Proof of Proposition~\ref{prop:poleantidiagonal}}
\label{app:proofpoleantidiagonal}

Our goal is to analyse the local behaviour of
$\omega^{\vee(x,Y)}_{m+1}(z,z_1,z_2,\ldots,z_m)$ 
at the anti-diagonal $z+z_i=0$. This amounts to study the pole of
$\omega^{\vee(x,Y)}_{m+1}(-z,z_1,z_2,\ldots,z_m)$ at the diagonal $z=z_i$.
We start from the reflection formula (\ref{reflection-formula}), namely (\ref{reflectionappendix-new})
with $\omega^{\vee(x,\tilde{Y})}_{1,m}(z;z_{\set{m}})=
-\omega^{\vee(x,Y)}_{1+m}(-z,z_{\set{m}})$. On the rhs,
$\mathcal{W}_{m+1}^{\vee(x,Y)}$ is given by
$\mathcal{W}_{m+1}^{\vee}$ in
(\ref{calWvee}) or 
Cor.~\ref{cor:1variableextended}. Hence, only multi-differentials of the form
\[
\omega^{\vee(x,Y)}_{k+l}
\bigl(
\underbrace{z,\ldots,z}_{k\,\text{times}},
z_{i_1},\ldots,z_{i_l}
\bigr)
\]
with pairwise different $z_{i_j}$ enter. These are all regular at the diagonal $z=z_i$, except for
$\omega_{0,2}(z,z_i)$. Thus, the only contribution to
\eqref{reflectionappendix-new} that is singular at $z=z_i$ comes from a term
containing $\omega_{0,2}(z,z_i)$ in the graphical description
of (\ref{calWvee}).
Therefore, we conclude that
\begin{align}
\frac{\omega^{\vee(x,Y)}_{g,m+1}(-z,z_{\set{m}})}{d\tilde{Y}(z)}
    &= -
    \sum_{s=0}^\infty
    \Big({-}\lambda \frac{\partial}{\partial \tilde{Y}(z)}\Big)^{s}[w^{s}]
    \bigg( \frac{dY(z)}{d\tilde{Y}(z)}
    \mathcal{W}_{m}^{\vee(x,Y)}
    \bigl(w;z;z_{\set{m}\setminus\{i\}}\bigr)   
    \nonumber
    \\
    &\qquad \times \hbar w
        \mathscr{S}\Big(\hbar w d_z\frac{1}{dY(z)}\Big)
        \omega_{0,2}(z,z_i) \bigg)
    +\mathcal{O}\bigl((z-z_i)^0\bigr),
  \label{app1-new}
  \end{align}
where
$\mathcal{W}_{m}^{\vee(x,Y)}
(w;z;z_{\llbracket m \rrbracket\setminus\{i\}})$
is independent of $z_i$ and, in particular, regular at $z=z_i$.\\

The next step consists of deriving a formula quadratic in the
$\omega_{m}$'s from the identity (\ref{calWVvee})
about $\mathcal{W}^{\vee(x,\tilde{Y})}_{1,m}$. As special case of
(\ref{bbW-mixed}), $\mathcal{W}^{\vee(x,\tilde{Y})}_{1,m}$ has a
graphical expansion
into a sum over tripartite graphs $\Gamma\in G_{\{0\},\set{m}}$
with one white vertex labelled $0$ and $m$ dual white vertices
labelled by $\set{m}$. Since $\mathcal{W}$ results from $\mathbb{W}$ by
taking $[w_i^0]$-coefficients, the $m$ dual white vertices are
all one-valent. Edges $e$ at black vertices incident to the white vertex $0$
are treated with the operator $\hbar v \mathscr{S}(\hbar v
\frac{\partial}{\partial \tilde{Y}_e})\frac{1}{d\tilde{Y}_e}$, and edges
incident to $\set{m}$ have
no further operator acting on them. Up to linear order in $v$ we thus have
\begin{align}
  \mathcal{W}^{\vee(x,\tilde{Y})}_{1,m}(v;z;z_{\set{m}}) -\frac{\delta_{m,0}}{\hbar v}
  &= -\frac{\omega_{1+m}^{\vee(x,Y)}(-z,z_{\set{m}})}{d\tilde{Y}(z)}
  +\frac{\hbar v}{2} \frac{\omega_{2+m}^{\vee(x,Y)}(-z,-z,z_{\set{m}} )}{
    d\tilde{Y}(z) d\tilde{Y}(z)}
\nonumber
  \\
  &+\frac{\hbar v}{2} \sum_{\set{m}=I\sqcup J}^{\text{ordered}}
\frac{  \omega^{\vee(x,Y)}_{1,|I|}(-z,z_{I})}{d\tilde{Y}(z)}
 \frac{ \omega^{\vee(x,Y)}_{1,|J|}(-z,z_{J})}{d\tilde{Y}(z)}
+\mathcal{O}(v^2)\,.
  \label{eq:onevariablechange-new}
\end{align}
The factor $\frac{1}{2}$ in the last term of the first line is the
symmetry factor of the two edges of the black vertex incident to
the white vertex $0$. In the last line, the additional factor
$\frac{1}{2}$ compensates the overcount in the ordered splitting
$\set{m}=I\sqcup J=J\sqcup I$. Moreover,
we have converted
$\omega^{\vee(x,\tilde{Y})}_{1,n}(z;z_1,...,z_n)
=-\omega^{\vee(x,Y)}_{1+n}(-z,z_{\llbracket n \rrbracket})$
and
$\omega^{\vee(x,\tilde{Y})}_{2,n}(z,z;z_1,\dots,z_n)
=\omega^{\vee(x,Y)}_{2+n}(-z,-z,z_{\llbracket n \rrbracket})$.

We compute the following expression:
\begin{align}
&(d\tilde{Y}(z))^2\Big(  [v^1]d^{-1}_i\Big(e^{-\frac{v}{\lambda}Y(z)}
\Big(\mathcal{W}^{\vee(x,\tilde{Y})}_{1,m}(v;z;z_{\set{m}})-\frac{\delta_{m,0}}{\hbar v}\Big)
\Big)
\nonumber\\
&\qquad +\frac{Y(z_i)}{\lambda}
  [v^0]d^{-1}_i
\Big(e^{-\frac{v}{\lambda}Y(z)}
\Big(\mathcal{W}^{\vee(x,\tilde{Y})}_{1,m}(v;z;z_{\set{m}})
-\frac{\delta_{m,0}}{\hbar v}\Big)
\Big)\Big)
  \label{app2-new}
\\
 &=
\frac{1}{\lambda} (Y(z)-Y(z_i))d\tilde{Y}(z) 
d_i^{-1}\omega_{1+m}^{\vee(x,Y)}(-z,z_{\set{m}})
\nonumber
\\
&  +\frac{\hbar}{2} d_i^{-1}\omega_{2+m}^{\vee(x,Y)}(-z,-z,z_{\set{m}} )
+\frac{\hbar}{2}  \sum_{\set{m}=I\sqcup J}^{\text{ordered}}
d_i^{-1}\Big( \omega^{\vee(x,Y)}_{1+|I|}(-z,z_{I})
\omega^{\vee(x,Y)}_{1+|J|}(-z,z_{J})\Big)
\nonumber
  \end{align}
when employing  (\ref{eq:onevariablechange-new}),
where $d^{-1}_i$ means that we integrate with respect to $z_i$. The ambiguity
by an integration constant will not play any r\^ole.
On the other hand, the same function evaluates with (\ref{calWVvee}) to 
\begin{align}
  &\eqref{app2-new}
  \\
  &= 
d\tilde{Y}(z) \sum_{s=0}^\infty 
\Big({-}\lambda d_z \frac{1}{d \tilde{Y}(z)}\Big)^{s} 
\Big(
{-}\frac{1}{\lambda} (Y(z){-}Y(z_i))
dY(z)[w^{s}] d_i^{-1}
\mathcal{W}^{\vee(x,Y)}_m(w;z;z_{\set{m}})\Big)
\nonumber
\end{align}
As before, we analyse the pole at $z=z_i$, whose only contribution comes from
a factor $\omega_{0,2}(z,z_i)$, as discussed previously. Hence, the above
expression is equal to
\begin{align}
\eqref{app2-new} & =
-\frac{d\tilde{Y}(z)}{\lambda} \sum_{s=0}^\infty 
\Big({-}\lambda d_z \frac{1}{d \tilde{Y}(z)}\Big)^{s} 
\Big(
(Y(z){-}Y(z_i))
dY(z)
\nonumber
\\
&\times [w^{s}]
\Big(
\mathcal{W}^{\vee(x,Y)}_m(w;z;z_{\set{m}\setminus \{i\}})
\hbar w \mathscr{S}\Big(\hbar w \frac{\partial}{\partial Y(z)}\Big)
\frac{d_i^{-1}\omega_{0,2}(z,z_i)}{dY(z)}
\Big)\Big)
\nonumber
\\
&+\mathcal{O}\bigl((z-z_i)^0\bigr).
\label{app3-new} 
\end{align}
We have $\frac{d_i^{-1}\omega_{0,2}(z,z_i)}{dY(z)}
= \frac{1}{(z-z_i)Y'(z)}+
\mathcal{O}((z-z_i)^0)= 
\frac{1}{Y(z)-Y(z_i)}+\mathcal{O}((z-z_i)^0)$
and thus
\begin{align*}
&(Y(z)-Y(z_i))\mathscr{S}\Big(\hbar w \frac{\partial}{\partial Y(z)}\Big)  
\frac{d_i^{-1}\omega_{0,2}(z,z_i)}{dY(z)}
\\
&= \sum_{k=0}^\infty \frac{(\hbar w)^{2k}}{(2k+1)!2^{2k}} 
\frac{(2k)!}{(Y(z)-Y(z_i))^{2k}}+\mathcal{O}((z-z_i)^0)
\\
&= \sum_{k=0}^\infty \frac{(\hbar w)^{2k+2}(2k+2)}{(2k+3)! 2^{2k+2}} 
\Big(\frac{\partial}{\partial Y(z_i) }\Big)^{2k}
\frac{1}{(Y(z)-Y(z_i))^{2}}+\mathcal{O}((z-z_i)^0)
\\
&= \hbar^2w^2
\tilde{\mathscr{S}}\Big(\hbar w \frac{\partial}{\partial Y(z_i)}\Big)
\mathscr{S}\Big(\hbar w \frac{\partial}{\partial Y(z_i)}\Big)
\frac{1}{(Y(z)-Y(z_i))^{2}}+\mathcal{O}((z-z_i)^0)\,,
\end{align*}
where ($B_k$ are the Bernoulli numbers)
\begin{align}
  \tilde{\mathscr{S}}(t):=\frac{\mathscr{S}'(t)}{t\mathscr{S}(t)}
  = \sum_{k=0}^\infty \frac{B_{2k+2}}{(2k+2)!} t^{2k} .
\end{align}

Next, we use the relation
$\mathscr{S}\Big(\hbar w \frac{\partial}{\partial Y(z_i)}\Big)
\frac{1}{(Y(z)-Y(z_i))^{2}}
=\mathscr{S}\Big(\hbar w \frac{\partial}{\partial Y(z)}\Big)
\frac{1}{(Y(z)-Y(z_i))^{2}}$, and finally
$\frac{1}{(Y(z)-Y(z_i))^{2}}=\frac{\omega_{0,2}(z,z_i)}{dY(z)dY(z_i)}
+\mathcal{O}((z-z_i)^0)$. This brings 
(\ref{app3-new}) into 
\begin{align}
\eqref{app2-new} & =
-\frac{(d\tilde{Y}(z))^2}{\lambda}
\sum_{s=0}^\infty 
\Big({-}\lambda \frac{\partial}{\partial \tilde{Y}(z)}\Big)^{s} 
[w^{s}]\Big(
\frac{dY(z)}{d\tilde{Y}(z)}\mathcal{W}^{\vee(x,Y)}_m(w;z;z_{\set{m}\setminus \{i\}})
\nonumber
\\
&\times 
\hbar^2 w^2\tilde{\mathscr{S}}\Big(\hbar w
\frac{\partial}{\partial Y(z_i)}\Big)
\hbar w \mathscr{S}\Big(\hbar w \frac{\partial}{\partial Y(z)}\Big)
\frac{\omega_{0,2}(z,z_i)}{dY(z)dY(z_i)}\Big)
\nonumber
\\
&+\mathcal{O}\bigl((z-z_i)^0\bigr).
\label{app4-new} 
\end{align}
Now, $\hbar^2\lambda^2 w^2\tilde{\mathscr{S}}(\hbar w
\frac{\partial}{\partial Y(z_i)})$ can be moved through 
the extraction $[w^s]$ and is turned into 
$\hbar^2 \lambda^2 \frac{\partial^2}{\partial \tilde{Y}^2} \tilde{\mathscr{S}}(\hbar 
\lambda \frac{\partial}{\partial \tilde{Y}(z)}
\frac{\partial}{\partial Y(z_i)})$
at the other side of $[w^s]$.
Comparing with (\ref{app1-new}),
the result is an operator acting on
  $\omega^{\vee(x,Y)}_{m+1}(-z,z_{\set{m}})$:
\begin{align}
\eqref{app2-new} & =
\lambda\hbar^2 (d\tilde{Y}(z))^2
\frac{\partial^2}{\partial (\tilde{Y}(z))^2}
 \tilde{\mathscr{S}}\Big(\hbar 
\lambda \frac{\partial}{\partial \tilde{Y}(z)}
\frac{\partial}{\partial Y(z_i)}\Big)
\frac{\omega^{\vee(x,Y)}_{m+1}(-z,z_{\set{m}})}{d\tilde{Y}(z)dY(z_i)}
\nonumber
\\
&+\mathcal{O}\bigl((z-z_i)^0\bigr).
\label{app5-new} 
\end{align}
  
We combine this with the rhs of (\ref{app2-new}),
multiply by $\frac{\lambda}{(Y(z)-Y(z_i))d\tilde{Y}(z) }$
and replace $z\mapsto -q$:
\begin{align*}
&d_i^{-1}\omega_{1+m}^{\vee(x,Y)}(q,z_{\set{m}})
\\
&=\frac{\lambda}{2(\tilde{Y}(q)-\tilde{Y}(-z_i))dY(q) }
\Big\{
-\hbar d_i^{-1}\omega_{2+m}^{\vee(x,Y)}(q,q,z_{\set{m}} )
\\
&\qquad -\hbar  \sum_{\set{m}=I\sqcup J}^{\text{ordered}}
d_i^{-1}\Big( \omega^{\vee(x,Y)}_{1+|I|}(q,z_{I})
\omega^{\vee(x,Y)}_{1+|J|}(q,z_{J})\Big)
\\
& \qquad +
2 \lambda \hbar^2 (dY(q))^2
\frac{\partial^2}{\partial (Y(q))^2}
 \tilde{\mathscr{S}}\Big(\hbar 
\lambda \frac{\partial}{\partial Y(q)}
\frac{\partial}{\partial Y(z_i)}\Big)
\frac{\omega^{\vee(x,Y)}_{m+1}(q,z_{\set{m}})}{dY(q)dY(z_i)}
\Big\}
\nonumber
\\
&+\mathcal{O}\bigl((q+z_i)^{-1}\bigr),
\end{align*}
where $\tilde{Y}(-q)=Y(q)$ was used.
We multiply by $\int_{-z_i}^q B^{TR}(z,\bullet)$
(which is of $\mathcal{O}\bigl((q+z_i)^{1}\bigr)$)
and take the residue $\Res_{q = -z_i}$. Note that $\Res_{q = -z_i} d_i^{-1}\omega_{1+m}^{\vee(x,Y)}(q,z_{\set{m}})=0$,
which can be shown from the $x$-$y$ ($=Y$-$\tilde{Y}$) duality formula ($=$ reflection formula) with respect to one variable $q$, after integrating with
respect to $z_i$. The first term is
$-d_i^{-1}\omega_{1+m}^{\vee(x,Y)}(-q,z_{\set{m}})$,
which is regular at $q=-z_i$, and the other terms are written as
total derivatives of locally meromorphic functions with respect
to $q$ and are therefore residue-free.
Finally, we apply $d_i$ and take the $[\hbar^{2g-2+(m+1)}]$-coefficient.
This establishes the desired formula.
By using the dictionary (\ref{dictionary})
we confirm \cite[eq.~(1.3)]{Hock:2023nki} with all signs and
rational coefficients.
\hspace*{\fill} $\square$%

\section{Proof of Proposition \ref{prop:poleatzero}}\label{app:proofpoleatzero}

This appendix proves a formula for the polar part of
$\omega_{1,n}^{\vee(x,Y)}(z_{\set{n}})\equiv \omega_{1,n}^{\vee(X,Y)}(z_{\set{n}})$
at $z_i=0$. The derivation uses the $x$-$Y$ duality of this family to the
family of differentials $\omega_{g,n}^{(x,Y)}$, which agrees with
the differentials generated by eq-TR for $2g-2+n\geq 0$, see
Fig.~\ref{fig:duality-network} for details.

The family $\omega_{g,n}^{(x,Y)}$ does not have poles at $z=0$ (except
for $\omega^{(0)}_{0,2}$ at the diagonal). We work with the mixed
multi-differentials $\omega^{(g)(x,Y)}_{n,m}$ with
$\omega^{(g)(x,Y)}_{n,0}=\omega_{g,n}^{(x,Y)}$ for $2g+n-2\geq 0$ and 
$\omega^{(g)(x,Y)}_{0,m}=\omega^{\vee(x,Y)}_{g,m}$.
Important for us is the
fact that $\omega^{(g)(x,Y)}_{n,m}(z_1,\ldots,z_n;\tilde{z}_1,\ldots,\tilde{z}_m)$ has no poles at $z_i=0$ (even after diagonalisation $z_j=z_i$), except for
$(g,n,m)=(0,2,0)$ at the diagonal, which is the extended Bergman
kernel.

The desired formula is derived from the first order of the
parametrised version of the $x$-$Y$ duality formula obtained by
swapping just one variable, which appears in
\cite[Proposition~4.8]{Alexandrov:2022ydc}. We have rederived it
in (\ref{calW-calVnew}) for the $(X,Y)$-duality. Adapted to $(x,Y)$,
the formula reads
\begin{align}\label{eq:onevariablechange2}
  &  e^{ w x(z)}\cW^{\vee(x,Y)}_{1+n}( w;z;z_{\set{n}})dY(z) - \delta_{n,0}e^{wx(z)} \frac{dY(z)}{\hbar w}
  \\
  &=
    -\sum_{r\geq0}
    \Big({-}d_z\frac{1}{dY(z)}\Big)^r
    \Bigl(
        e^{ w x(z)}
        [u^r]dx(z)
        \cW^{(x,Y)}_{1,n}(u;z;z_{\set{n}})
    \Bigr),
    \nonumber
  \end{align}
  where $\cW^{\vee(x,Y)}_{1+n}( w;z;z_{\set{n}})=
  \cW^{(x,Y)}_{0,1+n}( w;z;z_{\set{n}})$.
  The $[\hbar^{2g+n-1}w^1]$-coefficient of the lhs reads
\begin{align*}
[w^1\hbar^{2g+n-1}]\eqref{eq:onevariablechange2}_{LHS}
  &=x(z)\omega_{g,n+1}^{\vee(x,Y)}(z,z_{\set{n}}))+
  \frac{1}{2dY(z)}\omega_{g-1,n+2}^{\vee(x,Y)}(z,z,z_{\set{n}})
  \\
  & +\frac{1}{2dY(z)}
  \sum_{\substack{I_1\sqcup I_2=\set{n}\\ g_1+g_2=g\\
      (g,|I_i|)\neq (0,1)}}^{\text{ordered}}
 \omega_{g_1,n_1+1}^{\vee(x,Y)}(z,z_{I_1})\omega_{g_2,n_2+1}^{\vee(x,Y)}(z,z_{I_2}) \\
 & -\delta_{g,0} \frac{1}{2}(x(z))^2 dY(z)
  .
\end{align*}
The 1-variable extended multi-differential
$\cW^{(x,Y)}_{1,n}(w;z;z_{\set{n}})$ is a sum over connected tripartite graphs
with a single white vertex labelled $z$ of any valency and 1-valent dual
white vertices labelled $\set{n}$. The black vertices have weight
$\omega^{(g')(x,Y)}_{m',n'}(z,\ldots,z;z_{I'})$ with $m'\geq 1$ times the
argument $z$ and $\mathscr{S}$-action.
Since we are interested in the pole structure of
$\omega^{\vee(x,Y)}_{g,n+1}(z,z_1,\ldots,z_n)$ at $z=0$, it is enough
to consider on the rhs only contributions to this pole. 
The only
multi-differential with a pole at $z=0$ is
${:}\omega^{(0)(x,Y)}_{2,0}{:}(z,z)$ with $g'=n'=0$ and $m'=2$. At
sufficiently large $\hbar$-order, several such
${:}\omega^{(0)(x,Y)}_{2,0}{:}$ can appear. We restrict ourselves to
$g=1$ which is of order $[\hbar^{n+1}]$, as required for the
proposition. For $g=1$, there is just one
${:}\omega^{(0)(x,Y)}_{2,0} {:}(z,z)$ appearing, graphically giving a
non-trivial loop and thus $g=1$: 
\[
    \begin{tikzpicture}
        \node[draw,circle] (a) at (0,0) {};
        \node[circle,minimum size=30pt] (A) at (0,0) {};
        \node[fill,circle] (b) at (-0.5,0.5) {};
        \draw[bend left=60] (a) to (b);
        \draw[bend right=60] (a) to (b);
        \draw (a) -- (A.south);
        \draw (a) -- (A.south east);
        \draw (a) -- (A.east);
        \draw (a) -- (A.south west);
    \end{tikzpicture}
\]

If $n  \geq 1$, the remaining terms are sums over
partitions, and we can put all $\cS=1$. Thus, we get for the rhs of
(\ref{eq:onevariablechange2}) at $g=1$
\begin{align*}
  &[\hbar^{n+1}w^1]\eqref{eq:onevariablechange2}_{RHS}
  +\mathcal{O}(z^0)\\
  &=\sum_{r\geq0}\Big({-}d_z\frac{1}{dY(z)}\Big)^r[u^r]\bigg(
  {-}x(z)\frac{dx(z)}{u}\frac{\frac{1}{2}u^2\frac{(dz)^2}{(2z)^2}}{dx(z)^2}\sum_{I_1\sqcup...\sqcup I_l=\set{n}}\prod_{i=1}^l \frac{u \omega^{(0)(x,Y)}_{1,|I_i|}
    (z;z_{I_i})}{dx(z)}\bigg),
\end{align*}
where $\frac{(dz)^2}{(2z)^2}={:}\omega^{(0)(x,Y)}_{2,0}{:}(z,z)
+\mathcal{O}(z^0)$. For $n=0$ the terms in the big parenthesis
reduce to $\Big(
  {-}x(z)\frac{dx(z)}{u}\frac{\frac{1}{2}u^2\frac{(dz)^2}{(2z)^2}}{dx(z)^2}
\Big)$.
Applying Lemma~\ref{lem:om0n2z0identity}, we have in summary for $g=1$
\begin{align}
  &x(z)\omega^{\vee(x,Y)}_{g,n+1}(z,z_{\set{n}})+
  \frac{1}{2 dY(z)} \omega^{\vee(x,Y)}_{g-1,n+2}(z,z,z_{\set{n}})
\nonumber  \\
  &+\frac{1}{2dY(z)}  \sum_{\substack{I_1\sqcup I_2=\set{n}\\ g_1+g_2=g\\
      (g,|I_i|)\neq (0,1)}}^{\text{ordered}}
 \omega_{g_1,n_1+1}^{\vee(x,Y)}(z,z_{I_1})\omega_{g_2,n_2+1}^{\vee(x,Y)}(z,z_{I_2})
 \nonumber
 \\
 &=-\frac{1}{4}\Big(d_z\frac{1}{dY(z)}\Big)d^{-1}_{z'}
 {:}\omega^{\vee(x,Y)}_{g-1,n+2}(z,z',z_{\set{n}}){:}\Big|_{z'=z}+\mathcal{O}(z^0).
\end{align}
Last but not least, one can divide by $x(z)$, which actually generates a simple pole at $z=0$, because $R'(0)\neq 0$. However, $\omega^{(Y,x)}_{g,n+1}(z,z_1,\ldots,z_n)$ is residue-free, and acting on the resulting equation with the operator $Res_{z=0}\int^q_0B^{TR}(\bullet,z)$ projects the lhs onto the principal part of $\omega^{(Y,x)}_{g,n+1}(z,z_1,\ldots,z_n)$ and annihilates on the rhs any contribution from $\frac{\mathcal{O}(z^0)}{x(z)}$. We conclude the formula of Prop.~\ref{prop:poleatzero}.

\begin{lemma}\label{lem:om0n2z0identity}
    The following identity holds for $n\geq 1$:
    \begin{align}
      &\frac{1}{4}\bigg(-d_z\frac{1}{dY(z)}\bigg)d^{-1}_{z'}
      \omega^{(0)(x,Y)}_{0,n+2}(z,z',z_{\set{n}})\vert_{z'=z}
      \label{lemmaD1-n>0}
      \\
      &=\sum_{r\geq0}\Big({-}d_z\frac{1}{dY(z)}\Big)^r[u^r]dx(z)
      \bigg({-}x(z)\frac{1}{u}\frac{\frac{1}{2}u^2\frac{(dz)^2}{(2z)^2}}{
        dx(z)^2}\sum_{I_1\sqcup...\sqcup I_l={\llbracket n \rrbracket}}\prod_{i=1}^l
      \frac{u \omega^{(0)(x,Y)}_{1,|I_i|}(z;z_{I_i})}{dx(z)}\bigg)
      \nonumber
      \\
      &+\mathcal{O}(z^0).\nonumber
    \end{align}
It reduces for $n=0$ to 
    \begin{align}
      &\frac{1}{4}\bigg(-d_z\frac{1}{dY(z)}\bigg)d^{-1}_{z'}
      {:}\omega^{(0)(x,Y)}_{0,2}{:}(z,z')\vert_{z'=z}
\label{lemmaD1-n0}      
      \\
      &=\sum_{r\geq0}\Big({-}d_z\frac{1}{dY(z)}\Big)^r[u^r]dx(z)
      \bigg({-}x(z)\frac{1}{u}\frac{\frac{1}{2}u^2\frac{(dz)^2}{(2z)^2}}{
        dx(z)^2}
      \bigg)      +\mathcal{O}(z^0).
\nonumber
    \end{align}    
  \begin{proof} 
      One has 
      $d^{-1}_{z'}
      {:}\omega^{(0)(x,Y)}_{0,2}{:}(z,z')\vert_{z'=z}=-dz/(2z)+\mathcal{O}(z^0)$.
      Then (\ref{lemmaD1-n0}) follows from the following identity, valid
      for general
      $x(0)=0$ and $x'(0)\neq0$:
 \begin{align}
    \frac{dz}{2z}=2x(z)\frac{(dz)^2}{(2z)^2dx(z)}+\mathcal{O}(z^0).
     \label{appD-identity-x}
 \end{align}              
      
 To prove (\ref{lemmaD1-n>0}) for $n\geq 1$ we start from the mixed extended
 multi-differential  $\mathbb{W}^{(x,Y)}_{2,n}(u_1,u_2;w_{\set{n}};z,z';z_{\set{n}})$
with two variables moved to the other side.  We
 consider its $[w_1^0\cdots w_n^0]$-coefficient. Similar to
 Prop.~\ref{prop:bbW} it has a graphical expansion into tripartite
 graphs made of two white vertices of any valency labelled $a,b$ and
 $n$ dual white vertices labelled $\set{n}$, which are all one-valent
 because we restrict to $[w_1^0\cdots w_n^0]$. Moreover, we consider
 the leading order $[\hbar^{2+n-2}]$ corresponding to genus $g=0$,
 which means that the connected graphs are trees. Hence there will be
 \begin{itemize}
      \item   one black vertex incident to both $a,b$ and a subset
        $L\subset\set{n}$
        of the dual white vertices;
      \item $k_a$ black vertices each incident to $a$ and to a
        non-empty subset $I_i \subset \set{n}$,
        but not incident to $b$, for each $i=1,...,k_a$;
      \item $k_b$ black vertices each incident to $b$ and to a
        non-empty subset $J_j \subset \set{n}$, but not incident to $a$,
        for each $j=1,...,k_b$,        
 \end{itemize}
 with
 $L\sqcup I_1 \sqcup ...\sqcup I_{k_a} \sqcup J_1 \sqcup ...\sqcup
 J_{k_b}=\set{n}$. See Fig.~\ref{fig:bridge}. That the $I_i,J_j$ are non-empty follows from the
 fact that $\omega_{0,1}$ is excluded in one-valent black vertices. 

 \begin{figure}[ht]
    \centering
    \begin{tikzpicture}[scale=0.85]
    \begin{scope}[every node/.style={draw,circle, minimum size=20pt}]
        \node (a) at (-2,0) {$a$};
        \node (b) at (2,0) {$b$};
    \end{scope}
    \begin{scope}[every node/.style={fill,circle}]
        \node (1) at (0,0) {};
        \node (I1) at (-3,-1) {};
        \node (I2) at (-2,-1) {};
        \node (Ik) at (-0.5,-1) {};
        \node (J1) at (1,-1) {};
        \node (Jk) at (2.5,-1) {};
    \end{scope}
    \begin{scope}[every node/.style={draw,circle}]
        \node (w1) at (-3.5,-2) {};
        \node (w2) at (-3,-2) {};
        \node (w3) at (-2.25,-2) {};
        \node (w4) at (-1.75,-2) {};
        \node (w5) at (-1.25,-2) {};
        \node (w6) at (-0.5,-2) {};
        \node (w7) at (-0.5,1) {};
        \node (w8) at (0,1) {};
        \node (w9) at (0.5,1) {};
        \node (w10) at (0.5,-2) {};
        \node (w11) at (1,-2) {};
        \node (w12) at (1.5,-2) {};
        \node (w13) at (2.25,-2) {};
        \node (w14) at (2.75,-2) {};
    \end{scope}
        \path (a) edge (1);
        \path (b) edge (1);
        \path (a) edge (I1);
        \path (a) edge (I2);
        \path (a) edge (Ik);
        \path (b) edge (J1);
        \path (b) edge (Jk);
        \path (I2) -- node[midway] {$\dots$} (Ik);
        \path (J1) -- node[midway] {$\dots$} (Jk);
        \path (I1) edge (w1);
        \path (I1) edge (w2);
        \path (I2) edge (w3);
        \path (I2) edge (w4);
        \path (I2) edge (w5);
        \path (Ik) edge (w6);
        \path (1) edge (w7);
        \path (1) edge (w8);
        \path (1) edge (w9);
        \path (J1) edge (w10);
        \path (J1) edge (w11);
        \path (J1) edge (w12);
        \path (Jk) edge (w13);
        \path (Jk) edge (w14);
        \draw[decoration={brace,mirror,raise=8pt},decorate] (w1.west) -- node[below=10pt] {$I_1$} (w2.east);
        \draw[decoration={brace,mirror,raise=8pt},decorate] (w3.west) -- node[below=10pt] {$I_2$} (w5.east);
        \draw[decoration={brace,mirror,raise=8pt},decorate] (w6.west) -- node[below=10pt] {$I_{k_a}$} (w6.east);
        \draw[decoration={brace,raise=8pt},decorate] (w7.west) -- node[above=10pt] {$L$} (w9.east);
        \draw[decoration={brace,mirror,raise=8pt},decorate] (w10.west) -- node[below=10pt] {$J_1$} (w12.east);
        \draw[decoration={brace,mirror,raise=8pt},decorate] (w13.west) -- node[below=10pt] {$J_{k_b}$} (w14.east);
    \end{tikzpicture}
\caption{\label{fig:bridge}}
\end{figure}

 In
 leading order $[\hbar^{2+n-2}]$, the weights $\omega_{2,|L|}$,
 $\omega_{1,|I_i|}$ and $\omega_{1,|J_j|}$ corresponding to these
 black vertices restrict to genus 0, and all $\cS$-operators
 restrict to 1.  We thus have 
      \begin{align}
        &[w_1^0\cdots w_n^0 \hbar^{2+n-2}]
        \mathbb{W}^{(x,Y)}_{2,n}(u_1,u_2;w_{\set{n}};z,z';z_{\set{n}})
        \prod_{i=1}^n dY(z_i)
        \\
        &= \sum_{\substack{L\sqcup I_1\sqcup ...\sqcup I_{k_a} \sqcup \\
         J_1\sqcup ...\sqcup J_{k_b}=\set{n}\\
       I_1,...,I_{k_a},J_1,...,J_{k_b} \neq \emptyset}} \!\!\!\!
        \frac{\omega_{2,|L|}^{(0)(x,Y)}(z,z';z_{L})}{dx(z)dx(z')}
    \prod_{i=1}^{k_a} \frac{u_1\omega_{1,|I_i|}^{(0)(x,Y)}(z;z_{I_i})}{dx(z)}
    \prod_{j=1}^{k_b} \frac{u_2\omega_{1,|J_l|}^{(0)(x,Y)}(z';z_{J_j})}{dx(z')}.
        \nonumber
      \end{align}
      Note that there is no normal ordering for trees.
      From this we get $\omega^{(0)(x,Y)}_{0,n+2}(z,z',z_{\set{n}})$  by
      multiplication with $dx(z)dx(z')$ and turning
      $u_1$ into $-d_z\frac{1}{Y(z)}$ and
      $u_2$ into $-d_{z'}\frac{1}{Y(z')}$:
      \begin{align}
        &\omega^{(0)(x,Y)}_{0,n+2}(z,z',z_{\set{n}})
        \label{om0-n+2}
         \\
        &=\sum_{\substack{L\sqcup I_1\sqcup ...\sqcup I_{k_a} \sqcup \\
         J_1\sqcup ...\sqcup J_{k_b}=\set{n}\\
       I_1,...,I_{k_a},J_1,...,J_{k_b} \neq \emptyset}}
\Big(-d_z\frac{1}{dY(z)}\Big)^{k_a}
\Big(-d_{z'}\frac{1}{dY(z')}\Big)^{k_b}
\bigg(      \omega_{2,|L|}^{(0)(x,Y)}(z,z';z_{L})
\nonumber
\\*[-4ex]
&\hspace*{5cm} \times
\prod_{i=1}^{k_a} \frac{\omega_{1,|I_i|}^{(0)(x,Y)}(z;z_{I_i})}{dx(z)}
    \prod_{j=1}^{k_b} \frac{\omega_{1,|J_l|}^{(0)(x,Y)}(z';z_{J_j})}{dx(z')}
    \bigg).
    \nonumber
\end{align}
Note first that $\omega^{(0)(x,Y)}_{0,n+2}(z,z',z_{\set{n}})$ is of
genus $g=0$ and thus has no poles at $z,z'=0$ generically. However,
sending $z'=z$ generates a pole for
$\omega^{(0)(x,Y)}_{0,n+2}(z,z,z_{\set{n}})$ at $z=0$, coming purely
from the original pole of $\omega^{(0)(x,Y)}_{2,0}(z,z')$ at the
anti-diagonal $z+z'=0$, which remains present at $z'=z$. For
$L\neq \emptyset$, all $\omega^{(0)(x,Y)}_{2,|L|}(z',z;z_{L})$ are
regular when sending $z'\to z$ and afterwards $z\to 0$. So we restrict
to $L=\emptyset$.  Then
$\omega_{2,0}^{(0)(x,Y)}(z,z')=\frac{dzdz'}{(z-z')^2}+
\frac{dzdz'}{(z+z')^2}$ has first of all a pole at the diagonal.
However, Gen-TR proves that this pole cancels for
$n\geq 1$ \cite[Prop. 4.6]{Alexandrov:2022ydc}, and the regular remainder consists of combinations of
derivatives of $\frac{1}{Y'(z)}$ and the factors in the last line of
(\ref{om0-n+2}). These are all regular at $z=z'=0$.  Hence, the poles
of (\ref{om0-n+2}) at $z=z'=0$ comes exclusively from $L=\emptyset$
and $\omega_{2,0}^{(0)(x,Y)}(z,z') \mapsto
\frac{dzdz'}{(z+z')^2}$. Converting differentials to partial
derivatives, we thus get
 \begin{align}
        &\omega^{(0)(x,Y)}_{0,n+2}(z,z',z_{\set{n}})
        \label{om0-n+2-a}
         \\
         &=dY(z)dY(z') \hspace*{-0.7cm}
         \sum_{\substack{I_1\sqcup ...\sqcup I_{k_a} \sqcup \\
         J_1\sqcup ...\sqcup J_{k_b}=\set{n}}} \hspace*{-0.5cm}
\Big({-}\frac{\partial}{\partial Y(z)}\Big)^{k_a}
\Big({-}\frac{\partial}{\partial Y(z')}\Big)^{k_b}
\bigg(    \Big( -\frac{\partial}{\partial Y(z')}
\frac{1}{Y'(z)(z+z')} \Big)
\nonumber
\\*[-4ex]
&\hspace*{5cm} \times
\prod_{i=1}^{k_a} \frac{\omega_{1,|I_i|}^{(0)(x,Y)}(z;z_{I_i})}{dx(z)}
    \prod_{j=1}^{k_b} \frac{\omega_{1,|J_l|}^{(0)(x,Y)}(z';z_{J_j})}{dx(z')}
    \bigg)
    \nonumber
    \\
    &+\mathcal{O}((z+z')^0)+\mathcal{O}(z^0)+\mathcal{O}((z')^0).
    \nonumber
\end{align}
If $k_a>0$ or $k_b>0$, then the conditions $I_i\neq \emptyset$ and $J_j\neq \emptyset$, respectively, 
are from now on self-understood.
Observe that all terms have at least one
$\frac{\partial}{\partial Y(z')}$-derivative
so that $d_{z'}=dY(z')\frac{\partial}{\partial Y(z')}$ factors out.
We can thus integrate with $d_{z'}^{-1}$ to get
      \begin{align}
        &d_{z'}^{-1}\omega^{(0)(x,Y)}_{0,n+2}(z,z',z_{\set{n}})
        \label{om0-n+2-b}
         \\
         &=dY(z)\hspace*{-0.7cm}
         \sum_{\substack{I_1\sqcup ...\sqcup I_{k_a}=\set{n}}}
           \hspace*{-0.5cm}
   (-1)^{k_a+1}
  \Big(\frac{\partial}{\partial Y(z)}\Big)^{k_a}
\bigg(
\frac{1}{Y'(z)(z+z')}
\prod_{i=1}^{k_a} \frac{\omega_{1,|I_i|}^{(0)(x,Y)}(z;z_{I_i})}{dx(z)}
\bigg)
\nonumber
\\
 &+dY(z)\hspace*{-0.7cm}
         \sum_{\substack{I_1\sqcup ...\sqcup I_{k_a} \sqcup \\
         J_1\sqcup ...\sqcup J_{k_b}=\set{n}\\ k_b\geq 1}} \hspace*{-0.5cm}
(-1)^{k_a+b_b+1}\Big(\frac{\partial}{\partial Y(z)}\Big)^{k_a}
\Big(\frac{\partial}{\partial Y(z')}\Big)^{k_b-1}
\bigg(    \Big( \frac{\partial}{\partial Y(z')}
\frac{1}{Y'(z)(z+z')} \Big)
\nonumber
\\*[-2ex]
&\hspace*{5cm} \times
\prod_{i=1}^{k_a} \frac{\omega_{1,|I_i|}^{(0)(x,Y)}(z;z_{I_i})}{dx(z)}
    \prod_{j=1}^{k_b} \frac{\omega_{1,|J_l|}^{(0)(x,Y)}(z';z_{J_j})}{dx(z')}
    \bigg)
    \nonumber
    \\
   & +\mathcal{O}((z+z')^0)+\mathcal{O}(z^0)+\mathcal{O}((z')^0).
    \nonumber
  \end{align}
  At the diagonal $z=z'$, Lemma~\ref{conj-appendix} simplifies
  this expression to
        \begin{align}
        &d_{z'}^{-1}\omega^{(0)(x,Y)}_{0,n+2}(z,z',z_{\set{n}})\Big|_{z'=z}
        \label{appdeq1}
         \\
         &=\hspace*{-0.5cm}
         \sum_{\substack{I_1\sqcup ...\sqcup I_{k}=\set{n}}} \hspace*{-0.3cm}
  \Big(-d_z\frac{1}{dY(z)}\Big)^{k}
\bigg(
-\frac{dz}{2z}
\prod_{i=1}^{k} \frac{\omega_{1,|I_i|}^{(0)(x,Y)}(z;z_{I_i})}{dx(z)}
\bigg)
 +\mathcal{O}(z^0).
    \nonumber
  \end{align}

\noindent
The term $\frac{dz}{2z}$ is expressed by (\ref{appD-identity-x}).
We can ignore the $\mathcal{O}(z^0)$ contributions, which remain regular
$z=0$ 
when applying the $d_z\frac{1}{dY(z)}$ derivatives. The result is
  \begin{align}\label{appdeq2}
    &d^{-1}_{z'}\omega^{(0)}_{n+2,0}(z,z',z_{\set n})\vert_{z'=z}\\\nonumber
    &=-\sum_{r\geq0}  \Big({-}d_z\frac{1}{dY(z)}\Big)^r \!
     [u^r]\bigg(\! dx(z)  2x(z)\frac{(dz)^2}{(2z)^2 dx(z)^2} \!\!
     \sum_{\substack{I_1\sqcup ...\sqcup I_k=\set n}}
     \prod_{i=1}^{k}
     \frac{u\omega^{(0)(x,Y)}_{1,|I_i|}(z;z_{I_i})}{dx(z)}\bigg)
    \nonumber
    \\
    &+\mathcal{O}(z^0).\nonumber
  \end{align}
  Acting with one more differential $-d_z\frac{1}{dY(z)}$ literally
  adds a $u$ on the rhs, turning
  $\frac{(dz)^2}{(2z)^2 dx(z)^2}\to \frac{u^2(dz)^2}{u(2z)^2
    dx(z)^2}$, and we identify exactly the rhs of (\ref{lemmaD1-n>0}).
 This finishes the proof.
    \end{proof}
\end{lemma}

\begin{lemma} \label{conj-appendix}
  For functions $f_1,...,f_n$ of one variable and a function $g$ of two
  variables the following holds:
  \begin{align}
&\frac{\partial^{n}}{\partial x^{n}}  \Big(g(x,x)\prod_{i=1}^n f_i(x)\Big)
\\
&=      \bigg(\frac{\partial^{n}}{\partial x^{n}}  \Big[
    \prod_{i=1}^n f_i(x) \cdot g(x,x')\Big]\bigg)_{x'=x}
    \nonumber
    \\
    &+ \sum_{\substack{I\sqcup J=\set{n}\\ J\neq \emptyset}}\bigg(
    \frac{\partial^{|I|}}{\partial x^{|I|}}\frac{\partial^{|J|-1}}{\partial {x'}^{|J|-1}}
    \Big[\prod_{i \in I} f_i(x)
    \Big(\frac{\partial g(x,x')}{\partial x'}\Big)\prod_{j \in J} f_j(x')
    \Big]\bigg)_{x'=x}.
    \nonumber
 \end{align}

\end{lemma}
 We give a proof based on
the following classical theorem (e.g.\ \cite[Thm.~A.1]{Hock:2021tbl}),
which can be seen as a multivariate version of the
celebrated Cayley’s formula for the number of labelled trees:
\begin{lemma}
For any positive integer $n$ and family of indeterminates
$x_1,...,x_n,x_{n+1}$ the following formulae hold true:
\begin{align}
  (x_1+...+x_{n+1})^{n-1} &=\sum_{T\in T_{\set{n+1}} }
    \prod_{v\in \set{n+1}}
    x_v^{\text{deg}(v)-1},
    \label{Maciej-1}
      \\
  (x_1+...+x_{n+1})^n &=\sum_{T\in T^\bullet_{\set{n+1}} }
    x_{v_\bullet}^{\text{deg}(v_{\bullet})}
    \prod_{v\in \set{n+1}\setminus \{v_{\bullet}\}} 
    x_v^{\text{deg}(v)-1}.
    \label{Maciej-2}
\end{align}
Here $T_{\set{n+1}}$ is the set of labelled trees
whose vertices are labelled by $\set{n+1}$, and 
$T^\bullet_{\set{n+1}}$ is the set of labelled rooted trees
whose vertices are labelled by $\set{n+1}$.
In $T^\bullet_{\set{n+1}}$ we let $v_{\bullet}$ be the label of the root,
and $\deg(v)$ the valency of the vertex labelled $v$ in $T$.
\end{lemma}
  Alternatively, in the labelled trees we can just declare
  any labelled vertex (no matter which) as the root:
\begin{align}
  (x_1+...+x_{n+1}+x_{n+2})^{n}
  &=\sum_{\substack{T\in T^\bullet_{\set{n+2}} \\ n+1 \text{ is root}}}
    \prod_{v\in \set{n+2}}
    x_v^{\text{deg}(v)-1}.
    \label{Maciej-1b}
    \\
  &=\sum_{\substack{T\in T^\bullet_{\set{n+2}} \\ n+1 \text{ is root} \\
  n+2 \text{ is leaf }}}
    \prod_{v\in \set{n+2}}
    x_v^{\text{deg}(v)-1}.
+ \!\!\!\! \sum_{\substack{T\in T^\bullet_{\set{n+2}} \\ n+1 \text{ is root} \\
  n+2 \text{ not a leaf }}} \hspace*{-0.5em}
    \prod_{v\in \set{n+2}}
    x_v^{\text{deg}(v)-1},
    \nonumber
  \end{align}
  where in the second line we have distinguished trees where
  $n+2$ is a leaf. Since leaves provide a factor $1$, we have
\begin{align}
    \sum_{\substack{T\in T^\bullet_{I \cup \{n+1,n+2\}} \\ n+1 \text{ is root} \\
  n+2 \text{ is leaf }}}
    \prod_{v\in I\cup \{n+1,n+2\}}
    x_v^{\text{deg}(v)-1}
    &=   \Big(x_{n+1}+\sum_{i\in I} x_i\Big)^{|I|}.
    \label{Maciej-6}
  \end{align}
In the case that $n+2$ is not a leaf of a tree $T$, we cut the edge
below $n+2$ (i.e.\ the
edge incident to $n+2$ which belongs to the path to the root  $n+1$).
We obtain two trees one containing $n+1$, the other one
containing $n+2$,
where both have an additional half-edge (see Fig.\,\ref{fig:cut}):
\begin{itemize}
\item In the tree containing $n+2$, we remove the half-edge from $n+2$ and view
  $n+2$ as root of this tree $T_2$. The remaining vertices of $T_2$
  are labelled by a subset
  $J\subset \set{n}$, where 
  $J\neq \emptyset$ because $n+2$ is not a leaf.
  If in the original tree $T$ the vertex $n+2$ had
  $\mathrm{deg}_T(v_{n+2})=r$, then $r=\mathrm{deg}_{T_2}(v_{n+2})+1$
  from the point of view of $T_2$. 

\item 
In the tree containing $n+1$ (as root), we artificially add $n+2$ as a
leaf to the
half-edge to get a tree $T_1$. Its root is $n+1$; the other vertices are
labelled $I\cup \{n+2\}$ with $I=\set{n}\setminus J$ (possibly empty),
and $n+2$ is one of the leaves.
\end{itemize}

\begin{figure}[h]
    \centering
    \begin{tikzpicture}[scale=0.85]
        \begin{scope}[yscale=-1]
                    \node[fill,circle,label={[xshift=-15pt, yshift=-15pt]$\scriptstyle n+2$}] (n) at (0,0) {};
        \node[draw,circle] (3) at (0.25,1) {};
        \node[draw,circle] (4) at (-0.5,0.75) {};
        \node[draw,circle] (5) at (0.75,0) {};
        \node[draw,circle] (2) at (1,1.25) {};
        \node[fill,circle,label={[xshift=-15pt, yshift=-15pt]$\scriptstyle n+1$}] (1) at (0,1.75) {};
        \node[draw,circle] (7) at (0.25,-0.5) {};
        \node[draw,circle] (6) at (-1,-0.5) {};
        \node[draw,circle] (10) at (0.5,-1.25) {};
        \node[draw,circle] (8) at (1.5,-0.75) {};
        \node[draw,circle] (9) at (-0.5,-1) {};
        \draw (1) -- (2);
        \draw (1) -- (3);
        \draw (1) -- (4);
        \draw (3) -- node[midway,strike out,draw] {} (n);
        \draw (3) -- (5);
        \draw (n) -- (7);
        \draw (5) -- (8);
        \draw (7) -- (10);
        \draw (n) -- (9);
        \draw (9) -- (6);
        \draw[decoration={brace,mirror,raise=50pt},decorate] (-1,0) -- node[below=52pt] {$T$} (1.75,0);

        \node[xshift=65pt] (a) at (0,0) {$\Longrightarrow$};
        \end{scope}

        \begin{scope}[xshift=150pt,yscale=-1]
        \node[fill,circle] (n) at (0,0) {};
        \node[draw,circle] (7) at (0.25,-0.5) {};
        \node[draw,circle] (6) at (-1,-0.5) {};
        \node[draw,circle] (10) at (0.5,-1.25) {};
        \node[draw,circle] (9) at (-0.5,-1) {};
        \draw (n) -- (7);
        \draw (7) -- (10);
        \draw (n) -- (9);
        \draw (9) -- (6);
        \draw[decoration={brace,mirror,raise=50pt},decorate] (-1,0) -- node[below=52pt] {$T_2$} (0.5,0);
        \end{scope}

        \begin{scope}[xshift=190pt,yscale=-1]
        \node[fill,circle] (n) at (0,0) {};
        \node[draw,circle] (3) at (0.25,1) {};
        \node[draw,circle] (4) at (-0.5,0.75) {};
        \node[draw,circle] (5) at (0.75,0) {};
        \node[draw,circle] (2) at (1,1.25) {};
        \node[fill,circle] (1) at (0,1.75) {};
        \node[draw,circle] (8) at (1.5,-0.75) {};
        \draw (1) -- (2);
        \draw (1) -- (3);
        \draw (1) -- (4);
        \draw (3) -- (n);
        \draw (3) -- (5);
        \draw (5) -- (8);
        \draw[decoration={brace,mirror,raise=50pt},decorate] (-0.5,0) -- node[below=52pt] {$T_1$} (1.75,0);
        \end{scope}
    \end{tikzpicture}
    \caption{\label{fig:cut}}
\end{figure}

We produce all $T\in T^\bullet_{\set{n+1}}$ with $n+1$ root and $n+2$ not a leaf
and precisely these, if we independently vary
\begin{itemize}
\item the decompositions $I\sqcup J=\set{n}$ with $J\neq \emptyset$,
\item the trees $T_1 \in T^\bullet_{I\cup \{n+1,n+2\}}$ with $n+1$ the root and $n+2$ a leaf,
\item the trees $T_2 \in T^\bullet_{J\cup \{n+1\}}$ with $n+1$ the root.
\end{itemize}
With (\ref{Maciej-6}) for $T\mapsto T_1$ and
\[
\sum_{\substack{T_2 \in T^\bullet_{J\cup \{n+2\}}\\
    n+2 \text{ is root} }}
    \prod_{v\in T_1} x_v^{\mathrm{deg}(v)-1}
=\Big(x_{n+2}+\sum_{j\in J}x_i\Big)^{|J|-1}
\]
as well as the factor $x_{n+2}$ to account for the cut edge we get
the following result:

\begin{corollary}
For any positive integer $n$ and family of indeterminates
$x_1,...,x_n,x_{n+1},x_{n+2}$ the following formula holds true:
  \begin{align}
    (x_1{+}...{+}x_n{+}x_{n+1}{+}x_{n+2})^n &=
    (x_1+...+x_n+x_{n+1})^n
\label{Maciej-5}
    \\
    &+
\sum_{\substack{I\sqcup J=\set{n}  \\ J\neq \emptyset}} \!\!\!
      \Big(x_{n+1}+\sum_{i\in I} x_i\Big)^{|I|}
    x_{n+2} \Big(x_{n+2}+\sum_{j\in J} x_j\Big)^{|J|-1}.
    \nonumber
\end{align}
\end{corollary}
\noindent
\emph{Proof of Lemma \ref{conj-appendix}.}
We set $x_i=\frac{\partial}{\partial t_i}$ in (\ref{Maciej-5})
apply it to $g(t_{n+1},t_{n+2})\prod_{i=1}^nf_i(t_i)$ and go to the diagonal
 $t_1=...=t_n=t_{n+1}=t_{n+2}=x$:
\begin{align*}
  &\frac{d^n}{dx^n}\Big(g(x,x)\prod_{i=1}^n f_i(x)\Big)
  \\
  &\equiv \Big(\frac{\partial}{\partial t_1}+ ...+
  \frac{\partial}{\partial t_n}+\frac{\partial}{\partial t_{n+1}}
  +\frac{\partial}{\partial t_{n+2}}\Big)^n
  g(t_{n+1},t_{n+2})\prod_{i=1}^nf_i(t_i)\Big|_{t_1=...=t_n=t_{n+1}=t_{n+2}=x}
  \\
  &=\bigg(\frac{d^n}{dx^n}\Big(g(x,t_{n+2})\prod_{i=1}^n f_i(x)\Big)
  \bigg)_{t_{n+2}=x} +\sum_{\substack{I\sqcup J=\set{n}  \\ J\neq \emptyset}} F_{I,J}(x)
\end{align*}
where
\begin{align*}
F_{I,J}(x) &=
  \Big(\frac{\partial}{\partial t_{n+1}}+\sum_{i\in I}
  \frac{\partial}{\partial t_{i}}\Big)^{|I|} \Big[\prod_{i\in I} f_i(t_i)
  \\
  & \times \Big(\frac{\partial}{\partial t_{n+2}}+\sum_{j\in J}
\frac{\partial}{\partial t_{j}}\Big)^{|J|-1} \Big(
\frac{\partial g(t_{n+1},t_{n+2})}{\partial t_{n+2}}\prod_{j\in J} f_j(t_j)\Big)
\Big]_{t_1=...=t_n=t_{n+1}=t_{n+2}=x}
\\
 &=
  \Big(\frac{\partial}{\partial t_{n+1}}+\sum_{i\in I}
  \frac{\partial}{\partial t_{i}}\Big)^{|I|} \Big[\prod_{i\in I} f_i(t_i)
  G_J(t_{n+1},t_{n+2})\Big]_{t_1=...=t_n=t_{n+1}=t_{n+2}=x}
\\
 &=
  \Big(\frac{\partial}{\partial x}\Big)^{|I|} \Big[\prod_{i\in I} f_i(x)
  G_J(x,t_{n+2})\Big]_{t_{n+2}=x},
\end{align*}
for
\[
  G_J(t_{n+1},t_{n+2})  :=
  \Big(\frac{\partial}{\partial t_{n+2}}\Big)^{|J|-1} \Big(
  \frac{\partial g(t_{n+1},t_{n+2})}{\partial t_ {n+2}}
    \prod_{j\in I} f_j(t_{n+2})\Big).
\]
This finishes the proof.
\hfill $\square$%

\end{appendix}

\bibliographystyle{halpha-abbrv}
\bibliography{fock}

\end{document}